\documentclass[a4paper,11pt,twoside,onecolumn,openright,final,fleqn,openbib]{memoir}

\pdfoutput=1

\usepackage[latin1]{inputenc}
\usepackage[OT1]{fontenc}

\usepackage{cite}

\usepackage{import} 
\usepackage{graphicx}
\usepackage{color}
\usepackage{transparent} 

\usepackage{pdfpages}

\usepackage{tikz}
\usepackage{pgfplots}
\usepackage{standalone}
\pgfplotsset{compat=1.14}

\usepackage{pstricks}

\usepackage[american,arrowmos]{circuitikz}
\ctikzset{v/.append style={/tikz/european voltages}}
\usetikzlibrary{positioning}

\usepackage{tabu}
\usepackage{array}
\newcolumntype{P}[1]{>{\centering\arraybackslash}p{#1}}
\newcolumntype{M}[1]{>{\centering\arraybackslash}m{#1}}

\usepackage[UKenglish,USenglish,american]{babel}
\usepackage{anyfontsize} 
\usepackage{soul}

\usepackage{amssymb, amsmath, bm}
\usepackage{xfrac} 
\usepackage[LGRgreek]{mathastext} 

\usepackage{doi}
\usepackage{expl3}

\usepackage{color}

\usepackage[version=4]{mhchem}
\usepackage{xfrac}

\sloppybottom 

\setsecnumdepth{subsubsection}

\newcommand{\ndbd}{$0\nu\beta\beta\ $}

\setlrmargins{*}{*}{1}
\checkandfixthelayout

\copypagestyle{cgpage}{Ruled}
\captionnamefont{\small}
\captiontitlefont{\small}
\captiondelim{: }
\postcaption{\rule{\linewidth}{0.5pt}}

\AtBeginDocument{\addtocontents{toc}{\protect\thispagestyle{empty}}} 

\newcommand{\unitm}[1]{\ \textrm{#1}}

\newcommand{\mytilde}{\raise.17ex\hbox{$\mathtt{\sim}$}}

\newcommand{\degC}{^\circ C}

\title{Electronic Instrumentations for High Energy Particle Physics and \\ Neutrino Physics}
\author{Paolo Carniti}
\date{\today}

\begin{document}




\chapterstyle{verville}
\renewcommand*{\printchapternonum}{%
    \hrule \vskip 0.5\onelineskip
    \LARGE \bfseries \centering }

\makeevenhead{cgpage}{}{}{}
\makeoddhead{cgpage}{}{}{}
\makeevenfoot{cgpage}{}{}{}
\makeoddfoot{cgpage}{}{}{}


\thispagestyle{empty}


\begin{center}
{\LARGE{\textsc{University of Milano Bicocca }}} \\
\vspace{3mm}
\Large{\textsc{Doctorate School}}\\
\vspace{3mm}
\begin{figure}[h]
	\centering
	\includegraphics[width=.2\linewidth]{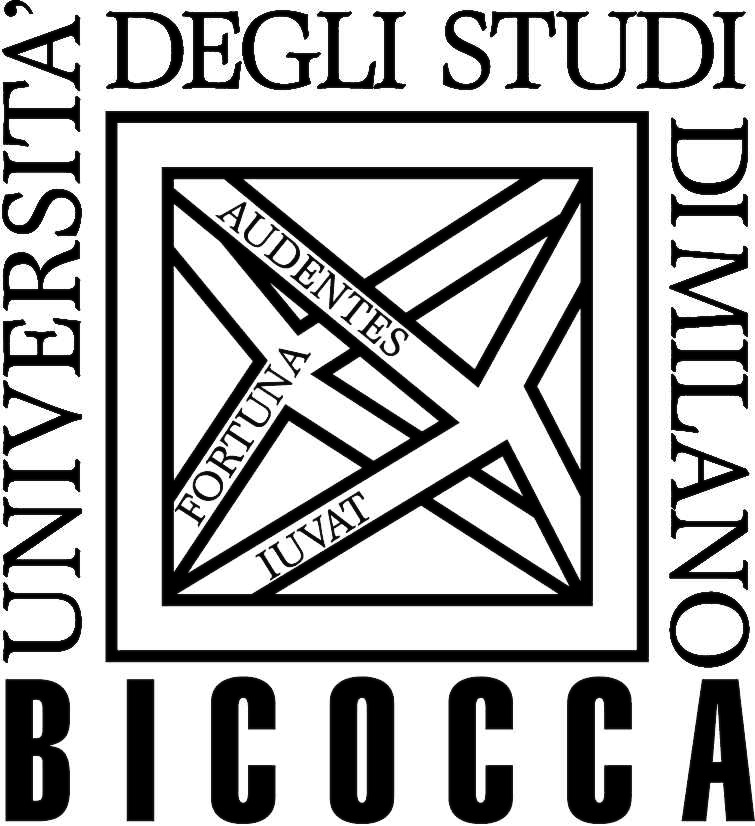}
\end{figure}
\vspace{3mm}
\normalsize{\textsc{Department of Physics "G. Occhialini"}}\\
\vspace{3mm}
\normalsize{\textsc{Doctorate Program in Physics and Astrophysics -- Cycle XXX}}\\
\vspace{3mm}
\normalsize{\textsc{Curriculum in Subnuclear Physics and Physics Technologies}}\\

\vspace{17mm}

\LARGE{\textbf{\thetitle}}\\
\vspace{17mm}

\end{center}



\begin{minipage}{0.4\linewidth}
\begin{center}
\textbf{Candidate}\\
Doct. Paolo \textsc{Carniti}\\
Reg. n. 709379\\
\vspace{1mm}
\phantom{\rule{\linewidth}{0.5pt}}
\phantom{\textbf{Co-tutor}}
\phantom{Prof. Gianluigi \textsc{Pessina}}
\vspace{1mm}
\phantom{\rule{\linewidth}{0.5pt}}
{\textbf{Coordinator}}\\
{Prof. Marta \textsc{Calvi}}
\vspace{1mm}
\phantom{\rule{\linewidth}{0.5pt}}
\end{center}
\end{minipage}
\hspace{18mm}
\begin{minipage}{0.4\linewidth}
\begin{center}
\textbf{Tutor}\\
Prof. Gianluigi \textsc{Pessina}\\
\phantom{Matr. 709379}
\vspace{1mm}
\phantom{\rule{\linewidth}{0.5pt}}
\phantom{\textbf{Co-tutor}}
\phantom{Prof. Marta \textsc{Calvi}}
\vspace{1mm}
\phantom{\rule{\linewidth}{0.5pt}}
\phantom{\textbf{Co-tutor}}
\phantom{Prof. Marta \textsc{Calvi}}
\vspace{1mm}
\phantom{\rule{\linewidth}{0.5pt}}
\end{center}
\end{minipage}

\vfill

\begin{center}
\textsc{Academic Year 2016 -- 2017}
\end{center}

\cleardoublepage


\thispagestyle{empty}
 
\vphantom{.} 
\vspace{0.4\textheight}
\epigraphtextposition{flushright}





\setlength{\epigraphwidth}{8.5cm} 

\epigraph{\textit{Da quass\`{u} il mondo degli uomini \\altro non sembra che follia, \\grigiore racchiuso dentro se stesso. \\E pensare che lo si reputa vivo \\soltanto perch\'{e} \`{e} caotico e rumoroso.}}{\textsc{Walter Bonatti}}


\cleardoublepage

\chapter*{Abstract}
\thispagestyle{empty}
\addcontentsline{toc}{chapter}{Abstract}

The present dissertation describes design, qualification and operation of several electronic instrumentations for High Energy Particle Physics experiments (LHCb) and Neutrino Physics experiments (CUORE and CUPID).
Starting from 2019, the LHCb experiment at the LHC accelerator will be upgraded to operate at higher luminosity and several of its detectors will be redesigned.
The RICH detector will require a completely new optoelectronic readout system.
The development of such system has already reached an advanced phase, and several tests at particle beam facilities allowed to qualify the performance of the entire system.
In order to achieve a higher stability and a better power supply regulation for the front-end chip, a rad-hard low dropout linear regulator, named ALDO, has been developed. Design strategies, performance tests and results from the irradiation campaign are presented.
In the Neutrino Physics field, large-scale bolometric detectors, like those adopted by CUORE and its future upgrade CUPID, offer unique opportunities for the study of neutrinoless double beta decay.
Their operation requires particular strategies in the readout instrumentation, which is described here in its entirety.
The qualification and optimization of the working parameters as well as the integration of the system in the experimental area are also thoroughly discussed, together with the latest upgrades of two electronic subsystems for the future CUPID experiment.

\cleardoublepage

\tableofcontents


\chapterstyle{southall} 
\renewcommand*{\chaptitlefont}{\LARGE\bfseries\memRTLraggedright}
\renewcommand*{\chapnumfont}{\HUGE \scshape\memRTLraggedright}

\makeevenhead{cgpage}{\textbf{\thepage}}{}{\leftmark}
\makeoddhead{cgpage}{\rightmark}{}{\textbf{\thepage}}
\makeevenfoot{cgpage}{}{}{}
\makeoddfoot{cgpage}{}{}{}


\chapter*{Introduction}
\addcontentsline{toc}{chapter}{Introduction}
\label{Chapter01}
\thispagestyle{empty}

The present thesis summarizes the work I carried out during the last three years in design, realization and operation of electronic instrumentations for particle physics experiments.

Being a physicist and an electronic designer enhance the development of the electronic instrumentations as they are not specification-driven but instead physics-driven.
This seems a subtle or trivial difference, but is often a requirement for the successful operation of complex detectors, for which electronics is not just an amplifier that can be plugged in but is part of the detector itself, like, for example, bolometric detectors.
For this reason, during the present dissertation, large sections were dedicated to the description of the physics motivations and their implications on front-end and back-end electronics. 

The work presented here is focused on the development of electronic instrumentation for two fields of particle physics that are currently at the frontier of the search of New Physics beyond the Standard Model and that lie at the opposites of the energy spectrum: CP-violation studies at high energy particle accelerators, and low-energy neutrino physics.

In the first part I will present my work for the LHCb experiment at the Large Hadron Collider, probably the most advanced machine that man ever made.
A machine that was built not for \emph{doing} something but just for \emph{understanding} something.
A machine that is basically useless, from a materialistic perspective.
LHCb is one of the four biggest detectors that are currently operating at the LHC and, from 2019-2020, it will undergo a large upgrade that will require a complete redesign of some sub-detectors.
After this upgrade, physicists at LHCb hope to shed a brighter light on the matter-antimatter asymmetry problem, one of the most intriguing open issues in contemporary physics.

In the second part I will describe the design and operation of the whole electronic system for two bolometric experiments for the search of the neutrinoless double beta decay, CUORE and CUPID.
These experiments are trying to observe an ultra-rare decay with an half-life in excess of $10^{25}$ years, which corresponds to \mbox{1,000,000,000,000,000} ($10^{15}$) times the age of the Universe.
For this reason such experiments require huge detector masses, long acquisition times, and the lowest background sources possible, in order to observe just a few candidates of the interesting decay.
Bolometers are a type of detector that operate at very low cryogenic temperatures within special machines (cryostats) that are able to reach temperatures as low as $10\ mK$ and even below.
For the CUORE experiment it was necessary to build an unprecedented large cryostat that created the coldest cubic meter in the known Universe.


\chapter{LHCb RICH upgrade}
\label{Chapter04}
\thispagestyle{empty}

\section{Introduction}

One of the most intriguing and unanswered questions of contemporary Physics, is related to the matter-antimatter asymmetry.
Why the known Universe is almost entirely made by matter, while experimental evidences and theoretical predictions do not seem to explain such a huge difference?

The Standard Model, in fact, already predicts a tiny asymmetry in the matter-antimatter production but its amount is not enough even for the matter that compose a single galaxy.
Experimental and theoretical physicists are thus looking past the Standard Model to understand this problem.

The mechanism that causes the imbalance in matter and antimatter production is the charge parity (CP) symmetry violation.
Charge-conjugation (C) and parity (P), together with the time-reversal (T), are the three fundamental discrete near-symmetries of the SM.
Table~\ref{tab:DiscreteSymmetry} shows the effect of such operators on some physical quantities.

\begin{table}[t]
\centering
\begin{tabular}{|c c |c c c| }
\multicolumn{2}{|c|}{\textbf{Observable}} & \textbf{P-transform} & \textbf{C-transform} & \textbf{T-transform}\\
\hline
Time & $t$ & $t$ & $t$ & $-t$\\ 
Position & $\vec{x}$ & $-\vec{x}$ & $\vec{x}$ & $\vec{x}$\\
Energy & $E$ & $E$ & $E$ & $E$\\ 
Momentum & $\vec{p}$ & $-\vec{p}$ & $\vec{p}$ & $-\vec{p}$\\
Angular momentum & $\vec{J}$ & $\vec{J}$ & $\vec{J}$ & $-\vec{J}$\\
Electric field & $\vec{E}$ & $-\vec{E}$ & $-\vec{E}$ & $\vec{E}$\\
Magnetic field & $\vec{B}$ & $\vec{B}$ & $-\vec{B}$ & $-\vec{B}$\\
Electric charge & $q$ & $q$ & $-q$ & $q$\\ 
Baryon number & $B$ & $B$ & $-B$ & $B$\\ 
Lepton number & $L$ & $L$ & $-L$ & $L$\\ 
\end{tabular}
\caption{Effects of parity, charge-conjugation and time-reversal transformations on some physical quantities.}
\label{tab:DiscreteSymmetry}
\end{table}

Perfect symmetries would transform the Universe into an identical copy of itself, however all these three are individually broken in some specific process.
The parity operator flips the sign of the three spatial coordinates.
Parity is conserved in strong, electromagnetic and gravitational interactions but it is violated in weak interactions~\cite{PV}.
The charge-conjugation operator C flips both the electric and magnetic field directions and converts each particle in its corresponding antiparticle by changing the sign of all the internal quantum numbers (charge, baryon number, lepton number).
C-symmetry is again violated in weak interactions~\cite{CV}.
Neutrino properties are the most evident example of P- and C-symmetry violations: SM predicts that only left-handed neutrinos and right-ended antineutrinos exist, hence their P-symmetric versions (right-handed neutrinos and left-handed antineutrinos, respectively), or C-symmetric versions (left-handed antineutrinos and right-handed neutrinos, respectively) do not exist, maximally violating these two symmetries.
The sequential combination of C and P operators (CP transformation) is also violated by the weak force~\cite{CPV}, and will be described with more detail in the following.
The only transformation that represents a perfect symmetry in the SM is the sequential combination of all the three operators, the so called CPT transformation, which means that physics laws are invariant for particle and antiparticles with opposite momenta in a mirrored space frame.

During Big Bang, it is natural to assume that an equal amount of positive and negative charged particles, as well as matter and antimatter, was produced.
Therefore it must exist a process that privileges the survival of matter, since our known Universe is mainly composed by matter.
This problem is often referred to as the baryon-antibaryon asymmetry problem.  

Andrei Sakharov first proposed the set of conditions that would explain the different rate of production of matter and antimatter in the early Universe~\cite{10}, and these are: the baryon number violation (not yet observed), C- and CP-symmetry violations (both already observed), and departure from thermal equilibrium in order to avoid re-annihilation.

CP violation is related to the different behavior of matter and antimatter with respect to the weak interaction and it was included in the theory of weak force provided by the SM, thanks to the work of Cabibbo~\cite{cabibbo}, Kobayashi and Maskawa~\cite{ckm}. In this model the mass eigenstates (d, s, b) of quarks do not coincide with the quark eigenstates which participate in the weak interaction ($d'$, $s'$, $b'$).
The mixing of the two sets of eigenstates is determined by the so called Cabibbo-Kobayashi-Maskawa matrix (CKM-matrix), as shown in the following equation:
\begin{equation}\label{Eq:CKM}
\left( \begin{array}{c} d' \\ s' \\ b' \end{array} \right)=%
\left( \begin{array}{ccc} V_{ud} & V_{us} & V_{ub} \\ V_{cd} & V_{cs} & V_{cb} \\ V_{td} & V_{ts} & V_{tb}  \end{array} \right)%
\left( \begin{array}{c} d \\ s \\ b \end{array} \right) \ .
\end{equation}

The CKM-matrix is a $3\times3$ complex matrix, but not all the elements $V_{ij}$ are independent from each other, since it is a unitary matrix ($V^{*}_{CKM}V_{CKM}=V_{CKM}V^{*}_{CKM}= 1$).
This reduces the number of free real parameters to nine.
Six phases can be absorbed in the mass and weak eigenstates, at the expense of adding a global phase.
This further parametrization reduces the number of independent parameters to four.

A convenient (approximate) parametrization of the CKM matrix was proposed by Wolfenstein~\cite{wolf}.
It consists in a series expansion of the CKM-matrix in the small expansion parameter $\lambda$ $(\lambda \simeq |V_{us}|)$, obtaining:
\begin{equation}\label{qq:CKM2}
V_{CMK}=\left( \begin{array}{ccc} 1-\lambda^2 & \lambda & A\lambda^3(\rho-i\eta) \\%
							        -\lambda & 1-\lambda^2 & A\lambda^2  \\%
						 A\lambda^3(1-\rho-i\eta) & -A\lambda^2 & 1  \end{array} \right) + \mathcal{O}(\lambda^4)%
\end{equation}
where $\lambda$ , $A$ , $\rho$ and $\eta$ are four real and independent parameters that are all of order 1.
The latest values~\cite{ckm_pdglbl} for these parameters are shown here:
\begin{equation*}
\begin{split}
\lambda &= 0.22506 \pm 0.00050	\\
A &= 0.811\pm 0.026				\\
\rho &= 0.124^{+0.019}_{-0.018}	\\
\eta &= 0.356 \pm 0.011 \ .
\end{split}
\end{equation*}

Up to date, three of the CKM parameters $A$, $\rho$ and $\eta$, all related to the third generation of quark eigenstates, are quite poorly determined.
The complex phase $\eta$ is the term that causes CP violation to occur and it only appears in $V_{ub}$ and $V_{td}$ terms.

The unitarity of the CKM matrix gives some constraints on its terms.
On the diagonal terms it is possible to write:
\begin{equation}\label{eq:CMK-Unity-diag}
\sum _{k}{\left|{V}_{ik}\right|}^{2} = \sum _{i}{\left|{V}_{ik}\right|}^{2} = 1 \ ,
\end{equation}
while, on the other components:
\begin{equation}\label{eq:CMK-Unity}
\sum _{k}{V}_{ki}{V}_{kj}^{*} = 0 \ .
\end{equation}

From the last equation, we can see that for any of the six pairs of $i$ and $j$, a triangle in the complex Gauss plane is defined, which is called unitarity triangle.
The area of the triangles is always the same and is determined by the CP violating phase $\eta$, while the shape can be different.

\begin{figure}
\centering
\def\svgwidth{\linewidth}
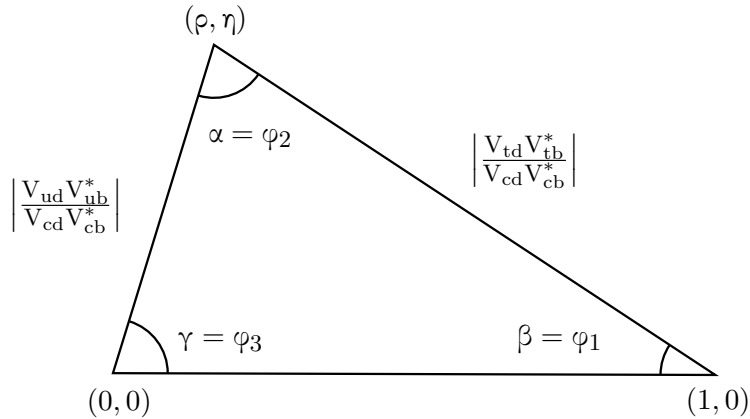
\caption{The CKM unitarity triangle with angles $\alpha$, $\beta$ and $\gamma$ and vertices $(0,0)$, $(1,0)$ and $(\rho,\eta)$.}
\label{fig:CMK} 
\end{figure}

CP violation can thus be studied by investigating the parameters that form the unitarity triangles.
In particular, applying equation \ref{eq:CMK-Unity} to the first and third columns ($i=d$ and $j=b$), we obtain:
\begin{equation}\label{eq:CMK-Unity-expl}
{V}_{ud}{V}_{ub}^{*} + {V}_{cd}{V}_{cb}^{*} + {V}_{td}{V}_{tb}^{*} = 0 \ .
\end{equation}
This equation can be represented as the unitarity triangle shown in Figure~\ref{fig:CMK}.
The triangle chosen above is particularly interesting since all the three sides have dimensions of the order of $\lambda$.

Precision measurements of the unitarity triangle parameters offer a number of opportunities to verify either CP violation with respect to the value predicted by the SM, or evidences of New Physics, like a fourth quark generation.

Several decays of B and D mesons are particularly interesting for such studies.
For instance, $B^0_d-\overline{B}^0_d$ mixing depends on the combination $|V_{td}V_{tb}^{*}|^2$ and thus provides information about the unitarity triangle of equation \ref{eq:CMK-Unity}.
Beyond the detailed test of the CKM description of CP violation, a wide spectrum of rare B decays can be studied, yielding independent information on the CKM parameters.

\section{Large Hadron Collider}

The Large Hadron Collider (LHC)~\cite{lhc} is the most powerful particle accelerator in the world.
This astonishing machine was built by a worldwide collaboration composed by 22 member states, as of 2017, at the CERN (European Organization for Nuclear Research) laboratories in Geneve, Switzerland.
The energy scale of 13~TeV in combination with the proton beam instantaneous luminosity in excess of $10^{34}\ cm^{-2}\: s^{-1}$ are unprecedented and allow for fundamental tests of the Standard Model of Particle Physics.

The LHC is built inside a $26.7\ km$ circular tunnel at a depth of about 100~m underground, at the border between Switzerland and France.
LHC is composed of two beam pipes where the proton bunches are accelerated in opposite directions in ultra-high vacuum.
A series of helium-cooled superconducting dipole magnets bend the beams, while quadrupole magnets squeeze the proton bunches in order to keep the beam well focused in the plane perpendicular to the beam axis.
The RF cavities accelerate the protons and allow them to be packed into bunches along the beam direction. The bunch crossing rate is $40\ MHz$ and two consecutive collisions are thus spaced by $25\ ns$.

Among the experiments that are currently exploiting the large possibilities offered by the LHC machine, the four biggest ones are ALICE, ATLAS, CMS and LHCb.
Each one is located in one of the four crossing points where the two proton beams are deviated from their parallel trajectories and focused on the collision point.

ATLAS~\cite{atlas} and CMS~\cite{cms} are the two general purpose detectors.
They cover the whole solid angle around the collision point and investigate the largest range of physics possible.
These two detectors have different design strategies which allow to cross-confirm any new discovery.
Their most groundbreaking result was without doubt the first observation of the Higgs boson made in 2012~\cite{higgs1, higgs2}.

ALICE and LHCb detectors are specialized for focusing on specific phenomena.
ALICE~\cite{alice} studies heavy-ion collisions and the physics of strongly interacting matter at extreme energy densities, where a phase of matter called quark-gluon plasma forms.

LHCb~\cite{lhcb, lhcbperf} is designed for high sensitivity searches of charge-parity (CP) violations in beauty and charm hadron decays and further high precision studies of the heavy flavor sector.
Its detector design is rather different from the other three experiments and its characteristics will be thoroughly discussed in the next sections.

In this chapter, firstly, I will introduce the LHCb detector with a brief description of all the sub-detectors that compose it, then I will explain with more detail the RICH sub-detectors and their working principle.
I will introduce the upcoming LHCb RICH upgrade and its implications in the optoelectronic readout system, describing the work done by our group about photodetectors and electronics.
The last and more detailed part of the chapter is dedicated to the work done during my Ph.D.: the participation at testbeams for the qualification of the optoelectronic readout system and the development of the ALDO ASIC.

\section{The LHCb experiment}

The LHCb experiment is dedicated to the study of CP-violation and rare decays of heavy hadrons containing a beauty or charm quark.
The experiment will test the components of the CKM matrix in order to find indirect evidence of new physics beyond the Standard Model.

Beauty and charmed hadrons are generated mainly in gluon-gluon interactions.
The gluons in this type of interactions carry a largely differing fraction of the beam particle momentum, causing a boost of the resulting particles along the beam line.

\begin{figure}
	\centering
	\includegraphics[width=.99\textwidth]{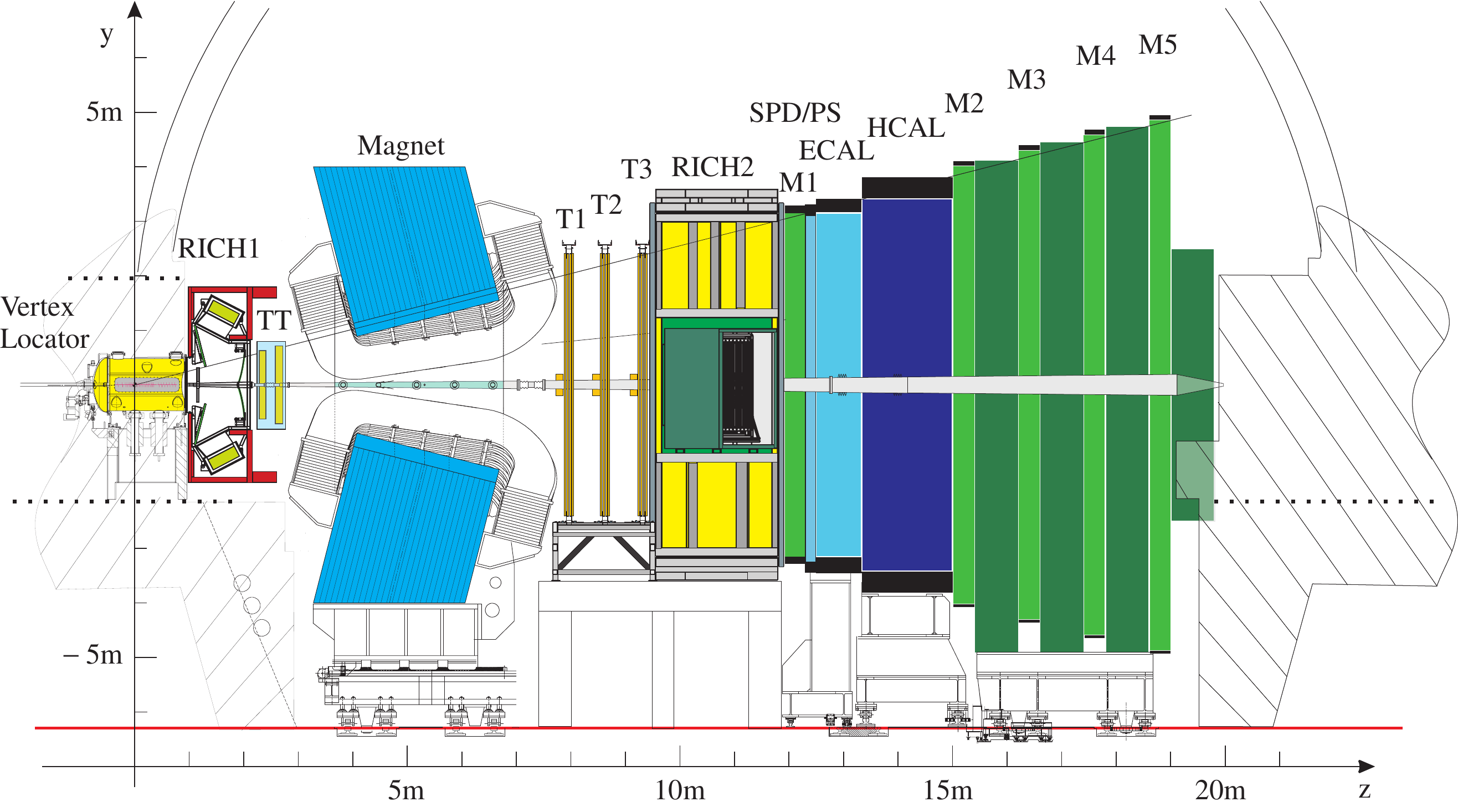}
	\caption{LHCb detector drawing.}
	\label{fig:lhcb_disegno}
\end{figure}

For this reason the layout of the LHCb detector is radically different than any of the other three main experiments and it is designed as a single-arm forward spectrometer.
Figure~\ref{fig:lhcb_disegno} shows a schematic side view of LHCb and its sub-detectors.

\begin{figure}
\begin{minipage}[b][][t]{.47\linewidth}
	\centering
	\includegraphics[width=\linewidth]{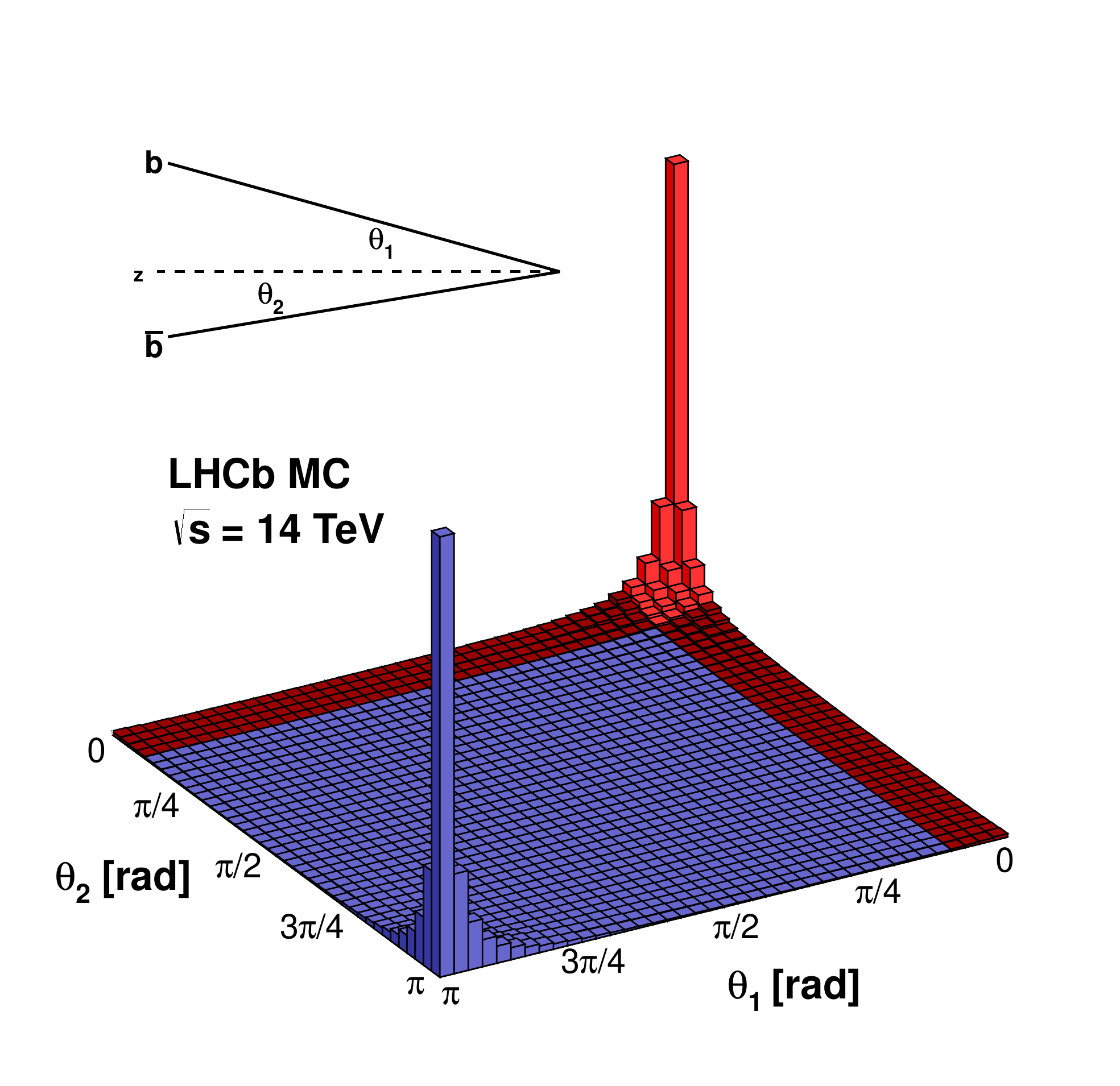}
\end{minipage}
\ \hspace{1mm} \
\begin{minipage}[b][][t]{.47\linewidth}
	\centering
	\includegraphics[width=\linewidth]{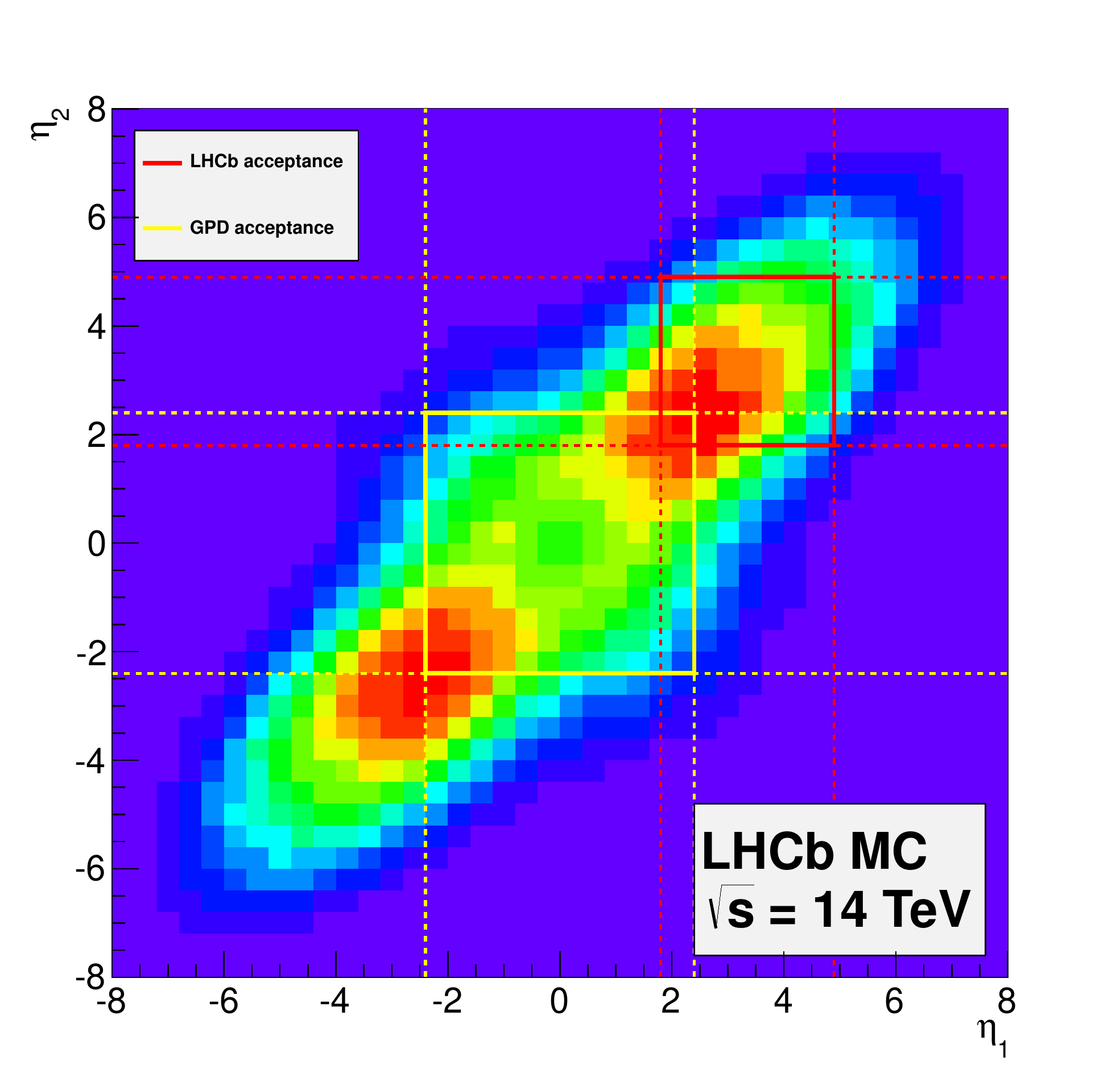} 
\end{minipage}
\caption{Plots for the $b\bar{b}$ pair production as a function of the angle (\textit{left}) and pseudorapididy (\textit{right}). The LHCb acceptance is highlighted in red, while the acceptance of general purpose experiments like ATLAS and CMS is highlighted in yellow.}
\label{fig:cross_section}
\end{figure}

The particle interaction point is located at the origin of the coordinate system.
The detector has full tracking coverage in the pseudorapidity range $1.8 < \eta < 4.9$ ($10\ mrad - 300\ mrad$) in the bending plane, which allows studying boosted hadrons of interest.
Figure~\ref{fig:cross_section} shows the produced number of $b\bar{b}$ pairs as a function of polar angle and pseudorapidity.
The LHCb acceptance is highlighted by the red square, while the acceptance of a general purpose detector, like ATLAS or CMS, is marked in yellow.

The LHCb detector operates at a lower luminosity with respect to the other experiments in order to have less interactions per bunch crossing, one proton-proton collision on average in the current operation phase.
The low pile-up is a fundamental requirement, since the studies performed at LHCb rely on the precise reconstruction of the decay chain, both geometrically, with high resolution primary and secondary vertex positioning, and topologically, with excellent particle identification efficiency.
This last requirement explains why the LHCb experiment is equipped with specific RICH detectors entirely dedicated to particle identification.
The design instantaneous luminosity of the LHCb experiment is $4 \cdot 10^{32}\ cm^{-2}\:s^{-1}$, a factor of about $50\times$ less than the machine peak value.

In the following sections, a brief overview of the LHCb detector and its various subdetectors is presented.

\subsubsection*{VELO}

The VErtex LOcator (VELO)~\cite{VELO_TDR} provides precise measurements of the track coordinates close to the interaction points.
This information is crucial in order to identify displaced secondary vertexes, which are typical for the short lived particles containing b and c quarks.

The VELO consists of a series of D-shaped planar silicon modules arranged along the beam direction.
Each unit provides a two-dimensional measure of the track coordinates expressed in polar units ($r$ is the radial distance from the beam axis and $\phi$ the angular displacement).
These two coordinates are acquired by silicon microstrip detectors with a thickness of 300 $\mu$m.
The third coordinate along the beam axis ($z$) is provided by the presence of multiple of such D-shaped detectors, with a higher granularity close to the expected interaction point.

In order to prevent detector damage during beam injection phase, when the beam is not fully focused, the VELO silicon modules are mounted on a retractable system that allows to move them 3 cm further from the beam axis.
As soon as the beam is collimated and stable, the modules are moved to 8 mm from the beam axis.

The VELO detector performance is essential for the precise detection of the vertex location. The detector is capable of a spatial resolution in the primary vertex reconstruction of $40$ $\mu$m and 10 $\mu$m in the parallel and perpendicular direction, respectively.
Two additional planes, located upstream of the main interaction point, constitute the pile-up detection system, acting as a trigger veto.
The pile-up system also estimates the number of primary proton-proton interactions for each bunch crossing.

\subsubsection*{Magnet}

The LHCb experiment requires precise measurements of charged particles momentum.
The bending magnet~\cite{Magnet_TDR} used in the experiment is a non superconductive dipole magnet with a total integrated magnetic field of $4\ Tm$ along a $10\ m$ distance.
The magnet is calibrated up to less than $10^{-4}$ relative precision in order to provide the most accurate estimation of the particle momentum.
To reduce the systematic effects of the detector, the direction of the magnetic field can be periodically flipped, so that data can be acquired with field polarity directed either upwards or downwards along the vertical y axis direction.
The magnet design is optimized in order to achieve low stray magnetic fields in the upstream direction, where sensitive detectors like the VELO and the RICH-1 are placed, while having a wide aperture, necessary to cover the whole detector acceptance.

\subsubsection*{Tracking}

The charged particle track reconstruction is a crucial feature necessary to identify the signatures of the decays and the particle momenta.
Beside the VELO, the LHCb tracking system is composed by TT (Tracker Turicensis), T1, T2 and T3 tracking stations~\cite{lhcb}.

The TT is a 150 cm wide and 130 cm high planar tracking station.
It is made of 4 layers of silicon microstrips and is large enough to cover the full angular acceptance of the experiment.
Unlike the TT station, T1-T2-T3 are located downstream the magnet, as illustrated in Figure~\ref{fig:lhcb_disegno}.
The T1-T3 are divided in two parts built with different technologies called Inner Tracker (IT) and Outer Tracker (OT).
The IT\cite{IT_TDR}, composed by silicon microstrip detectors, covers a $120\ cm$ wide and $40\ cm$ high cross-shaped region in the center of the T1-T2-T3 tracking stations.
The OT \cite{OT_TDR}, which covers the most peripheral areas, is a drift-time detector designed as an array of individual, gas-tight straw-tube modules, composed of two staggered layers of 64 drift tubes each, filled by a mixture of Argon and CO$_2$.
This choice ensures fast drift time (below $50\ ns$) and a sufficient spatial resolution ($200\ \mu m$).

From the position of the particle interaction in the tracking system, the trajectories of the charged particles can be reconstructed with an efficiency larger than 96\% in the momentum range going from $5\ GeV/c$ up to $200\ GeV/c$, and in the pseudorapidity ranging from 2 to 5.

\subsubsection*{RICH detectors}

One of the peculiar differences that characterizes the LHCb detector with respect to the other experiments at LHC, is the presence of two detectors dedicated to particle identification (PID).
At LHCb, in fact, the accurate topological reconstruction of every decay channels is one of the main requirements for the study of rare CP-violating processes.
For this purpose, LHCb includes two RICH (Ring Imaging CHerenkov) detectors, mainly responsible for the proton, kaon and pion identification~\cite{rich_TDR}.

The RICH1 detector is located upstream of the LHCb dipole magnet, between the VELO and the TT, and provides particle identification in the low momentum range (from $\sim1$ to $\sim 60\ GeV/c$) over a wide angular acceptance ranging from $\pm 25\ mrad$ to $\pm 300\ mrad$ ($\pm 250\ mrad$), along the horizontal (vertical) direction.
The RICH2 detector is located downstream of the magnet, between the last tracking station (T3) an the first muon station (M1).
It is capable to operate in the high momentum range, going from $\mytilde15$ to $\mytilde100\ GeV/c$, in a limited angular acceptance ranging from $15\ mrad$ to $120\ mrad$ in the horizontal direction and to $100\ mrad$ in the vertical direction, large enough to cover the region where the high momentum particles are produced.

A more detailed description of this detector will follow.

\subsubsection*{Calorimeters}

The calorimetry system at LHCb~\cite{calo_TDR} consists of two main detectors, the electromagnetic (ECAL) and the hadronic (HCAL) calorimeter.
On the upstream side of the ECAL, a Scintillating Pad Detector (SPD) and a Pre-Shower detector (PS) are added to the calorimetry system for faster discrimination between electrons, photons and heavier charged particles.

Both calorimeters are made with a shashlik structure, with interleaved absorbing material and active scintillating tiles.
They allow to measure position and energy of the incident charged and neutral particles by generating electromagnetic or hadronic showers in the absorbing material.
Interleaved layers of scintillating material produce an amount of light proportional to the charge released during the electromagnetic shower and thus to the energy of the impinging particle.
Light is captured by wavelength shifting fibers (WLS) that guide the photons to photomultiplier tubes for detection and allow a better spectral matching to the photocathode quantum efficiency.

The measured transverse energy is passed to the first level trigger (L0) and is used to trigger on particles with high transverse energy, typical for events containing a quark b or c. 

The ECAL is capable of an energy resolution of $\sigma_{E}/E = 9\% / \sqrt{E/GeV}$, while the HCAL provides a resolution of $\sigma_{E}/E = 69\% / \sqrt{E/GeV}$.

\subsubsection*{Muons}

The muon system~\cite{muon_TDR} consists of five Multi-Wire Proportional Chamber layers.
The first one is located upstream of the calorimeters and features a Gas Electron Multiplier (GEM) in the high occupancy region near the beam pipe.
The other four muon detection layers are located on the downstream side of the calorimeters, in the outermost part of the LHCb detector.

These layers are interleaved with 80~cm thick iron absorbers, reaching up to 20 hadronic interaction lengths in combination with the HCAL.
The muon system provides essential information to the trigger, since muons appear in many final states of CP-sensitive decays.
Muons are identified with 95\% efficiency, with a misidentification rate of about 2\%.

\subsubsection*{Trigger}

The LHCb detector operates at a luminosity such that, on average, only one interaction per bunch crossing happens.
Among all the possible type of interactions, only a fraction of them is interesting for the research performed at LHCb.
The trigger system~\cite{trigger_TDR} is responsible for this task of selection of interesting events.

To select events containing beauty and charmed hadron decays, a multi level trigger system is used, consisting of the L0 hardware trigger, and the High Level Trigger (HLT) implemented in software.

The L0 trigger accesses the data from the calorimeters and muon systems in order to select events producing particles with high transverse energy and momentum.
These detectors are the only that are accessed at the full bunch crossing rate of $40\ MHz$ and L0 is able to reduce the rate of interesting events down to about $1\ MHz$.
The L0 also uses the information from the pile-up system in the VELO, as veto for events with too high multiplicity. 

The software-based HLT, further reduces the data flow from 1 MHz to a few kHz.
It consists of a C++ application which initially performs a partial reconstruction of the physical properties of the particles using the information from the VELO and the tracking system to confirm the L0 decision.
To the remaining candidates, the HLT performs a full event reconstruction and applies a series of event selection criteria, reducing the data flow to a rate of a few kHz.
Selected events are stored in the dedicated server farm.

\subsection{LHCb upgrade}

During the next planned long shutdown (LS2), starting from 2019, the LHCb detector will undergo a substantial upgrade~\cite{LOI,Framework_TDR} aimed to significantly increase the luminosity at which the experiment is operating.
The goal is to increase the instantaneous luminosity up to $2 \cdot 10^{33}\ cm^{-2} \: s^{-1}$, a factor 5 above the current level.
Over the course of the following 10 years of operation LHCb will aim to collect $50\ fb^{-1}$ of data, ten times the integrated luminosity acquired so far, since the LHC construction.

The limiting factor to fully exploit the higher luminosity already offered by the LHC machine is the L0 hardware trigger which relies on the information from few detectors, since most of the existing front-end electronics is limited to $1\ MHz$ readout.

The LHCb detector upgrade avoids this limitation by removing the L0 hardware trigger and by reading the detector information at the $40\ MHz$ bunch crossing rate~\cite{trigger_upg_TDR}.
Events are recorded and sent to the LHCb data acquisition farm at the full event rate of the LHC and
processed by a software trigger.
The software trigger features a simplified version of the offline reconstruction to increase its efficiency.
The upgrade imposes two major challenges to the sub-detectors of LHCb.
Their front-end electronics has to be able to read out at $40\ MHz$ without any dead time, and the detectors have to withstand a five times higher luminosity in comparison to Run 2, which means increased pile-up, higher detector occupancy and harsher radiation levels.

The upgraded LHCb will feature a new Vertex Locator~\cite{VELO_upg_TDR}, modifications in the tracking systems~\cite{tracker_upg_TDR} like the Upstream Tracker (UT) and the Scintillating Fibre Tracker (SciFi), changes to the calorimetry and muon systems, and a major upgrade of the RICH detectors~\cite{pid_upg}.

\subsection{Cherenkov effect}

In the previous section, I briefly introduced Ring Imaging Cherenkov (RICH) detectors.
These kind of detectors are responsible for particle identification (PID) of charged hadrons, exploiting the Cherenkov effect to distinguish between $\pi$, K, and p.

Cherenkov effect consists in the emission of photons when a charged particle traverse a medium with a velocity $v_p$ higher than the velocity of light in the medium.
This effect was first observed by P. A. Cherenkov in 1934~\cite{cherenkov} and theoretically formulated by I. Frank and I. Tamm in 1937. Cherenkov, Frank, and Tamm were collectively awarded the Nobel prize in 1958.

The speed of light in a medium ($c_m$) with refractive index $n$ is given by
\begin{equation}
c_m = \frac{c}{n} \ .
\end{equation}
As an example, the refractive index for a typical silicate glass is about 1.4, which means that the speed of light in that material is \mytilde$0.7c$.
Ultra-relativistic particles like the ones present in high energy physics experiments can easily surpass that speed.

\begin{figure}
\centering
\includegraphics[width=.6\linewidth]{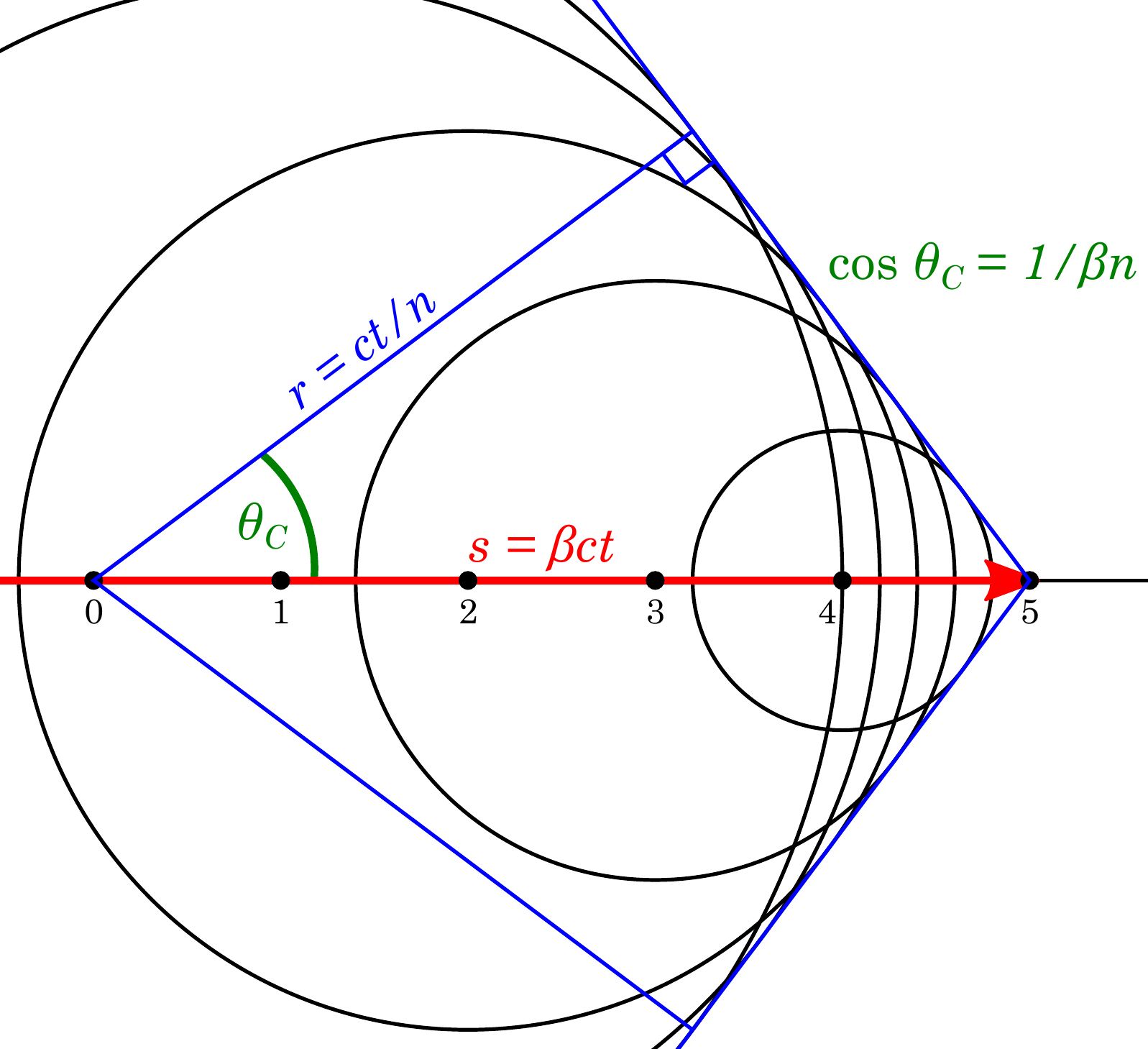} 
\caption{Schematic representation of the Cherenkov radiation emission.}
\label{fig:Cherenkov_Wavefront}
\end{figure}

When a charged particle travels through the medium, the particle's electric field locally superimposes to that of the medium lattice.
For a dielectric radiator, this time-dependent local field results in a dipole momentum induced on the atoms of the medium.
If the particle is slow enough, then the relaxation of the dipoles does not lead to any electromagnetic emission since the dipoles are symmetrically distributed around the particle and the wavelets emitted by the relaxation exhibit a destructive interference.
If the high energy particle passes through the medium with $v_p>\frac{c}{n}$, then the dipoles does not reach a symmetrical distribution and, as the dipoles relax, the wavelets generated interfere constructively.
The wavefront of the resulting Cherenkov radiation is emitted on the surface of a cone.
A schematic representation of the phenomenon is shown in Figure~\ref{fig:Cherenkov_Wavefront}.

The light is emitted at a constant angle $\theta_{C}$ relative to the direction of propagation of the charged particle, given by the following expression:
\begin{equation}
\cos \theta_{C} = \frac{c_m}{v_p} = \frac{1}{n\beta} \ ,
\end{equation}
where $\beta = \frac{v_p}{c}$ is the ratio between the velocity of the particle and the speed of light.

The condition $v_p>\frac{c}{n}$ represent the lower threshold for the Cherenkov effect.
For high energy particles when $\beta$ approaches unity, then the Cherenkov angle $\theta_{C}$ saturates at $\theta_{C,max}=\arccos\frac{1}{n}$.
These two values set a range of momenta where a particular RICH detector (i.e. radiating material) is effective for PID.

RICH detectors~\cite{rich_det} are able to measure the radius of the emitted light cone and, combining this information with the momentum measurement provided by the tracking system, can measure the mass of the incident particle, allowing its identification.

Since particles are ultra-relativistic, their $\beta$ is close to unity and thus also $n$ must be chosen close to 1.
For the same reasons, the Cherenkov angles are quite small (few mrad) and long focal lengths together with high pixel resolutions are required in order to detect the Cherenkov rings.

The energy spectrum of the Cherenkov radiation can be calculated from the Maxwell's equations and is expressed by the Frank-Tamm formula:
\begin{equation}\label{eq:Frank}
\frac{d^2E}{d\omega \: dx}=\frac{q^2}{4\pi}\mu(\omega)\omega \left(  1-\frac{1}{\beta^2 n(\omega)^2} \right) \ ,
\end{equation}
where $q$ is the charge of the high energy particle, $\omega$ is the angular velocity, $n\left(\omega\right)$ and $\mu\left(\omega\right)$ are the refractive index and the magnetic permeability of the dielectric medium, dependent on $\omega$.

Assuming $\mu\left(\omega\right)$ and $n\left(\omega\right)$ constant with respect to $\omega$, equation \ref{eq:Frank} can be expressed in term of number of photons emitted per unit of length and wavelength. Photon energy is given by $E=h\nu$ and the total energy can be written as $E=Nh\nu = Nhc/\lambda$ obtaining:
\begin{equation}\label{eq:PhotonYield}
\frac{d^2N}{d\lambda \: dx} \propto \frac{q^2 }{\lambda^2} \left(  1-\frac{1}{\beta^2 n^2} \right) \ .
\end{equation}

The photon production yield is proportional to the thickness of the radiator material and inversely proportional to the radiation wavelength.
Cherenkov photons are typically emitted in the optical part of the spectrum towards the violet and ultraviolet.
Photon production does not depend on the energy of the particles, but solely on their $\beta$ .
In order to operate successfully, RICH detectors have to provide and adequate number of emitted photons in the sensitive wavelength range of the photodetectors, so radiator length has to be adequately high.
Typically, a set of focusing and planar mirrors is used to focus and project the Cherenkov cone as a ring onto the photon detection plane.

\subsection{LHCb RICH detector and its upgrade}

\begin{figure}
\begin{minipage}[b][][t]{.47\linewidth}
	\centering
	\includegraphics[height=8 cm]{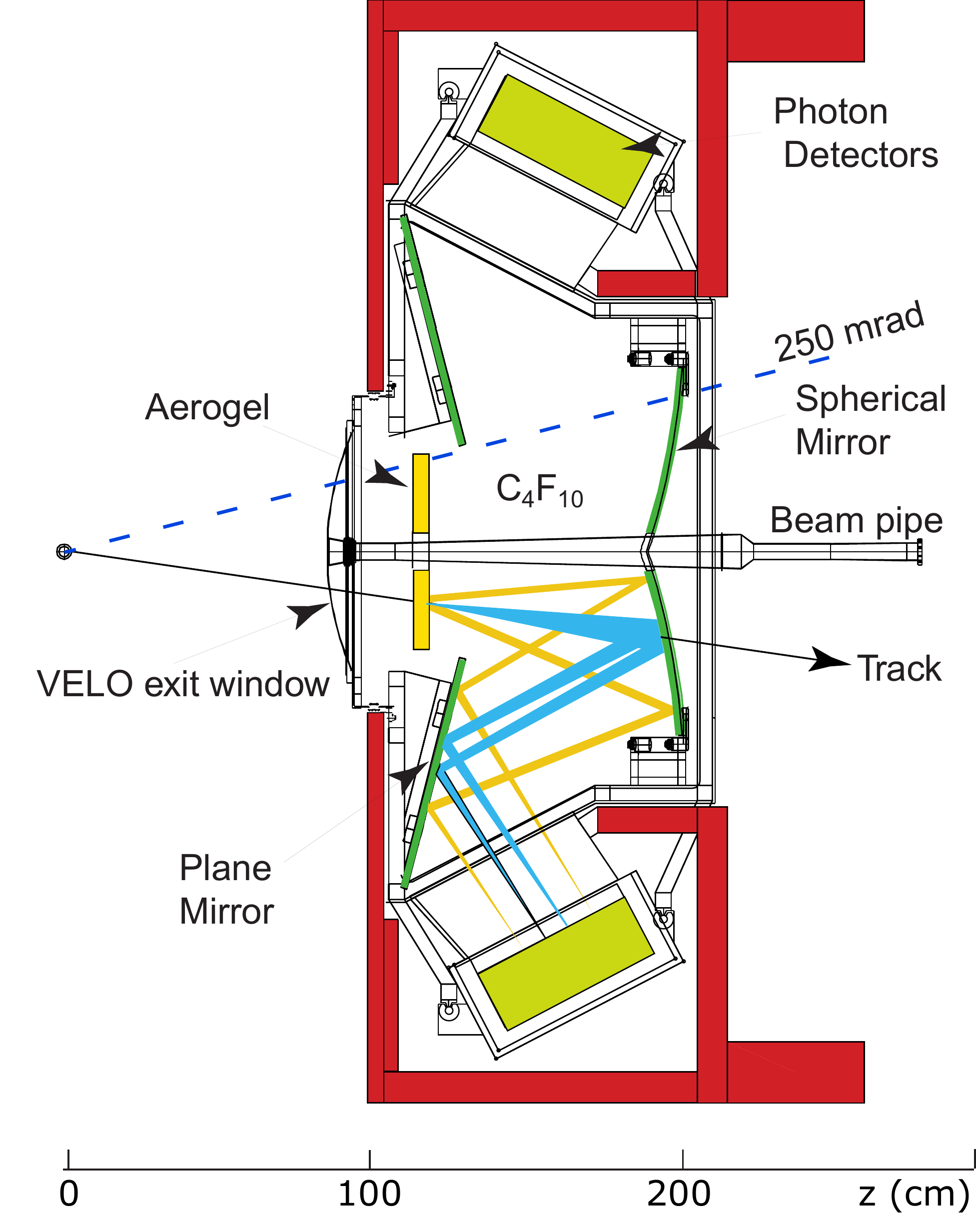}
\end{minipage}
\ \hspace{3mm} \
\begin{minipage}[b][][t]{.47\linewidth}
	\centering
	\includegraphics[height=8 cm]{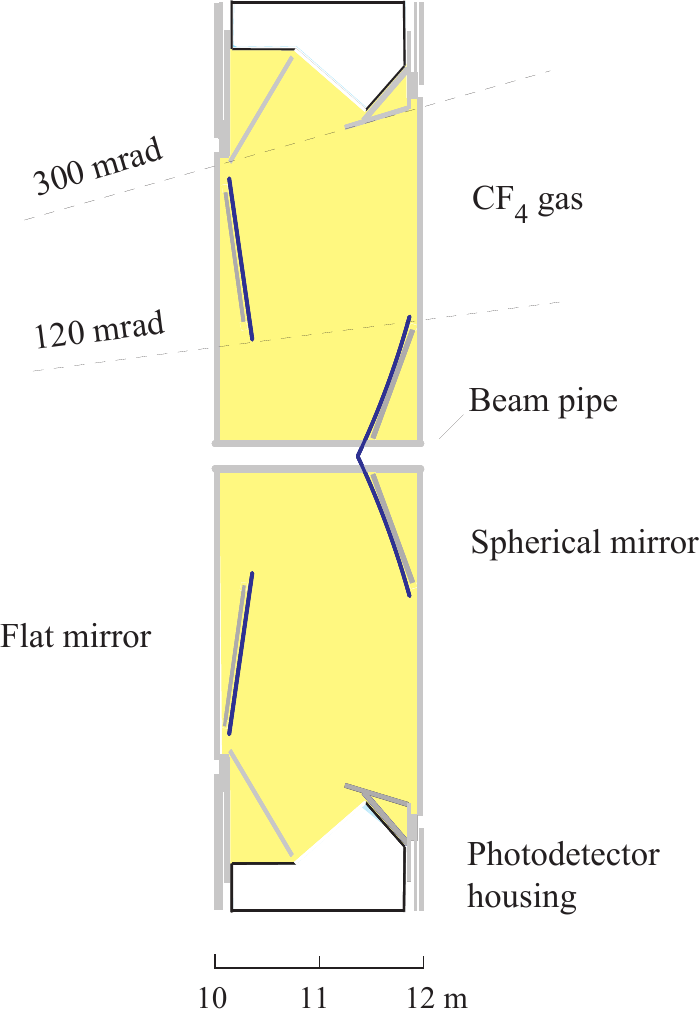} 
\end{minipage}
\caption{Schematics of the two RICH detectors. RICH1 is on the left, while RICH2 is on the right.}
\label{fig:rich_sch}
\end{figure}

The RICH detectors in LHCb are two, in order to cover the whole momentum range required by the experiment~\cite{lhcb,rich_perf}.
The schematic of the two detectors is shown in Figure~\ref{fig:rich_sch}.

The first station, called RICH1, is located upstream of the magnet and is responsible for the PID in the momentum range below 60~GeV/c.
RICH1 used two radiating materials, a solid aerogel radiator, few cm thick, and a gaseous radiator (fluorobutane, $C_4 F_10$, $n = 1.0014$).
From Run 2, the aerogel was removed due to his low yield (low number of Cherenkov photons emitted and tendency to absorb the gas, thus changing its refractive index).
Its removal allows to slightly increase the performance of the gaseous radiator and reduce the material within the acceptance, at the expense of poorer PID efficiency in low momentum range.

RICH2 is located after the tracking stations, just before the first muon station.
Its gaseous radiator is made by $CF_4$ ($n = 1.00048$) and allow PID up to 100~GeV/c.

Both detectors feature two sets of mirrors, one spherical mirror for focusing the ring, plus a flat mirror outside of the acceptance, which project the Cherenkov light further onto the photodetection plane.
Due to the small number of Cherenkov photons, the photodetectors have to be sensitive to single photons and should allow pixellated readout with a resolution high enough to reconstruct the Cherenkov rings.

Currently, the photosensitive plane is made with a matrix of Hybrid Photodetectors (HPDs)~\cite{hpd}, arranged in two planes for each RICH.
The planes cover a large area of $1.2\ m^2$ in RICH1 and $2.6\ m^2$ in RICH2 with an active area of 64\% and quantum efficiency of about 30\%.
The photodetectors are enclosed in a magnetic shielding box that reduces the stray fields due to the dipole magnet and allow the proper operation of the photodetectors.
Each photodetector is individually shielded with a wrap of mu-Metal for additional magnetic shielding.

HPDs feature a quite low gain of about 5000 electrons for each incoming photon, thus the readout electronics have to be installed as close as possible to the anode, in order to maximize signal to noise ratio.
In the current LHCb RICH, the readout chip is directly bump-bonded to the anode and it is encapsulated within the vacuum tube of the detector.
The front-end chip exhibits a noise of about $130\ e^-$ and is able to sustain rates up to $1\ MHz$.

As briefly introduced in previous sections, the LHCb detector will undergo a large upgrade in order to operate at higher luminosity~\cite{pid_upg}.
For the RICH detector, the upgrade will require a major redesign since the readout rate is limited by the electronics, which cannot be replaced.
Furthermore, the increased occupancy (i.e. the number of lighted pixels for each bunch crossing) will require a change in the optical scheme in order to reduce the number of hits per pixel.

To summarize, the upgraded RICH detectors will feature:
\begin{itemize}
\item \emph{New optics:} the photodetector plane is moved backwards and the mirror position is optimized for reducing aberrations in order to improve the ring angular resolution and keep the occupancy at about $25\%-30\%$ in the hottest regions of RICH1.
\item \emph{New photodetectors:} the HPDs will be substituted with multi-anode photomultipliers tubes (MaPMT) which exhibit a higher active area, higher quantum efficiency, smaller pixel size and higher gain. This allow to use external electronics, while still having a very compact system.
\item \emph{New electronics:} the new custom electronics will be able to resist to the higher radiation levels and to read out signals at the full 40~MHz interaction, with a power consumption of less than 1~mW per channel, avoiding the need of a dedicated cooling.
\end{itemize}

The system is based on a smaller modular element, called Elementary Cell (EC), which will be described in the following section.

\subsection{The Elementary Cell}

\begin{figure}
\centering
\includegraphics[height=5 cm]{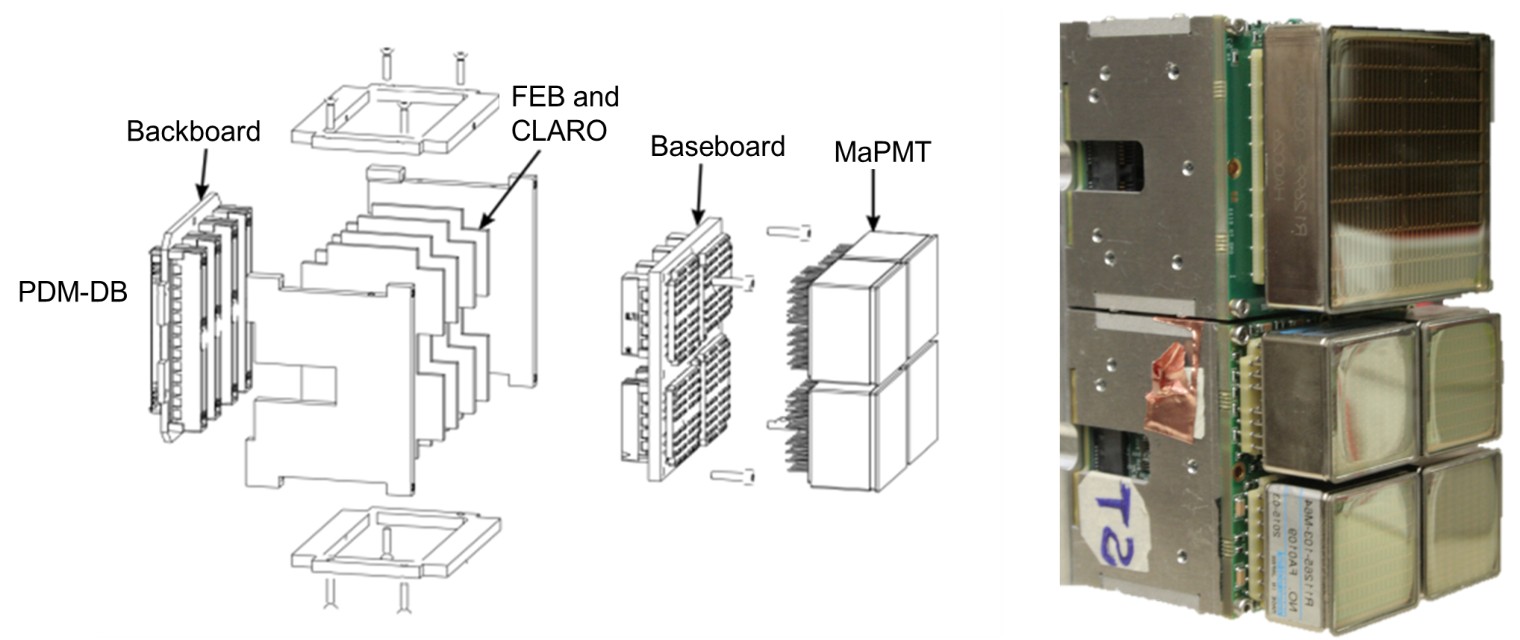} 
\caption{On the left, an exploded schematic sketch of the Elementary Cell. On the right a photo of the two types of Elementary Cells.}
\label{fig:foto_ec}
\end{figure}

The photon detector system and its electronics will be composed of autonomous functional units, called Elementary Cells (ECs).
This facilitates detector assembly and maintenance, and provides flexibility of the geometrical arrangement of the photon detection planes. 
With such a modular system also MaPMT geometrical acceptance is maximized.
An exploded drawing of the EC is shown in Figure~\ref{fig:foto_ec}, together with a photograph of the two types of EC.

\begin{figure}
\centering
\includegraphics[width=.8\linewidth]{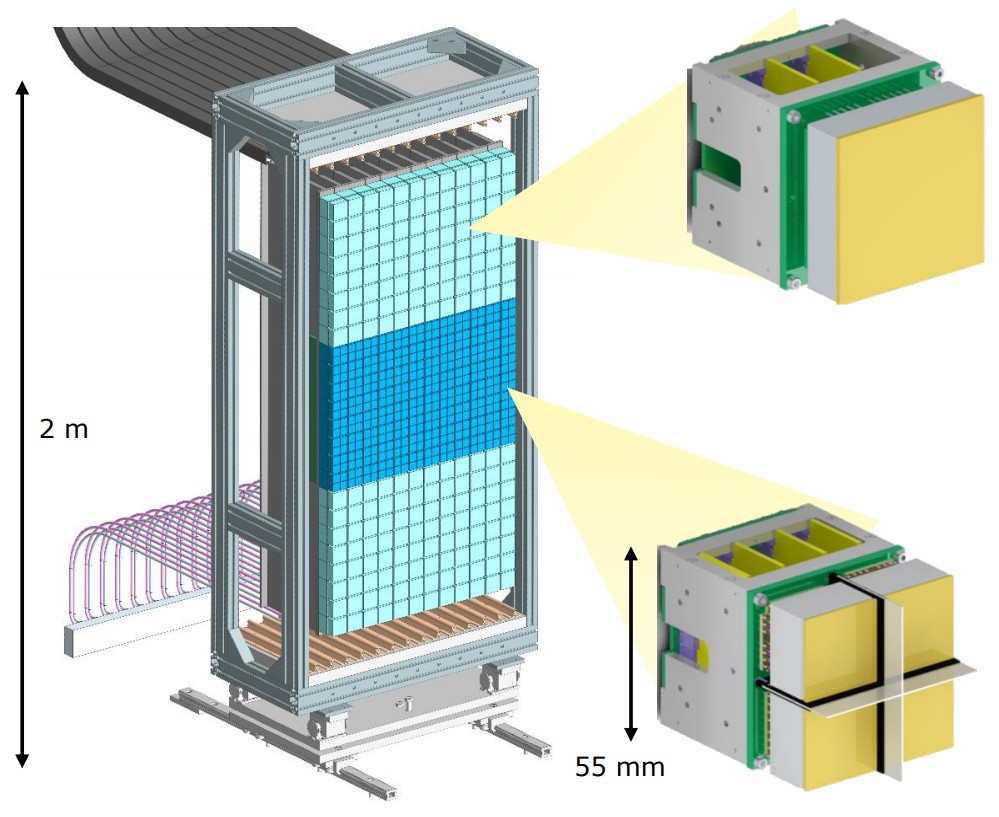} 
\caption{Graphical picture of the upgraded RICH2 photosensitive plane. The central area with higher pixel occupancy adopts the R-type EC, with smaller pixels, while the external parts adopts the H-type EC.}
\label{fig:rich2_upgrade}
\end{figure}

One EC contains either four R-type MaPMTs (Hamamatsu R11265~\cite{R11265}, 64-channel, $2.9\times2.9\ mm^2$ pixel size) or a single H-type MaPMT (Hamamatsu H12700~\cite{H12700}, 64-channel, $6\times6\ mm^2$ pixel size), plugging into custom designed sockets which are located on a PCB, the baseboard, which also hosts the voltage divider for the HV biasing of the sensor and provides a cross-shaped thermal mass that behaves as a low thermal impedance path driving the heat dissipated by the divider towards the metallic structure surrounding the EC.
The R-type and H-type ECs are shown respectively on the bottom and on the top of the photograph in Figure~\ref{fig:foto_ec}.
The aluminum case provides both structural rigidity and heat flow to a cooling bar.
The H-type ECs are used in the outer areas of the RICH2 detector, where pixel size is not limiting the Cherenkov angle resolution.
Figure~\ref{fig:rich2_upgrade} shows a graphical sketch of RICH2 photosensitive plane, where it is possible to appreciate the different segmentation in the inner and outer areas.

\begin{figure}
\centering
\includegraphics[width=.7\linewidth]{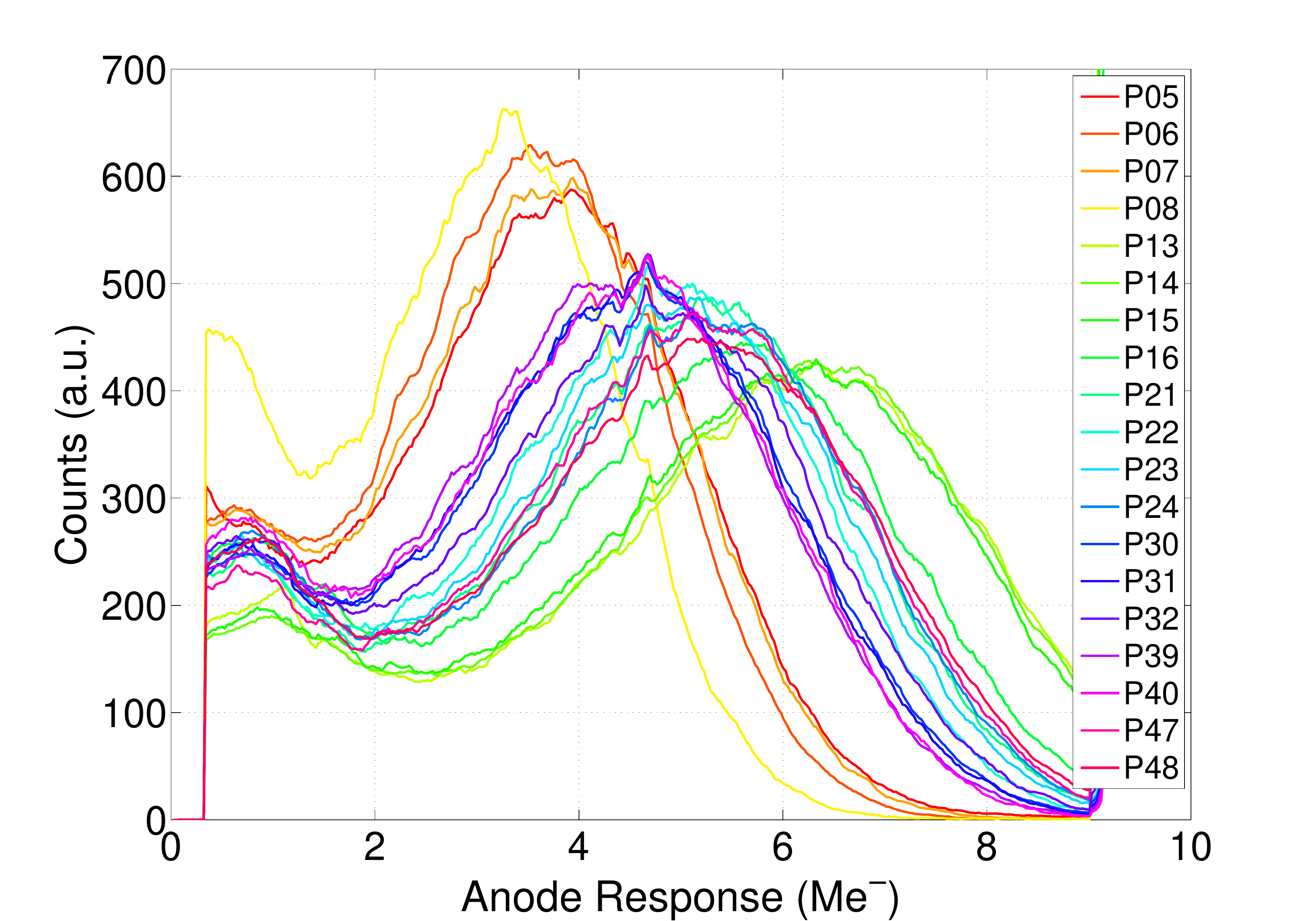} 
\caption{Single photon spectra for several pixels of a R11265 MaPMT.}
\label{fig:r11265_spectra}
\end{figure}

MaPMTs offer several improvements with respect to HPDs.
Geometrical acceptance is much higher, more than 85\%, with respect to about 67\%.
The photon conversion process is always performed in a photocathode, but the quantum efficiency of the MaPMT's photocathode has improved from about 30\% for the HPD's one, to about 40\%, thanks to the adoption of the latest ultra bi-alkali (UBA) materials.
Output signal is in the $Me^-$ range ($\sim160\ fC$), due to the multi-stage amplification chain.
The typical output capacitance is of the order of $0.5\ pF$.
The plot in Figure~\ref{fig:r11265_spectra} shows the typical single photon spectra for several pixels on a R11265 MaPMT.

The baseboard interfaces with 4 (or 2 in the case of the EC-H) front-end boards (FEBs) that host the readout chip, named CLARO and developed by our group in Milano Bicocca.
This device generates a digital pulse when a photon hits the corresponding MaPMT pixel.
A detailed description of the chip is provided in the following section.
FEBs interface with a backboard that routes the output signals to the digital board (PDMDB).
The core of the PDMDB is an FPGA that latches the binary signals coming from the CLAROs, packages them and sends the packets out of the detector.
The PDMDB is also responsible for the configuration of the operational parameters of the chips and hosts the low voltage power supply circuitry.

The power supply of the front-end chips will be provided by a rad-hard DC/DC converter module designed at CERN, named FEASTMP~\cite{feast}.
This switching device is able to operate in presence of high magnetic fields up to $4\ T$ and up to a total ionizing dose of $200\ Mrad$ with high efficiency ($>75 \%$).
The noise is quite good for such a compact device, about $2\ mV$ peak-to-peak.
Stability over temperature is about $400\ ppm/^{\circ}C$, which correspond to $1\ mV/^{\circ}C$ at a power supply of $2.5\ V$.
This last specification must be considered carefully, since the temperature excursion of this type of devices can be very high due to changing loads.

\subsection{The CLARO ASIC}

\begin{figure}
\centering
\begin{minipage}[b][][t]{.40\linewidth}
	\centering
	\includegraphics[height=5 cm]{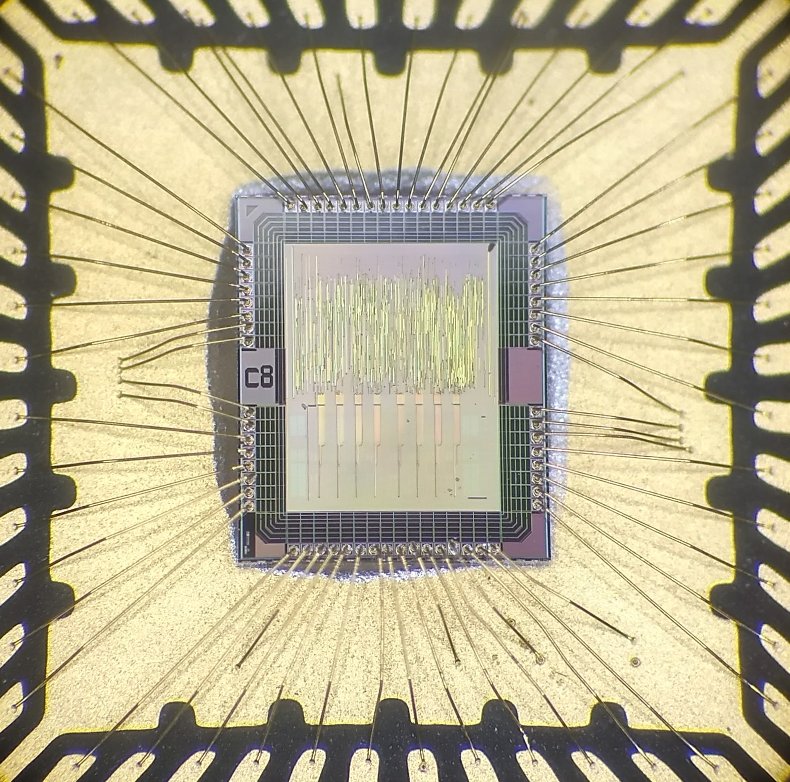} 
\end{minipage}
\ \hspace{3.25mm} \
\begin{minipage}[b][][t]{.54\linewidth}
	\centering
	\includegraphics[height=5 cm]{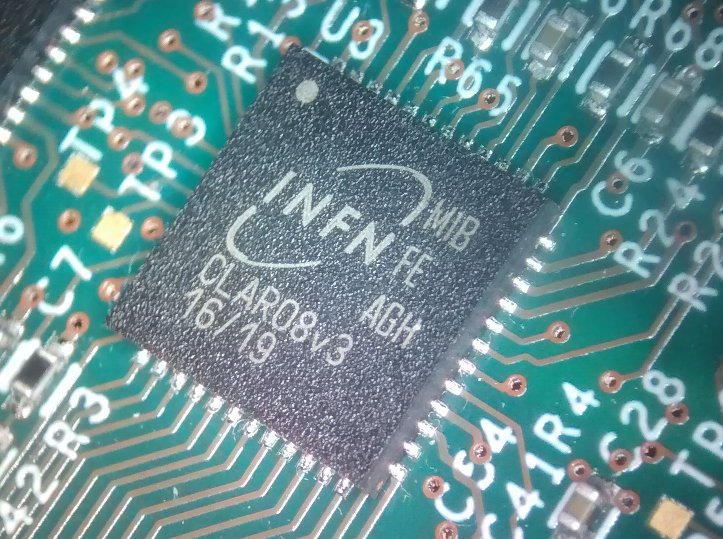}
\end{minipage}
\caption{Photographs of the CLARO ASIC die (on the left) and packaged (on the right).}
\label{fig:claro_photo}
\end{figure}

The CLARO~\cite{claro} is a radiation-hard 8-channel application specific integrated circuit (ASIC) specifically designed for single photon counting with MaPMTs.
It is made in $0.35\ \mu m$ CMOS technology from ams (ex Austria Microsystems), a reliable and inexpensive technology that proved to meet the LHCb requirements while also ensuring a very high yield and a good tolerance to radiation.
Two photos of the chip are shown in Figure~\ref{fig:claro_photo}, one in bare die within the package, and the other in the QFN56 package.

\begin{figure}
	\centering
		\begin{minipage}[b][][t]{.48\textwidth}
			\centering
			\includegraphics[width=\linewidth]{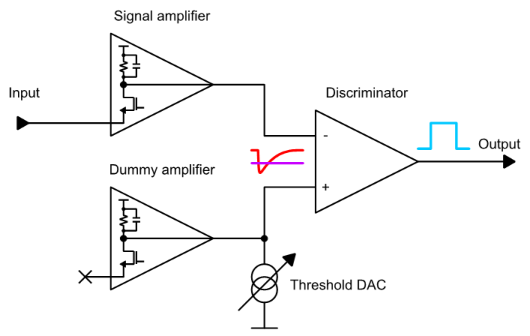}
			\caption{Simplified electrical scheme of the CLARO analog block.}
			\label{fig:CLARO_Schema}
		\end{minipage}
		\ \hspace{3mm} \
		\begin{minipage}[b][][t]{.46\textwidth}
			\centering
			\includegraphics[width=\linewidth]{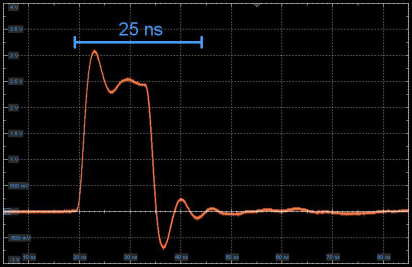}
			\caption{Typical output signal. The baseline is restored in 25 ns.}
			\label{fig:CLARO_pulse}
		\end{minipage}
\end{figure}

In a typical single photon counting application, a charge amplifier reads the signals from the PMT on a low input impedance, and produces voltage signals with amplitude proportional to the deposited charge.
The signals feed the input of a comparator (discriminator), which generates a digital output signal if their amplitude exceeds a given threshold.
The gain of the PMT pixels in general is not uniform, therefore the gain of the amplifier or the threshold of the comparator should be adjusted on a channel by channel basis.
This approach is used by the CLARO and by other ASICs in the field \cite{nino, maroc}, including an early 4-channel CLARO prototype~\cite{claro-cmos}.
A simple block schematic of the chip is shown in Figure~\ref{fig:CLARO_Schema}.
The first prototype dates back to 2011, and it evolved through few revisions up to the final version, named CLARO8v3.

The main features of this ASIC are the fast baseline recovery, within $25\ ns$ for typical working conditions, and below $50\ ns$ even for large input signals with negligible spillover, and the low power consumption, below $1\ mW$ per channel in idle and $2.5\ mW$ at high rate ($10^7$ hits per second, the forseen maximum occupancy).

The fast baseline recovery allow to count up to a rate of $40\ MHz$, as required by the LHCb upgrade specifications, even if the maximum average data rate is expected to be lower, since the occupancy in the hottest regions of RICH1 will be slightly below 30\%.
A typical output signal is shown in Figure~\ref{fig:CLARO_pulse}.

The low power consumption requirement comes from the fact that the EC is very tight packed and the cooling of the PMTs is a critical aspect to be considered, since the dark current depends exponentially on the operating temperature.
Any additional heating source has to be minimized and the readout circuit can usually become a major contributor due to its wide bandwidth.
With our solution, only $45\ mW$ per 8 chips -- the number that is needed to read out one full MaPMT -- are consumed in idle, a contribution that is negligible with respect to the power dissipated by the MaPMT bias circuitry (about $320\ mW$).

The analog block is designed with a pseudo-differential layout, adopting a dummy amplifier with floating input, completely identical to the main CSA.
When no current is generated from the threshold DAC current source, the output value of the dummy amplifier compensate any offset of the input CSA allowing its DC coupling to the discriminator stage.
Moreover, the symmetrical configuration obtained using the dummy amplifier improves the power supply rejection ratio of the ASIC.

MaPMTs have a quite high spread in their characteristics: the gain of two pixels in the same device can differ up to a factor 4, and also depends strongly on the HV bias voltage.
The spread can be even larger between different devices.
The CLARO chip must be able to compensate for such a high spread in the input signal characteristics, therefore both the gain and the discriminator threshold are digitally programmable.

The gain (attenuation) can be adjusted with 4 values (2 bits), and the available settings are $1\ V/V$, $1/2\ V/V$, $1/4\ V/V$ and $1/8\ V/V$.
Threshold can be set with 6 bits (64 values) by means of a programmable current generator.
The full range can be adjusted by changing an external resistor.
In the nominal configuration and with gain $1\ V/V$, the threshold step is $30\ ke^-$ ($5\ fC$) and the maximum threshold is $2\ Me^-$ ($320\ fC$).
At the lowest gain, minimum and maximum thresholds are respectively $240\ ke^-$ ($38\ fC$) and $15\ Me^-$ ($2.5\ pC$).
An additional bit is able to lower all the thresholds by half-scale, in order to compensate for channels with particularly high threshold offset.

All settings are stored in a 128-bit configuration and status register, accessible with serial peripheral interface (SPI).
The digital register allow also to enable many useful debug features like input test signal injection, SEU counter read back, and SEU injection.
The signal injection is particularly useful for mass production tests, since it can be used to quickly verify the functionality of each channel in its entirety.

Integrated circuits are sensitive to the effects of the environment radiation.
Impinging radiation causes charge deposits in oxides, electron-hole production in the channel, and damage to the silicon lattice.
In digital circuitry these effects are particularly relevant since they could cause bit-flips, often referred to as single event upset (SEU).
Another possible effect caused by radiation is single event latchup (SEL), which consist in the formation of a parasitic structure like a BJT transistors in the bulk or a CMOS transistor under passivation oxides, that can create low impedance paths between critical lines and compromise the device functionality.

Both the analog and digital parts of the chip adopt design and layout techniques in order to minimize the effect of radiation damage.

The configuration block, in particular, is protected against radiation-indu-ced SEUs and SELs thanks to a triple modular redundancy architecture with additional circuitry that can self-correct a flipped cell, and custom hardened digital cells~\cite{Siviglia1,Siviglia2}.

Table~\ref{tab:rich_fluences} shows the expected radiation levels in the RICH environment for its expected lifetime ($50\ fb^{-1}$ of total integrated luminosity).

\begin{table}
\centering
{\tabulinesep=1.5mm
\begin{tabu}{c|P{27mm} P{27mm} P{27mm}} 
 & \textbf{Neutrons (1~MeV~eq.)} & \textbf{Total ionizing \newline dose} & \textbf{Hadrons $\bm{\left(>20\ MeV\right)}$} \\ 
\hline 
\textbf{RICH-1} & $6.1 \cdot 10^{12}\ cm^{-2}$ & 400 krad & $2.3 \cdot 10^{12}\ cm^{-2}$ \\ 
\textbf{RICH-2} & $3.1 \cdot 10^{12}\ cm^{-2}$ &  & $1.0 \cdot 10^{12}\ cm^{-2}$ \\ 
\end{tabu}}
\caption{LHCb RICH radiation levels during the whole detector life.}
\label{tab:rich_fluences}
\end{table}

The radiation test campaign~\cite{claro_irr1, claro_irr2} was performed mainly by the INFN Ferrara group and demonstrated that the chip could operate successfully with equivalent radiation levels of 100 years of operation in LHCb environment.
The performance degradation of the analog block was minor, with a threshold drift of  about 5\%, without any relevant noise or power consumption increase.
Tests with protons and high energy ions caused several SEUs in the chip, all of them were corrected by the voter and detected in the SEU counter.
No SEL were observed.

\section{Testbeams}

Tests of the full optoelectronic chain have been performed in multiple testbeam sessions starting from 2014 at the SPS facility at CERN. In these tests, the performance of the proposed photon detectors and the feasibility of the front-end readout and DAQ chain were assessed in a particle beam environment.

The beam tests~\cite{testbeam} were performed in the North Area of the Prevessin site at CERN. A beam consisting mainly of pions and protons with momentum of $180\ GeV/c$  was obtained from the SPS facility and guided through a light-tight box containing a borosilicate planoconvex lens, used as the Cherenkov radiator. In the first tests the Cherenkov light was detected with two half populated Elementary Cells (4 MaPMTs, 256 channels), while the latest tests featured four fully populated ECs (1024 channels). A beam trigger is generated by two scintillator planes located upstream and downstream of the light-tight box. The box was also placed downstream of a tracking telescope, which is able to record and reconstruct track information for the particles.

\subsection{System setup}

\begin{figure}
\centering
\includegraphics[angle=90, width=.5\linewidth]{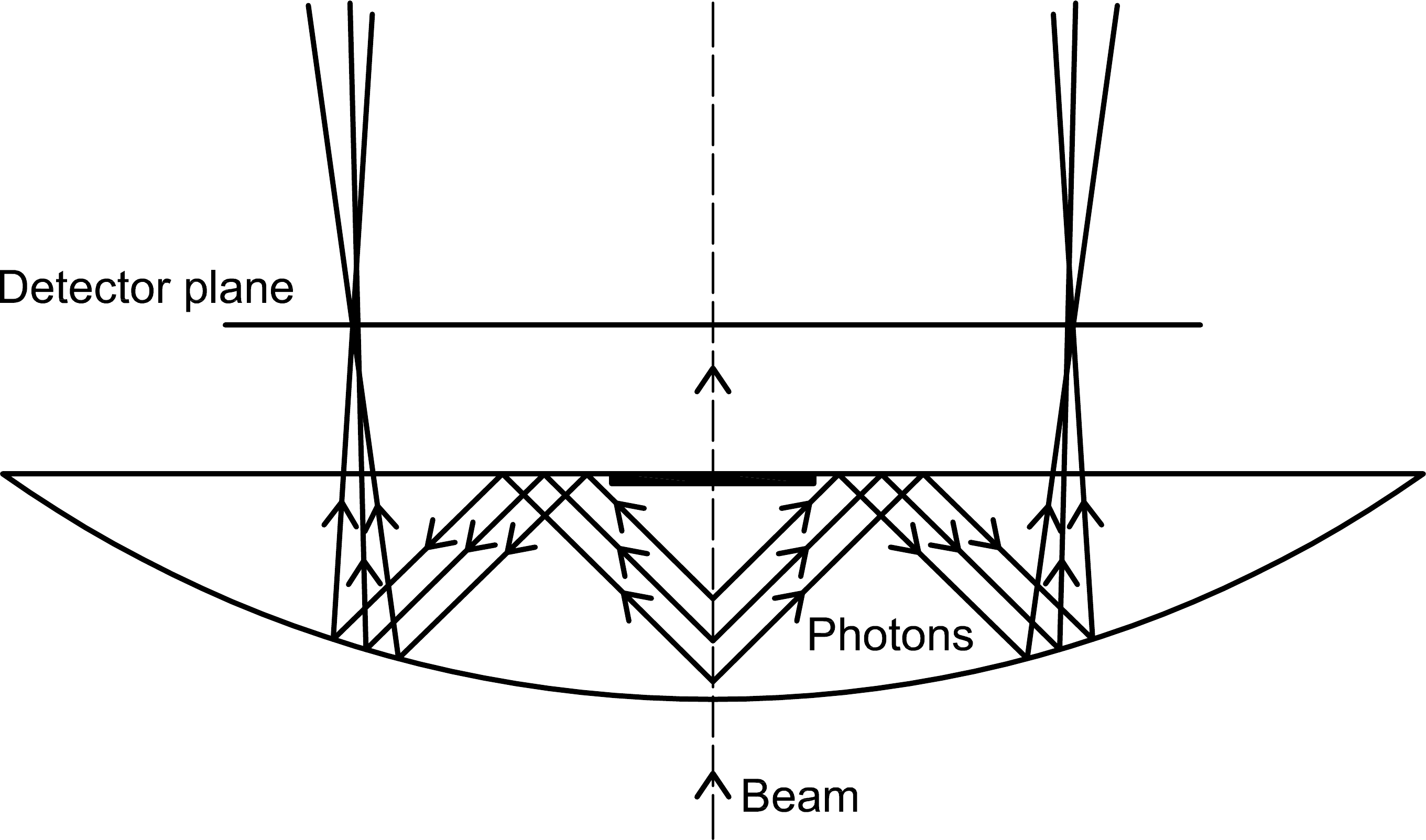}
\caption{Schematic of the optical setup.}
\label{fig:setup}
\end{figure}

The optical setup consisted of a planoconvex lens made of borosilicate glass which acts both as Cherenkov radiator and as focusing device. Figure~\ref{fig:setup} shows a sketch of the optical setup and its working principle. 


Charged particles enter the lens in the centre of the spherical side and they produce Cherenkov photons which are reflected at the flat surface due to total internal reflection. On the spherical surface, a $20\ mm$ ring-shaped reflective layer is deposited, which focuses the incident photons on the photodetector plane, forming the Cherenkov ring. The radius of the lens has been measured to be $R=144.6\unitm{mm}$, it has a diameter of $151.7\unitm{mm}$ and a thickness at the centre of $27\unitm{mm}$. The centre of the lens, on the flat surface, has been blackened to reduce the length of the path where detectable photons are produced and increase the spatial resolution.

Extensive simulations with particle simulation software (\textsc{Geant}) and optical CAD software (Optica) were performed. The expected Cherenkov ring radius is $\sim 60\ mm$, with a spatial resolution of $0.6\ mm$, mainly limited by the optical system and by the MaPMT pixel size. In the test beam configuration, the Cherenkov angle is always saturated to $\theta_{C,max}=\arccos\frac{1}{n}$, since the refractive index of the borosilicate glass radiator is large ($n\approx1.4$), thus it is not possible to discriminate the incident particles.

\begin{figure}
\centering
\begin{minipage}[b][][t]{.65\linewidth}
	\centering
	\includegraphics[height=5.5cm]{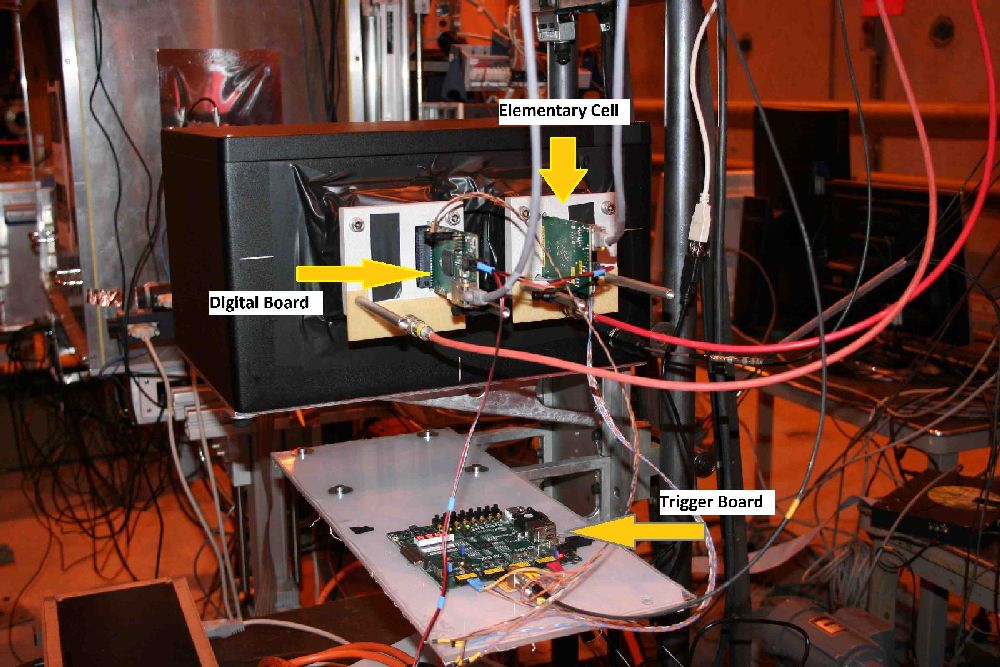}
	\caption{Photograph of the experimental setup during 2015 testbeam (two half ECs installed).}
	\label{fig:box}
\end{minipage}
\ \hspace{0.3mm} \
\begin{minipage}[b][][t]{.3\linewidth}
	\centering
	\includegraphics[height=5.5cm]{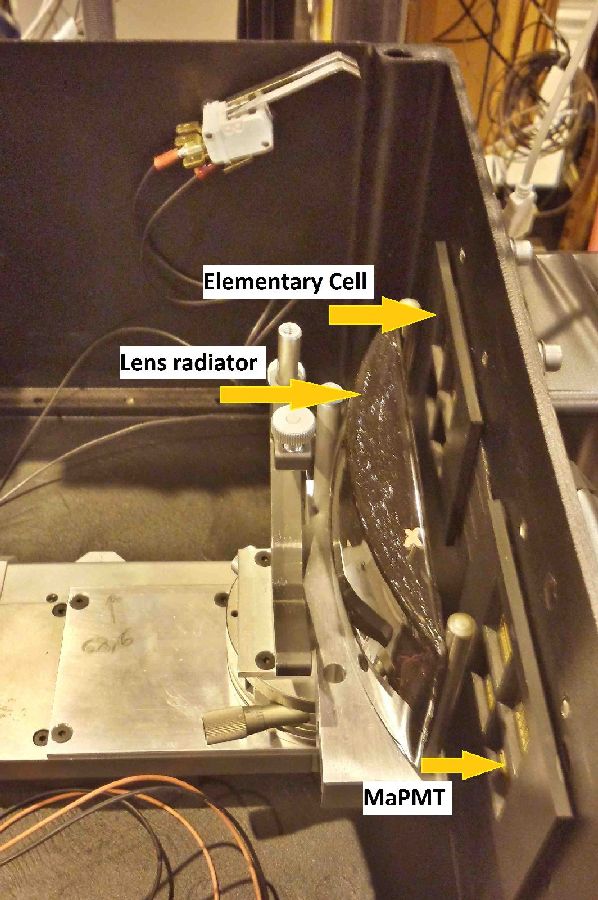}
	\caption{View of the inside of the box.}
	\label{fig:inside}
\end{minipage} \\
\end{figure}

The ECs are installed symmetrically around the beam, each capturing an arc of the Cherenkov ring. When more then two ECs are installed, they are stacked in two groups.
The setup was installed on a remotely movable support in order to align it to the beam.
Two photographs of the setup installed in the experimental area are shown in Figure~\ref{fig:box} and Figure~\ref{fig:inside}.

The digital outputs of the CLAROs are read out by the digital board (DB), which is still in a prototypical phase. The DB formats the data into event packets and transmits them over Gigabit Ethernet to a PC where they are stored. The DB is also used to configure the CLARO chips. Graphical user interfaces were developed to configure, monitor and control all the instrumentations.

The online data monitor is particularly crucial since it allows to validate the data quality during acquisition. A decoding program is able to read the acquired data and to display monitoring histograms and a map of the accumulated hits on the MaPMTs.

\subsection{DAC scan studies and CLARO threshold calibration}
\label{sec:dac_scan}

Before the latest testbeams, all the 1024 channels were individually calibrated and, for each CLARO channel, we characterized the threshold step and offset at threshold zero.
A 10-bit DAC installed on the backboard allow to set a programmable voltage at the input of the CLARO test circuit, which injects current pulses with amplitude proportional to the applied voltage into the charge sensitive amplifier.

\begin{figure}
	\centering
 	\includegraphics[width=.6\linewidth]{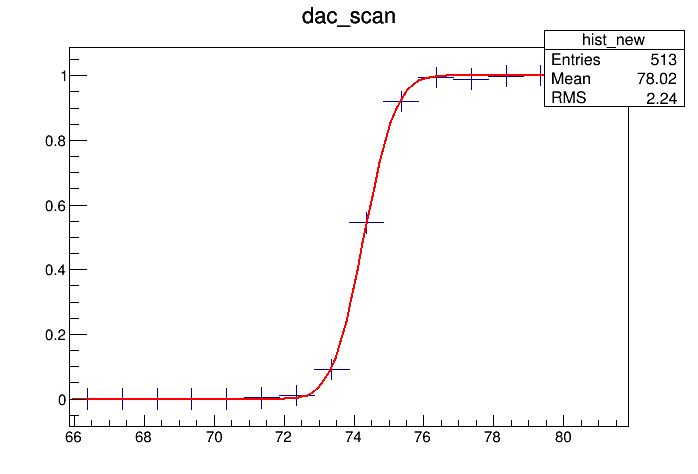}
    \caption{S-curve from a DAC scan. The curve is fitted with an erf(). The mean value is the threshold value and the sigma represents the RMS noise.}
    \label{fig:claro_dac_scan_s_curve}
\end{figure}

\begin{figure}
	\centering
	 \input{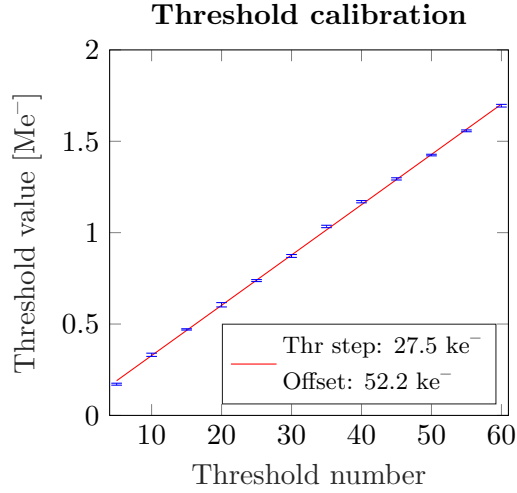} 
	\caption{Multiple DAC scans at different threshold are combined into this plot which allow to calculate the threshold step and offset at threshold zero.}
	\label{fig:claro_dac_scan_thr}
\end{figure}

A sweep in DAC voltage at fixed threshold allow to calibrate the threshold, as shown in Figure~\ref{fig:claro_dac_scan_s_curve}.
The signal amplitude at which the output starts to count is the calibrated value for that specific threshold, while the slope of the transition gives the noise of the device.
By repeating the same procedure for different thresholds, a linear curve like the one in Figure~\ref{fig:claro_dac_scan_thr} is reconstructed.
The slope of the linear fit of the curve gives the threshold step, while the intercept gives the offset at threshold zero.

We performed DAC scans at three attenuation (gain) settings (1, 1/2 and 1/4) and also with the offset bit enabled, in order to test the effectiveness of this functionality.

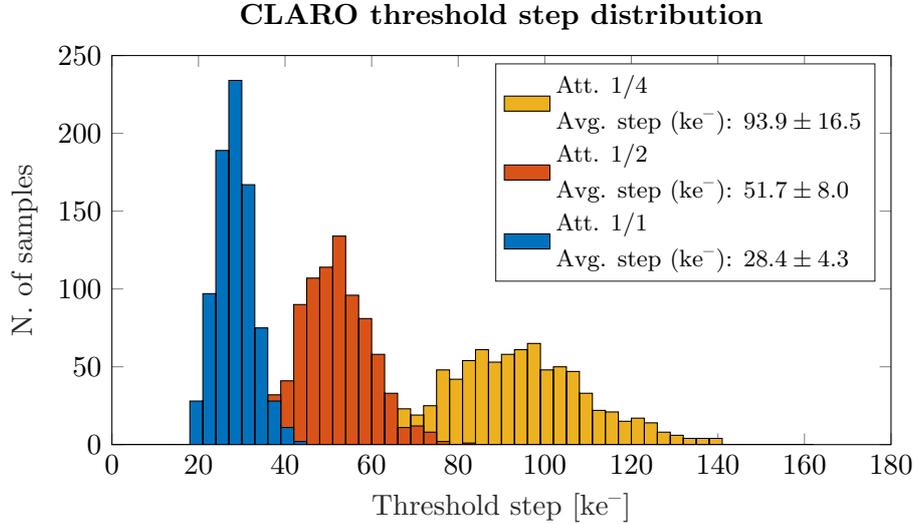
\begin{figure}
	\centering
	 \definecolor{mycolor1}{rgb}{0.92900,0.69400,0.12500}%
\definecolor{mycolor2}{rgb}{0.85000,0.32500,0.09800}%
\definecolor{mycolor3}{rgb}{0.00000,0.44700,0.74100}%
\begin{tikzpicture}

\begin{axis}[%
width=4.036in,
height=2.031in,
at={(0.677in,0.426in)},
scale only axis,
xmin=0,
xmax=180,
xlabel style={font=\color{white!15!black}},
xlabel={Threshold step [$ke^-$]},
ymin=0,
ymax=250,
ylabel style={font=\color{white!15!black}},
ylabel={N. of samples},
axis background/.style={fill=white},
title style={font=\bfseries},
title={CLARO threshold step distribution},
legend style={legend cell align=left, align=left, draw=white!15!black}
]
\addplot[ybar interval, fill=mycolor1, fill opacity=0.6, draw=black, area legend] table[row sep=crcr] {%
x	y\\
0	0\\
3	0\\
6	0\\
9	0\\
12	0\\
15	0\\
18	0\\
21	0\\
24	0\\
27	0\\
30	0\\
33	0\\
36	0\\
39	0\\
42	1\\
45	0\\
48	1\\
51	0\\
54	1\\
57	4\\
60	10\\
63	10\\
66	23\\
69	19\\
72	25\\
75	48\\
78	42\\
81	54\\
84	61\\
87	53\\
90	58\\
93	61\\
96	65\\
99	48\\
102	50\\
105	47\\
108	33\\
111	22\\
114	21\\
117	15\\
120	17\\
123	14\\
126	8\\
129	6\\
132	4\\
135	4\\
138	4\\
141	0\\
144	0\\
147	0\\
150	0\\
153	0\\
156	0\\
159	0\\
162	0\\
};
\addlegendentry{\footnotesize Att. 1/4\\ \footnotesize Avg. step ($ke^-$): $93.9 \pm 16.5$}

\addplot[ybar interval, fill=mycolor2, fill opacity=0.6, draw=black, area legend] table[row sep=crcr] {%
x	y\\
0	0\\
3	0\\
6	0\\
9	0\\
12	0\\
15	0\\
18	0\\
21	0\\
24	0\\
27	1\\
30	0\\
33	9\\
36	32\\
39	41\\
42	90\\
45	107\\
48	114\\
51	134\\
54	96\\
57	81\\
60	58\\
63	33\\
66	11\\
69	12\\
72	8\\
75	2\\
78	0\\
81	1\\
84	0\\
87	0\\
90	0\\
93	0\\
96	0\\
99	0\\
102	0\\
105	0\\
108	0\\
111	0\\
114	0\\
117	0\\
120	0\\
123	0\\
126	0\\
129	0\\
132	0\\
135	0\\
138	0\\
141	0\\
144	0\\
147	0\\
150	0\\
153	0\\
156	0\\
159	0\\
162	0\\
};
\addlegendentry{\footnotesize Att. 1/2\\ \footnotesize Avg. step ($ke^-$): $51.7 \pm 8.0$}

\addplot[ybar interval, fill=mycolor3, fill opacity=0.6, draw=black, area legend] table[row sep=crcr] {%
x	y\\
0	0\\
3	0\\
6	0\\
9	0\\
12	0\\
15	0\\
18	28\\
21	97\\
24	189\\
27	234\\
30	167\\
33	75\\
36	28\\
39	11\\
42	2\\
45	0\\
48	0\\
51	0\\
54	0\\
57	0\\
60	0\\
63	0\\
66	0\\
69	0\\
72	0\\
75	0\\
78	0\\
81	0\\
84	0\\
87	0\\
90	0\\
93	0\\
96	0\\
99	0\\
102	0\\
105	0\\
108	0\\
111	0\\
114	0\\
117	0\\
120	0\\
123	0\\
126	0\\
129	0\\
132	0\\
135	0\\
138	0\\
141	0\\
144	0\\
147	0\\
150	0\\
153	0\\
156	0\\
159	0\\
162	0\\
};
\addlegendentry{\footnotesize Att. 1/1\\ \footnotesize Avg. step ($ke^-$): $28.4 \pm 4.3$}

\end{axis}
\end{tikzpicture}%
	\caption{Threshold step distribution at different attenuation settings for 1024 channels.}
	\label{fig:claro_thr_step_distr}
\end{figure}

The distribution of the threshold steps at different attenuation is plotted in Figure~\ref{fig:claro_thr_step_distr}. For typical MaPMT gains of a few $Me^-$, the attenuation 1 is the preferred settings, since the average threshold step is $28.4\ ke^-$, with a full scale of about $1.8\ Me^-$. At higher attenuations the average step increases to $51.7\ ke^-$ (attenuation 1/2) and $93.9\ ke^-$ (attenuation 1/4). The standard deviation of the threshold steps is about 16\%, as expected from the simulations performed during the CLARO design phase.

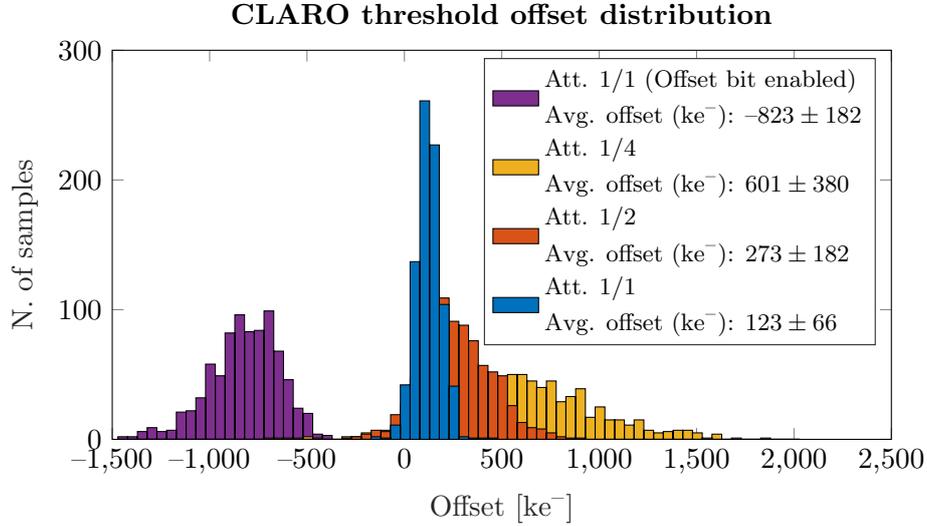
\begin{figure}
	\centering
	 \definecolor{mycolor1}{rgb}{0.49400,0.18400,0.55600}%
\definecolor{mycolor2}{rgb}{0.92900,0.69400,0.12500}%
\definecolor{mycolor3}{rgb}{0.85000,0.32500,0.09800}%
\definecolor{mycolor4}{rgb}{0.00000,0.44700,0.74100}%
\begin{tikzpicture}

\begin{axis}[%
width=4.036in,
height=2.031in,
at={(0.677in,0.426in)},
scale only axis,
xmin=-1500,
xmax=2500,
xlabel style={font=\color{white!15!black}},
xlabel={Offset [$ke^-$]},
ymin=0,
ymax=300,
ylabel style={font=\color{white!15!black}},
ylabel={N. of samples},
axis background/.style={fill=white},
title style={font=\bfseries},
title={CLARO threshold offset distribution},
legend style={legend cell align=left, align=left, draw=white!15!black}
]
\addplot[ybar interval, fill=mycolor1, fill opacity=0.6, draw=black, area legend] table[row sep=crcr] {%
x	y\\
-1470.85054421244	2\\
-1420.85054421244	2\\
-1370.85054421244	6\\
-1320.85054421244	9\\
-1270.85054421244	6\\
-1220.85054421244	7\\
-1170.85054421244	21\\
-1120.85054421244	22\\
-1070.85054421244	32\\
-1020.85054421244	58\\
-970.850544212438	49\\
-920.850544212438	82\\
-870.850544212438	96\\
-820.850544212438	83\\
-770.850544212438	84\\
-720.850544212438	99\\
-670.850544212438	68\\
-620.850544212438	46\\
-570.850544212438	24\\
-520.850544212438	20\\
-470.850544212438	4\\
-420.850544212438	3\\
-370.850544212438	0\\
-320.850544212438	0\\
-270.850544212438	0\\
-220.850544212438	0\\
-170.850544212438	0\\
-120.850544212438	0\\
-70.8505442124383	0\\
-20.8505442124383	0\\
29.1494557875617	0\\
79.1494557875617	0\\
129.149455787562	0\\
179.149455787562	0\\
229.149455787562	0\\
279.149455787562	0\\
329.149455787562	0\\
379.149455787562	0\\
429.149455787562	0\\
479.149455787562	0\\
529.149455787562	0\\
579.149455787562	0\\
629.149455787562	0\\
679.149455787562	0\\
729.149455787562	0\\
779.149455787562	0\\
829.149455787562	0\\
879.149455787562	0\\
929.149455787562	0\\
979.149455787562	0\\
1029.14945578756	0\\
1079.14945578756	0\\
1129.14945578756	0\\
1179.14945578756	0\\
1229.14945578756	0\\
1279.14945578756	0\\
1329.14945578756	0\\
1379.14945578756	0\\
1429.14945578756	0\\
1479.14945578756	0\\
1529.14945578756	0\\
1579.14945578756	0\\
1629.14945578756	0\\
1679.14945578756	0\\
1729.14945578756	0\\
1779.14945578756	0\\
1829.14945578756	0\\
1879.14945578756	0\\
1929.14945578756	0\\
1979.14945578756	0\\
2029.14945578756	0\\
};
\addlegendentry{\footnotesize Att. 1/1 (Offset bit enabled)\\\footnotesize Avg. offset ($ke^-$): $-823 \pm 182$}

\addplot[ybar interval, fill=mycolor2, fill opacity=0.6, draw=black, area legend] table[row sep=crcr] {%
x	y\\
-1470.85054421244	0\\
-1420.85054421244	0\\
-1370.85054421244	0\\
-1320.85054421244	0\\
-1270.85054421244	0\\
-1220.85054421244	0\\
-1170.85054421244	0\\
-1120.85054421244	0\\
-1070.85054421244	0\\
-1020.85054421244	0\\
-970.850544212438	0\\
-920.850544212438	0\\
-870.850544212438	0\\
-820.850544212438	0\\
-770.850544212438	0\\
-720.850544212438	1\\
-670.850544212438	1\\
-620.850544212438	1\\
-570.850544212438	1\\
-520.850544212438	2\\
-470.850544212438	1\\
-420.850544212438	0\\
-370.850544212438	0\\
-320.850544212438	2\\
-270.850544212438	2\\
-220.850544212438	5\\
-170.850544212438	3\\
-120.850544212438	7\\
-70.8505442124383	10\\
-20.8505442124383	14\\
29.1494557875617	16\\
79.1494557875617	17\\
129.149455787562	17\\
179.149455787562	20\\
229.149455787562	27\\
279.149455787562	30\\
329.149455787562	44\\
379.149455787562	50\\
429.149455787562	40\\
479.149455787562	46\\
529.149455787562	50\\
579.149455787562	50\\
629.149455787562	45\\
679.149455787562	40\\
729.149455787562	45\\
779.149455787562	29\\
829.149455787562	33\\
879.149455787562	39\\
929.149455787562	17\\
979.149455787562	25\\
1029.14945578756	15\\
1079.14945578756	15\\
1129.14945578756	11\\
1179.14945578756	15\\
1229.14945578756	7\\
1279.14945578756	5\\
1329.14945578756	6\\
1379.14945578756	7\\
1429.14945578756	7\\
1479.14945578756	5\\
1529.14945578756	1\\
1579.14945578756	4\\
1629.14945578756	0\\
1679.14945578756	1\\
1729.14945578756	0\\
1779.14945578756	0\\
1829.14945578756	1\\
1879.14945578756	0\\
1929.14945578756	0\\
1979.14945578756	0\\
2029.14945578756	0\\
};
\addlegendentry{\footnotesize Att. 1/4\\\footnotesize Avg. offset ($ke^-$): $601 \pm 380$}

\addplot[ybar interval, fill=mycolor3, fill opacity=0.6, draw=black, area legend] table[row sep=crcr] {%
x	y\\
-1470.85054421244	0\\
-1420.85054421244	0\\
-1370.85054421244	0\\
-1320.85054421244	0\\
-1270.85054421244	0\\
-1220.85054421244	0\\
-1170.85054421244	0\\
-1120.85054421244	0\\
-1070.85054421244	0\\
-1020.85054421244	0\\
-970.850544212438	0\\
-920.850544212438	0\\
-870.850544212438	0\\
-820.850544212438	0\\
-770.850544212438	0\\
-720.850544212438	0\\
-670.850544212438	0\\
-620.850544212438	0\\
-570.850544212438	0\\
-520.850544212438	0\\
-470.850544212438	0\\
-420.850544212438	0\\
-370.850544212438	0\\
-320.850544212438	1\\
-270.850544212438	2\\
-220.850544212438	4\\
-170.850544212438	7\\
-120.850544212438	6\\
-70.8505442124383	19\\
-20.8505442124383	27\\
29.1494557875617	47\\
79.1494557875617	63\\
129.149455787562	67\\
179.149455787562	109\\
229.149455787562	91\\
279.149455787562	88\\
329.149455787562	76\\
379.149455787562	57\\
429.149455787562	52\\
479.149455787562	49\\
529.149455787562	26\\
579.149455787562	13\\
629.149455787562	9\\
679.149455787562	8\\
729.149455787562	5\\
779.149455787562	2\\
829.149455787562	1\\
879.149455787562	1\\
929.149455787562	0\\
979.149455787562	0\\
1029.14945578756	0\\
1079.14945578756	0\\
1129.14945578756	0\\
1179.14945578756	0\\
1229.14945578756	0\\
1279.14945578756	0\\
1329.14945578756	0\\
1379.14945578756	0\\
1429.14945578756	0\\
1479.14945578756	0\\
1529.14945578756	0\\
1579.14945578756	0\\
1629.14945578756	0\\
1679.14945578756	0\\
1729.14945578756	0\\
1779.14945578756	0\\
1829.14945578756	0\\
1879.14945578756	0\\
1929.14945578756	0\\
1979.14945578756	0\\
2029.14945578756	0\\
};
\addlegendentry{\footnotesize Att. 1/2\\\footnotesize Avg. offset ($ke^-$): $273 \pm 182$}

\addplot[ybar interval, fill=mycolor4, fill opacity=0.6, draw=black, area legend] table[row sep=crcr] {%
x	y\\
-1470.85054421244	0\\
-1420.85054421244	0\\
-1370.85054421244	0\\
-1320.85054421244	0\\
-1270.85054421244	0\\
-1220.85054421244	0\\
-1170.85054421244	0\\
-1120.85054421244	0\\
-1070.85054421244	0\\
-1020.85054421244	0\\
-970.850544212438	0\\
-920.850544212438	0\\
-870.850544212438	0\\
-820.850544212438	0\\
-770.850544212438	0\\
-720.850544212438	0\\
-670.850544212438	0\\
-620.850544212438	0\\
-570.850544212438	0\\
-520.850544212438	0\\
-470.850544212438	0\\
-420.850544212438	0\\
-370.850544212438	0\\
-320.850544212438	0\\
-270.850544212438	0\\
-220.850544212438	0\\
-170.850544212438	2\\
-120.850544212438	1\\
-70.8505442124383	11\\
-20.8505442124383	42\\
29.1494557875617	137\\
79.1494557875617	261\\
129.149455787562	227\\
179.149455787562	104\\
229.149455787562	41\\
279.149455787562	2\\
329.149455787562	1\\
379.149455787562	1\\
429.149455787562	1\\
479.149455787562	0\\
529.149455787562	0\\
579.149455787562	0\\
629.149455787562	0\\
679.149455787562	0\\
729.149455787562	0\\
779.149455787562	0\\
829.149455787562	0\\
879.149455787562	0\\
929.149455787562	0\\
979.149455787562	0\\
1029.14945578756	0\\
1079.14945578756	0\\
1129.14945578756	0\\
1179.14945578756	0\\
1229.14945578756	0\\
1279.14945578756	0\\
1329.14945578756	0\\
1379.14945578756	0\\
1429.14945578756	0\\
1479.14945578756	0\\
1529.14945578756	0\\
1579.14945578756	0\\
1629.14945578756	0\\
1679.14945578756	0\\
1729.14945578756	0\\
1779.14945578756	0\\
1829.14945578756	0\\
1879.14945578756	0\\
1929.14945578756	0\\
1979.14945578756	0\\
2029.14945578756	0\\
};
\addlegendentry{\footnotesize Att. 1/1\\\footnotesize Avg. offset ($ke^-$): $123 \pm 66$}

\end{axis}
\end{tikzpicture}%
	\caption{Offset distribution at different attenuation and offset bit settings for 1024 channels.}
	\label{fig:claro_thr_offset_distr}
\end{figure}

Figure~\ref{fig:claro_thr_offset_distr} shows the distribution of the offset at threshold zero.
The average offset is not zero, but is shifted in the positive half-plane.
Average values are $120\ ke^-$ at attenuation 1, $270\ ke^-$ at attenuation 1/2 and $600\ ke^-$ at attenuation 1/4.
For the channels that need a threshold lower than the offset, it is possible to compensate this effect by enabling an offset correction bit, which shifts the thresholds by 32 units.
With this correction enabled, all the channels are shifted to a negative offset with an average value of $-820\ ke^-$.
This measurement demonstrated the effectiveness of the offset correction. 

\subsection{Threshold scan studies}
\label{subsec:threshold}

\begin{figure}
\centering
\includegraphics[width=\linewidth]{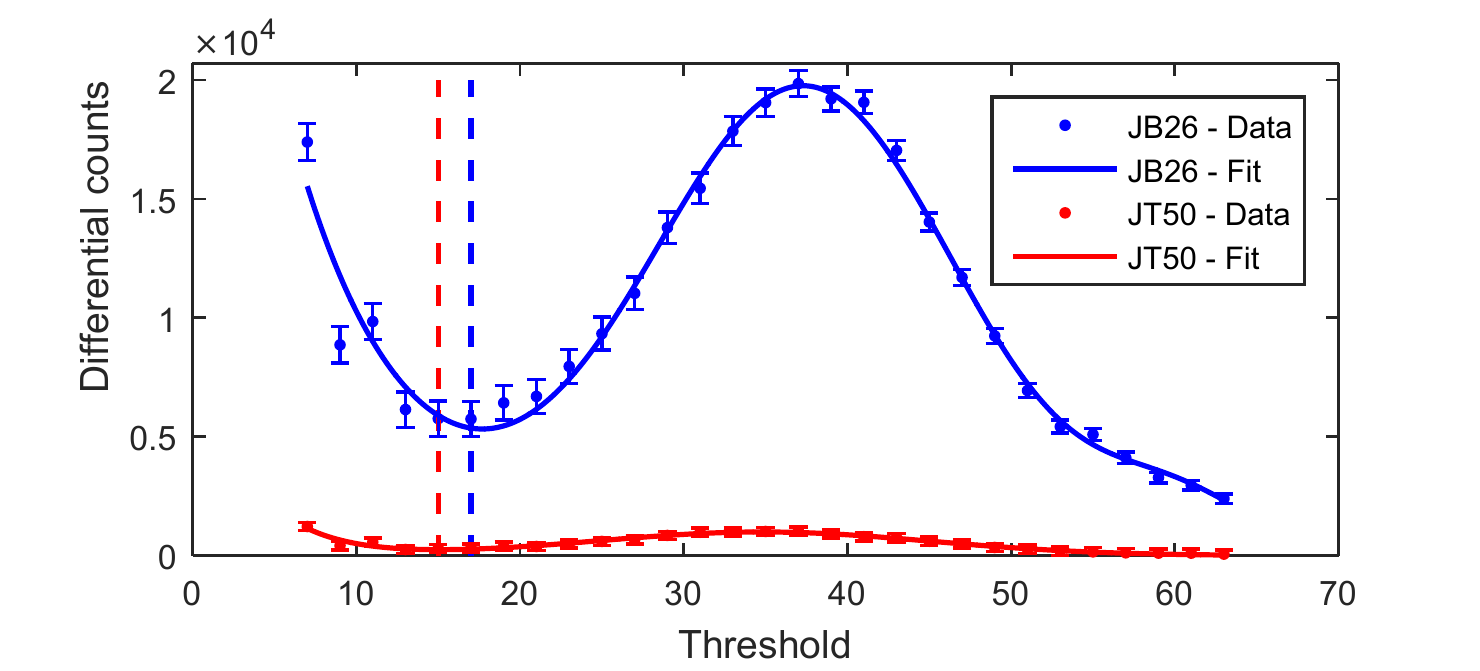}
\caption{Single photon spectrum for a pixel on the ring in \textit{blue} (pixel JB26) and for a pixel off the ring in \textit{red} (pixel JT50). The dashed lines highlight the optimal threshold value for each pixel (17 for pixel JB26 and 15 for pixel JT50).}
\label{thrscn:spectrum}
\end{figure}

Threshold scans allow to reconstruct the single photon spectrum of each pixel.
With constant MaPMT illumination, an increasing threshold is set into the CLARO channel and the same amount of triggers is acquired for each threshold (usually 1 million).
By differentiating the number of acquired events for each threshold it is possible to reconstruct single photon spectra like those shown in Figure~\ref{thrscn:spectrum}.

Several threshold scan runs were performed through the test period, at various MaPMT bias voltages.
In the measurement shown here we used a MaPMT voltage of 1000~V and threshold settings spanned from threshold 7 to threshold 63 in steps of 2 units.
Each nominal unit is about $30\ ke^-$, as shown in section \ref{sec:dac_scan}.
The differential spectra for each pixel were fitted with three gaussians (noise pedestal, single photon peak and double photon peak).
Threshold scans at testbeams were performed using Cherenkov photons and this implies that off-ring pixels were illuminated only by stray photons, hence their event rate is much lower than that for pixels on the Cherenkov ring.
Nevertheless, the spectra of off-ring pixels could be reconstructed adequately well despite the lower statistics.

In Figure~\ref{thrscn:spectrum}, two typical spectra for a pixel on the ring (blue curve) and off the ring (red curve) are shown.

\begin{figure}
\centering
\includegraphics[width=.8\linewidth]{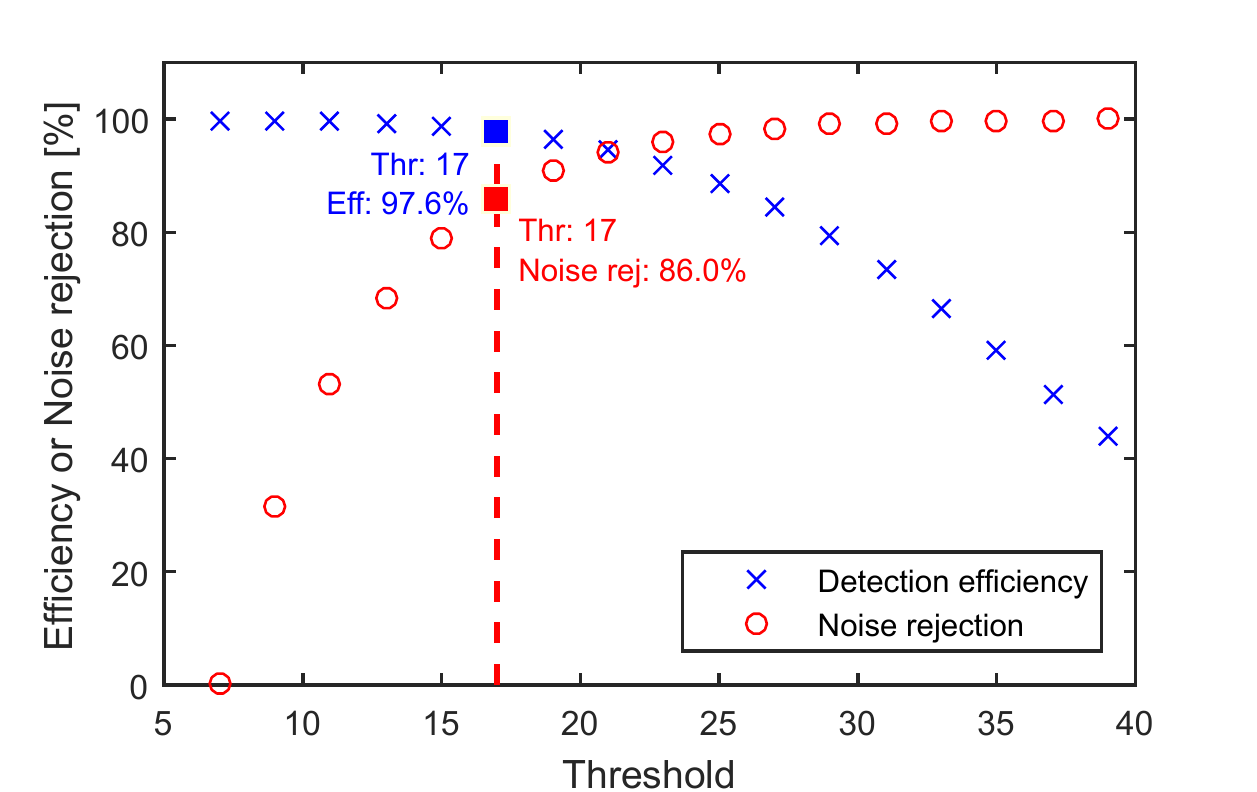}
\caption{Single photon peak detection efficiency and noise rejection versus threshold setting for pixel JB26. The red line and the datatips highlight the optimal threshold value.}
\label{thrscn:efficiency}
\end{figure}

The spectra were used to compute the optimum threshold value for each pixel and in each working condition.
The optimum threshold is defined as the threshold in the valley between the MaPMT noise pedestal and the single photon peak.
By selecting the optimal threshold it is possible to reject most of the noise while maintaining an adequate photon detection efficiency.
Figure~\ref{thrscn:efficiency} shows a plot of the photon detection efficiency and the pedestal noise rejection with respect to the threshold settings for one of the pixels of Figure~\ref{thrscn:spectrum}.
The efficiency is calculated by the ratio of the area above threshold and the area of full single photon peak, while noise rejection is calculated as the ratio of the noise pedestal below threshold and the area of the entire pedestal from the lowest threshold.

The optimized thresholds were used throughout the testbeam, while acquiring physics data for the Cherenkov angle reconstruction.

\subsection{Crosstalk}

Another interesting measurement that can be extracted from the data acquired in testbeams is the crosstalk of the entire optoelectronic chain.

Crosstalk can be evaluated by looking at pairs or clusters of neighboring active pixels during the same trigger, in a region of the MaPMT far from the Cherenkov ring.
On the ring, in fact, the rate of Cherenkov photons is quite high (about 4 photons per event per MaPMT are expected) so there would be a non negligible probability of two real photons hitting neighboring pixels, and thus triggering a false crosstalk count. For this reason all the pixels that are located on the Cherenkov ring and their first nearest pixels are masked out in this analysis.

MaPMT dark count rate is a few tens of Hz per pixel and it is much lower than the trigger rate (about $500\ kHz$), hence its contribution to the counts can be neglected.
Furthermore, since the illumination rate of pixels off-ring is less than 1\% of the total events, the probability of accidental coincidence between two neighboring pixels can also be neglected.
Therefore, coincidences on off-ring pixels can only be attributed to crosstalk.

Using the binary data coming from the readout system it is not possible to distinguish which of the two neighboring pixels induced crosstalk on the other one, as there is no information about signal amplitudes, thus the number of crosstalk events $\left(N_{AB}\right)$ was evenly split between the two pixels.

After these considerations, the crosstalk probability $p_{CT}$ is calculated as:
\begin{equation}
p_{CT} \left(A\rightarrow B\right) = \frac{N_{AB}/2}{N_A} \ .
\end{equation}
$N_{AB}$ is the number of events where both pixels are switched on and $N_{A}$ is the number of events where at least pixel A is on.
The same calculation is done exchanging pixel B with pixel A and the final crosstalk value between the pair is the average of the two values.


\begin{figure}
\centering
\includegraphics[width=.9\linewidth]{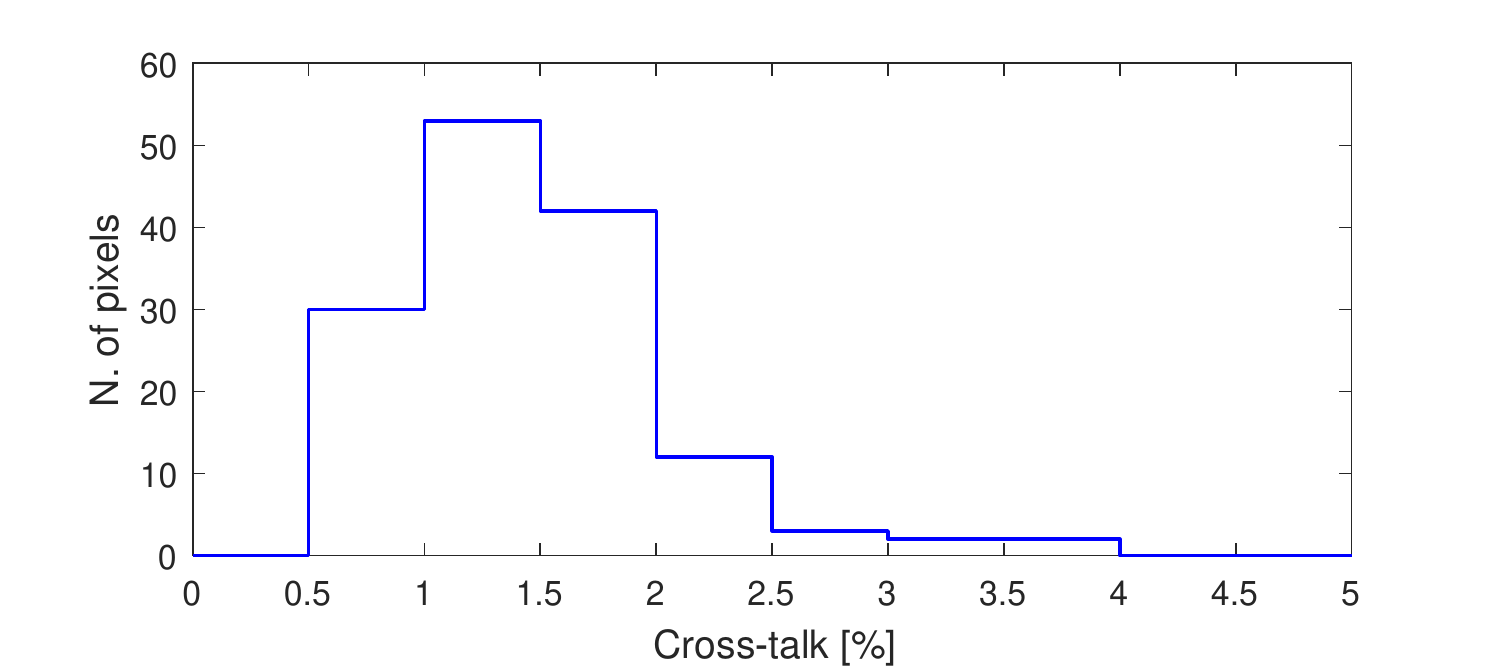}
\caption{Histogram of the crosstalk probability with optimized thresholds. The mean value is 1.48\%.}
\label{ct:histogram}
\end{figure}

The histogram of the computed values for the run with optimized thresholds is plotted in Figure~\ref{ct:histogram}.
This analysis was performed on two MaPMTs (128 channels) and the average value for the crosstalk is 1.48\%.
These results are obtained with the complete system and therefore include the contributions from the whole optoelectronic chain: charge sharing, inter-pixel capacitance, PCB parasitic capacitance, CLARO parasitic capacitance, etc.
This remarkably low crosstalk was possible thanks to the careful design of the entire system, starting from the excellent quality of the MaPMT selected, to front-end board's layout and CLARO preamplifier design.
This result is in good agreement with previous test bench measurements, presented in the CLARO papers.

\subsection{Summary}

The testbeam campaigns represented crucial milestones in the development of the upgraded LHCb RICH detectors.
First testbeams allowed to test the main concept of the upgraded system and verify the integration of the early prototypes, providing inputs for modifications and improvements.
Latest testbeam campaigns already featured pre-production components and they were a perfect showcase for production readiness reviews and to demonstrate maturity, reliability and performance of the optoelectronic chain for the upgraded LHCb RICH.

\section{The ALDO ASIC}

Most of existing particle and radiation sensors give analog signals at their output electrodes, and therefore require analog front-end electronics.
In some cases (e.g. calorimeters, etc.) the analog information is preserved all the way into the digital domain, where digitized analog waveforms are recorded.
In this case the signal to noise ratio clearly plays a central role, and the noise performance of the front-end amplifiers should not be affected by noise injected through the supply rails.
In other cases (e.g. RICH detectors, trackers, etc.) not all the analog information is of interest, and some is disregarded at an earlier stage, for instance by discriminating the analog signals against a threshold, like in the CLARO.
The resulting binary signals are much more robust against noise, and this could, at a superficial glance, hide the importance of keeping noise low before discrimination takes place: on the contrary, noise in the discrimination threshold, especially at low frequency, can impact counting precision and ultimately compromise the measurement of the quantities of interest (i.e. Cherenkov angle, track hits, etc.).

The power supply noise level that can be tolerated by a given system depends on many factors, among which the most important is the power supply rejection ratio of the front-end circuitry.
Power supply noise can also couple into sensitive analog signals through parasitic capacitances at the PCB level.
The above examples make the case that having clean and stable power supplies can be adopted as a general safe guideline for electronic design.
And this is true also if the signals are digitized at an early stage, as in the case of systems based on a binary readout scheme.

Filtering high frequency noise from power supplies is feasible with a few passive components.
However, in particle physics experiments space constraints are usually tight, thus active circuits (voltage regulators) can usually provide more compact solutions.
This becomes increasingly evident at lower frequencies, where the size of large value inductors and capacitors defines the lower frequency limit for effective filtering with passive components.
Stray magnetic fields are also a limiting factor for the adoption of large inductors.

Stability against time and temperature variations also plays an important role.
Using a voltage regulator to stabilize the power supplies close to the front-end circuitry helps in eliminating voltage drifts that may affect performance.
Stable supply voltages also reduce or eliminate the need for reference voltage generators as stable reference voltages can be derived by simply dividing the power supply rail.
Moreover, by using local voltage regulators to decouple the power supplies of small groups of front-end channels from their neighbors, undesired interactions are minimized.
This also allow to selectively switch-off or restart smaller clusters of devices in case of a malfunctioning of one part, decreasing its impact on the detector operation.

\begin{figure}
\centering
\def\svgwidth{\linewidth}
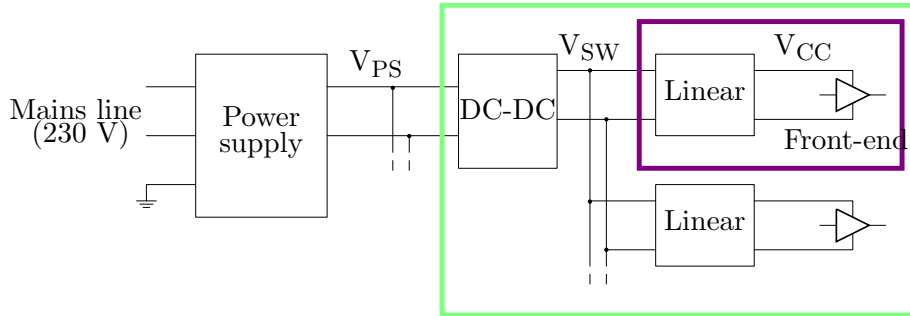
\caption{Power supply scheme that employs both switching and linear regulators to provide clean and stable power to the front-end circuitry with high efficiency.}
\label{fig:PowerScheme}
\end{figure}

For what concerns digital circuitry, switching (DC-DC) power converters are commonplace, as they offer high efficiency (80\% and higher) even with a large voltage drop between input and output.
Their principle of operation is based on clocked switching, with frequency of operation usually in the range $100\ kHz - 1\ MHz$.
In the case of analog circuitry, the situation is more delicate.
Eliminating the switching noise from the output of a switching regulator is not trivial, and noise levels below $1\ mV$ peak to peak are hard to obtain.
Therefore, when it comes to sensitive analog circuitry, linear voltage regulators still play a role.
To combine the best of both worlds, the general scheme depicted in Figure \ref{fig:PowerScheme} is often employed.
Switching regulators can be used to convert with high efficiency the input voltage $V_{PS}$ ($5\ V$, $12\ V$, $24\ V$, \dots) to a lower voltage $V_{SW}$ (e.g. $1.5\ V$, $2.8\ V$), with just enough headroom for a low dropout linear regulator to regulate the rail to the clean and stable voltage $V_{CC}$ that the front-end circuitry requires (e.g. $1.2\ V$, $2.5\ V$).
In the case of high energy physics experiments, the green area of Figure \ref{fig:PowerScheme} is located on the detector, and therefore must typically operate in presence of radiation and of strong magnetic fields.

With these motivations, I developed the first prototype of ALDO, in the framework of the LHCb RICH upgrade.
ALDO is a radiation-hard low dropout linear regulator in standard voltage 0.35 $\mu m$ CMOS technology from ams, the same technology used by the CLARO ASIC.
It can be used to generate a stable, low noise and adjustable supply voltage for powering the CLARO chips.
ALDO will be able to greatly enhance noise filtering, thermal stability and load regulation of the power supply generated by the FEASTMP DC/DC.
Its integration inside the Elementary Cell is easy, through an optional add-in board. 

\subsection{Theory of operation}
\label{sec:ldo_theory}

\begin{figure}
\centering
\begin{tikzpicture}
\draw (0,2) to[V, l_=$V_i$] (0,0) -- (5,0) to[R, l_=$R_L$] (5,2) -- (3,2) to[vR, l_=$R_S$] (1,2) -- (0,2)
node[ocirc, scale = 1.5] (A) at (3.5,0) {}
node[ocirc, scale = 1.5] (B) at (3.5,2) {}
(A) to[open, v_>=$V_o$ ] (B)
;
\node[draw,minimum width=2cm,minimum height=3.4cm,anchor=south west,dashed] at (1,-0.2){};
\node[minimum width=2cm,minimum height=1.8cm,anchor=south west,text width=1.7cm,align=center] at (1,0){Voltage regulator};
\end{tikzpicture}
\caption{Basic schematic of a linear voltage regulator.}
\label{fig:voltage_ref}
\end{figure}
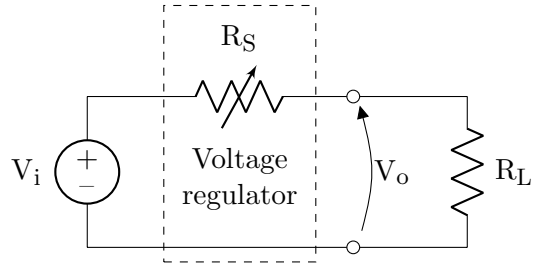

A voltage regulator is a device which is able to generate a specified voltage, independently of the applied load.
The basic schematic of such a device is shown in Figure~\ref{fig:voltage_ref}.

From Figure~\ref{fig:voltage_ref} we can write
\begin{equation}
V_o = \frac{R_L}{R_L+R_S}V_i\ ,
\end{equation}
with $V_o$ the output voltage of the regulator, $V_i$ the input voltage, $R_L$ the load resistance, and $R_S$ the source resistance of the regulator.
If the device has to keep $V_o$ stable with respect to the changing load $R_L$, then it has to regulate $R_S$ so that
\begin{equation}
R_S = \frac{V_i-V_o}{V_o} R_L = \alpha R_L\ ,
\end{equation}
hence the name linear regulator.

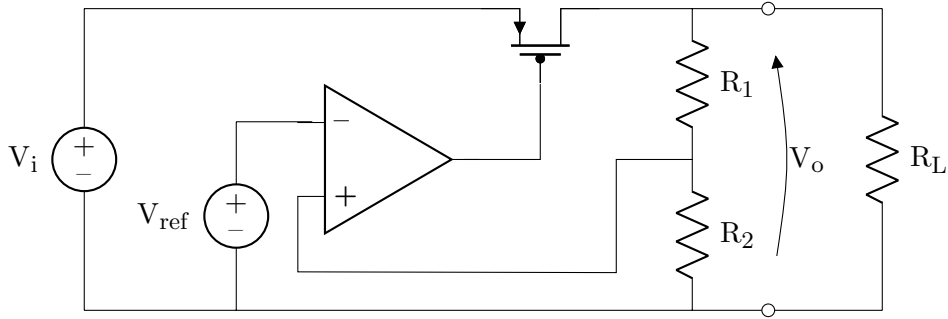
\begin{figure}
\centering
\begin{tikzpicture} \draw
(0,4) to[V, l_=$V_i$] (0,0) -- (8,0) to[R, l_=$R_2$] (8,2) to[R, l_=$R_1$] (8,4)
(8,4) -- (7,4) to[Tpmos] (5,4) -- (0,4)
(4,2) node[op amp,yscale=1] (opamp) {}
(6,3) -- (6,2) -- (5,2)
(3,2.5) -- (2,2.5) to [V, l_=$V_{ref}$] (2,0)
(opamp.+) -- ($(opamp.+)+(0,-1)$) -- (7,0.5) -- (7,2) -- (8,2)
(8,0) -- (10.5,0) to[R, l_=$R_L$] (10.5,4) -- (8,4)
node[ocirc, scale = 1.5] (A) at (9,0) {}
node[ocirc, scale = 1.5] (B) at (9,4) {}
(A) to[open, v_>=$V_o$ ] (B)
;
\end{tikzpicture}
\caption{Circuital implementation of a linear voltage regulator.}
\label{fig:ldo}
\end{figure}

One of the simplest circuit to implement such a device, is depicted in Figure~\ref{fig:ldo}.

The output current is controlled by the PMOS transistor, often referred to as pass transistor, connected between the input and the output. The gate of the PMOS is driven by the operational amplifier, which act as an error amplifier since it amplifies the error between the feedback node and a reference voltage. When more current is drawn by the load, the voltage at the feedback node tends to decrease. As a response, the error amplifier pulls down the gate of the MOS, allowing more current to flow into the load in order to keep the output voltage stable.

The output voltage $V_o$ can be adjusted arbitrarily by choosing the most appropriate feedback resistors, as from the following equation:
\begin{equation}
V_o = \left(1+\frac{R_1}{R_2}\right)V_{ref}\ .
\label{eq:vo_ldo}
\end{equation}

The reference voltage that is fed into the inverting input of the error amplifier is often included in the regulator and implemented as a Zener diode or as a bandgap reference voltage, with the latter being the most common between the two, due to the superior performance it offers.

\begin{figure}
\centering
\def\svgwidth{\linewidth}
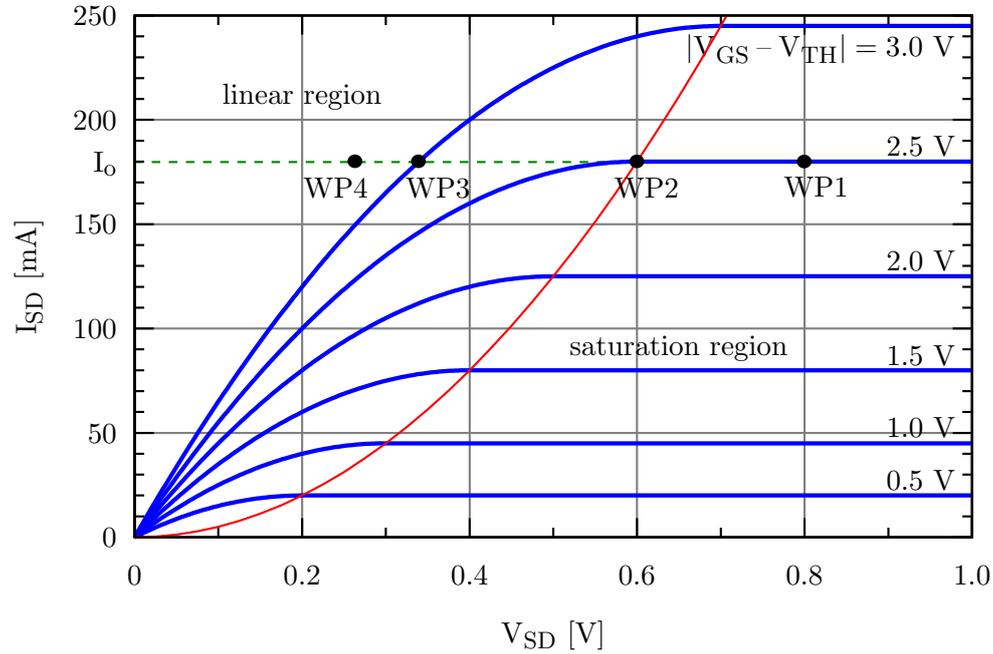
\caption{PMOS characteristic curves.}
\label{fig:pmos_curve}
\end{figure}

An important figure of merit of linear regulators is the dropout voltage. The dropout voltage is the minimum voltage required across the pass transistor ($V_{SD}$) in order to maintain the output voltage regulation. Dropout voltage can be explained by looking at Figure~\ref{fig:pmos_curve}. In working point WP1, a dropout of 0.8 V and an output current $I_o$ are set. The pass transistor is in saturation region and the error amplifier sets the overdrive voltage $V_{ov}=\left|V_{GS}-V_{TH}\right|$ at 2.5 V. When the input voltage is reduced by 0.2 V, the dropout goes to 0.6 V (WP2) and the transistor is approaching linear region, but it is still able to source $I_o$ with the same $V_{ov}$. In WP3 input voltage is further reduced and the overdrive of 2.5~V is not anymore sufficient for the desired output current. The error amplifier thus drives the pass transistor to $V_{ov}=3\ V$ and the dropout goes to 330 mV. If input voltage is reduced furthermore, like in WP4, and the error amplifier is not able to drive $V_{ov}$ to a higher voltage (for example because the gate is saturated to ground and $V_{ov}=V_i-\left|V_{TH}\right|$), then the output voltage cannot be regulated anymore and drops to a lower voltage in order to maintain the minimum dropout like in WP3. The slope of the I-V curve in WP3 is the reciprocal of the minimum series resistance of the regulator ($R_{DS\left(on\right)})$), hence a lower series resistance implies a lower dropout.

Low dropout regulators (LDO) are a specific class of linear regulator with a dropout voltage lower than 1 V, usually achieved with the adoption of specialized large-area power MOSFETs as pass element, in order to have a low $R_{DS\left(on\right)}$.

Another important characteristic of linear regulators is their capability to filter changes of the input voltage (noise, spikes, etc.).
This property can be quantified using the power supply rejection ratio (PSRR).
When the error amplifier is fast enough and its output is far from saturation voltage, then the amplifier is able to regulate the gate of the pass element in order to reject any fluctuation in the input voltage.
If the regulator is in dropout region, then the error amplifier is close to saturation and thus unable to reject power supply changes.

Another crucial property of LDOs is the high thermal stability and low noise.
Stability and noise of the error amplifier can be easily minimized by an accurate design of the input stage, and usually their contribution can be made negligible with respect to other parts of the LDO, like the voltage reference.
From equation~\ref{eq:vo_ldo}, it is possible to see that the output voltage is directly derived from the input reference, hence its thermal stability is typically the same of the reference, while the noise of the reference gets amplified by the gain of the feedback.
Voltage references are widely implemented with bandgap circuits, which will be described later.

\begin{figure}
\centering
\begin{tikzpicture}[scale=\textwidth/16.3cm] \draw
(-1.5,4) -- (-1.5,-1.5) -- (8.5,-1.5) to[R, l_=$R_2$] (8.5,1.25) to[R, l_=$R_1$] (8.5,4)
(2,1.5) node[buffer,yscale=1,xscale=1.4,scale=2.3] (opamp) {}
(opamp.out) -- ($(opamp.out)+(0,0)$) to[C, l^=$C_G$] ($(opamp.out)+(0,2.5)$)
(opamp.out) -- ($(opamp.out)+(-2,0)$) to[/tikz/circuitikz/bipoles/length=1cm,R, l_=$R_{OA}$] ($(opamp.out)+(-3,0)$) -- ($(opamp.out)+(-4.2,0)$) to[/tikz/circuitikz/bipoles/length=1cm,V, l^=$A_0 V_F$] ($(opamp.out)+(-4.2,-1.2)$) -- ($(opamp.out)+(-4.2,-3)$)
(opamp.in) -- ($(opamp.in)+(0,-2.5)$) -- (7,-1) -- (7,1.25) -- (8.5,1.25)
(-1.5,4) -- (4.5,4) to[american current source, l^=$-g_{m}V_{GS}$] (8.5,4)
(5.5,4) -- (5.5,3) to[R, l_=$R_{DS}$] (7.5,3) -- (7.5,4)
(8.5,-1.5) -- (10,-1.5) to[R, l^=$R_L$] (10,4) -- (8.5,4)
(10,-1.5) -- (11.5,-1.5) to[C, l_=$C_L$] (11.5,4) -- (10,4)
(11.5,-1.5) -- (13,-1.5) to[R, l_=$R_{ESR}$] (13,1.25) to[C, l_=$C_{O}$] (13,4) -- (11.5,4)
(9,4) node [above] {$V_{O}$}
(7.5,1.25) node [above] {$V_{F}$}
(opamp.out) node [right] {$V_{G}$}
;
\end{tikzpicture}
\caption{Small-signal model of circuit in Figure~\ref{fig:ldo}.}
\label{fig:ldo_smallsignal}
\end{figure}
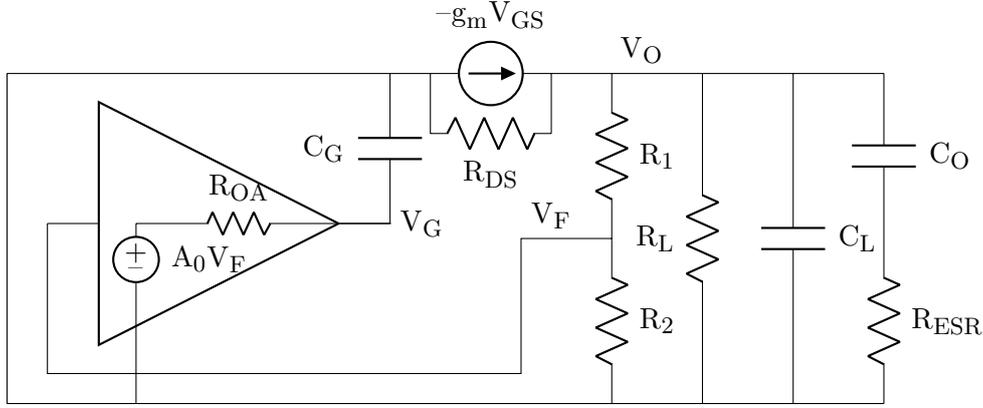

The operation of the linear regulator is based on a negative feedback and, like any other circuit with feedback, an accurate study of frequency response is needed. The small-signal equivalent circuit is shown in Figure~\ref{fig:ldo_smallsignal}. The PMOS transistor is substituted by its small-signal model with gate capacitance $C_G$ and drain-source resistance $R_{DS}$. The output is loaded with a capacitive load $C_L$ (e.g. ceramic bypass capacitors) and a stabilizing capacitor $C_{O}$ with an equivalent series resistance $R_{ESR}$, whose importance will be shown later.

We can proceed with the usual feedback analysis by interrupting the feedback at the voltage generator $A_0 V_F$. If we apply a voltage $V_X$ in that node we have that
\begin{equation}
V_G=\frac{1}{1+s R_{OA} C_G}V_X\ .
\end{equation}
At the node $V_O$ we can write the following equation for the currents:
\begin{equation}
-g_m V_{GS} = \left(R_{DS} \parallel \left(R_1 + R_2\right) \parallel R_L \parallel \frac{1}{sC_L} \parallel \left(R_{ESR}+\frac{1}{sC_{O}}\right)\right)^{-1} V_O\ .
\label{eq:ldo_eq2}
\end{equation}
We can define $R_O=R_{DS} \parallel \left(R_1 + R_2\right) \parallel R_L$ and resolve equation \ref{eq:ldo_eq2}:
\begin{equation}
V_{O}= -g_m V_{GS} R_O \frac{1+sR_{ESR} C_{O}}{1+s\left(R_{ESR} C_{O} +R_O C_{O} + R_O C_L\right)+s^2 R_O R_{ESR} C_L C_{O}}\ .
\label{eq:ldo_eq3}
\end{equation}
We can then assume that $C_{O} \gg C_L$ and approximate equation \ref{eq:ldo_eq3} to:
\begin{equation}
V_{O} = -g_m V_{GS} R_O \frac{1+sR_{ESR} C_{O} }{\left(1+s\left(R_{ESR}+R_O \right)C_{O}\right) \left(1 + s \left(R_{ESR} \parallel R_O\right) C_{L}\right)}\ .
\end{equation}
The loop closes itself at
\begin{equation}
V_{X,ret}=A_0 V_F = A_0 V_O \frac{R_2}{R_1+R_2}\ .
\end{equation}
The loop gain becomes therefore:
\begin{equation}
T=\frac{V_{X,ret}}{V_X} = -g_m R_O A_0 \frac{R_2}{R_1+R_2} \frac{1+\frac{s}{z_{ESR}}}{\left(1+\frac{s}{p_O}\right) \left(1+\frac{s}{p_{OA}}\right) \left(1+\frac{s}{p_{ESR}}\right)}\ ,
\label{eq:ldo_stability}
\end{equation}
with \[
z_{ESR} = \frac{1}{2\pi R_{ESR}C_{O}}\ ,
\]
\[
p_{O} = \frac{1}{2\pi \left(R_{ESR}+R_{O}\right)C_{O}}\ ,
\]
\[
p_{OA} = \frac{1}{2\pi R_{OA}C_{G}}\
\]
and
\[
p_{ESR} = \frac{1}{2\pi \left(R_{ESR} \parallel R_{O}\right)C_{L}}\ .
\]
Without the stabilizing capacitor $C_O$ ($R_{ESR}=0$), the zero is not present and the system behaves like a 2-pole system, thus susceptible to instability.
With the stabilizing capacitor, 3 poles and 1 zero are present. 
The poles relative positions usually satisfy $p_O < p_{OA} \ll p_{ESR}$. 
This system can be easily stabilized if an appropriate stabilizing capacitor is selected, so that the zero is able to cancel the mid-frequency pole $p_{OA}$. 
However, when $R_{ESR}$ is increased, also $p_{ESR}$ goes to lower frequencies, hence typically there is a defined range of $R_{ESR}$ which stabilizes the linear regulator.

\subsubsection{Bandgap circuit}

In the previous section it was highlighted the importance of the voltage reference for thermal stability and noise of linear regulators.
For this purpose most of the recent LDO designs adopt bandgap references, that exhibit the highest performance.

The working principle of bandgap circuits is based on the sum of two references voltages, one with a negative temperature coefficient and one with a positive temperature coefficient, in order to generate a reference with zero temperature coefficient.
Usually, these coefficients have different absolute values, so they have to be summed with an adequate weight, but this can be accomplished easily as it will be shown later.

The two references with opposite temperature coefficient can be obtained with bipolar junction transistors (BJT), which are available in almost every integrated technology. The negative reference voltage can be obtained from the base-emitter voltage $V_{be}$, while the positive one can be obtained from the thermal voltage $V_t=\frac{k_B T}{q}$.

The emitter current is defined as 
\begin{equation}
I_C = I_{S}\:e^{\frac{V_{be}}{V_t}}
\end{equation}
with saturation current $I_{S}$
\begin{equation}
I_{S} = I_0\:e^{\frac{-E_{g0}}{\alpha\:q\:V_t}}\
\end{equation}
where $I_0$ and  $\alpha$ are process-depending constants, $E_{g0}$ is the band gap energy at $0\ ^{\circ}C$, and $q$ is the electron charge. 
Then we can write $V_{be}$ as
\begin{equation}
V_{be} = V_g - V_t \ln \left(\frac{I_0}{I_C}\right)
\label{eq:vbe}
\end{equation}
with $V_g = \frac{E_{g0}}{\alpha\:q}$.

Its thermal dependence is
\begin{equation}
\frac{\partial V_{be}}{\partial T} = -\frac{k_B}{q} \ln \left(\frac{I_0}{I_C}\right) \approx -2\ mV/^{\circ}C \ .
\label{eq:vbebrift}
\end{equation}
Since $I_0 \gg I_C$, the logarithm does not change much with a change of $I_C$, hence any temperature dependance of $I_C$ can be neglected.

Unlike the $V_{be}$, the thermal voltage is not directly usable as a voltage reference. However, if we take the difference of the $V_{be}$ voltages of two transistors biased with the same $I_C$, we get:
\begin{equation}
\Delta V_{be}=V_{be1}-V_{be2} = V_t \ln \left(\frac{I_{C}}{I_{S1}}\right) - V_t \ln \left(\frac{I_{C}}{I_{S2}}\right) = V_t \ln \left(\frac{I_{S2}}{I_{S1}}\right)
\label{eq:vt}
\end{equation}
and
\begin{equation}
\frac{\partial \Delta V_{be}}{\partial T} = \frac{k_B}{q} \ln \left(\frac{I_{S2}}{I_{S1}}\right) \ ,
\end{equation}
which can be made positive if $I_{S2}$ is greater than $I_{S1}$ by choosing a larger area size for $Q_2$.

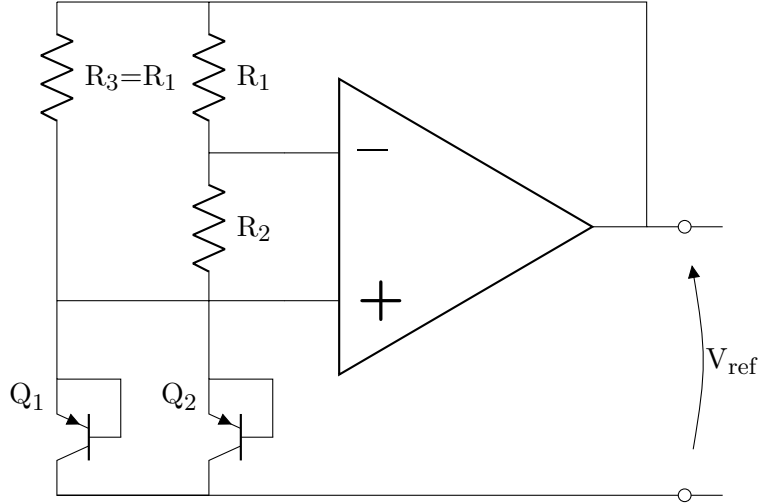
\begin{figure}
\centering
\begin{tikzpicture} \draw
(0,0) node[pnp, xscale=-1, anchor=C] (pnp1) {} 
		(pnp1.text) node[left,inner sep=0pt] {$Q_1$}
		(pnp1.B) -- (pnp1.B |- pnp1.E) -- (pnp1.E)
		(pnp1.E) -- ++(0,3) to[R, l_=$R_3{=}R_1$] ++(0,2)
		(pnp1.C) -- ++(2,0)
		($(pnp1.C)+(2,0)$) node[pnp, xscale=-1, anchor=C] (pnp2) {} 
		(pnp2.text) node[left,inner sep=0pt] {$Q_2$}
		(pnp2.B) -- (pnp2.B |- pnp2.E) -- (pnp2.E)
		(pnp2.E) -- ($(pnp2.E)+(0,1)$) to[R, l_=$R_2$] ($(pnp2.E)+(0,3)$) -- ($(pnp2.E)+(1,3)$)
		($(pnp2.E)+(1,3)$) node[op amp, anchor=-, scale = 2] (opamp) {}
		($(pnp2.E)+(0,3)$) to[R, l_=$R_1$] ++(0,2)
		(opamp.+) -- (pnp1.E |- opamp.+)
		($(pnp1.E)+(0,5)$) -- ++(opamp.out |- pnp1.C) -- (opamp.out)
		(opamp.out) -- ++(1,0)
		(pnp1.C) -- ($(opamp.out)-(pnp1.C |- opamp.out)+(1,0)$ |- pnp1.C)
		node[ocirc, scale = 1.5] (A) at ($(opamp.out)+(0.5,0)$) {}
		node[ocirc, scale = 1.5] (B) at ($(opamp.out)-(pnp1.C |- opamp.out)+(0.5,0)$ |- pnp1.C) {}
		(B) to[open, v_>=$V_{ref}$ ] (A)
;
\end{tikzpicture}
\caption{Circuital implementation of the bandgap voltage reference.}
\label{fig:bandgap}
\end{figure}

There are several circuit topologies for detecting and summing these references.
In our design we choose the topology shown in Figure~\ref{fig:bandgap}.

The opamp forces its inputs at the same potential, thus the currents through $R_1$ and $R_3$ are the same since they are of equal value. The inverting input is forced to $V_{be1}$, so we can write:
\begin{equation}
V_{be1} = V_{be2} + I\:R_2 \ .
\end{equation}
Then $V_{ref}$ voltage can be defined as:
\begin{equation}
V_{ref} = V_{be2} + I\:R_2 + I\:R_1 \ .
\end{equation}
We can then resolve the system and find that:
\begin{equation}
V_{ref} = V_{be1} + \frac{R_1}{R_2}\left(V_{be1}-V_{be2}\right) \ .
\end{equation}
Using equations \ref{eq:vbe} and \ref{eq:vt} we have that:
\begin{equation}
V_{ref} = V_{be1} + V_t \frac{R_1}{R_2} \ln \left(\frac{I_{S2}}{I_{S1}}\right) \ .
\label{eq:bandgap_bjt3}
\end{equation}

If the two transistors are well matched and placed close to each other, then ratio between $I_{S2}$ and $I_{S1}$ can be approximated by the ratio between the two areas $A_2$ and $A_1$.

The drift of the reference voltage is
\begin{equation}
\frac{\partial V_{ref}}{\partial T} = \frac{\partial V_{be}}{\partial T}+m\frac{\partial V_{t}}{\partial T}
\label{eq:bandgap_bjt1}
\end{equation}
with
\begin{equation}
m = \frac{R_1}{R_2} \ln \left(\frac{A_{2}}{A_{1}}\right) \ .
\label{eq:bandgap_bjt2}
\end{equation}

In order to have zero drift, a specific m has to be set:
\begin{equation}
m = \frac{-\frac{\partial V_{be}}{\partial T}}{\frac{\partial V_{t}}{\partial T}} = \frac{2\ mV/^{\circ}C}{0.086\ mV/^{\circ}C} \approx 23 \ .
\end{equation}

The voltage $V_{ref}$ becomes
\begin{equation}
V_{ref} = V_{be}+m\:V_t= 0.6\ V + 23 \cdot 0.026\ V \approx 1.2\ V \ .
\end{equation}
The obtained voltage reference has a value close to the band gap energy of Silicon, hence the name of the bandgap voltage reference.

BJT transistors made with parasitic vertical or planar structures are available in almost all CMOS technologies.
Bipolar devices, however, are known for being more sensitive to displacement damage in radiation environments~\cite{danno_radiazione}, and their use in such scenarios should be avoided.

MOS transistors operated in sub-threshold region have a similar characteristic behavior to standard diode-connected BJT transistors.

The drain current for a sub-threshold MOS can be written as:
\begin{equation}
I_{D,subthr}=\frac{W}{L} \mu C_{ox} \left(\eta-1\right) V_t^2 \: e^{\frac{V_{GS}-V_{th}}{\eta \: V_t}} \left(1-e^{-\frac{V_{DS}}{V_t}}\right)
\label{eq:idsubthr}
\end{equation}
with $\frac{W}{L}$ the transistor aspect ratio, $\mu$ the charge mobility, $C_{ox}$ the gate-oxide capacitance, $\eta$ the sub-threshold slope factor (process dependent), $V_t=\frac{K_B T}{q}$, $V_{GS}$ the gate-source voltage, $V_{th}$ the threshold voltage, and $V_{DS}$ the drain-source voltage.

If $V_{DS} \gg V_t$ we can approximate the content of the last parentheses to unity and calculate $V_{GS}$ voltage as
\begin{equation}
V_{GS}= V_{th}-\eta V_t \ln \left( \frac{\frac{W}{L} \mu C_{ox} \left(\eta-1\right) V_t^2}{I_D}\right)
\end{equation}
and then the thermal drift
\begin{equation}
\frac{\partial V_{GS}}{\partial T}= \frac{\partial V_{th}}{\partial T}-\eta \frac{k_B}{q} \ln \left( \frac{\frac{W}{L} \mu C_{ox} \left(\eta-1\right) V_t^2}{I_D}\right) \ .
\label{eq:vgs_drift}
\end{equation}
This is an approximation, since some of the terms within the logarithm and $\eta$ itself also depend on temperature. $\frac{\partial V_{th}}{\partial T}$ is negative (a parameter that is usually provided in the datasheet of the CMOS technology) hence the gate-source thermal drift of a MOS in sub-threshold region has a trend similar to the $V_{be}$ voltage in BJT transistors shown in equation \ref{eq:vbebrift}.

We can repeat the same calculation as for the BJT and generate a reference with positive thermal coefficient by taking the difference of two $V_{GS}$. We assume that the channel length and the drain current of the two transistors are the same. We can write
\begin{equation}
\Delta V_{GS} = V_{GS1}-V_{GS2} = V_{th1}-V_{th2}+\eta V_t \ln \left(\frac{W_2}{W_1}\right)
\end{equation}
and 
\begin{equation}
\frac{\partial \Delta V_{GS}}{\partial T} = \eta \frac{k_B}{q} \ln \left(\frac{W_2}{W_1}\right) \ ,
\end{equation}
which is positive as long as $W_2 > W_1$.

Using the same circuit topology as in Figure~\ref{fig:bandgap}, we have
\begin{equation}
V_{ref,MOS} = V_{GS1} + \eta V_t \frac{R_1}{R_2} \ln \left(\frac{W_{2}}{W_{1}}\right) \ .
\label{eq:bandgap_mos1}
\end{equation}
and
\begin{equation}
\frac{\partial V_{ref,MOS}}{\partial T} = \frac{\partial V_{GS}}{\partial T}+m_{MOS}\frac{\partial V_{t}}{\partial T}
\label{eq:bandgap_mos2}
\end{equation}
with
\begin{equation}
m_{MOS} = \eta \frac{R_1}{R_2} \ln \left(\frac{W_{2}}{W_{1}}\right) \ .
\label{eq:bandgap_mos3}
\end{equation}

\subsection{Specifications}

\begin{table}
\centering
{ 
\begin{tabu}{r r@{ }l l} 
\multicolumn{4}{c}{\textbf{Design specifications}} 								\\ 
Specification 			& \multicolumn{2}{c}{Value}					& Condition				\\ 
\hline
Output voltage			& $1.2-3$				& V					& 						\\
Output current			& 120 (200)				& mA				& 						\\
Dropout voltage			& 400					& mV				& (200 mA)				\\
Thermal drift 			& 15					& $ppm/^{\circ}C$	& ($20-50^{\circ}C$)	\\
Ripple rejection 		& 30					& dB 				& (200 mA)				\\
Output noise RMS 		& 15					& $\mu V$			& ($10\ Hz-1\ MHz$)		\\
Total ionizating dose	& 1.2					& Mrad				& 						\\
Neutron fluence			& $2\cdot 10^{13}$		& $n_{eq}\ cm^{-2}$	& 						\\ 
\hline
\end{tabu}}
\caption{ALDO design specifications.}
\label{tab:aldo_design_specs}
\end{table}

The ALDO ASIC has to be integrated in the EC on an add-in card placed between the FEBs.
The add-in card hosts two ALDOs, each one powering half EC, which corresponds to 16 CLARO chips or 128 channels.
This arrangement defines the minimum specifications required for the ALDO.

The supply current of each CLARO is 2.3 mA in idle, which corresponds to about 36 mA for 16 chips. At full load in high occupancy regions, the CLARO consumes 7 mA per chip, 112 mA for 16 chips. The ALDO was thus designed for 120 mA maximum output current in normal mode. In order to have some additional headroom, an optional high current mode with 200 mA maximum output current was also included.

The ALDO chip has to operate with the lowest possible energy loss in order to contribute marginally to the total power budget with respect to the front-ends. As a goal we fixed an efficiency of about 85\%. For a linear regulator, we can define the efficiency as
\[
\epsilon = \frac{P_{load}-P_{LDO}}{P_{load}} = 1-\frac{V_{dropout}}{V_{out}}
\]
and in our case we find that
\[
V_{dropout} = \left(1-\epsilon\right)V_{out} = 0.15 \cdot 2.5\ V \approx 400\ mV \ .
\]

The CLARO chip generates its threshold voltage by mirroring a reference current which is provided by an external bias resistor, referred directly to the power supply. The threshold sensibility to power supply changes was measured to $\frac{1}{V_{thr}}\frac{\partial V_{thr}}{\partial V_{CC}}=120\ ppm/mV$. In order to have a threshold thermal stability $\frac{1}{V_{thr}}\frac{\partial V_{thr}}{\partial T}$ better than $5\ ppm/^\circ C$, then the power supply have to be as stable as
\[
\frac{1}{V_{CC}}\frac{\partial V_{CC}}{\partial T} = \frac{1}{V_{CC}} \cdot \frac{1}{V_{thr}}\frac{\partial V_{thr}}{\partial T} \cdot \left(\frac{1}{V_{thr}}\frac{\partial V_{thr}}{\partial V_{CC}}\right)^{-1} \approx 16\ ppm/^\circ C
\]
in the typical range of operation between $20\ ^\circ C$ and $50\ ^\circ C$. This value is our specification for the thermal stability of the bandgap reference.

Switching noise due to the DC/DC spikes has to be attenuated by more than 30 dB at any load at frequency higher than 100 kHz.

The radiation hardness specifications are the same of the CLARO, defined as 10 years of operation in the LHCb RICH environment, with an additional safety margin ($\times 3$).

Table~\ref{tab:aldo_design_specs} summarizes the ALDO design specifications.

\subsection{Circuit design}

\begin{figure}
\centering
\def\svgwidth{\linewidth}
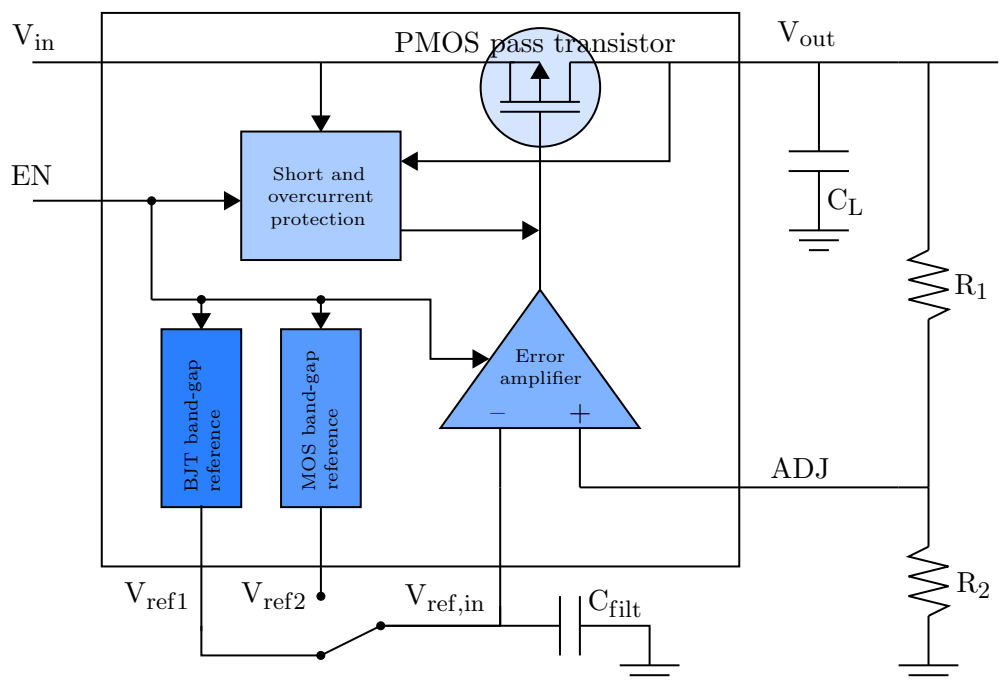
\caption{ALDO block scheme.}
\label{fig:aldo_block}
\end{figure}

Figure~\ref{fig:aldo_block} shows the block schematic of the ALDO chip.
The chosen architecture, shown in the previous section, is quite standard and widely discussed in literature.
The innovative aspect of this design is related to its radiation hardness.

Two bandgap references were included, one based on a traditional scheme with BJT transistors and the other one based on sub-threshold MOSFETs.
The choice of including two bandgaps is driven by the fact that the BJT-based one is expected to be more stable and less process-dependent, but more susceptible to displacement damage due to radiation.
On the contrary, MOS-based reference is much more resistant to displacement damage but more sensitive to process variations and total ionizing dose effects.
In this first prototype the two references are simultaneously active and each one can be used as the regulator reference, by shorting its output to the corresponding error amplifier input ($V_{ref,in}$ in Figure ~\ref{fig:aldo_block}).
This configuration gives the versatility to independently test both the solutions on the same chip, without the need to manufacture different versions.

The regulator is fully adjustable and the output voltage can be properly set by selecting the values of the feedback resistors between the output of the regulator and the ADJ pin of the error amplifier.

A foldback protection circuitry protects the regulator against shorts and overcurrents, closing the gate of the PMOS pass transistor and preventing the regulator from damaging itself. 

\subsubsection{BJT-based bandgap reference}

The bandgap reference based on the vertical BJT offered by the ams 0.35 $\mu m$ technology adopts the same circuit topology shown in Figure~\ref{fig:ldo}. We decided to adopt a BJT area ratio of 19:1 in order to create a matrix of 20 contiguous BJTs arranged in 4 rows and 5 columns, for improved matching. The bias current flowing into each of the two transistors is set to 13.6 $\mu A$, by choosing $R_1 = R_3 = 32.5\ k\Omega$ . The drift of the $V_{be}$ from simulation was found to be $-1.74\ mV/^{\circ}C$. From equations \ref{eq:bandgap_bjt1} and \ref{eq:bandgap_bjt2}, we calculated the value of the coefficient $m_{BJT}$ which nullifies the thermal drift:
\begin{equation}
m_{BJT} = \frac{-\frac{\partial V_{be}}{\partial T}}{\frac{\partial V_{t}}{\partial T}} = \frac{1.74\ mV/^{\circ}C}{0.086\ mV/^{\circ}C} \approx 20.23 \
\end{equation}
and then the required resistor ratio
\begin{equation}
\frac{R_1}{R_2} = \frac{m}{\ln \left(\frac{A_2}{A_1}\right)} = \frac{20.23}{\ln \left(19\right)} = 6.87 \ .
\end{equation}
Finally we can calculate $R_2$:
\begin{equation}
R_2 = \frac{32.5\ k\Omega}{6.87} = 4.73\ k\Omega \ .
\end{equation}

\begin{figure}
	\centering
 	\includegraphics[width=\linewidth]{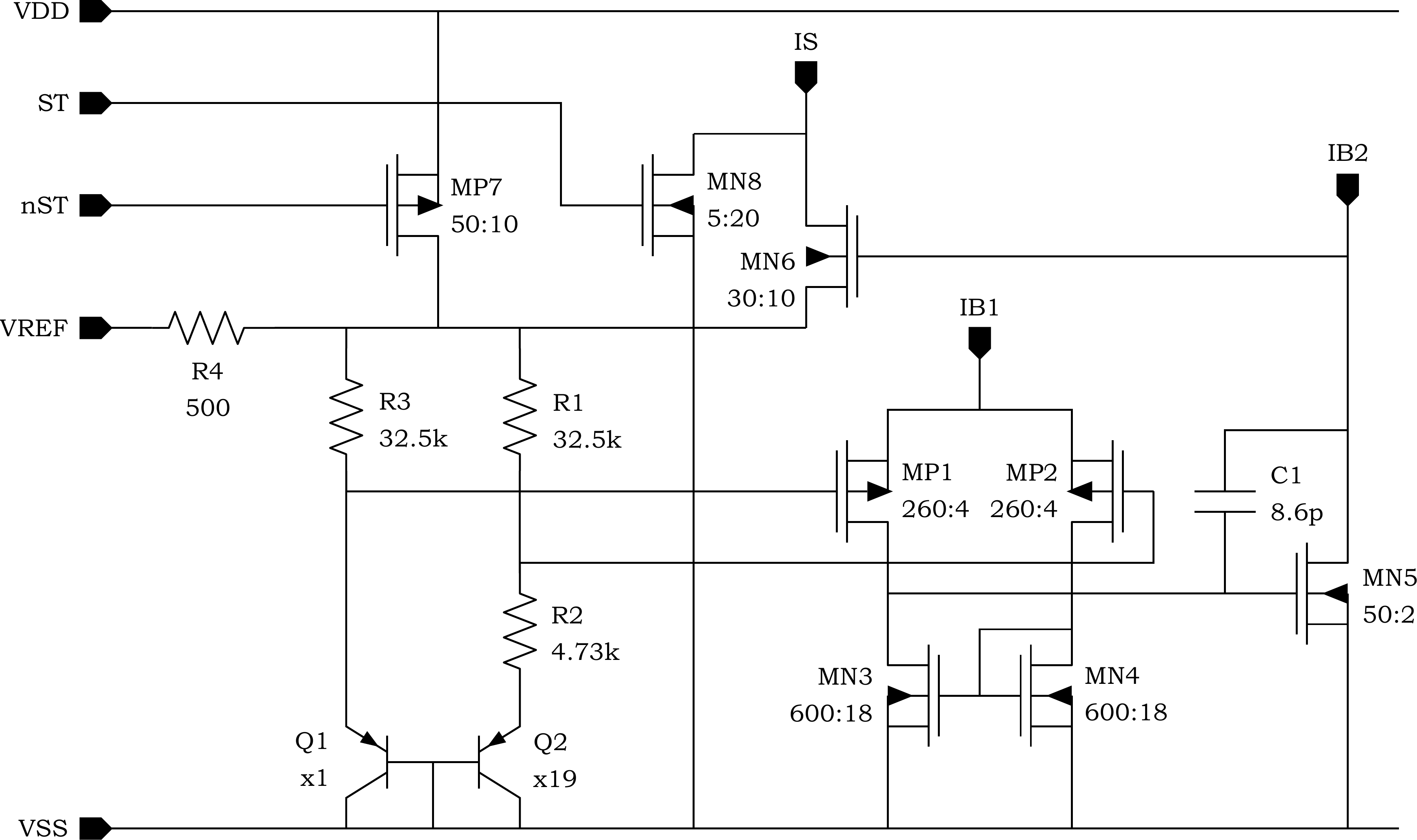}
    \caption{Schematic of the BJT-based bandgap, dimensions W:L are in $\mu m$.}
    \label{fig:aldo_prog_bandgap}
\end{figure}

The output of the bandgap can be calculated using equation \ref{eq:bandgap_bjt3}:
\begin{equation}
V_{ref} = V_{be1} + m_{BJT}\:V_t = 0.698\ V + 20.23 \cdot 0.0259\ V = 1.22\ V
\end{equation}

The amplifier adopts a standard two-stage design with an input PMOS pair and NMOS second stage. The PMOS input was chosen due to the superior 1/f noise performance with respect to NMOS input pair. Transistor sizes were also optimized to minimize noise and offset, choosing larger area transistors and making the two $V_{DS}$ of the active load as equal as possible to minimize systematic offset due to different current gains.
The amplifier is stabilized with a 8.6 pF Miller capacitor across the second gain stage.

The bias current is mirrored on node IS and provides the current biases of the amplifier on nodes IB1 and IB2.

The startup circuitry was omitted from the schematic shown here. It consists in a set of inverters whose input is IB2 and whose output activates transistors MP7 and MN8 in order to bias the BJTs and the amplifier. When the circuit is powered up, the inverters switch off these transistors.

\subsubsection{MOS-based bandgap reference}

\begin{figure}
	\centering
 	\includegraphics[width=\linewidth]{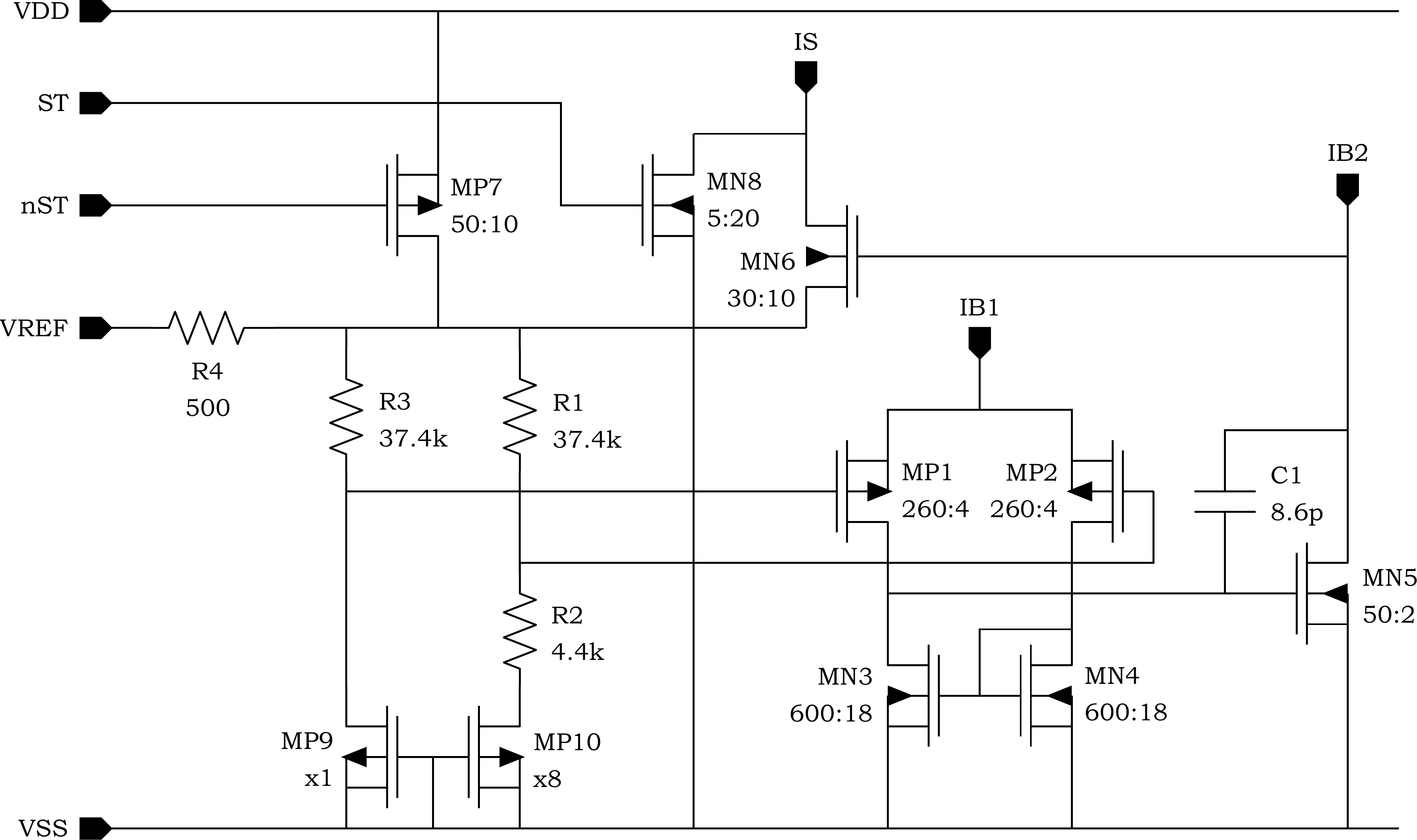}
    \caption{Schematic of the MOS-based bandgap.}
    \label{fig:aldo_prog_bandgap_mos}
\end{figure}

The second bandgap reference adopts the same circuit topology of the BJT-based one, except for having two PMOS transistors in sub-threshold region in place of the BJTs.
Figure~\ref{fig:aldo_prog_bandgap_mos} shows the schematic of the circuit.

Transistors have to be biased in sub-threshold region, thus very large W/L ratio was chosen in order to keep a similar bias current as for the BJT-based bandgap.
W/L was set to $2160\ \mu m / 0.35\ \mu m = 6170$ that gives, according to simulations, a $V_{GS}$ of $477\ mV$ with a bias current of $7.9\ \mu A$.

Combining equations \ref{eq:bandgap_mos2} and \ref{eq:bandgap_mos3}, and imposing $\frac{\partial V_{ref}}{\partial T} = 0$ we find that
\begin{equation}
m_{MOS} = \frac{-\frac{\partial V_{GS}}{\partial T}}{\frac{\partial V_{t}}{\partial T}}=\eta \frac{R_1}{R_2} \ln \left(\frac{W_2}{W_1}\right) \ .
\label{eq:mmos}
\end{equation}
$\frac{\partial V_t}{\partial T}=\frac{k_b}{q}$, while $\frac{\partial V_{GS}}{\partial T}$ is defined by equation \ref{eq:vgs_drift}:
\begin{equation}
\frac{\partial V_{GS}}{\partial T}= \frac{\partial V_{th}}{\partial T}-\eta \frac{k_B}{q} \ln \left( \frac{\frac{W}{L} \mu C_{ox} \left(\eta-1\right) V_t^2}{I_D}\right) \ .
\label{eq:dvgs_dt}
\end{equation}
The datasheet of the 0.35 $\mu m$ CMOS technology by ams defines the following parameters for a standard PMOS transistor:
\begin{equation}
\frac{\partial V_{th}}{\partial T}=-1.83\ mV/^{\circ}C 
\end{equation}
and
\begin{equation}
\mu C_{ox}= 58\ \mu A/V^2 \ ,
\end{equation}
but there is no specification for $\eta$, so we simulated the $I_D\left(V_{GS}\right)$ curve for the PMOS MP9. The results of this simulation are shown in Figures \ref{fig:aldo_ids_vgs_simu} and \ref{fig:aldo_ids_vgs_simu_fit}. The fitted value of $\eta$ is 1.36 and the $V_{GS}$ at 300~K is 470~mV.

\begin{figure}
\begin{minipage}[b][][t]{.47\linewidth}
	\centering
	 \definecolor{mycolor1}{rgb}{0.00000,0.44700,0.74100}%
\begin{tikzpicture}[thick,scale=0.850, every node/.style={scale=0.850}]

\begin{axis}[%
width=2.126in,
height=1.432in,
at={(0.514in,0.43in)},
scale only axis,
xmin=0.2,
xmax=0.6,
xlabel style={font=\color{white!15!black}},
xlabel={$V_{GS}\ \left[V\right]$},
ymin=0,
ymax=100,
ylabel style={font=\color{white!15!black}},
ylabel={$I_D\ \left[\mu A\right]$},
axis background/.style={fill=white},
title style={font=\bfseries},
title={Simulated $\mathbf{I_D\left(V_{GS}\right)}$}
]
\addplot [color=mycolor1, forget plot]
  table[row sep=crcr]{%
0.2013	0.001585\\
0.2026	0.00166\\
0.204	0.001738\\
0.2053	0.00182\\
0.2067	0.001905\\
0.208	0.001995\\
0.2094	0.002089\\
0.2107	0.002188\\
0.2121	0.002291\\
0.2134	0.002399\\
0.2148	0.002512\\
0.2162	0.00263\\
0.2175	0.002754\\
0.2189	0.002884\\
0.2202	0.00302\\
0.2216	0.003162\\
0.223	0.003311\\
0.2243	0.003467\\
0.2257	0.003631\\
0.2271	0.003802\\
0.2284	0.003981\\
0.2298	0.004169\\
0.2312	0.004365\\
0.2326	0.004571\\
0.2339	0.004786\\
0.2353	0.005012\\
0.2367	0.005248\\
0.238	0.005495\\
0.2394	0.005754\\
0.2408	0.006026\\
0.2422	0.00631\\
0.2436	0.006607\\
0.2449	0.006918\\
0.2463	0.007244\\
0.2477	0.007586\\
0.2491	0.007943\\
0.2505	0.008318\\
0.2518	0.00871\\
0.2532	0.00912\\
0.2546	0.00955\\
0.256	0.01\\
0.2574	0.01047\\
0.2588	0.01096\\
0.2602	0.01148\\
0.2615	0.01202\\
0.2629	0.01259\\
0.2643	0.01318\\
0.2657	0.0138\\
0.2671	0.01445\\
0.2685	0.01514\\
0.2699	0.01585\\
0.2713	0.0166\\
0.2727	0.01738\\
0.2741	0.0182\\
0.2755	0.01905\\
0.2769	0.01995\\
0.2783	0.02089\\
0.2797	0.02188\\
0.2811	0.02291\\
0.2825	0.02399\\
0.2839	0.02512\\
0.2853	0.0263\\
0.2867	0.02754\\
0.2881	0.02884\\
0.2895	0.0302\\
0.2909	0.03162\\
0.2923	0.03311\\
0.2938	0.03467\\
0.2952	0.03631\\
0.2966	0.03802\\
0.298	0.03981\\
0.2994	0.04169\\
0.3008	0.04365\\
0.3022	0.04571\\
0.3037	0.04786\\
0.3051	0.05012\\
0.3065	0.05248\\
0.3079	0.05495\\
0.3093	0.05754\\
0.3108	0.06026\\
0.3122	0.0631\\
0.3136	0.06607\\
0.315	0.06918\\
0.3165	0.07244\\
0.3179	0.07586\\
0.3193	0.07943\\
0.3208	0.08318\\
0.3222	0.0871\\
0.3236	0.0912\\
0.325	0.0955\\
0.3265	0.1\\
0.3279	0.1047\\
0.3294	0.1096\\
0.3308	0.1148\\
0.3322	0.1202\\
0.3337	0.1259\\
0.3351	0.1318\\
0.3366	0.138\\
0.338	0.1445\\
0.3395	0.1514\\
0.3409	0.1585\\
0.3423	0.166\\
0.3438	0.1738\\
0.3452	0.182\\
0.3467	0.1905\\
0.3482	0.1995\\
0.3496	0.2089\\
0.3511	0.2188\\
0.3525	0.2291\\
0.354	0.2399\\
0.3554	0.2512\\
0.3569	0.263\\
0.3584	0.2754\\
0.3598	0.2884\\
0.3613	0.302\\
0.3628	0.3162\\
0.3642	0.3311\\
0.3657	0.3467\\
0.3672	0.3631\\
0.3686	0.3802\\
0.3701	0.3981\\
0.3716	0.4169\\
0.3731	0.4365\\
0.3745	0.4571\\
0.376	0.4786\\
0.3775	0.5012\\
0.379	0.5248\\
0.3805	0.5495\\
0.382	0.5754\\
0.3835	0.6026\\
0.3849	0.631\\
0.3864	0.6607\\
0.3879	0.6918\\
0.3894	0.7244\\
0.3909	0.7586\\
0.3924	0.7943\\
0.3939	0.8318\\
0.3954	0.871\\
0.3969	0.912\\
0.3984	0.955\\
0.3999	1\\
0.4015	1.047\\
0.403	1.096\\
0.4045	1.148\\
0.406	1.202\\
0.4075	1.259\\
0.409	1.318\\
0.4106	1.38\\
0.4121	1.445\\
0.4136	1.514\\
0.4151	1.585\\
0.4167	1.66\\
0.4182	1.738\\
0.4197	1.82\\
0.4213	1.905\\
0.4228	1.995\\
0.4244	2.089\\
0.4259	2.188\\
0.4275	2.291\\
0.429	2.399\\
0.4306	2.512\\
0.4321	2.63\\
0.4337	2.754\\
0.4352	2.884\\
0.4368	3.02\\
0.4384	3.162\\
0.4399	3.311\\
0.4415	3.467\\
0.4431	3.631\\
0.4446	3.802\\
0.4462	3.981\\
0.4478	4.169\\
0.4494	4.365\\
0.451	4.571\\
0.4526	4.786\\
0.4542	5.012\\
0.4558	5.248\\
0.4574	5.495\\
0.459	5.754\\
0.4606	6.026\\
0.4622	6.31\\
0.4638	6.607\\
0.4654	6.918\\
0.467	7.244\\
0.4686	7.586\\
0.4703	7.943\\
0.4719	8.318\\
0.4735	8.71\\
0.4752	9.12\\
0.4768	9.55\\
0.4784	10\\
0.4801	10.47\\
0.4817	10.96\\
0.4834	11.48\\
0.485	12.02\\
0.4867	12.59\\
0.4884	13.18\\
0.49	13.8\\
0.4917	14.45\\
0.4934	15.14\\
0.495	15.85\\
0.4967	16.6\\
0.4984	17.38\\
0.5001	18.2\\
0.5018	19.05\\
0.5035	19.95\\
0.5052	20.89\\
0.5069	21.88\\
0.5086	22.91\\
0.5103	23.99\\
0.512	25.12\\
0.5137	26.3\\
0.5154	27.54\\
0.5172	28.84\\
0.5189	30.2\\
0.5206	31.62\\
0.5224	33.11\\
0.5241	34.67\\
0.5258	36.31\\
0.5276	38.02\\
0.5293	39.81\\
0.5311	41.69\\
0.5328	43.65\\
0.5346	45.71\\
0.5364	47.86\\
0.5381	50.12\\
0.5399	52.48\\
0.5417	54.95\\
0.5435	57.54\\
0.5453	60.26\\
0.5471	63.1\\
0.5489	66.07\\
0.5507	69.18\\
0.5525	72.44\\
0.5543	75.86\\
0.5561	79.43\\
0.5579	83.18\\
0.5597	87.1\\
0.5615	91.2\\
0.5634	95.5\\
0.5652	100\\
};
\end{axis}
\end{tikzpicture}%
	\caption{Characteristic curve of the sub-threshold PMOS transistor simulated by the design tool.}
	\label{fig:aldo_ids_vgs_simu}
\end{minipage}
\ \hspace{1mm} \
\begin{minipage}[b][][t]{.47\linewidth}
	\centering
	 \input{04_Chapter04/Figures/aldo_ids_vgs_simu_fit.tex} 
	\caption{The curve was fitted near the working point ($7.9\ \mu A$) to calculate the sub-threshold slope $\eta$ from \ref{eq:idsubthr}.}
	\label{fig:aldo_ids_vgs_simu_fit}
\end{minipage}
\end{figure}

Using these data, we can calculate
\begin{equation}
\frac{\partial V_{GS}}{\partial T}= -2.11\ mV/^{\circ}C \ .
\end{equation}

We chose a MOS area ratio of 8 and from \ref{eq:mmos} we can find the required resistor ratio in order to have zero drift:
\begin{equation}
\frac{R_1}{R_2} \approx 8.5 \ .
\end{equation}
$R_1$ was set to $37.4\ k\Omega$ for biasing the MOS, hence $R_2$ was set to $4.4\ k\Omega$.
The output voltage is given by equation \ref{eq:bandgap_mos1}:
\begin{equation}
V_{ref,MOS} = V_{GS1} + m_{MOS}\:V_t = 0.470\ V + 24.04 \cdot 0.0259\ V = 1.09\ V \ .
\end{equation}

The amplifier schematic is exactly the same as for the BJT band gap.

\subsubsection{Output PMOS transistor}

In section \ref{sec:ldo_theory}, we have qualitatively shown that dropout depends on the series resistance of the pass transistor in linear region. The expression that gives $R_{DS\left(on\right)}$ is given by
\begin{equation}
R_{DS\left(on\right)} = \frac{1}{\mu C_{ox} \frac{W}{L} \left(V_{GS}-V_{TH}-V_{DS}\right) }
\end{equation}
from which we can easily find out that the output PMOS transistor has to be made as wide as possible in order to lower its output resistance.

Our specifications determined that the maximum dropout allowed has to be 400~mV at full load of 120~mA, which means that the maximum series impedance has to be $3.3\ \Omega$. In high current mode (200~mA) the maximum series resistance has to be less than $2\ \Omega$.
In order to match the specifications for these two operating modes, we decided to split the transistor into two independent transistor each one with $W = 5000\ \mu m$ and $L = 0.35\ \mu m$ (the minimum channel length offered by the technology).
At this short transistor length, however, the effective length is $0.5\ \mu m$.
From the previous formula, we can find out that in the worst case (when the gate is pulled as low as possible so that $V_{SG} = V_{in}$ while $V_{SD} = V_{in}-V_{out}$ and the $V_{TH}$ is the highest at 0.78 V), we have that
\begin{equation}
R_{DS\left(on\right)} = \frac{1}{48\ \frac{\mu A}{V^2} \frac{10000}{0.5} \left(3\ V-0.78\ V\right)} = 0.47\  \Omega \ .
\end{equation}
In addition to this contribution we have to add the bonding wire resistance (made by 2 gold wires with a diameter of $25\ \mu m$ for each pad) of $0.1\ \Omega$, the total package pin resistance of $0.05\ \Omega$, and the in-chip trace and contact resistances of about $0.15\ \Omega$. These last contribution was specifically minimized during layout phase by using large ($200\ \mu m$) metal traces across the top three metal layers provided by the technology. With these estimations, total series resistance should not exceed $0.8\ \Omega$, well lower than the maximum $2\ \Omega$ from the high current specification.

\begin{figure}
	\centering
 	\includegraphics[width=.6\linewidth]{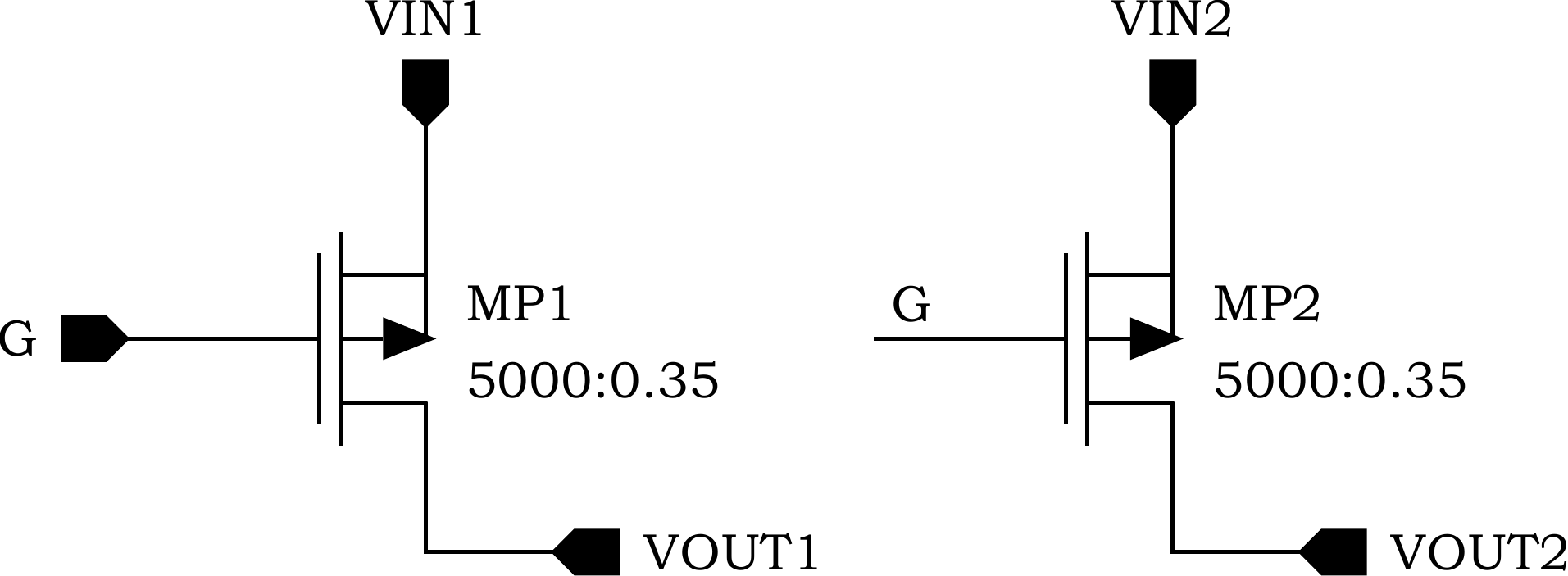}
    \caption{Schematic block of the PMOS pass transistor. Total transistor length is $2 \times 5\ mm$.}
    \label{fig:aldo_prog_pmos}
\end{figure}

Figure \ref{fig:aldo_prog_pmos} shows the schematic of the pass transistors. The gate of the two MOS is shorted together, however source and drain (i.e. input and output voltages of the regulator, respectively) are routed to independent pins. This has two advantages:
the first is that current densities can be lowered since currents are split between separate bonding wires, pads and metal traces, the second is that, in case less output current is required, only one transistor can be connected, leaving the other one floating and thus improving stability due to the reduced capacitive load at the output of the error amplifier. From simulations, the total capacitive load at the gate of the pass transistor is about 9 pF for each PMOS, so 18 pF when both are active.

\subsubsection{Error amplifier}

\begin{figure}
	\centering
 	\includegraphics[width=.6\linewidth]{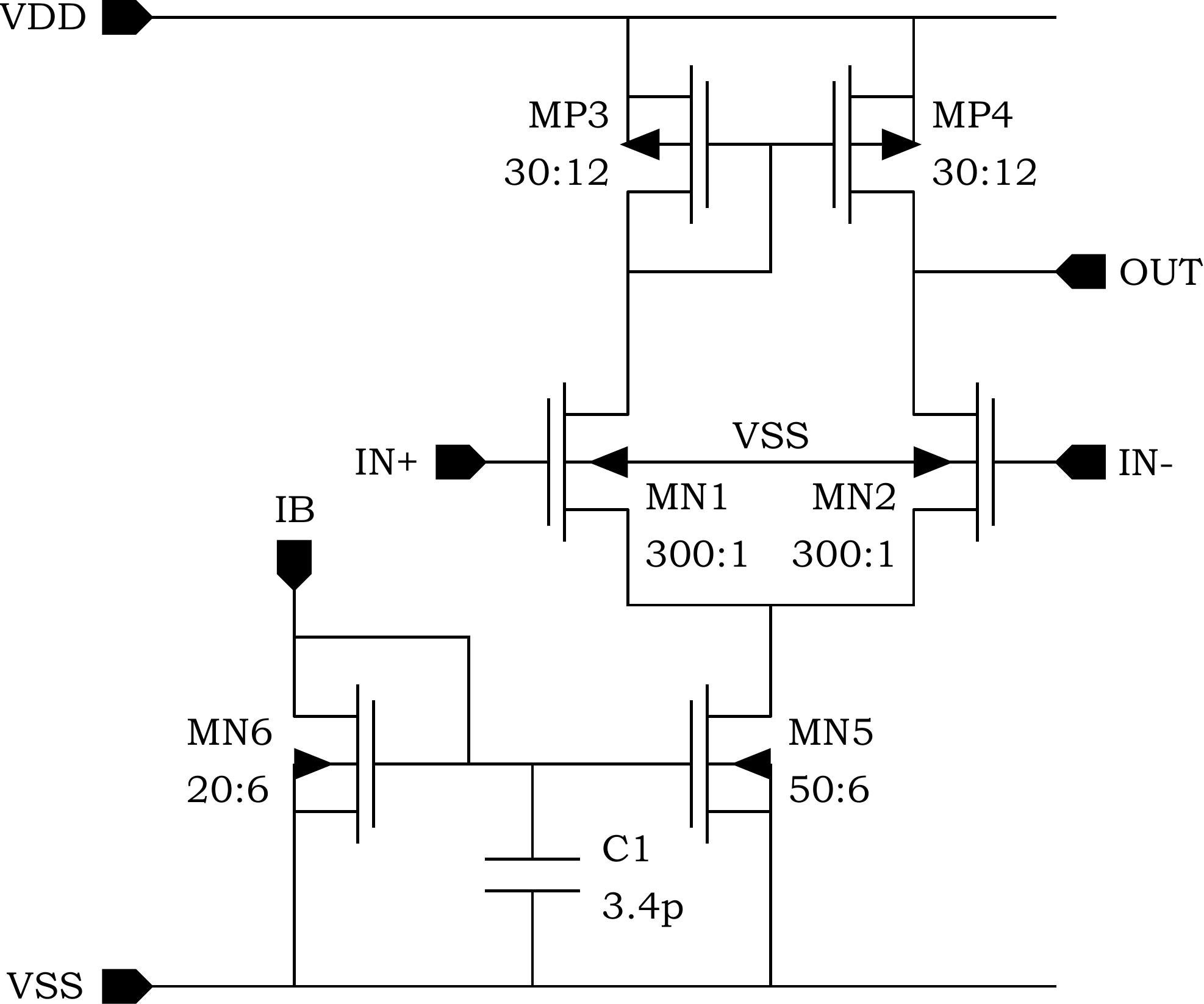}
    \caption{Schematic block of the error amplifier.}
    \label{fig:aldo_prog_erroramp}
\end{figure}

The error amplifier is a single stage amplifier implemented with an NMOS input pair. The NMOS pair provides an higher gain with respect to a PMOS pair, while the single stage solution offers a lower output impedance which is preferable for stability concerns.
A single stage NMOS input also offers a compromise between output dynamics and the required compatibility to positive rail.
The schematic of this block is shown in Figure~\ref{fig:aldo_prog_erroramp}.

The open loop gain of the amplifier is $53\ dB$, with a dominant pole at $4.9\ kHz$ due to the output capacitive load at the pass element gate ($18\ pF$) and the amplifier output impedance ($1.8\ M\Omega$).
The second pole due to the active load is placed at much higher frequency ($3.2\ MHz$), well above the unity gain bandwidth.

From the stability analysis carried out in section \ref{sec:ldo_theory} and referring in particular to equation \ref{eq:ldo_stability}, we have that the zero determined by the output capacitor and its equivalent series resistance must compensate the mid-frequency pole due to the error amplifier that we have just calculated.
In addition to that, the high frequency pole due to the load capacitor and the equivalent series resistance ($p_{ESR}$) has to be placed at a frequency such that the gain has already fallen below 0 dB in any load condition.

In order to find the optimum working point, a multi-dimensional sweep with varying $C_L$, $C_O$ and $R_{ESR}$ was performed with the simulation tool.
The optimum value for the output capacitor was found in the range $22\ \mu F - 100\ \mu F$ with an ESR of $50\ m\Omega - 300\ m\Omega$.
For the final board, a tantalum capacitor of $68\ \mu F$ with $100\ m\Omega$ ESR was selected.

\subsubsection{Protections}
\label{sec:aldo_protections}

\begin{figure}
	\centering
 	\includegraphics[width=\linewidth]{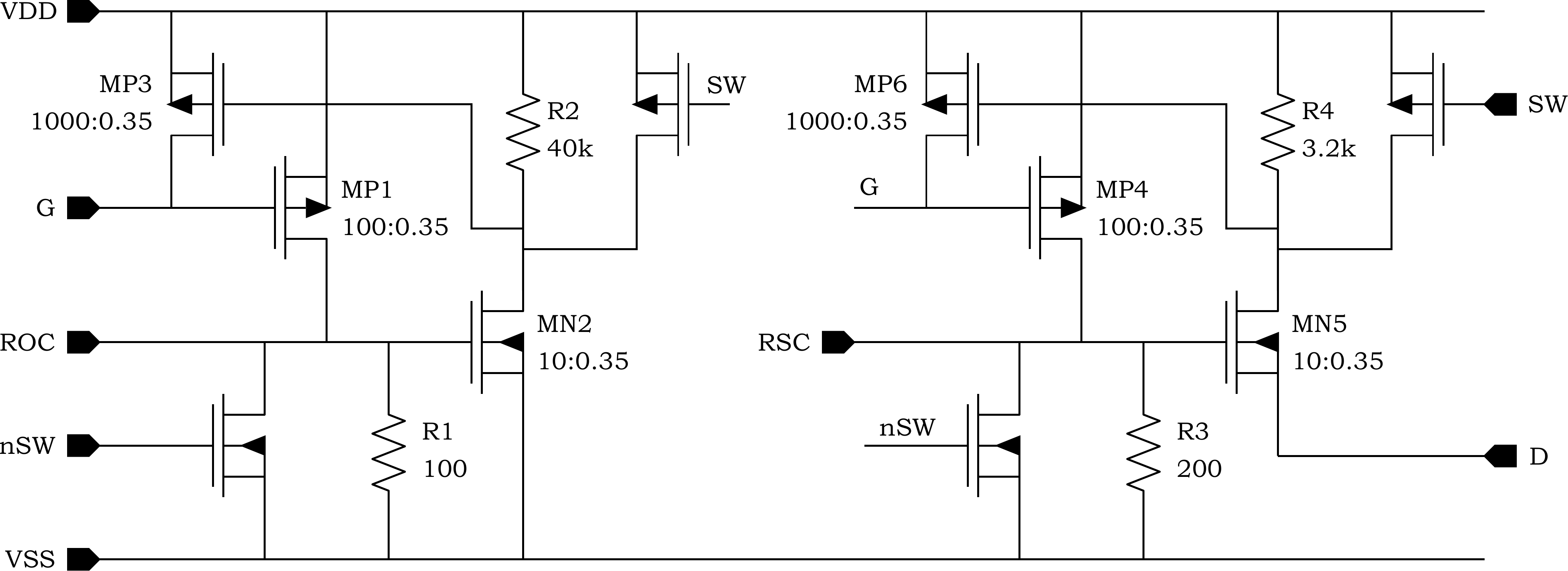}
    \caption{Block schematic of the overcurrent and short protections.}
    \label{fig:aldo_prog_overcurrent}
\end{figure}

For a safe integration in complex and tight systems, the regulator has to be protected against common failures like excessive load and shorts at the output.
The circuit that perform this task is shown in Figure~\ref{fig:aldo_prog_overcurrent}.

The protection is made by two different blocks that implement the fold-back protections.
The overcurrent block (on the left of the schematic) mirrors the output current with a 1:100 ratio, converts it to a voltage on resistor R1 and compares it to the threshold of MN2.
If the load current exceeds the maximum value, then the gate of the pass transistor (G node) is pulled up and the current is reduced.

The short protection block (on the right of the schematic) adopts a similar concept, with the difference that the source of MN5 is connected to the output voltage (node D), thus the threshold of the protection becomes effective only when a low voltage (short) is present at the output.
In order to prevent the circuit from triggering at the start-up, an external bypass capacitor has to be applied on the SW node, so that the startup of the protection circuit is delayed.
The nSW signal is generated by inverting SW. 

In the regulator we also included a dummy diode-connected MOS with the same dimensions and layout of the one used in the MOS-based bandgap. This device will allow us to accurately characterize its characteristics and eventually design a thermal protection circuit based on this device, which could be included in future revisions of the chip. 

The chip also has an enable pin which can be used to switch down the output voltage and perform a power cycle on the devices supplied by the regulator.

\subsection{Circuit layout and radiation hardening techniques}

Circuit layout with power devices have some additional challenges due to the high current involved. In addition to that, radiation hardening techniques should be adopted in order to maximize the ASIC tolerance to radiation damage.

All the high current paths (i.e. input and output paths to the source and drain of the output pass transistor) were enlarged to keep the series resistance as low as possible and minimize the risk of electro-migration damage. For the same reasons multiple vias arrays were also used.

Radiation hardening techniques consist mainly in avoiding that parasitic structures can be triggered by the charges deposited by incident radiation. This phenomenon is called single-event latchup (SEL). The parasitic structures (i.e. parasitic MOS or bipolar transistors) can be established, for example, under the shallow trench insulation (STI) between complementary PMOS and NMOS structures, or in the bulk of the device in the form of a bipolar transistor between the source and drain, through the silicon bulk.

\begin{figure}
\centering
\def\svgwidth{.7\linewidth}
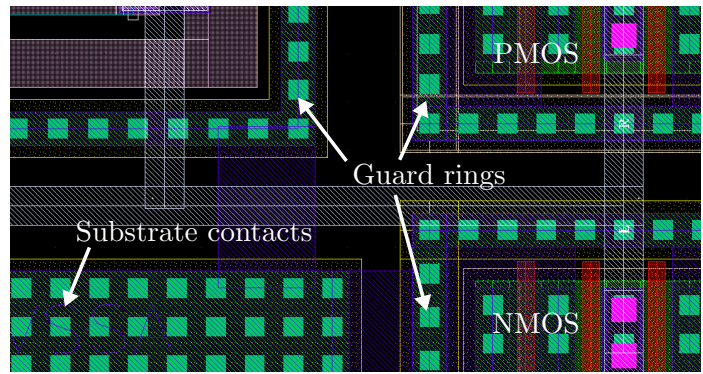
\caption{Detail of the layout with guard rings and substrate contacts.}
\label{fig:aldo_layout2}
\end{figure}

Mitigation strategies consist in the adoption of guard rings (substrate or N-well contacts) that surround each transistor or group of transistor of the same type. These contacts are very effective in removing any unwanted free charge and prevent the formation of these parasitic structures. Figure~\ref{fig:aldo_layout2} shows a screenshot of the design tool where the application of such technique is highlighted.

\begin{figure}
\centering
\def\svgwidth{\linewidth}
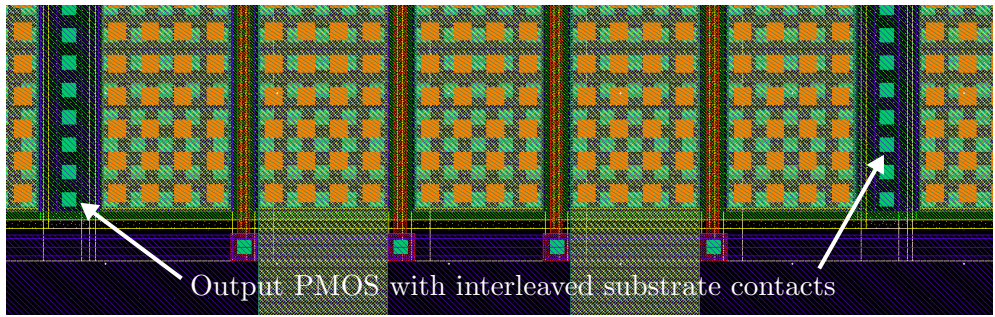
\caption{Detail of the output PMOS layout with interleaved substrate contacts in order to prevent latchup in the bulk.}
\label{fig:aldo_layout1}
\end{figure}

In order to avoid the formation of parasitic bipolar transistor underneath the very large pass transistor, substrate contacts were also interleaved every $25\ \mu m$, as shown in Figure~\ref{fig:aldo_layout1}.

Apart from these special techniques, all the standard and advanced techniques for improving device matching were adopted. These included, for example, interleaved differential pairs, dummy structures around critical devices, strategical positioning of temperature sensitive blocks (bandgaps) far from the output pass transistor where more heat will be dissipated, and many others.

\subsection{ASIC production and test boards}

\begin{figure}
	\centering
 	\includegraphics[width=.5\linewidth]{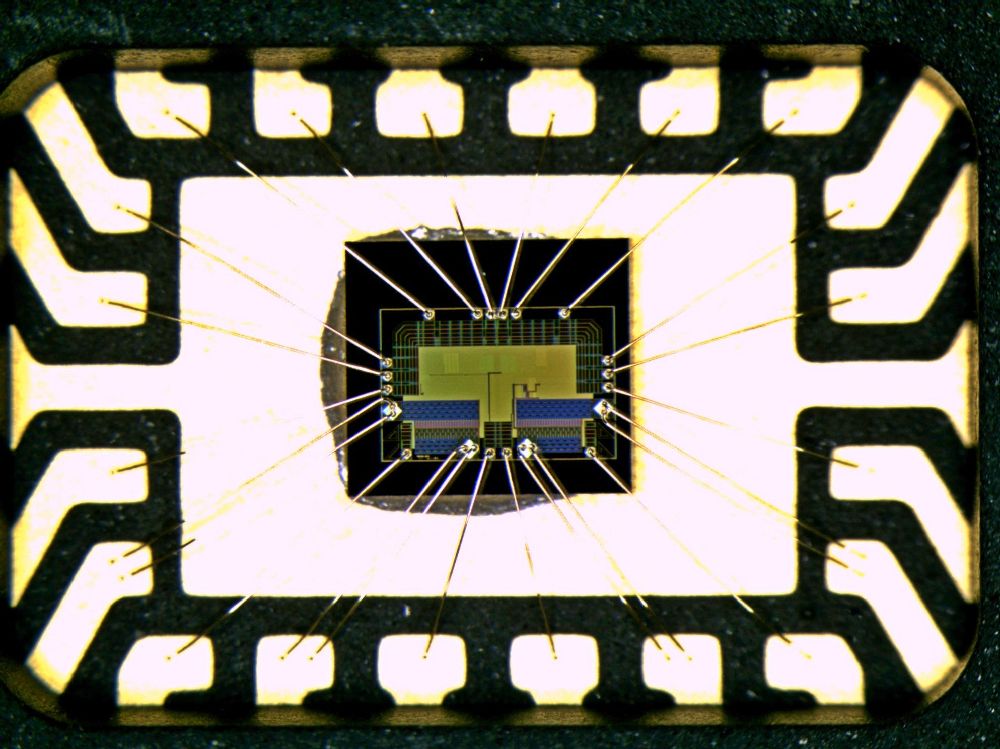}
    \caption{Photograph of the ALDO chip die using a microscope. Die area is $2.1 \times 1.4\ mm^2$.}
    \label{fig:aldo_photo}
\end{figure}

A first prototype production run with 40 chips was commissioned to ams.
The chips were mounted in a SOIC20 package for easiness of mounting and better debug capabilities, but it could also fit in a more compact QFN16, if space constraints will require it.
A few dies were placed in a special package with a lid that can be opened for irradiation purposes.
A photo of the bonded chip is shown in Figure~\ref{fig:aldo_photo}.

\begin{figure}
	\centering
 	\includegraphics[width=.6\linewidth]{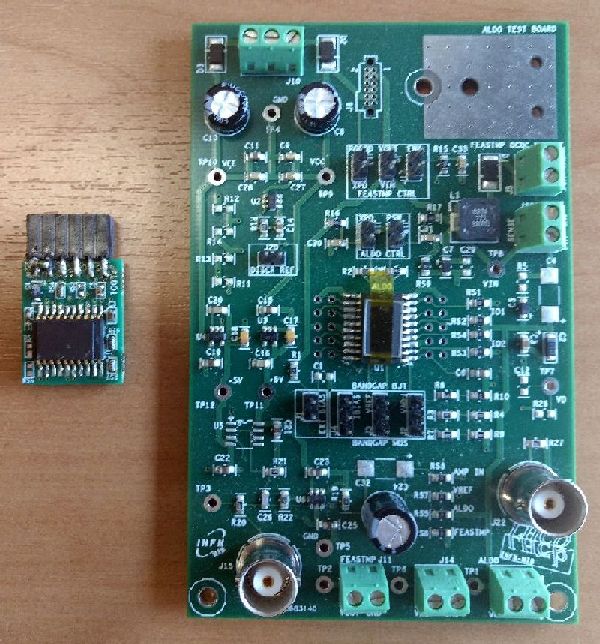}
    \caption{Photographs of the board for device characterization (on the right) and of the smaller board for EC integration and irradiation tests (on the left).}
    \label{fig:aldo_testboard}
\end{figure}

For the tests of the chips we developed a custom testboard, depicted on the left in Figure~\ref{fig:aldo_testboard}, which included additional circuitry for hosting the FEASTMP DC/DC, as well as for performing power supply rejection and noise measurements.

A much more compact board, $18.9 \times 23.4 mm^2$ large, was also designed. A photograph of the board is visible on the left in Figure~\ref{fig:aldo_testboard}.
This board can be easily integrated in the EC of the LHCb RICH.
All the control pins and reference voltage are available on the upper connector, so this board can be also used for monitoring the chip in irradiation facilities, where space constraints can also be very tight.

\subsection{Performance}

After receiving the first prototypes, we performed a thorough test campaign in order to qualify the performance of the ASIC. 
The tests started with circuit and stability optimizations, then continued with thermal stability, noise and power supply rejection measurements.
The results will be discussed with more detail in the next sections.

\subsubsection{Circuit optimization}

The ASIC offers the capability to use the single-size output MOS for low current mode (120~mA) and the double-size output MOS for an output current up to 200~mA.
In normal mode the ASIC worked as expected and it was able to source 125~mA with a dropout of 300~mV.
When the double-size MOS was used, the ALDO showed some instability during startup at high output current, using a small value output resistor as load.
This particular condition is the very worst case, since, during normal operation with the CLAROs or any other active device, the output load is not purely resistive. In addition to that, the load of the CLAROs also depends on the rate of incoming photons, which is expected to increase only after some time from the startup of the regulator.

In order to improve the startup with low resistive loads, we tried to increase the stabilizing capacitor.
This slightly mitigated the effect but did not lead to a definitive solution of the issue.
This unexpected behavior was not simulated by the development tools in neither of the worst case corners.

Given the fact that the issue is present only in high-current mode and in a worst case scenario of a low resistive load, it was decided to postpone further studies on this issue and continue the characterization of the ASIC in normal mode, which already provides the required output current for the LHCb RICH operation.

In section~\ref{sec:aldo_protections} it was already discussed the need of a bypass capacitor in order to slow down the protection circuit. The minimum value for this capacitor is $1\ \mu F$.

\subsubsection{Thermal stability}

Three samples of the ALDO chip were separately placed in a climatic chamber and temperature were swept from $20\: ^{\circ}C$ up to $50\: ^{\circ}C$, $70\: ^{\circ}C$ or $90\: ^{\circ}C$, depending on the sample. The various voltages (bandgaps, input and output voltages, dummy diode, etc.) were monitored with a multi-channel 6.5-digit multimeter. The system was completely remotely controlled using a MATLAB GUI that recorded the voltages and saved log files to disk.

\begin{figure}
\centering
\input{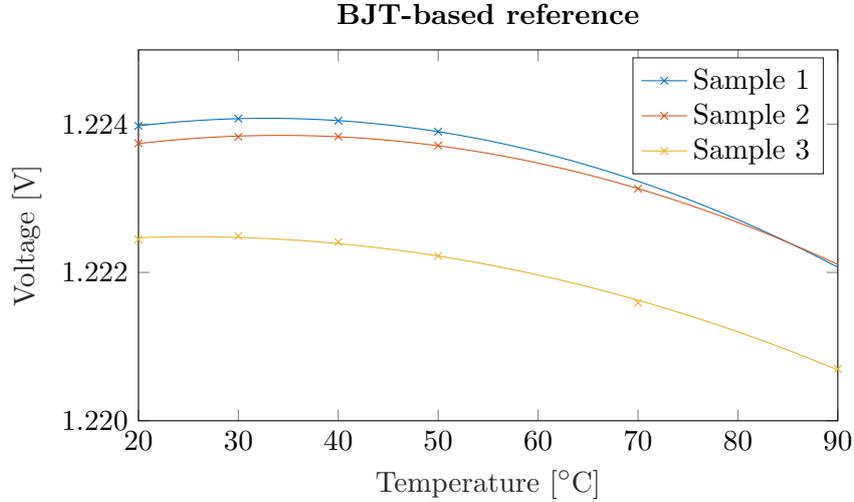} 
\caption{Temperature drift of the BJT-based bandgap reference for three ALDO chips. The spread between the references is less than 1.6~mV (0.13\%). The thermal drift is less than 15~ppm/$^{\circ}C$ in the range between 20 $^{\circ}C$ and 50~$^{\circ}C$.}
\label{fig:aldo_temp_vref_bjt}
\end{figure}

Figure~\ref{fig:aldo_temp_vref_bjt} shows the results for the BJT-based bandgap. The three samples have a spread of 1.6~mV and the thermal drift is below 15~ppm/$^{\circ}C$ in the range between 20 $^{\circ}C$ and 50~$^{\circ}C$ for all the three samples. The reference voltage is about 1.223~V, as expected from the simulations.

\begin{figure}
\centering
\input{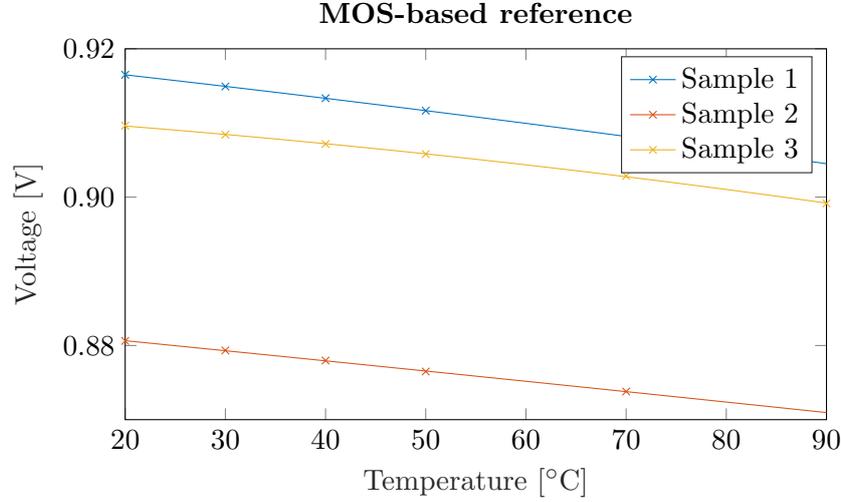} 
\caption{Temperature drift of the MOS-based bandgap reference for three ALDO chips.}
\label{fig:aldo_temp_mos_ref}
\end{figure}

Figure~\ref{fig:aldo_temp_mos_ref} shows the results for the MOS-based bandgap.
In this case the ASIC exhibits a significant deviation from the results given by simulations, both in the reference voltage, 0.9~V versus 1.1~V, and in the thermal drift, -160~ppm/$^{\circ}C$ versus few ppm/$^{\circ}C$.
The three devices tested also have a higher spread than expected between them, hinting that the issue could also be related to some process mismatch.
Montecarlo simulations using the development tools did not lead to a satisfying explanation of the observed behavior, hinting to a poor modelization of the thermal characteristics in sub-threshold region.

\begin{figure}
	\centering
	 \input{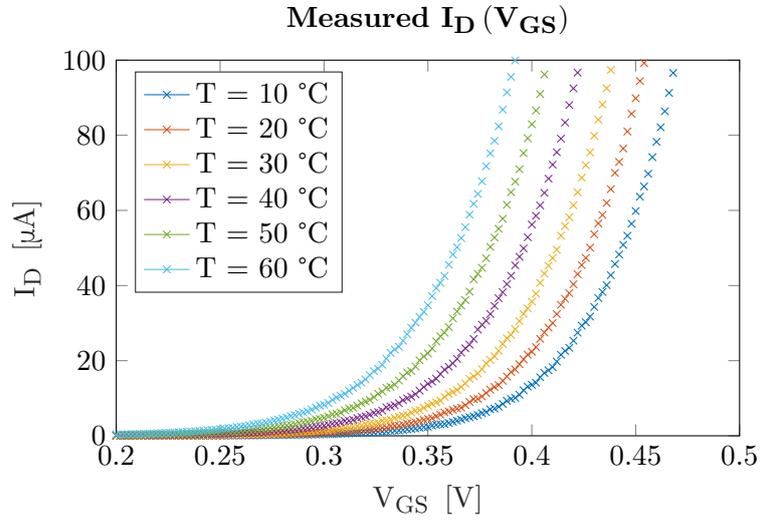} 
	\caption{Characteristic curve of the sub-threshold MOS measured at different temperatures.}
	\label{fig:aldo_ids_vgs_temp}
\end{figure}

In order to understand the cause of such an unexpected behavior, we measured the characteristic I-V curve of the reference MOS at different temperatures, thanks to the dummy device that was included specifically for debug purposes.
The measurements were performed with a Keithley 4200A semiconductor analyzer.
The results are shown in Figure~\ref{fig:aldo_ids_vgs_temp}.

The drain voltage $I_{D}$ as a function of $V_{GS}$ is given by equation~\ref{eq:idsubthr}.
In our case we can simplify the equation to
\begin{equation}
I_{D}=\frac{W}{L} \mu C_{ox} \left(\eta-1\right) V_t^2 \: e^{-\frac{V_{th}}{\eta \: V_t}}e^{\frac{V_{GS}}{\eta \: V_t}} = I_0 \: e^{\frac{V_{GS}}{\eta \: V_t}}
\label{eq:id_misure}
\end{equation}
since $V_{DS} \gg V_t$.

In order to fit the curve, we can calculate the logarithm
\begin{equation}
\log\left(I_{D}\right)= \log\left(I_0\right) +  \frac{V_{GS}}{\eta \: V_t}
\end{equation}
and proceed with a linear fit.

\begin{figure}
	\centering
	 \input{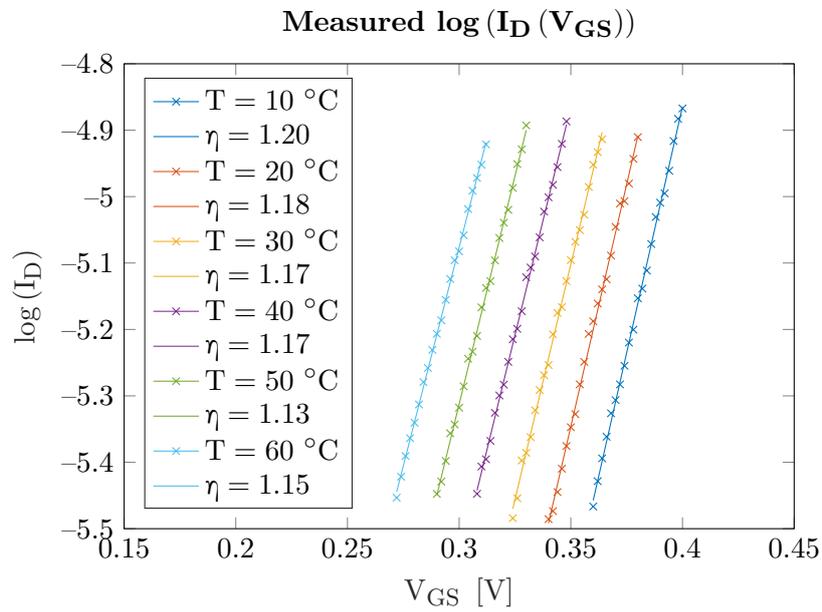} 
	\caption{The logarithm of the I-V curve was fitted around the working point of the sub-threshold MOS.}
	\label{fig:aldo_log_ids_vgs_temp}
\end{figure}

The fit plot is shown in Figure~\ref{fig:aldo_log_ids_vgs_temp}, while the results are summarized in Table~\ref{tab:fit_id}.
The threshold voltage $V_{th}$ was calculated from $I_0$, as from equation \ref{eq:id_misure}, taking into account the thermal drift of the hole mobility ($\mu \propto T^{-1.5}$).
The measured values of the sub-threshold slope $\eta$ are considerably different from the one obtained in simulation, which was 1.36 (from Figure~\ref{fig:aldo_ids_vgs_simu_fit}).

\begin{table}
\centering
{\tabulinesep=1.2mm
\begin{tabu}{c c c c c} 
\textbf{Temperature} 	& $\bm{\eta}$		& $\bm{I_0}$	& $\bm{V_{th}}$ & \textbf{$\bm{V_{GS}}$ @ $\bm{I_D=6.7\ \mu A}$}	\\ 
$[^{\circ} C]$ 			&  					& [nA]				& [mV]				& [mV]	\\ 
\hline
10 						& 1.199 			& 0.016				& 425				& 379	\\
20 						& 1.178 			& 0.036 			& 406				& 362	\\
30 						& 1.172 			& 0.086 			& 389				& 345	\\
40 						& 1.169 			& 0.209 			& 373 				& 327	\\
50 						& 1.134 			& 0.368 			& 349				& 310 	\\
60 						& 1.150 			& 0.956 			& 338				& 293	\\
\end{tabu}}
\caption{Fit results from Figure~\ref{fig:aldo_log_ids_vgs_temp}.}
\label{tab:fit_id}
\end{table}

Using the measured values, we can also calculate the threshold voltage $V_{th}$ and the forward voltage $V_{GS}$ in the working point, plotted in Figures~\ref{fig:aldo_vgs_vs_temp} and \ref{fig:aldo_vth_vs_temp}.

We can then calculate the thermal drifts $\frac{\partial V_{GS}}{\partial T}= -1.73\ mV/^{\circ}C$ and $\frac{\partial V_{th}}{\partial T}= -1.78\ mV/^{\circ}C$.

\begin{figure}
\begin{minipage}[b][][t]{.47\linewidth}
	\centering
		 \input{04_Chapter04/Figures/aldo_vgs_vs_temp.tex} 
		\caption{Temperature drift of the $V_{GS}$ voltage. The data was fitted with a linear curve and the calculated drift is $-1.73\ mV/^{\circ}C$.}
		\label{fig:aldo_vgs_vs_temp}
\end{minipage}
\ \hspace{1mm} \
\begin{minipage}[b][][t]{.47\linewidth}
	\centering
		 \input{04_Chapter04/Figures/aldo_vth_vs_temp.tex} 
		\caption{Temperature drift of the threshold voltage $V_{th}$. These values are calculated from the fit of the I-V curves. The drift is $-1.78\ mV/^{\circ}C$.}
		\label{fig:aldo_vth_vs_temp}
\end{minipage}
\end{figure}

Using equations \ref{eq:bandgap_mos1} and \ref{eq:bandgap_mos2} with the chosen design parameters, we obtain the expected reference voltage of $V_{ref,MOS} = 903\ mV$ at $20\ ^{\circ}C$ and thermal drift $\frac{1}{V_{ref,MOS}}\frac{\partial V_{ref,MOS}}{\partial T} = -210\ ppm/^{\circ}C$, which are in much better agreement with the measured values.
 
%

\begin{figure}
\centering
\input{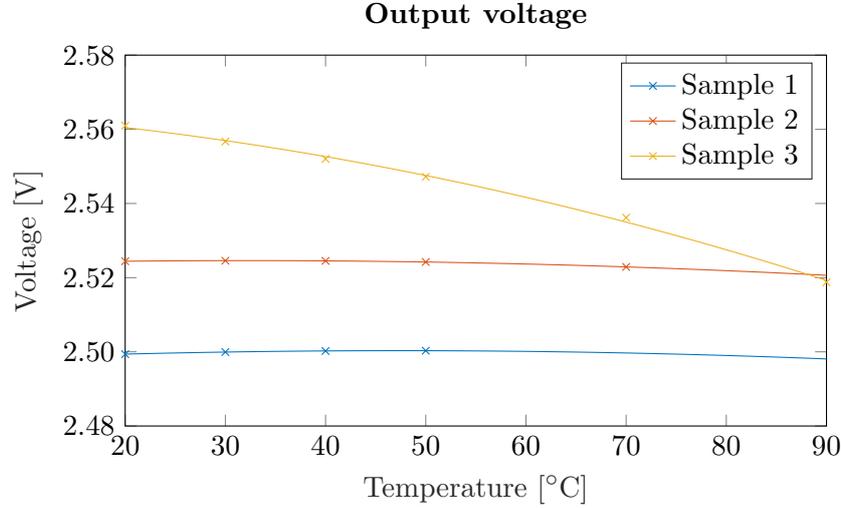} 
\caption{Temperature drift of the ALDO output voltage. Samples 1 and 2 adopt the BJT-based bandgap as voltage reference, while Sample 3 adopts the MOS-based reference.}
\label{fig:aldo_temp_output}
\end{figure}

The measured thermal drift of the output voltage is shown in Figure~\ref{fig:aldo_temp_output}.
Samples 1 and 2 adopt the BJT-based bandgap as the voltage reference of the error amplifier and in fact are more stable than sample 3, which adopts the MOS-based bandgap.
The thermal drift of the output voltage is also sensible to the relative thermal coefficient of the resistances that form the feedback network.
In this measurement, in particular, standard non-matched resistors were used, so the thermal coefficient of the output voltage is slightly worse than the bandgap references.

The measurements performed in this section demonstrated an excellent performance for the BJT-based bandgap, which exhibit a thermal drift of less than $15\ ppm/^{\circ} C$ in the desired operation range between $20\ ^{\circ}C$ and $50\ ^{\circ}C$, satisfying the specifications.
The MOS-based bandgap, on the contrary, showed a performance slightly worse than the simulated one.
The measurement of the reference diode, however, allowed to characterize its behavior and the results obtained will be used for future revisions of the chip.
In future revisions we will also evaluate the use of trimmable or external resistances which could be used to compensate for process variations.

\subsubsection{Power supply rejection}

\begin{figure}
\centering
\input{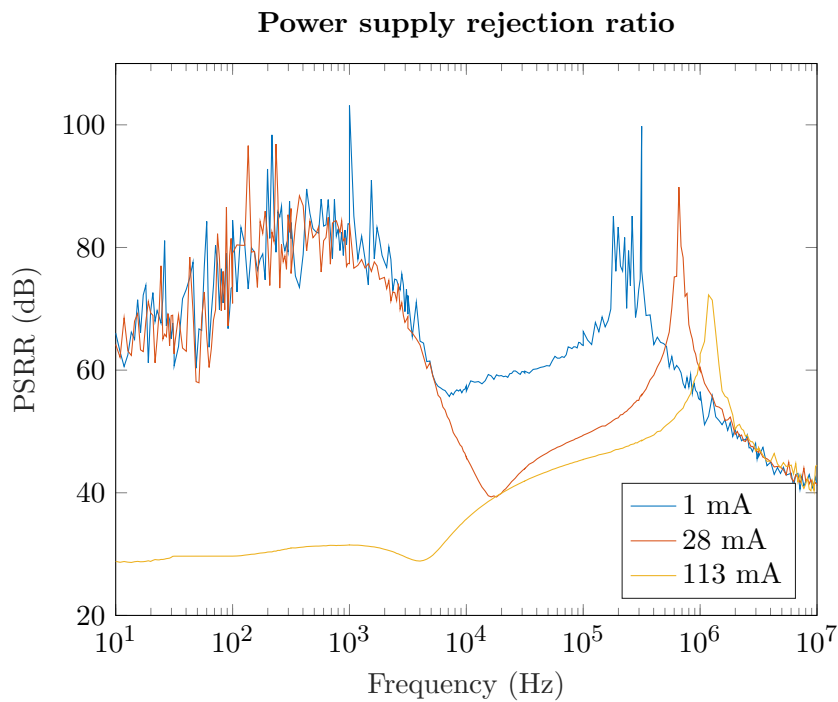} 
\caption{Power supply rejection ratio at different output currents. The dropout voltage during this measurement was set to 400 mV.}
\label{fig:aldo_psrr}
\end{figure}

Power supply rejection ratio (PSRR) was measured by injecting a sinusoidal wave of known amplitude and frequency at the input of the regulator and measuring the amplitude of the same sinusoid at the regulated output of the ALDO.
The ratio between the two amplitudes is the PSRR.
By sweeping all the frequencies of interests, in our case from 10 Hz up to 10 MHz, we reconstructed the plot shown in Figure~\ref{fig:aldo_psrr}.

The PSRR decreases with an higher output load or with lower dropout, since the output pass transistor is forced to work in linear region, where its capability to reject changes in the input voltage are reduced.
For the ALDO we measured a PSRR in excess of 40~dB at any frequency above 10~kHz, independently from the applied load.
At lower frequencies with a load of about 110~mA, the PSRR drops under 40 dB but it is still about 30~dB, which means a factor 32.

Such PSRR will allow to properly filter noisy DC/DC power supplies, as shown in the next section.

\subsubsection{Noise and FEASTMP integration}

Noise measurements are very important to qualify the performance of the ALDO ASIC. Low noise power supplies can be used to directly generate voltage references for biasing front-end chips or for generating threshold voltages. 
During this measurement the ALDO was also tested with the FEASTMP, the rad-hard DC/DC regulator that will be used in  most of the next generation detectors, in order to evaluate the integration and noise rejection of the ALDO.

\begin{figure}
\centering
\input{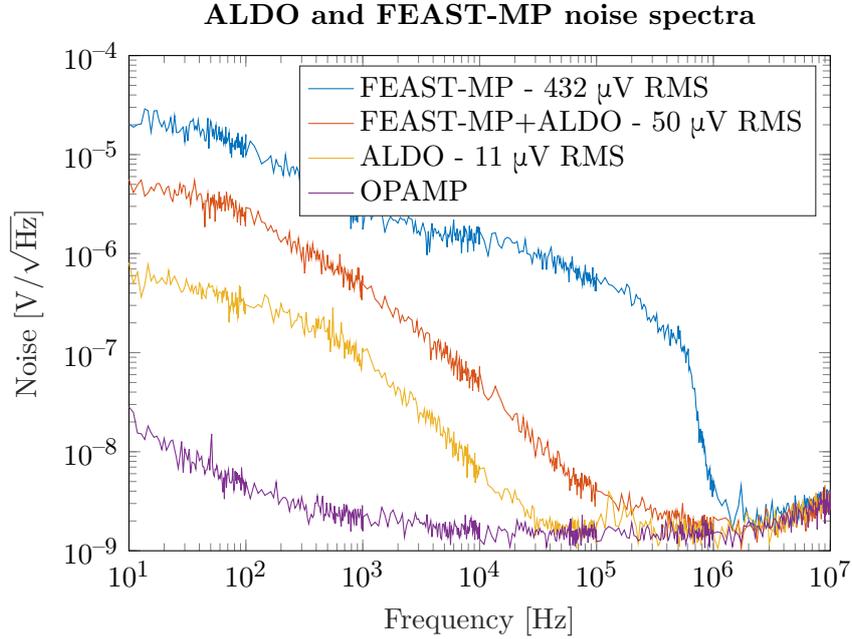} 
\caption{Noise spectra in different configurations. In \textit{blue} the noise of the FEASTMP DC/DC regulator alone. In \textit{red} the noise of the ALDO powered by the FEASTMP, where it is possible to appreciate that the ALDO is actively filtering the FEASTMP noise due to its high PSRR. In \textit{orange} the noise of the ALDO when powered by a low noise commercial power supply. In \textit{purple} the noise of the operational amplifier used for noise measurements. The input voltage of the FEASTMP is 5~V and its load is 200~mA. The input voltage of the ALDO is 2.85~V and its load is 50~mA.}
\label{fig:aldo_noise}
\end{figure}

Figure~\ref{fig:aldo_noise} shows the results of the noise measurements. A low noise operational amplifier was used to interface the device with the measurement instrumentation and its noise represent the lower measurable limit. As it can be seen, the noise of the DC/DC is quite high, even if the ripple due to the switching operation of the regulator is well filtered and is only barely visible as a small peak at 1.8~MHz. The RMS noise of the FEASTMP is $430\ \mu V$ in a bandwidth from 10~Hz to 10~MHz. As it is clearly visible from the plot, the ALDO is able to effectively filter the DC/DC noise down to $50\ \mu V$ RMS in the same frequency range. The noise of the ALDO regulator alone is even lower and was measured using a low noise power supply as input voltage. In this case the noise on the output voltage is $11\ \mu V$ RMS.

\subsection{Irradiation}

The ALDO has to withstand the radiation environment expected in the LHCb RICH detectors.
The amount of the simulated radiation levels per year in the two RICH detectors is summarized in Table~\ref{tab:rich_fluences}, already shown before.

One of the ALDO bandgap reference voltages is based on diode-connected vertical BJT transistors, which are known to be sensitive to displacement damage in the substrate.
An irradiation test with neutrons was performed in order to evaluate the radiation hardness of the ASIC with respect to this type of radiation damage.

\begin{figure}
\centering
 \input{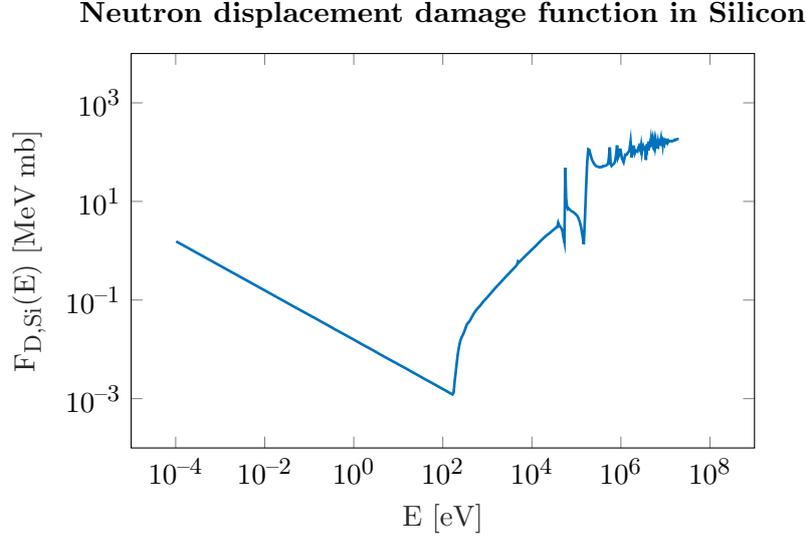} 
\caption{Neutron displacement damage function as a function of neutron energy, from~\cite{e722}.}
\label{fig:aldo_irr_damage_function}
\end{figure}

In radiation hardness tests it is common practice to quantify the neutron displacement damage in silicon with the equivalent mono-energetic (usually 1 MeV) neutron fluence.
Literature provides a standard practice for characterizing this parameter~\cite{e722}.
The equivalent mono-energetic neutron fluence $\Phi_{eq,E_{ref},mat}$ is given by the following formula:
\begin{equation}
\Phi_{eq,E_{ref},mat} = \frac{\int_{0}^{\infty} \Phi\left(E\right)F_{D,mat}\left(E\right)dE}{F_{D,mat}\left(E_{ref}\right)}
\label{eq:eq_fluence}
\end{equation}
where $\Phi\left(E\right)$ is the neutron fluence spectral distribution, and $F_{D,mat}\left(E\right)$ is the neutron displacement damage function for the irradiated material.
The actual values for $F_{D,Si}\left(E\right)$ in Silicon are provided by literature and its plot is shown in Figure~\ref{fig:aldo_irr_damage_function}. 

As it can be seen, most of the displacement damage is determined by neutrons at energies higher than 0.2 MeV.
This is why nowadays most of the neutron irradiation tests are performed with high energy neutron beams ($>10\ MeV$) in particle accelerator facilities.

\begin{figure}
\begin{minipage}[b][][t]{.35\linewidth}
	\centering
 	\includegraphics[width=\linewidth]{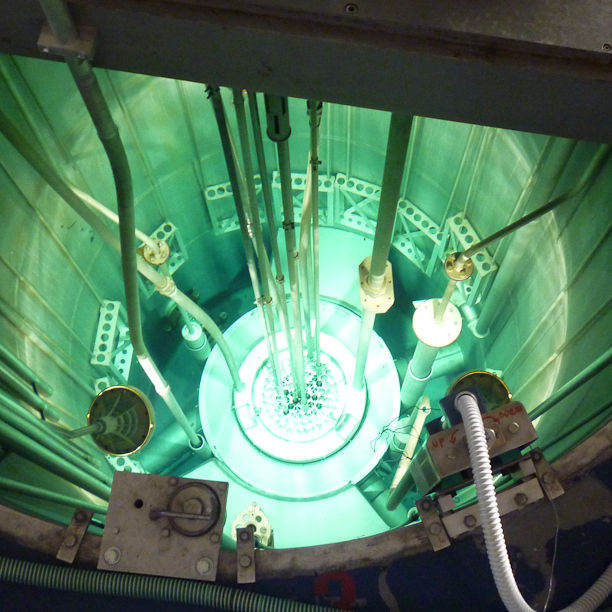}
    \caption{Photograph of the Triga Mark II reactor at LENA.}
    \label{fig:lena_reattore}
\end{minipage}
\ \hspace{4mm} \
\begin{minipage}[b][][t]{.58\linewidth}
	\centering
	\def\svgwidth{.85\linewidth}
	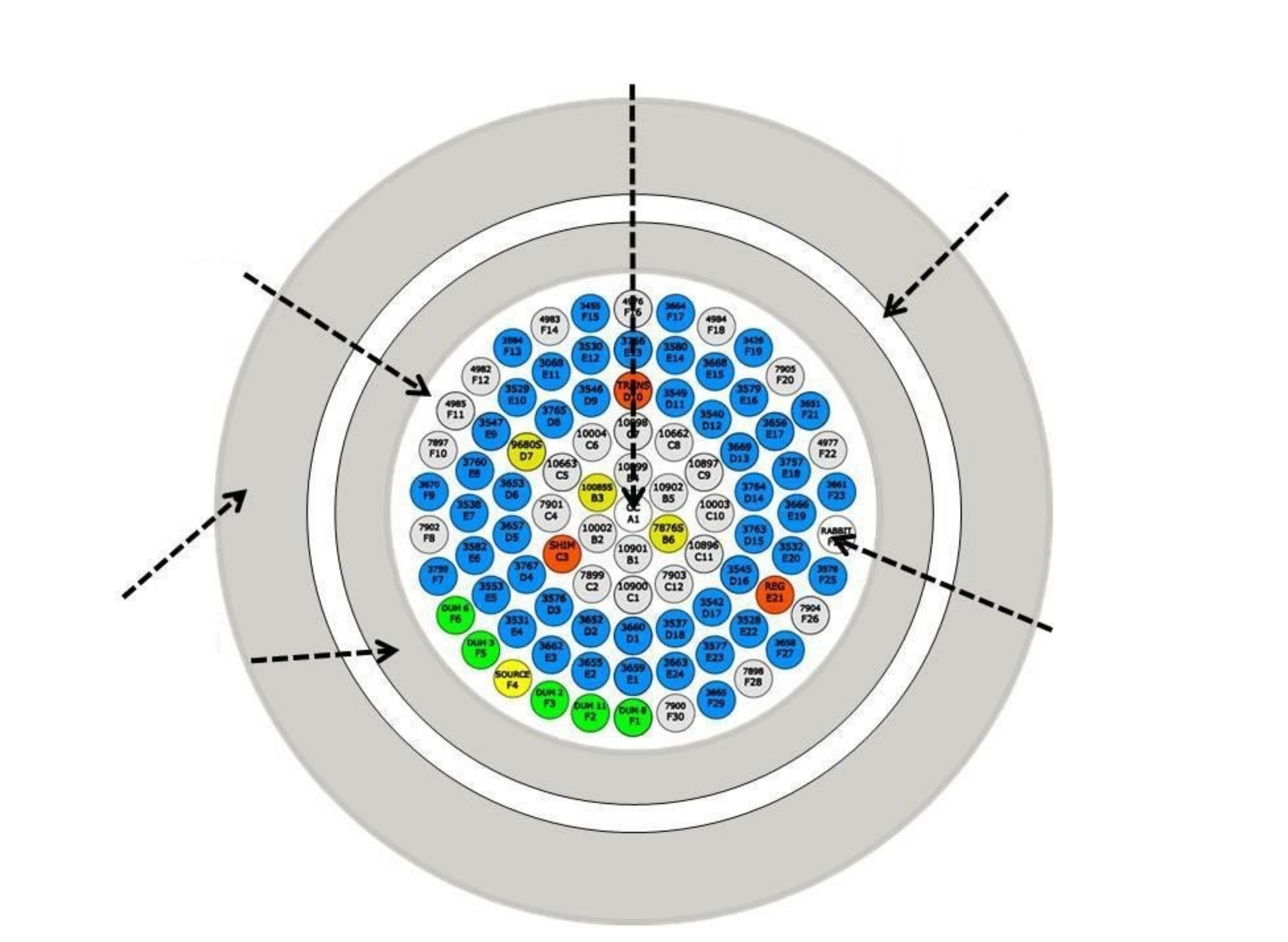
	\caption{Drawing of the reactor core with the available measurement channels (Central Thimble, Rabbit Channel, Lazy Susan).}
	\label{fig:lena_schema}
\end{minipage}
\end{figure}

We decided to follow a more classic approach by using the neutron source provided by the Laboratorio Energia Nucleare Applicata (LENA) in Pavia, which operates a 250 kW TRIGA Mark II research nuclear reactor.
Figure~\ref{fig:lena_reattore} shows a photograph of the reactor. 
The reactor offers several channels for irradiation purposes, each one characterized by different neutron and gamma spectra~\cite{di2014triga}.
For the ALDO irradiation, the Central Thimble that runs directly inside the core of the reactor was chosen, since it offers the hardest spectrum, with a large contribution of high energy neutrons.
Figure~\ref{fig:lena_schema} illustrates the scheme of the reactor core, showing the channels that can be used for measurement purposes.

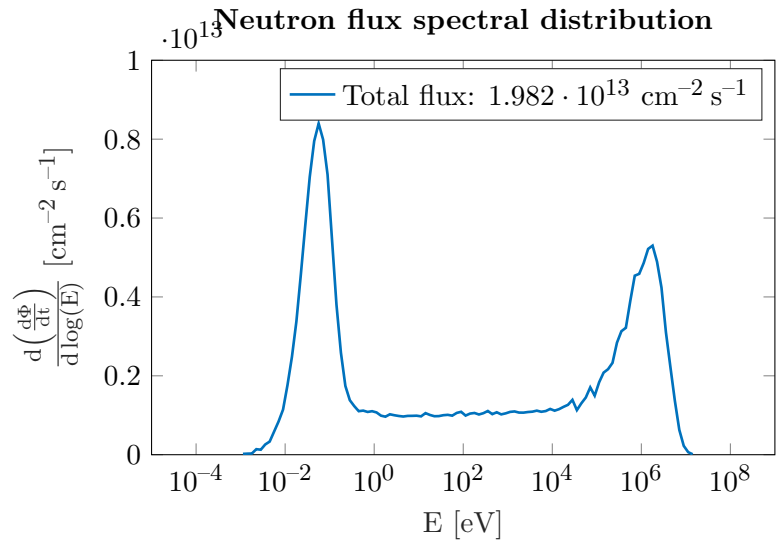
\begin{figure}
\centering
\definecolor{mycolor1}{rgb}{0.00000,0.44700,0.74100}%
\begin{tikzpicture}

\begin{axis}[%
width=3.229in,
height=2.059in,
at={(0.542in,0.475in)},
scale only axis,
xmode=log,
xmin=1e-05,
xmax=1000000000,
xminorticks=true,
xlabel style={font=\color{white!15!black}},
xlabel={E [eV]},
ymin=0,
ymax=10000000000000,
ylabel style={font=\color{white!15!black}},
ylabel={$\frac{d\left(\frac{d\Phi}{dt}\right)}{d\log(E)}\ [cm^{-2}\:s^{-1}]$},
axis background/.style={fill=white},
title style={font=\bfseries},
title={Neutron flux spectral distribution},
legend style={legend cell align=left, align=left, draw=white!15!black}
]
\addplot [color=mycolor1, line width=1.0pt]
  table[row sep=crcr]{%
0.00112945	21237481774.3499\\
0.0014219	26922637622.3435\\
0.0017901	30529573295.9973\\
0.0022536	140226808559.529\\
0.0028371	126600973784.317\\
0.0035717	259517671791.875\\
0.0044965	332967095759.8\\
0.00566075	587817146585.767\\
0.00712645	842860191055.647\\
0.00897165	1143128112940.64\\
0.0112945	1765181783872.32\\
0.014219	2482154500760.9\\
0.017901	3378296712364.1\\
0.022536	4569537220698.42\\
0.028371	5823470747335.41\\
0.035717	7044532691404.28\\
0.044965	7952637232279.52\\
0.0566075	8396099801035.16\\
0.0712645	7988891717332.73\\
0.0897165	7100070526577.02\\
0.112945	5452061152504.71\\
0.14219	3804563104104.21\\
0.17901	2586056117816.85\\
0.22536	1743092846426.35\\
0.28371	1381619502997.61\\
0.35717	1231565771223.76\\
0.44965	1103428807981.48\\
0.566075	1117841404142.58\\
0.712645	1084467555316.61\\
0.897165	1102996725611.17\\
1.12945	1071794518387.53\\
1.4219	990024022013.03\\
1.7901	963870838086.063\\
2.2536	1024714693705.71\\
2.8371	1000598254442.49\\
3.5717	981822617616.168\\
4.4965	968319196420.321\\
5.66075	983882481481.906\\
7.12645	985343620209.606\\
8.97165	989231347600.673\\
11.2945	972327558660.951\\
14.219	1054095490023.02\\
17.901	1007503553016.69\\
22.536	975849031826.724\\
28.371	978858091433.369\\
35.717	1001352053988.37\\
44.965	1010706448674.28\\
56.6075	989982036208.149\\
71.2645	1058785453766.23\\
89.7165	1088133902162.11\\
112.945	991884193084.571\\
142.19	1042383775196.37\\
179.01	1057856682064.78\\
225.36	1019734042007.58\\
283.71	1052816197364.33\\
357.17	1108230834134.52\\
449.65	1028160369016.52\\
566.075	1075961845712.08\\
712.645	1022674102246.42\\
897.165	1048255357272.95\\
1129.45	1085517681422.48\\
1421.9	1097024725845.95\\
1790.1	1068012758459.26\\
2253.6	1066435737151.13\\
2837.1	1081984706731.6\\
3571.7	1090988717337.42\\
4496.5	1116751019836.2\\
5660.75	1088513757173.16\\
7126.45	1106390344436.73\\
8971.65	1160372359374.09\\
11294.5	1116014034871.69\\
14219	1154836619365.26\\
17901	1213398305163.15\\
22536	1264587662168\\
28371	1390768716928.96\\
35717	1130492509999.18\\
44965	1302822942775.11\\
56607.5	1449953534937.6\\
71264.5	1707215887545.39\\
89716.5	1496997768325.9\\
112945	1837616013786.72\\
142190	2081200729612.2\\
179010	2167897017174.64\\
225360	2328929996716.37\\
283710	2839843092818.92\\
357170	3131177818350.94\\
449650	3217205382164.19\\
566075	3903009420780.48\\
712645	4538693553698.02\\
897165	4585405456199.85\\
1129450	4859324790536.41\\
1421900	5216984937365.2\\
1790100	5300327310455.67\\
2253600	4892368127696.77\\
2837100	4234608220888.85\\
3571700	3101209423917.88\\
4496500	2224830939886.44\\
5660750	1369257915517.84\\
7126450	630447278644.86\\
8971650	224518206945.264\\
11294500	64115848687.2255\\
14228500	11444337345.8542\\
};
\addlegendentry{Total flux: $1.982\cdot10^{13}\ cm^{-2}\:s^{-1}$}

\end{axis}
\end{tikzpicture}%
\caption{Neutron spectrum in the Central Thimble channel of the LENA reactor.}
\label{fig:aldo_irr_neutron_spectrum_log}
\end{figure}

The total neutron flux ($\frac{d\Phi}{dt}$) in the Central Thimble at full 250 kW power is $1.982 \cdot 10^{13}\ cm^{-2}\ s^{-1}$ and its spectral distribution is shown in Figure~\ref{fig:aldo_irr_neutron_spectrum_log}.
It can be seen that the fast neutron contribution is almost as relevant as the thermal neutron contribution, and is as high as $5 \cdot 10^{12}\ cm^{-2}\ s^{-1}$ at 1 MeV.
This is not true for the other channels, where a large fraction of the fast neutrons has been already thermalised by the moderators.
Another important motivation that drove the choice of the Central Thimble is that this is the channel where simulations are more accurate, with statistical and systematic errors of about 10\% for the simulated flux.

\begin{figure}
\centering
 \input{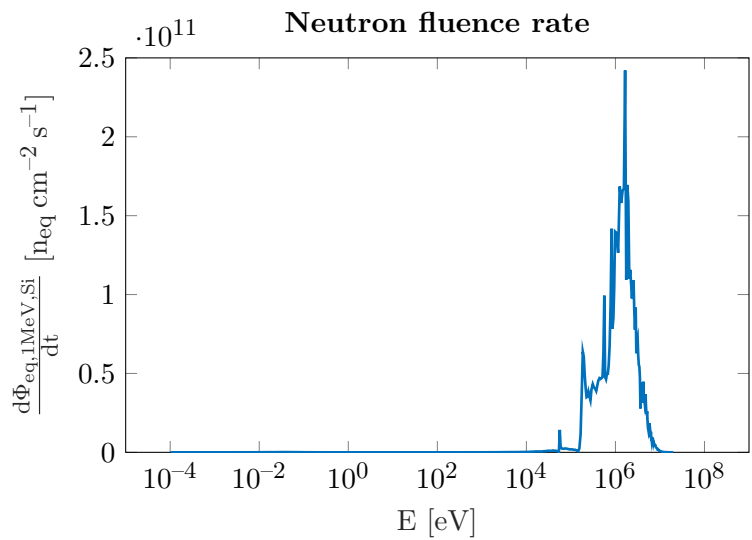} 
\caption{Fluence rate as a function of neutron energy. As it is possible to see, most of the damage comes from the hardest part of the spectrum, with a peak at about 1 MeV.}
\label{fig:aldo_irr_fluencem}
\end{figure}

Using Equation \ref{eq:eq_fluence}, the 1-MeV-equivalent neutron fluence can be calculated. Its rate is shown in Figure~\ref{fig:aldo_irr_fluencem} as a function of energy. At full 250 kW power, we obtain:
\begin{equation}
\frac{d\Phi_{eq,1MeV,Si}}{dt} = 5.943 \cdot 10^{12}\ n_{eq}\ cm^{-2}\ s^{-1}.
\end{equation}

Since the neutron flux scales proportionally to the reactor power, we have two degrees of freedom, power and time of irradiation, in order to reach the desired fluence, which can be set to the expected detector lifetime (10 years) plus a safety factor.

Three steps of irradiation were chosen, with a final accumulated fluence of $4.3 \cdot 10^{13} cm^{-2}$ corresponding to about 70 years of operation in the hottest region of the RICH-1.
The irradiation was performed in three steps, with increasing reactor power, in order to better study the behavior of the ASIC during the irradiation.
More details about the steps performed can be found in Table \ref{tab:irr_steps}.

\begin{table}
\centering
{\tabulinesep=1.2mm
\begin{tabu}{c|c c c M{13mm} c M{13mm}} 
	 			& \textbf{Power} 	& \textbf{Duration}	& \multicolumn{2}{c}{\textbf{Neutrons}}	&	\multicolumn{2}{c}{\textbf{TID}}	\\
 				& [W] 				& [hours] 			& [$n_{eq}\ cm^{-2}$]	& [LHCb years]		& [krad]	& [LHCb years]	\\
\hline 
\textbf{Step 1} & 50 				& 1 				& $0.43 \cdot 10^{13}$ 	& 7					& 5.0		& 0.13	\\ 
\textbf{Step 2} & 100 				& 2 				& $1.71 \cdot 10^{13}$ 	& 28				& 20.2		& 0.51	\\ 
\textbf{Step 3} & 250 				& 1 				& $2.14 \cdot 10^{13}$ 	& 35				& 25.2		& 0.63	\\ 
\hline
\textbf{Total} 	& --	 			& 4 				& $4.28 \cdot 10^{13}$ 	& 70				& 50.4		& 1.26	\\ 
\end{tabu}}
\caption{Irradiation steps performed.}
\label{tab:irr_steps}
\end{table}

During the nuclear reaction, a large quantity of high energy gammas is also produced.
LENA did not provide any simulation model or spectrum, but only stated that the total ionizing dose (TID) at full power in the Central Thimble is 70 Gy/s (7 krad/s).
Scaling this number to the chosen powers and durations, gives a total TID of 50.4 krad, corresponding to about 1.26 years of operation in RICH-1.
This dose is not enough to qualify the chip for the whole detector operation, thus an irradiation with a Cobalt-60 source or other ionizing source is planned for the following months.

\begin{figure}
\begin{minipage}[b][][t]{.3\linewidth}
	\centering
 	\includegraphics[width=.7\linewidth]{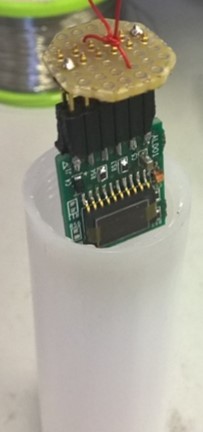}
    \caption{Photograph of the ALDO test board.}
    \label{fig:aldo_irr_photo}
\end{minipage}
\ \hspace{3mm} \
\begin{minipage}[b][][t]{.65\linewidth}
	\centering
	\includegraphics[width=\linewidth]{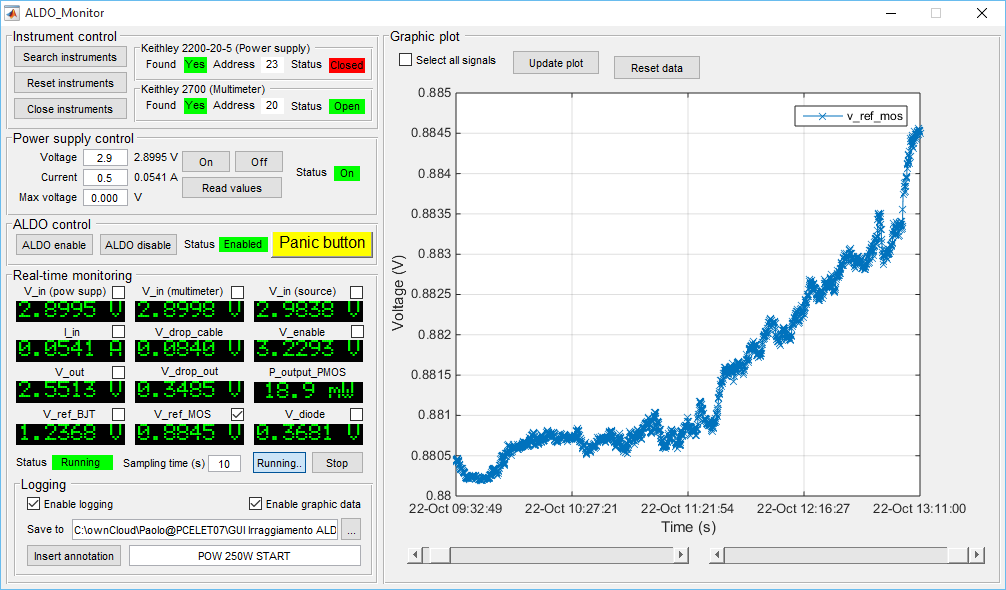}
	\caption{MATLAB GUI for remote control, online monitoring and data taking.}
	\label{fig:aldo_irr_gui}
\end{minipage}
\end{figure}

The Central Thimble can be accessed through a \mytilde 6 m tube with a diameter of a few centimetres. The ASIC board was inserted in a specific plastic container provided by LENA, shown in Figure~\ref{fig:aldo_irr_photo}. During the irradiation the chip was fully operational. Power supplies, output and monitor voltages were connected to the ASIC through a long cable with 8 wires. The ALDO was operating with a 400 mV dropout and with an output current of 50 mA, to simulate an average load at LHCb RICH. A Matlab GUI was developed for instrument remote control, online monitoring and data taking, shown in Figure~\ref{fig:aldo_irr_gui}.

\begin{figure}
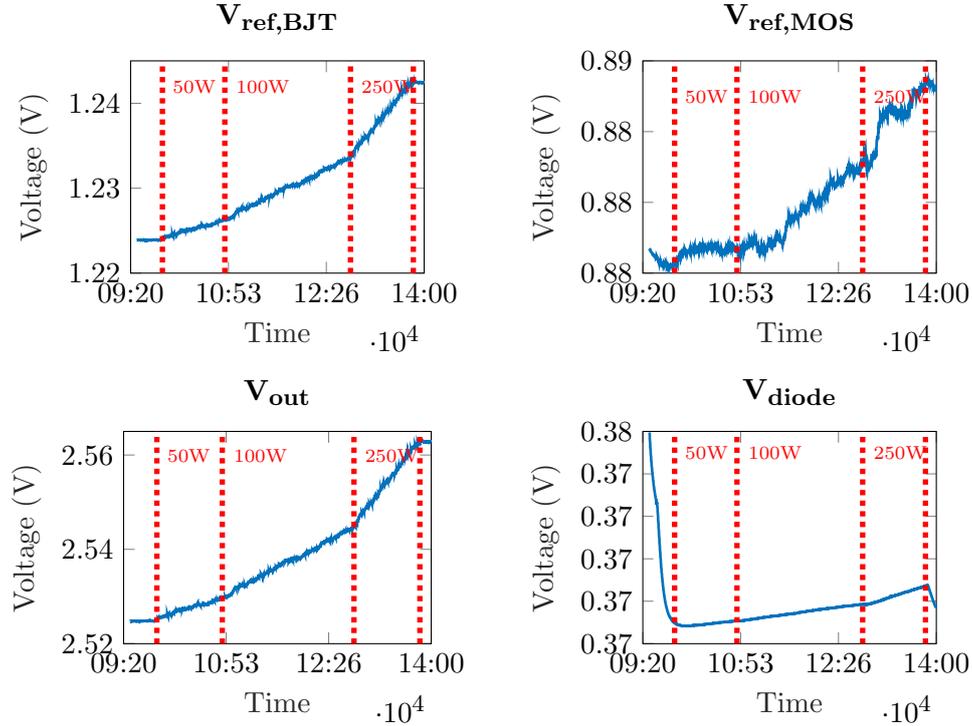

\centering
\begin{minipage}[b][][t]{.48\linewidth}
	\centering
 	\input{04_Chapter04/Figures/v_ref_bjt.tex}
\end{minipage}
\ \hspace{0.01mm} \
\begin{minipage}[b][][t]{.48\linewidth}
	\centering
	\input{04_Chapter04/Figures/v_ref_mos.tex}
\end{minipage} \\
\vspace{2mm}
\begin{minipage}[b][][t]{.48\linewidth}
	\centering
 	\input{04_Chapter04/Figures/v_out.tex}
\end{minipage}
\ \hspace{0.01mm} \
\begin{minipage}[b][][t]{.48\linewidth}
	\centering
	\input{04_Chapter04/Figures/v_diode.tex}
\end{minipage}
\caption{Behavior of different voltages during irradiation. Top-left plot shows the bipolar-based voltage reference, top-right shows the MOS-based voltage reference, bottom-left shows the output voltage of the regulator, and bottom-right shows the MOS diode forward voltage when biased with 10 $\mu A$.}
\label{fig:aldo_irr_mon_plots}
\end{figure}

\begin{figure}
\centering
\input{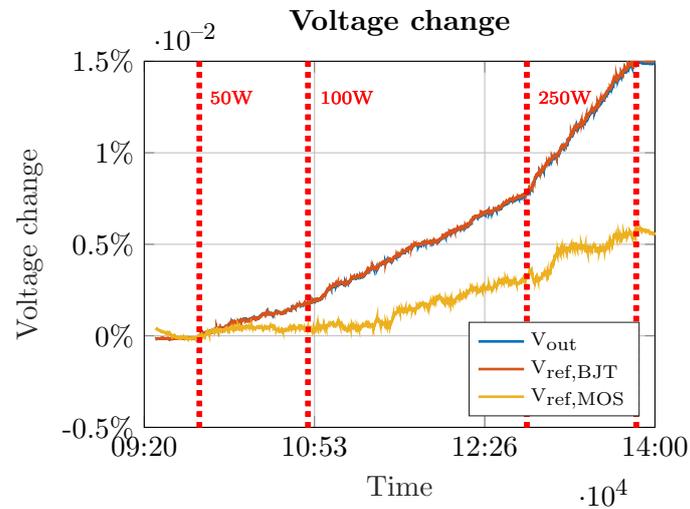} 
\caption{Voltage change during irradiation. $V_{out}$ is generated using $V_{ref,BJT}$, so the voltage change is the same and the two curves are indistinguishable).}
\label{fig:aldo_irr_volt_change}
\end{figure}

The data recorded during irradiation is shown in Figure~\ref{fig:aldo_irr_mon_plots}, while Figure~\ref{fig:aldo_irr_volt_change} show the fractional change from the start of the irradiation. The bipolar-based voltage reference exhibits a voltage change of 1.5\% over the whole irradiation period. The output voltage has a similar behaviour since it is generated using this reference. This means that no appreciable changes of the input voltage offset and bias current are present. The MOS-based voltage reference, as expected, shows a lower voltage change of 0.59\%, due to the lower sensitivity to displacement damage. Neither of the two bandgaps exhibit unacceptably high performance degradation, and both of them behave well, with a voltage reference degradation in LHCb RICH hottest regions of 270 ppm/y for the bipolar-based reference and 95 ppm/y for the MOS-based reference. Low dose effects can be appreciated, with a slightly higher degradation at lower irradiation fluences.

The on-chip diode was also biased and read out during irradiation, in order to monitor any temperature change during the irradiation. On the diode, temperature changes should appear as exponentially shaped slopes over time, while radiation damage should appear as a linear slope. As it can be seen in the bottom-right plot in Figure~\ref{fig:aldo_irr_mon_plots}, apart from the initial temperature stabilization when the device was inserted in the reactor, the only changes are due to the radiation damage.

No single event effects were detected during the whole irradiation process and the ALDO worked flawlessly without the need of any power cycle.

\begin{figure}
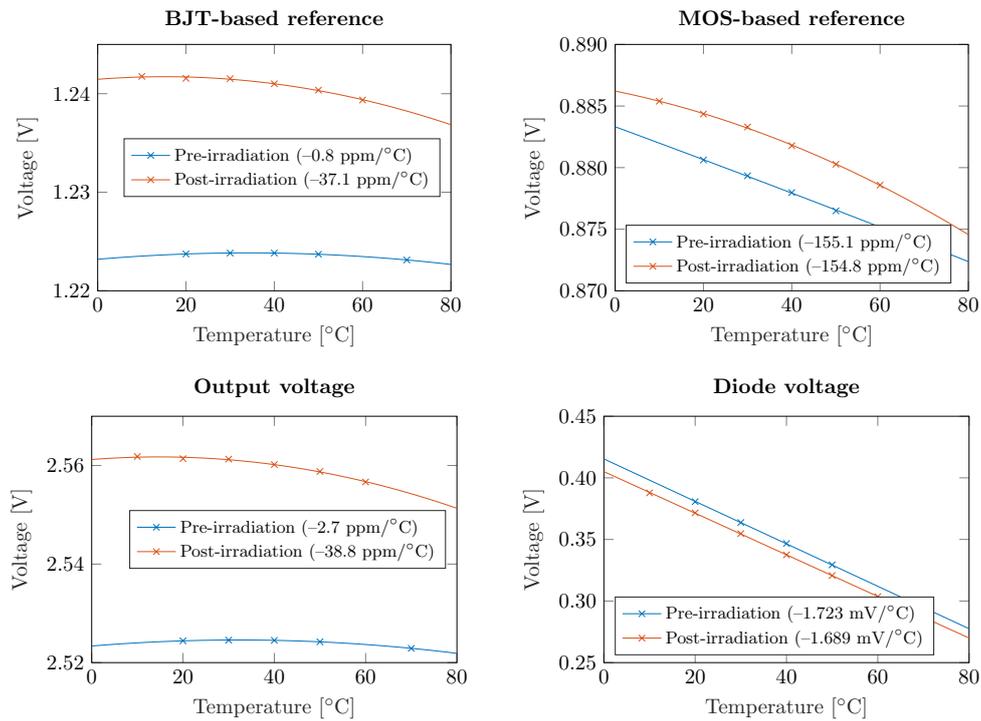

\centering
\begin{minipage}[b][][t]{.48\linewidth}
	\centering
 	\input{04_Chapter04/Figures/aldo_temp_vref_bjt_irr.tex}
\end{minipage}
\ \hspace{0.01mm} \
\begin{minipage}[b][][t]{.48\linewidth}
	\centering
	\input{04_Chapter04/Figures/aldo_temp_mos_ref_irr.tex}
\end{minipage} \\
\vspace{2mm}
\begin{minipage}[b][][t]{.48\linewidth}
	\centering
 	\input{04_Chapter04/Figures/aldo_temp_output_irr.tex}
\end{minipage}
\ \hspace{0.01mm} \
\begin{minipage}[b][][t]{.48\linewidth}
	\centering
	\input{04_Chapter04/Figures/aldo_temp_diode_irr.tex}
\end{minipage}
\caption{Pre and post irradiation thermal drifts. The numbers in brackets are calculated from 20~$^{\circ}C$ to 50~$^{\circ}C$.}
\label{fig:aldo_irr_temp_plots}
\end{figure}

After the irradiation took place, the device was re-tested in the climatic chamber. A comparison between pre and post irradiation behaviors are shown in Figure~\ref{fig:aldo_irr_temp_plots}. As it can be seen, the MOS-based reference exhibits the least change in both the output voltage and thermal drift, while BJT-based reference and thus the output voltage show a degradation of the thermal drift from 20~$^{\circ}C$ to 50~$^{\circ}C$ from a few ppm/$^{\circ}C$ to almost 40~ppm/$^{\circ}C$. This drift, however, is still better than the drift of the MOS-based bandgap. The diode reflects the behavior of the MOS-based bandgap with almost no change in its voltage drift. 

From these results we can conclude that the radiation hardness to non-ionizing energy losses is satisfactory for both the references. The MOS-based bandgap offers the smaller voltage change, while the BJT-based one still offers the smaller thermal drift. For future revisions, when the thermal drift of the MOS-based reference will be optimized, this seems to be the best choice.

\subsection{Summary and future prospects}

The ALDO ASIC was fully characterized both on the testbench for electrical and thermal performance, and in radiation environment with neutrons, up to a fluence of $4.3 \cdot 10^{13}\ n_{eq}\ cm^{-2}$. The ALDO proved to improve both the power supply noise and the thermal stability of the FEASTMP DC/DC regulator which is the baseline option for the supply voltage generation for the CLARO chips in the upgraded LHCb RICH detectors.

A final decision on the inclusion of the ALDO is still pending until the final revision of the PDMDB digital board will be available and the CLARO will be tested at the full 40~MHz rate. ALDO will be used as a backup solution in case power supply noise and thermal stability will not satisfy the required standards.

Other groups, like the CMS barrel timing layer group, have expressed their interest for the adoption of the ALDO and for the development of new customized and improved versions, which are currently under development.


\chapter{The CUORE and CUPID experiments}
\label{Chapter03}
\thispagestyle{empty}

\section{Introduction}

Neutrinos are probably the most elusive, and thus least ``understood'', among the fundamental particles of the Standard Model.
Neutrinos are neutral fermions that do not interact either with electromagnetic or strong forces and
they exist in three flavors, each one associated to a charged lepton: electronic neutrino ($\nu_e$), muonic neutrino ($\nu_\mu$) and tauonic neutrino ($\nu_\tau$).
The original Standard Model formulation classified neutrinos as massless particles, the only massless particle apart from the two vector bosons of the electromagnetic and strong forces, the photon and the gluon.
Nevertheless, several experiments (Super-Kamiokande~\cite{superkamiokande}, SNO~\cite{sno}, KamLAND~\cite{kamland}, K2K~\cite{k2k}) have demonstrated the existence of neutrino oscillations between the three flavors, which can only be explained by massive and mixed neutrinos.

The formalism that explains such scenario was pioneered in the work by Gribov and Pontecorvo~\cite{gribov} where they first theorized that the three neutrino flavor eigenstates ($\nu_e$, $\nu_\mu$ and $\nu_\tau$) are not identical to the mass eigenstates, but are instead a superposition of the three mass eigenstates ($\nu_1$, $\nu_2$ and $\nu_3$).
The Pontecorvo-Maki-Nakagawa-Sakata (PMNS) matrix~\cite{mns} determines the mixing of the three mass eigenstates so that
\begin{equation}\label{eq:Pontecorvo}
|\nu_{\alpha} \rangle=\sum_{i}U_{\alpha ,i}|\nu_{i} \rangle \ ,
\end{equation}
where $\alpha$ represents the flavor family ($e$, $\mu$, $\tau$), $i$ represents the mass family (1, 2, 3), and $U_{\alpha ,i}$ is one of the element of the PMNS matrix.

The probability that a flavored neutrino converts to another flavor is not zero, but oscillates with respect to time and space.
The wavelength of such an oscillation for ultra-relativistic neutrinos is proportional to $\Delta m_{i,j}^2$.

The aforementioned experiments confirmed the existence of this phenome-non beyond the SM, and provided an estimation of $\Delta m_{i,j}^2$~\cite{neutrino_pdg}:

\begin{equation} \label{eq:masse}
\Delta m^{2}_{1,2} \approx 7.54 \cdot 10^{-5} eV^{2} \ll \Delta m^{2}_{2,3} \approx 2.42 \cdot 10^{-3} eV^{2} \ .
\end{equation}

The observation of neutrino oscillations has been one of the most groundbreaking discoveries of the latest decades and it was awarded the Nobel Prize in 2015.

Neutrino oscillations, however, have opened much more questions than they have answered, since they do not allow to determine neither the mass absolute value nor the sign of the $\Delta m$ coefficients, so that two neutrino mass hierarchies can be hypothesized: direct hierarchy, for which the neutrino mass scales proportionally to that of the charged lepton masses, or inverse hierarchy.
Furthermore, neutrino oscillations cannot tell us if the neutrino is a Dirac or Majorana, (i.e. differs or coincides with its own antiparticle, respectively).

It is thus clear than neutrino physics represent an exciting and promising field for the discovery of new physics beyond the Standard Model.

\subsection{Neutrinoless double beta decay}
\label{sec:ndbd}

Beta decay ($\beta$) is a type of radioactive decay in which an atomic nucleus emits a fast energetic electron or positron.
In $\beta^-$ decay, a neutron transforms into a proton, with an emission of an electron and an electronic antineutrino, while in a $\beta^+$ decay, a proton transforms into a neutron, with an emission of a positron and an electronic neutrino.
The former increases the atomic number Z of one unity, while the latter decreases it.
Both the total baryon and lepton numbers are conserved.
\begin{equation}
\begin{split}
\beta^- &:\ \ \ ^{A}_{Z}X \longrightarrow \ce{^{A}_{Z+1}X}' + e^{-} + \bar{\nu_e}    \\
\beta^+ &:\ \ \ ^{A}_{Z}X \longrightarrow \ce{^{A}_{Z-1}X}' + e^{+} + {\nu_e}  \ .  \\
\end{split}
\end{equation}

\begin{figure}
\centering
\includegraphics[width=\linewidth]{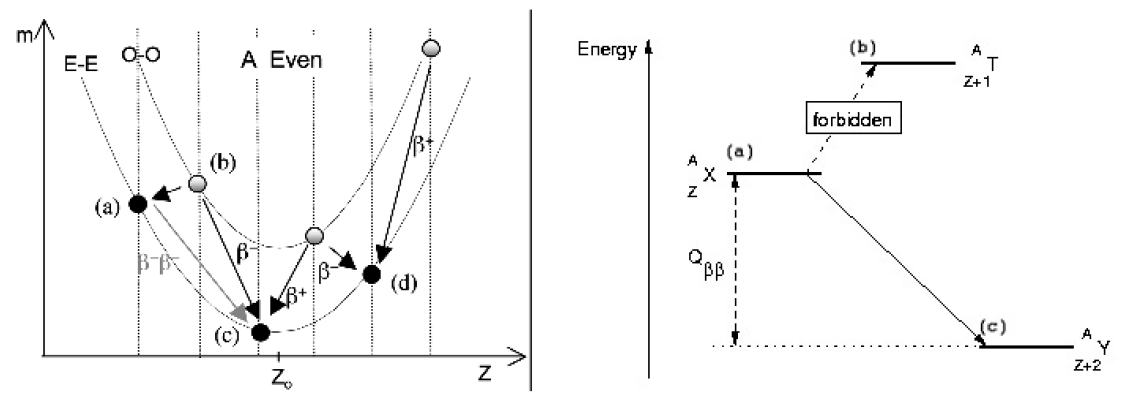}
\caption{On the left, a plot of the nucleus mass as a function of atomic number $Z$ in a beta decay chain for an even mass number $A$. On the right, a zoom on the binding energy for cases a (local minimum), b (forbidden state) and c (global minimum).}
\label{fig:dbd}
\end{figure}

With this process, unstable atoms can reach a more stable energetic configuration and the probability of a nuclide decaying due to beta decay is determined by its nuclear binding energy.
For a given mass number $A$, only one particular configuration of nuclides have the highest binding energy.
The expression for the binding energy $E_{binding}$ is given by the semi-empirical mass formula (SEMF or Weizs\"{a}cker's formula)~\cite{weizsacker} and it has a parabolic dependence on the atomic number $Z$:
\begin{equation}
-E_{binding} = a(A)\:Z^2 - b(A)\:Z + \delta\left(A,Z\right) + c(A) \ ,
\end{equation}
with $a(A)$, $b(A)$ and $c(A)$ being constants dependent on $A$.
The $\delta$ term is worth to be explicitly shown:
\begin{equation}
\delta = \begin{cases}
            0  & A\ odd\\
            +d(A) & A,Z\ odd\\
            -d(A) & A,Z\ even
         \end{cases} \ .
\end{equation}
Therefore, for even mass numbers $A$, there are two distinct parabolas that could lead to local minima in the beta decay chain.
An example of such an eventuality is shown in Figure \ref{fig:dbd}, where a local minimum can be seen in the (a) and (d) cases.

\begin{figure}
\centering
\includegraphics[width=.8\linewidth]{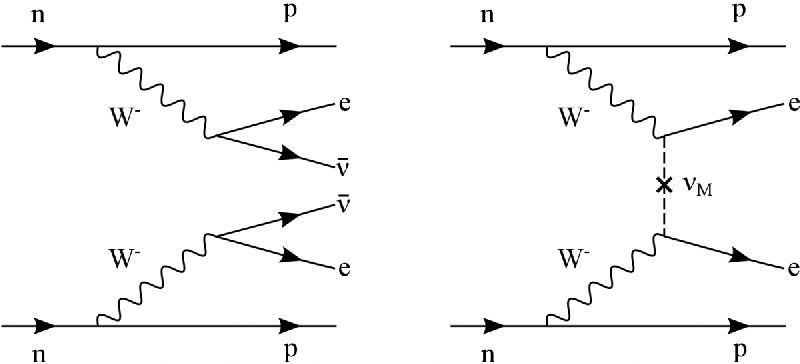}
\caption{Feynman diagrams of $2\nu\beta\beta$ (on the left) and $0\nu\beta\beta$ (on the right).}
\label{fig:dbd_feynman}
\end{figure}

When a local minima is reached (a), no spontaneous beta decay can happen since such a process would lead to a lower binding energy state (b).
However, in 1935, Maria Goeppert-Mayer~\cite{mgm} first proposed the existence of double beta decay ($\beta\beta$), a rare second-order weak process that consists in the simultaneous decay of two protons or neutrons, which can convert an atom in (a) state directly to the most stable nuclide configuration (c).
This process was directly observed for the first time in 1987~\cite{2nudbd} and since then several experiments observed it for different nuclei~\cite{2nudbd_nuclei} with very long half-lives, typically in the range $10^{18} - 10^{22} \ y$.
The Feynman diagram for such decay is shown on the left part of Figure~\ref{fig:dbd_feynman}.

An alternative process, the neutrinoless double beta decay ($0\nu\beta\beta$), was proposed by Furry~\cite{furry}, after the Majorana theory of the neutrino.
The neutrinoless decay can only take place if the neutrino is a massive Majorana particle and demands an extension of the Standard Model of the electroweak interactions, because it violates the lepton number conservation.
Therefore, the observation of the double beta decay without emission of neutrinos will sign the Majorana character of the neutrino.
The Feynman diagram for \ndbd is shown in Figure~\ref{fig:dbd_feynman}, on the right.

The corresponding nuclear reactions are the following:
\begin{equation}
\begin{split}
2\nu\beta\beta &:\ \ \ ^{A}_{Z}{X}_{N} \longrightarrow \ce{^{A}_{Z+2}X}'_{N-2} + 2e^{-} + 2\bar{\nu_e} \ ,   \\
0\nu\beta\beta &:\ \ \ ^{A}_{Z}{X}_{N} \longrightarrow \ce{^{A}_{Z+2}X}'_{N-2} + 2e^{-} \ .\\
\end{split}
\end{equation}

The theoretical expression of the half-life of the process can be written as
\begin{equation}
\left(T^{0\nu}_{\sfrac{1}{2}}\right)^{-1} = G_{0\nu} \left|\mathcal{M}\right|^2 \frac{\left|m_{\beta\beta}\right|^2}{m_e} \ .
\end{equation}
$G_{0\nu}$ is the phase space factor (PSF) of the transition (it represents the pure kinematic
contribution to the \ndbd), $\mathcal{M}$ is the nuclear matrix element (NME) (it takes into account the nuclear structure aspects), $m_e$ is the electron mass at rest, and $m_{\beta\beta}$ is the effective Majorana mass, defined as 
\begin{equation}
m_{\beta\beta} = \left| \sum_{i=1,2,3}U_{e,i}^2 m_i \right| \ .
\end{equation}

A measure of the half-life of \ndbd would therefore provide a measurement of the effective Majorana mass, and also provide constraints on the neutrino mass hierarchy.

Currently, there is a number of experiments either taking place or planned for the near future devoted to the search of this process, using different experimental techniques.
Most stringent limits on the lifetime are of the order of $10^{22}\ y$ to $10^{25}\ y$, depending on the chosen isotope~\cite{henning}.
A discussed claim for the existence of \ndbd decay in $^{76}Ge$ declares that its half-life is $2.2 \cdot 10^{25} \ y$~\cite{klapdor}.

In the case of a \ndbd peak showing up in the energy spectrum, the half-life can be evaluated from the law of radioactive decay:
\begin{equation}
T^{0\nu}_{\sfrac{1}{2}} = \ln 2 \: T \: \epsilon \: \frac{N_{\beta\beta}}{N_{peak}} \ ,
\end{equation}
where $T$ is the measuring time, $\epsilon$ is the detection efficiency, $N_{\beta\beta}$ is the number of decaying nuclei under observation, and $N_{peak}$ is the number of observed decays in the peak.

If no peak is detected, it is still possible to introduce a physical quantity that describes the sensitivity of the experiment.
The sensitivity $S^{0\nu}$ to the \ndbd is defined as the corresponding process half-life that could be hidden by the background fluctuations ($\sqrt{N_B}$), with a given confidence level ($n_\sigma$):
\begin{equation}
\label{eq:sensitivity0}
S^{0\nu} = \ln 2 \: T \: \epsilon \: \frac{N_{\beta\beta}}{n_\sigma \sqrt{N_{B}}} \ .
\end{equation}
Sensitivity can be rewritten in order to explicitly show some of the experimental characteristics, assuming a background that scales proportionally with detector mass:
\begin{equation}
\label{eq:sensitivity}
S^{0\nu} = \ln 2 \: \epsilon \: \frac{1}{n_\sigma} \frac{x \: \eta \: N_A }{\mathcal{M}_A} \sqrt{\frac{M \: T}{B\: \Delta E}} \ ,
\end{equation}
where $B$ is the background level per unit mass, energy,
and time, $M$ is the detector mass, $\Delta E$ is the FWHM energy resolution, x is the stoichiometric multiplicity of the element containing the candidate, $\eta$ is the candidate isotopic abundance, $N_A$ is the Avogadro number and $\mathcal{M}_A$ is the compound molecular mass.

This formula clearly emphasizes many of the most critical experimental parameters, which can be summarized in the following list.
\begin{itemize}
\item \emph{Large mass.} Present experiments have masses of the order of some tens of kg up to a few hundreds kg. Next generation experiments will need masses of the order of tons, hence scalable experimental techniques are advantaged.
\item \emph{Long measurement time.} Experiments have to operate over long times, usually several years, with excellent stability and low downtimes.
\item \emph{Low background.} Background minimization is particularly crucial since it is one of the parameters that can be greatly optimized with proper procedures and techniques. $0\nu\beta\beta$ experiments are located underground, in order to shield against cosmic rays. Moreover, radio-pure materials for the detector and the surrounding parts, as well as proper passive and/or active shielding, are mandatory to protect against environmental radioactivity. The specific target of the background rejection depends on the isotope and on the detector technique. A high Q-value ($\Delta E_{binding}$) is beneficial since it automatically rejects the natural radioactivity background which rolls off at the $^{208}Tl$ line ($2615\ keV$).
\item \emph{High energy resolution.} This is a fundamental requirement to identify the sharp \ndbd peak over an almost flat background and against the intrinsic background induced by the tail of the $2\nu\beta\beta$ peak. Energy resolution can be optimized but it is often limited by the specific readout technique adopted.
\item \emph{High isotopic abundance.} Most of the \ndbd candidates have a natural isotopic abundance of a few percent, with the exception of $^{130}Te$ that is present in natural tellurium with an abundance of about 34\%. For most candidates, enrichment techniques are essential, thus the easiness or cost-effectiveness of the process is crucial.
\end{itemize}

When the background level $B$ is so low that the expected number of background events in the region of interest of the \ndbd during the experiment life is close to zero, then the expression of equation~\ref{eq:sensitivity} is not valid anymore.
This is the so-called ``zero background'' experimental condition.
In this case $N_B$ in equation \ref{eq:sensitivity0} is constant and the expression for the sensitivity becomes simpler:
\begin{equation}
\label{eq:sensitivity2}
S^{0\nu}_{zero-background} = \ln 2 \: \epsilon \: \frac{1}{n_\sigma} \frac{x \: \eta \: N_A }{\mathcal{M}_A} \frac{M \: T}{N_S} \ ,
\end{equation}
where $N_S$ is now the number of observed events in the region of interest.
Reaching the zero background condition is the ultimate goal of the experimental searches since, in this case, the sensitivity depends linearly on detector mass and measurement time.
For this reason many future experiments will implement active techniques for background rejection.

In the next section I will describe the bolometric technique, one of the most promising for the search of neutrinoless double beta decay.
This technique is the one that is used by the CUORE and CUPID experiments, for which I have developed the readout electronics during my Ph.D. work.


\subsection{Bolometric technique}

Bolometers are calorimetric sensors that operate at cryogenic temperatures.
Any energy release due to an interacting particle is detected as a temperature rise in the bolometer crystal.
In this kind of detectors, the source material for the \ndbd is coincident with the detector material since the crystal is composed by the candidate isotope.

Bolometers are used in low energy particle physics experiments since the 1980s~\cite{fiorini, mibeta, cuoricino, cuore0}, and their excellent energy resolution (from the eV for microbolometers, up to a few keV for macrobolometers) makes them perfectly suitable for the search of the neutrinoless double beta decay and even the direct kinematic measurement of neutrino mass.

\begin{figure}
\centering
\def\svgwidth{\linewidth}
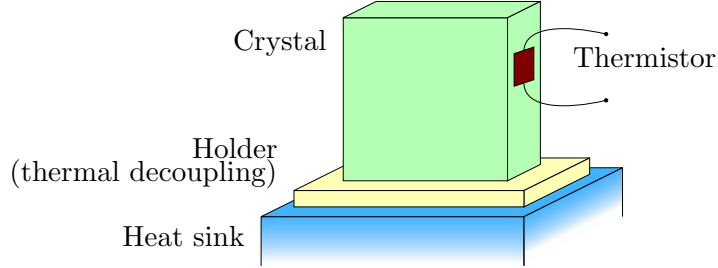
\caption{Typical setup of a bolometric detector.}
\label{fig:TypicalBolometricSensor}
\end{figure}

Figure~\ref{fig:TypicalBolometricSensor} shows the typical setup of a bolometric detector.

Crystals operate at a very low temperatures (around $10\ mK$) so that their heat capacity $\mathcal{C}$ is minimized and thus the temperature rise is maximized, since we can write
\begin{equation}
\Delta T = \frac{\Delta E}{\mathcal{C}} \ .
\label{eq:ch4DeltaT}
\end{equation}

The heat capacity of a dielectric and diamagnetic material is given by the Debye model:
\begin{equation}\label{eq:Debye}
\mathcal{C}(T)=k_BN_T\frac{12\pi^4}{5}\left(  \frac{T}{\theta_D} \right)^3 \ ,
\end{equation}
where $k_B$ is the Boltzmann constant, $N_T$ is the total number of atoms and $\theta_D$ is the characteristic Debye temperature.
At temperatures below the Debye temperature, the heat capacity quickly decreases, and for a typical bolometer crystal kept at about $10\ mK$ it can reach values of about one $nJ/K$.

The sensitive crystal is thermally coupled to a heat sink via a weak link with thermal conductivity $\mathcal{G}$.
This link provides the thermalization power after each energy deposition due to a particle interaction.
Usual values are in the range of $nW/K$

The ratio between the heat capacity $\mathcal{C}$ and the thermal conductance $\mathcal{K}$ gives the typical time constant of the bolometric detector:
\begin{equation}\label{eq:TauB}
 \tau_B = \frac{\mathcal{C}}{\mathcal{K}} \approx \frac{1\ nJ/K}{1\ nW/K} = 1\ s \ .
\end{equation}

In order to detect the thermal increase in the crystal, each bolometer must be equipped with a sensor able to convert the temperature increase into a detectable electric signal.
For the experiments discussed in this work, Neutron Transmutation Doping (NTD) thermistors~\cite{ntd} are adopted. Other solution used elsewhere are transition edge sensors (TES)~\cite{tes}, magnetic microcalorimeters (MMCs)~\cite{mmc}, kinetic inductance devices (KIDs)~\cite{kid} and others.

Each of our crystals is also usually equipped with a heater element (i.e. a resistor) thermally coupled to the bolometer, which is used for calibration and stabilization purposes.

NTD thermistors are obtained by irradiating wavers of germanium or silicon with thermal neutrons.
These neutrons determine nuclear reactions which are able to create a perfectly uniform doping in the semiconductor.
The wafers are doped up to a specific level, which causes the devices to have an exponential dependence of resistivity at very low temperatures.
The resistance of the NTD can be described by the following formula:
\begin{equation}
R_B (T) = R_0 \exp \left( \sqrt{\frac{T_0}{T}}\right)
\label{eq:ch4RT1}
\end{equation}
where $R_0$ and $T_0$ are technological parameters determined by the doping concentration.
Typical values are $R_0 \approx 10\ \Omega$ and $T_0 \approx 3\ K$, that give an NTD resistance at 10 mK of a few hundred $\ M \Omega$ with an inversely proportional dependence on temperature (resistance increases at lower temperatures).
This last feature is very important for proper operation of the detector and for the study of the electro-thermal dynamic of the bolometer.

Temperature signals are detected by a change in the resistance given by equation \ref{eq:ch4RT1}.
Thermistors can be biased either with a constant current $I_B$, usually generated from a voltage source $V_{bias}$ through very large load resistors $R_L$, or with an AC current.
In the following we will only consider the case of DC bias.
When thermistors are biased with constant current and the resistance of the thermistor decreases, a voltage signal is generated across the thermistor and the dissipated power also reduces.
This behavior configures a negative feedback.
This qualitatively working principle, however, must be studied more carefully since instabilities can occur at specific bias conditions and at different working points.
A thorough analysis will be continued in the following.


\begin{figure}
\centering
\def\svgwidth{\linewidth}
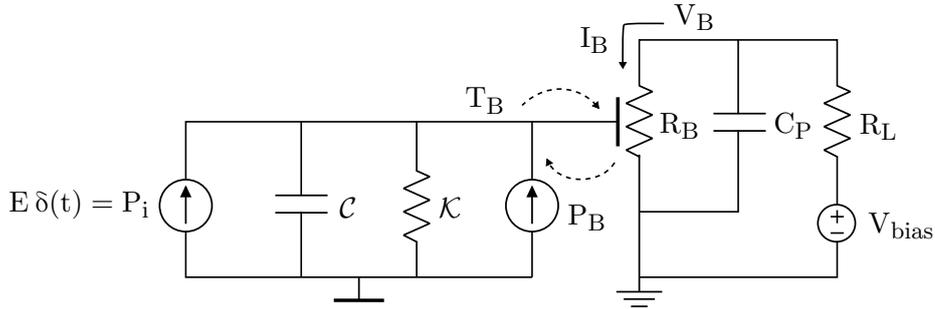
\caption{Schematic of the electro-thermal circuit of the bolometer.}
\label{fig:bolometer_scheme}
\end{figure}

Let us consider the case shown in Figure~\ref{fig:bolometer_scheme}.
For thermal stability we would like that, when the temperature $T_B$ of the bolometer increases, then the dissipated power $P_B$ by the electric part decreases, namely:
\begin{equation}
\frac{dP_B}{dT_B}<0 \ .
\end{equation}
The previous equation can be rewritten as:
\begin{equation}
\label{eq:bolo_stab}
\frac{dP_B}{dT_B} = \frac{dP_B}{dR_B} \frac{dR_B}{dT_B} < 0\ .
\end{equation}
From equation \ref{eq:ch4RT1} we can calculate that:
\begin{equation}
\label{eq:RBdT}
\frac{dR_B}{dT_B} = -\frac{1}{2}\sqrt{\frac{T_0}{T^3}} R_B = -\alpha
\end{equation}
with $\alpha > 0$.
NTDs (and other semiconductor based thermistors) always have a negative thermal coefficient, unlike metallic conductors that have a positive one.

Equation \ref{eq:bolo_stab} thus reduces to:
\begin{equation}
\frac{dP_B}{dR_B} > 0\ .
\end{equation}
We can rewrite the previous as:
\begin{equation}
\label{eq:dPBdRB}
\frac{dP_B}{dR_B} = \frac{d\left(V_B I_B\right)}{dR_B} = I_B\left(\frac{dV_B}{dR_B}+R_B\frac{dI_B}{dR_B}\right) \ .
\end{equation}
$I_B$ and $V_B$ are defined by the electrical bias circuit on the right of Figure~\ref{fig:bolometer_scheme}, as:
\begin{equation}
I_B = \frac{V_{bias}}{R_B+Z_L} \ ,
\end{equation}
with $Z_L$ the parallel of the bias resistance $R_L$ and the parasitic capacitance $C_P$, and:
\begin{equation}
V_B = \frac{R_B}{R_B+Z_L} V_{bias}\ .
\end{equation}
We can calculate the derivatives of the two with respect to $R_B$, substitute them into equation \ref{eq:dPBdRB}, and put the result into \ref{eq:bolo_stab} together with \ref{eq:RBdT}.
Finally we obtain:
\begin{equation}
\frac{dP_B}{dT_B} = -\alpha\: I_B^2 \frac{Z_L-R_B}{Z_L+R_B} \ .
\end{equation}

The feedback is negative if $Z_L > R_B$ but it can also become positive if $Z_L$ becomes too small.
For this reason (another reason is parallel noise of RL, as we saw already) the load resistance $R_L$ is chosen as high as possible, typically in the order of a few tens of $G\Omega$ and the parasitic capacitance $C_L$ has to be kept low.
However, in experiments with large array of bolometers with room temperature readout, reducing the parasitic capacitance of the link can be difficult and its value in our case is in the order of $500\ pF$.
In this case, also the $R_B$ value of the thermistor have to be kept low, by selecting an higher working point temperature, making a compromise between stability and the heat capacity $\mathcal{C}$ of the crystal (i.e. signal amplitude).
Optimal values for the NTD resistance are in the range of a few hundred $M\Omega$, also because higher values would excessively reduce the bandwidth of the signal below $1\ Hz$.
Biasing the detectors with constant voltage ($R_L = 0$) would lead to maximum instability.
This is why charge and current amplifiers cannot be used in these type of applications.

Let us study the system when an input signal is applied.
The input signal is represented by an incident particle that releases an instantaneous power $P_i(t) = E \: \delta(t)$ or, in frequency domain, $P_i(s)=E$.
The system includes and electro-thermal feedback that can be treated using the classical feedback theory approach.

The canonical equation $V_o = \frac{1}{\beta}\frac{-T_{loop}}{1-T_{loop}}V_i+\frac{A_{dir}}{1-T_{loop}}V_i$ with $T_{loop}=-A\beta$ becomes, in this case:
\begin{equation}
\label{eq:loop1}
dP_B = \frac{1}{\beta}\frac{-T_{loop}}{1-T_{loop}}E
\end{equation}
with $T_{loop} = -\frac{\alpha \beta}{s\mathcal{C}+\mathcal{K}}$ and $\alpha$ the gain coefficient defined in \ref{eq:RBdT}.
The NTD takes place of the amplifier in the canonical case.
The direct transmission term is zero, since, if $\alpha=0$, the thermal circuit cannot induce any change in the electrical one.

If the gain is increased, up to the limit $\alpha \rightarrow +\infty$, then the system is able to react immediately to an energy release and keep temperature constant.
We have thus:
\begin{equation}
\label{eq:loop2}
dP_B = \frac{1}{\beta}E = -E \ .
\end{equation}

Now we can evaluate the loop gain by switching off any external source ($P_i$ in this case), by breaking the loop and by applying an external temperature rise $dT_i$.
The feedback power becomes:
\begin{equation}
dP_B = -\alpha\: I_B^2 \frac{Z_L-R_B}{Z_L+R_B} dT_i \ .
\end{equation}
The power generates a temperature rise $dT_{ret}$ on the heat capacity $\mathcal{C}$ and thermal conductance $\mathcal{K}$:
\begin{equation}
dT_{ret} = \frac{dP_B}{s\mathcal{C}+\mathcal{K}} \ ,
\end{equation}
and the loop gain can be evaluated:
\begin{equation}
\label{eq:loop3}
T_{loop} = \frac{dT_{ret}}{dT_i} = -\alpha\: I_B^2 \frac{1}{s\mathcal{C}+\mathcal{K}} \frac{Z_L-R_B}{Z_L+R_B} \ .
\end{equation}
Putting equations \ref{eq:loop1}, \ref{eq:loop2} and \ref{eq:loop3} together:
\begin{equation}
dP_B = - \alpha \: I_B^2  \frac{Z_L-R_B}{Z_L+R_B} \frac{1}{s\mathcal{C}+\mathcal{K}+\alpha \: I_B^2  \frac{Z_L-R_B}{Z_L+R_B}} E \ .
\end{equation}
Which can be rewritten in term of the voltage $V_B$:
\begin{equation}
V_B(s) = \frac{- \alpha \: I_B}{s\mathcal{C}+\mathcal{K}+\alpha \: I_B^2  \frac{Z_L-R_B}{Z_L+R_B}} E \ .
\end{equation}

By calculating the inverse Laplace transform, the expression can be converted
to the time domain, obtaining:
\begin{equation}
\label{eq:bol_signal}
V_B\left(t\right) = -\alpha I_B \frac{E}{\mathcal{C}}\: \Theta(t)\: e^{-\frac{t}{\tau_B}} \ ,
\end{equation}
with
\begin{equation}
\tau_B = \frac{\mathcal{C}}{\mathcal{K} + \alpha R_B I_B^2\frac{Z_L-R_B}{Z_L+R_B}} \ .
\end{equation}
Equation \ref{eq:bol_signal} represent a voltage step, $\Theta(t)$, with amplitude $-\alpha I_B \frac{E}{\mathcal{C}}$ and decay time given by $\tau_B$.
From the point of view of the signal amplitude it is beneficial to maximize $\alpha$ and minimize the detector heat capacity by operating at lower temperatures.
However, as it was shown before, this would induce instabilities in the electro-thermal feedback and a compromise must be chosen.

The intrinsic energy resolution of bolometric detectors is related to the phonon energy ($E_0 = k_B T_B$), and is given by the following formula~\cite{moseley}:
\begin{equation}\label{eq:dEBol}
\sigma_E=\sqrt{k_B T_B^2 \mathcal{C}} \ .
\end{equation}
In macrobolometers the intrinsic resolution is worsened by other sources of disturbance coming mainly from mechanical friction and vibrational noise of the crystals~\cite{CUOREvibration1,CUOREvibration2}.

For what concerns the noise contribution of the electronic readout, we will briefly calculate the main contributions in the following.

\begin{figure}
\centering
\def\svgwidth{\linewidth}
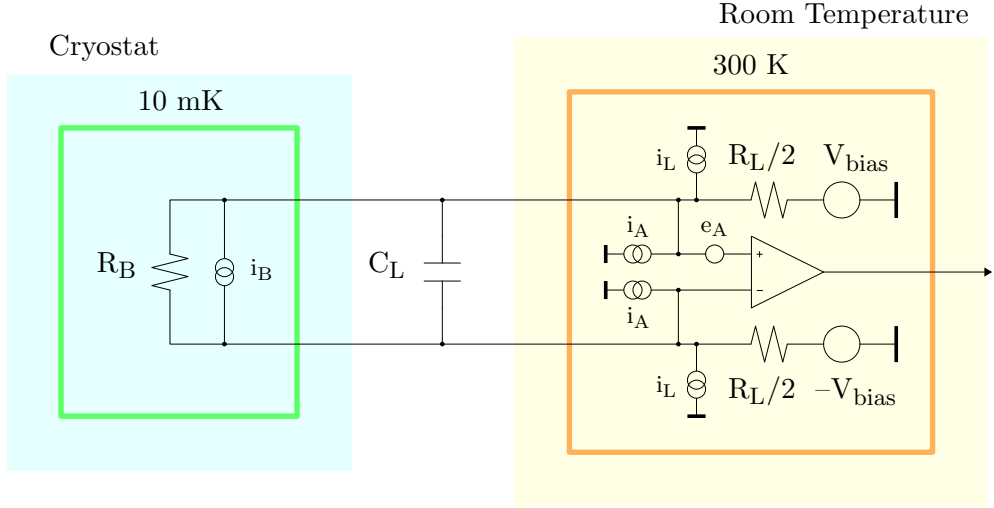
\caption{Readout of a bolometric sensor at room temperature.}
\label{fig:RoomTemperatureReadout}
\end{figure}

Let us consider the typical case depicted in Figure~\ref{fig:RoomTemperatureReadout} with differential readout and differential biasing through $R_L/2$ resistors.

The first source of noise to be considered is the thermistor noise, given by the following expression, with $R_B$ the dynamic impedance of the thermistor:
\begin{equation}
v_{B} = \sqrt{4 k_B T_B R_B} \ .
\end{equation}
For a typical dynamic impedance of $100\ M\Omega$ at $15\ mK$, the thermistor series noise can be estimated to about $10\ nV\sqrt{Hz}$, which correspond to a parallel noise ($i_{B}$) of $0.1\ fA\sqrt{Hz}$.
This value poses a limit on the other sources of parallel noise of the readout electronics.

For what concerns the input series noise of the preamplifier circuit, this is not the limiting contribution: if electronics is placed at room temperature, then large area transistor with high bias current can be selected, and both the white and flicker series noise can be kept at values lower than the intrinsic NTD noise.
The maximum limit on the transistor size is determined by the input capacitance, which has to be small with respect to the parasitic capacitance of the link.

Since high source impedances are involved, parallel noise is the most crucial aspect that must be taken into account.
The two main sources of parallel noise are the input current noise of the preamplifier and the current noise of the load resistors.

The former can be, again, kept low by accurately selecting the input transistors of the readout electronics.
The shot noise $i_A$ for each input is defined as $i_A = \sqrt{2qI_A}$ with q the electron charge and $I_A$ the input bias current of the transistor.
Since the transistor are two, the total input current noise is $i_A \sqrt{2}$.
However, since $R_L$ is much higher than $R_B$, the current is divided equally between the two branches $R_L/2$ and $R_L/2+R_B$, thus only half of the current develops a voltage difference on $R_B$.
The final contribution is therefore
\begin{equation}
i_{A,tot} = \sqrt{qI_A} \ .
\end{equation}

The contribution due to the load resistors can be evaluated in a similar fashion, with the final differential contribution being
\begin{equation}
\label{eq:LoadResNoise}
i_{L,tot} = \frac{\sqrt{2}}{2}\sqrt{\frac{4 k_B T_L}{R_L/2}} = \sqrt{\frac{4 k_B T_L}{R_L}} \ ,
\end{equation}
with $T_L$ the ambient temperature of the load resistors.

The current noise for a total load resistor of $60\ G\Omega$ at $50\: ^\circ C$ (typical operating temperature of the instrumentation without active cooling) is about $0.5\ fA/\sqrt{Hz}$, five times larger than the intrinsic noise of the thermistor.
This is clearly the dominant contribution to the input noise due to the electronic system and it highlights even more the necessity to keep the thermistor impedance low, since total RMS noise is proportional it.

For the bias current it is sufficient to require that $i_{A,tot} \ll i_{L,tot}$, which can be written as
\begin{equation}
I_{A} \ll \frac{4 k_B T_L}{q R_L} \approx 2\ pA \ .
\end{equation}
This value is a reasonable amount for high quality large-area JFETs even at a working temperature of $50\ ^\circ C$.
The transistors used in our system, for example, have a typical input current of less than $100\ fA$.

In the following sections I will present the two experiments for which we have developed the front-end electronics: CUORE and its upgrade CUPID.

\subsection{The CUORE experiment}

CUORE (Cryogenic Underground Observatory for Rare Events)~\cite{cuore} is a large scale experiment that searches the neutrinoless double beta decay in $^{130}Te$ using the bolometric technique described above.
As discussed in section~\ref{sec:ndbd}, the observation of this rare decay would represent the first experimental evidence of leptonic number violation, demonstrate the Majorana nature of the electronic neutrinos, and estimate the effective mass of the electronic neutrino and the neutrino mass hierarchy.

The CUORE experiment is the first bolometric experiment to reach the ton-scale, with a total detector weight of $742\ kg$ and a \ndbd candidate weight of $206\ kg$, thanks to a large array of macro-bolometers.
The CUORE detector will consist of an array of 988 $TeO_2$ crystals arranged in a cylindrical configuration of 19 towers, each made from the assembly of 4 columns of 13 crystals each, so that a single tower hosts 52 cubic bolometers, $5\times5\times5\ cm^3$ each.

The excellent performance of the bolometric technique with such crystal was already demonstrated by the pilot experiments CUORICINO~\cite{cuoricino} and CUORE-0~\cite{cuore0}, which were able to set the best limit on the \ndbd in $^{130}Te$.
CUORE aims to push this limit even further, thanks to its larger mass, longer operation time, and lower background.
The final goal, after 5 years of data taking, is a sensitivity $> 9.5\cdot10^{25} y$, which would test the Majorana effective mass in the $50\ meV$-$130\ meV$ range.

Tellurium-130 has a Q-value for the $\beta\beta$ decay of $2527.515 \pm 0.013\ keV$~\cite{te130qvalore}, which is quite close to the $^{208}Tl$ peak at $2615\ keV$.
However this is counterbalanced by the excellent energy resolution achievable (about $5\ keV$), as demonstrated by both CUORICINO and CUORE-0.
Tellurium dioxide also has several other excellent characteristics: the highest isotopic abundance among all the \ndbd candidates (34.167\%~\cite{te130abundance}), good mechanical and thermal properties, high level of cleanliness and radiopurity, and others.
All these properties make tellurium dioxide a good choice for the realization of a bolometric detector.

\begin{figure}
\centering
\includegraphics[width=.7\linewidth]{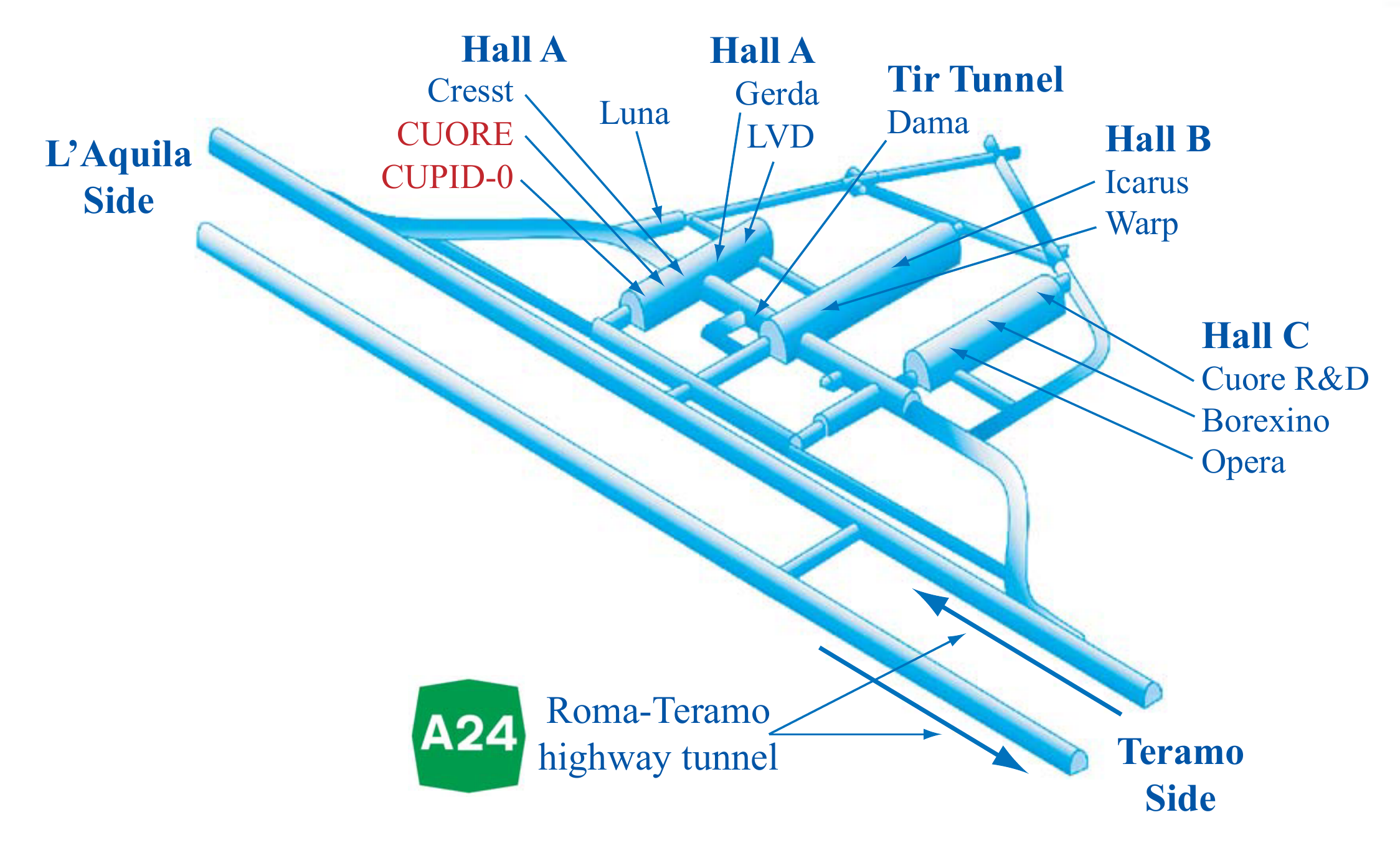}
\caption{Picture of the LNGS cave with the main experiments.}
\label{fig:lngs}
\end{figure}

CUORE is located at the underground National Laboratories of Gran Sasso (LNGS, L'Aquila, Italy), where the highest limestone mountain in continental Appennini range, Gran Sasso, is able to shield against cosmic rays with about $3600\ m$ of equivalent water.
In the experimental area, the measured fluxes of muons, gamma and thermal neutrons are: $\Phi_\mu = (1.5\pm0.06)\cdot10^{-6}\ s^{-1}\: cm^{-2}$, $\Phi_\gamma = 1\ s^{-1}\:cm^{-2}$ and $\Phi_n = (1.08 \pm 0.02) \cdot 10^{-6}\ s^{-1}\: cm^{-2}$.
Unlike other underground laboratories, LNGS can be easily accessed by means of a $10\ km$ highway tunnel, the longest two-tubes tunnel in Europe.
The cave hosts several experiments dedicated to low energy particle physics and dark matter searches.
A picture of the underground laboratories is shown in Figure~\ref{fig:lngs}.

\begin{figure}
\begin{minipage}[b][][t]{.37\linewidth}
	\centering
	\includegraphics[width=\linewidth]{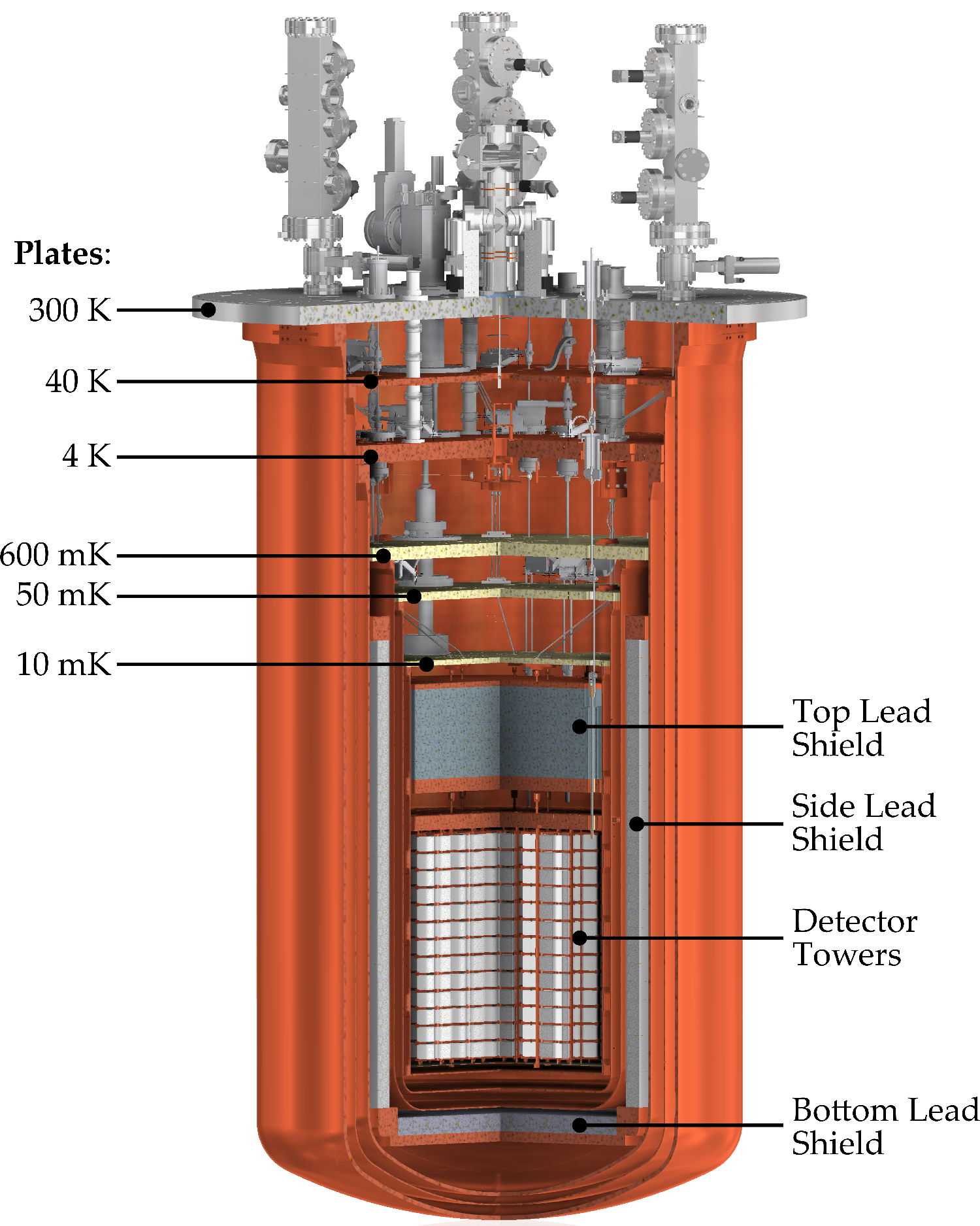}
\end{minipage}
\ \hspace{.1mm} \
\begin{minipage}[b][][t]{.6\linewidth}
	\centering
	\includegraphics[width=\linewidth]{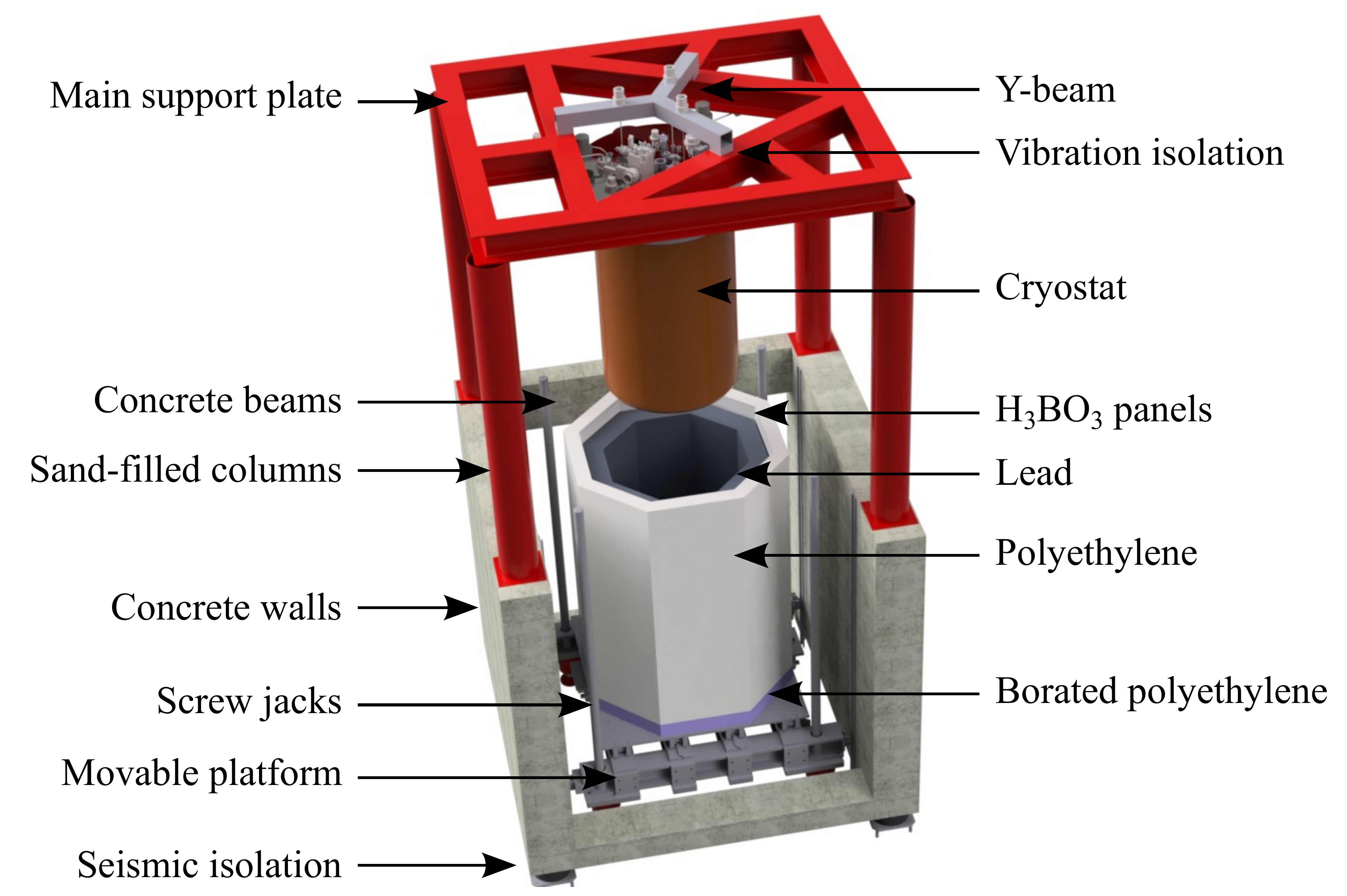} 
\end{minipage}
\caption{Schematic pictures of the CUORE cryostat (on the left) and of the suspension and shielding system (on the right).}
\label{fig:cuore_disegni}
\end{figure}

Building such a large cryogenic system, able to cool about 15 tons of material at temperatures below $4\ K$, with strict radio-purity requirements, is indeed a great challenge.
In Figure~\ref{fig:cuore_disegni} two pictures of the CUORE experiment are shown.

The CUORE cryostat~\cite{CUOREcryostat} is composed by 6 nested vessels corresponding to different temperature stages.
The outermost shield, Outer Vacuum Chamber (OVC), is kept at room temperature, while the Inner Vacuum Chamber (IVC), is at $4\ K$.
They are both vacuum-tight and they are separated by an intermediate radiation shield maintained at a temperature of $40\ K$.
The cooling of these stages is performed by a system of 5 Pulse Tubes (PTs), mounted on the OVC top plate.
Since PTs are cryocoolers, the cryostat is cryogen-free and thus no refills are required, maximising the data-taking duty cycle.
Inside the IVC there are three further stages at $600\ mK$, at $50\ mK$ and the coldest one at about $10\ mK$.
The coldest temperature is achieved through a $^{3}He/^{4}He$ dilution refrigerator (DU), specifically designed for CUORE.
A system of temperature stabilization, based on a PID (proportional-integral-differential) algorithm, has been developed to set the desired working temperature and stabilize possible drifts.
To attenuate neutron and $\gamma$-ray backgrounds, the cryostat is surrounded by layers of borated polyethylene, boric-acid powder, and lead bricks.
A further suppression of $\gamma$-rays from the cryostat materials is obtained with additional lead layers inside the cryostat, made with radio-pure ancient Roman lead~\cite{piombo_romano}.

Another crucial issue that must be taken into account is mechanical noise.
Vibrations are a source of background that generates energy dissipation into the crystals, worsening their energy resolution.
In this sense, pulse tubes are particularly critical due to their working principle, based on pressure waves.
The detector is equipped with an advanced suspension system that provides a mechanical decoupling of the detector from the outside environment.
The system minimizes the transmission of mechanical vibrations due to seismic noise and operations of the cryocoolers and pumps.

\begin{figure}
\centering
\includegraphics[width=.6\linewidth]{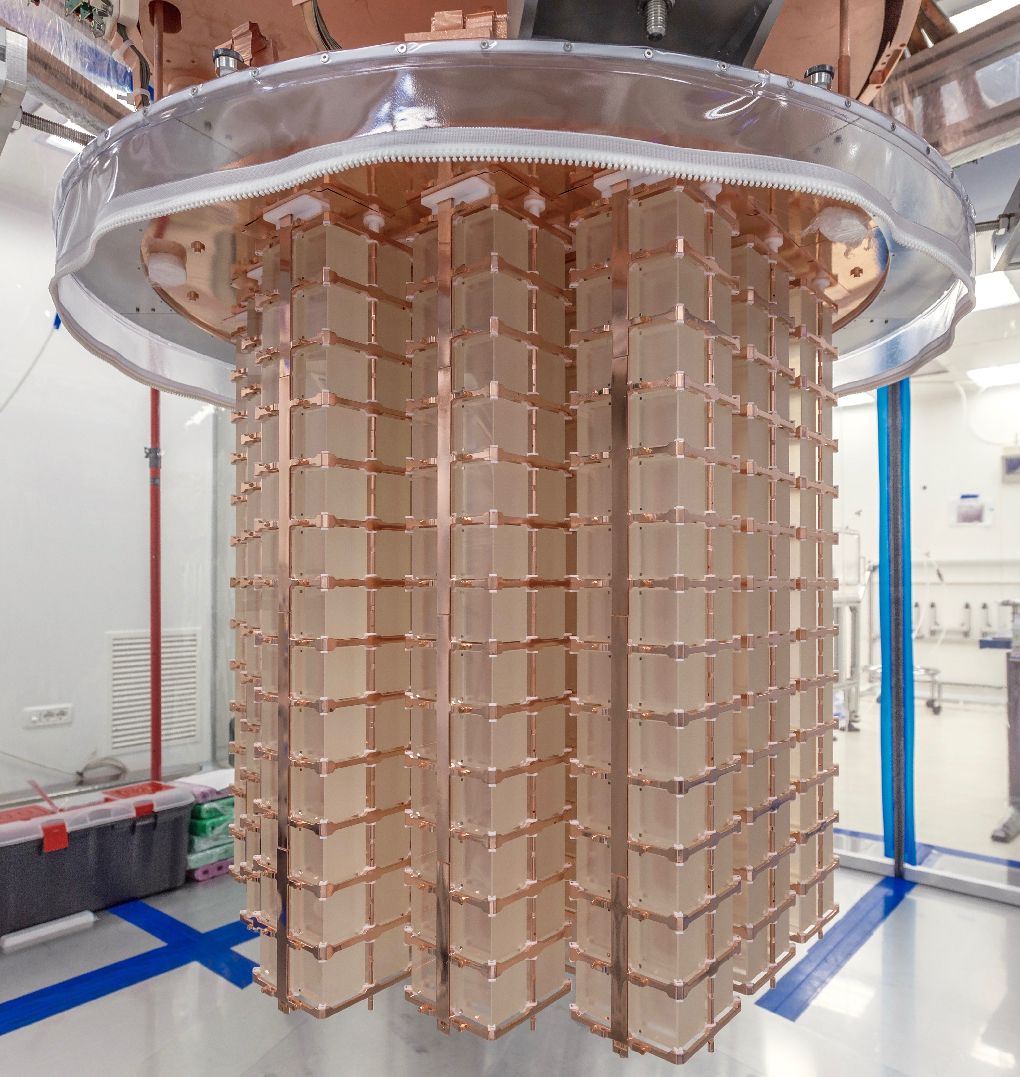}
\caption{Photograph of the complete detector, right after the assembly of the last tower.}
\label{fig:cuore_foto}
\end{figure}

The goal of low background ($10^{-2}\ counts/keV/kg/y$)~\cite{background_budget} was achieved with a wide effort in material selection, cleaning and careful detector assembly.
Excellent radio-purity is indeed one of the characteristics of $TeO_2$ crystals selected for CUORE, but it could be spoiled if surface contaminants are deposited during ``dirty'' phases like the detector assembly, sensor gluing, transportation, etc.
Strict protocols were adopted for this purpose, using radon-free clean rooms and constantly flushing the tower enclosures with nitrogen.
A photograph of the assembled towers is presented in Figure~\ref{fig:cuore_foto}.

The periodical energy calibration of the energy response is made with a low activity source that can be lowered directly within the cryostat thanks to deployable strings containing $^{232}Th$.
The decay chain of the thorium-232 includes the strong thallium-208 peak at $2615\ keV$, which allows precise energy calibration close to the Q-value of the \ndbd in tellurium-130, as well as other peaks at lower energies that can be used to test the linearity of the energy response.
The response of the detectors is also continuously monitored during data taking (with a period of about $300\ s$), by injecting heat pulses in the crystals by means of a heater resistor glued on one face of the crystal itself.
The pulse generation is provided by an ultra-stable and low noise pulser board designed by our group~\cite{CUOREpulsers1, CUOREpulsers2}.

\begin{figure}
\centering
\includegraphics[width=.6\linewidth]{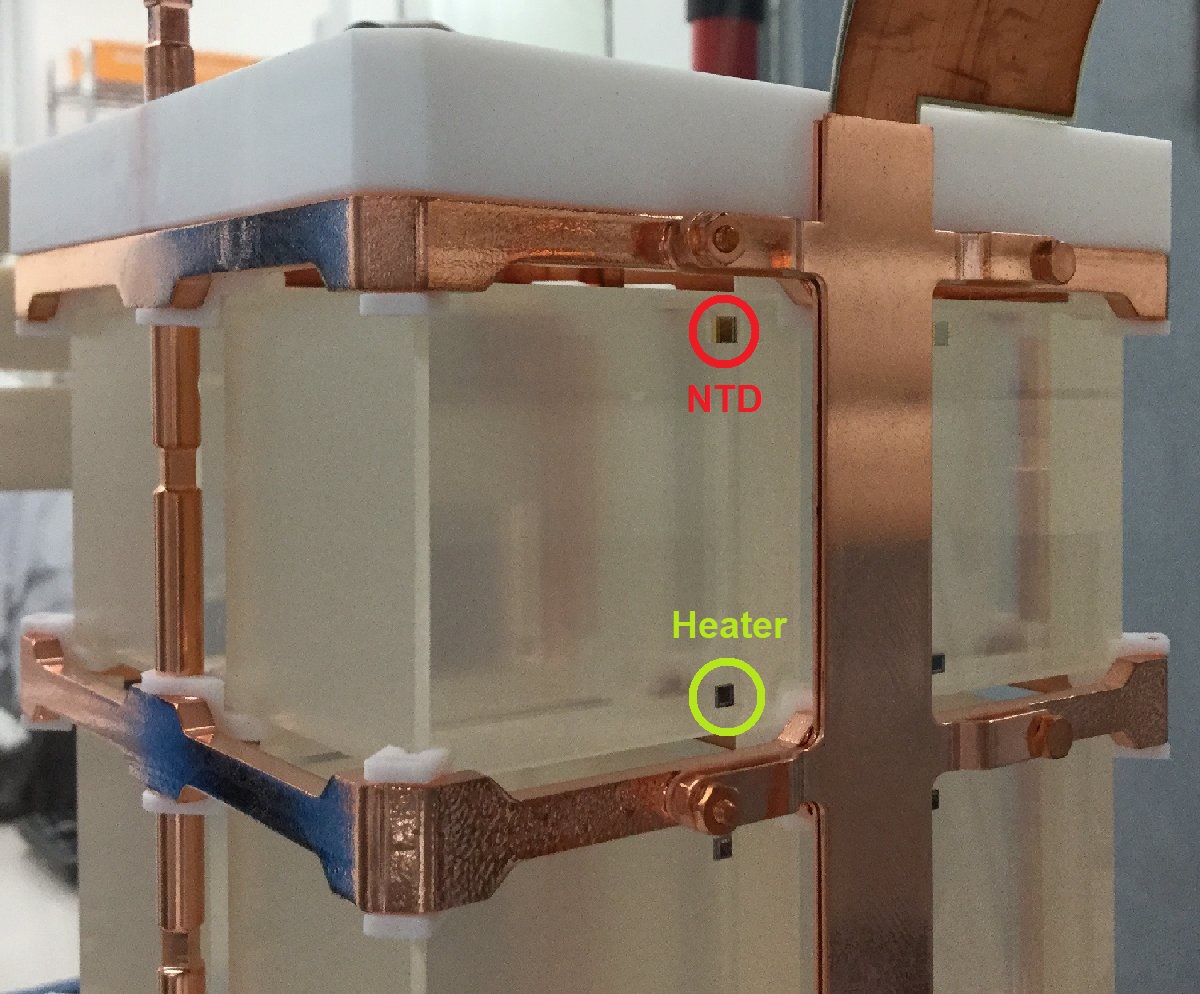}
\caption{Close view of a $TeO_2$ crystal. NTD thermistor and heater resistor are clearly visible.}
\label{fig:cuore_riv}
\end{figure}

A zoomed photograph of a CUORE $TeO_2$ crystal is shown in Figure~\ref{fig:cuore_riv}, where both the NTD thermistor and the heater resistor are clearly visible.

CUORE has started the data taking in late Q2 2017, with the first physics results presented in summer 2017.
The most stringent limit on the \ndbd half-life in $^{130}Te$ has been achieved after only a few weeks of operation~\cite{cuoreprl}.

\subsection{The CUPID experiment}

CUPID (CUORE Upgrade with Particle IDentification)~\cite{cupid_arxiv} represents the upgrade of the CUORE experiment.
The aim of this experiment is to expand the ton-scale bolometric technique pioneered by CUORE, by adding particle identification (PID) capabilities that would allow to greatly suppress background sources, rejecting the alpha particle background from the interesting beta signal.

The most promising techniques for particle identification in bolometric experiments are mainly two: scintillating detectors and Cherenkov detectors.
In scintillating detectors~\cite{ALESSANDRELLO1998109,Pirro2006}, in addition to the heat signal, the interacting particles also produce an amount of light that is different between each type of incident particle due to the different stopping power of the crystal.
Scintillating detectors have a much higher light yield (usually several $keV/MeV$) with respect to Cherenkov detector and this light can be detected by means of germanium or silicon bolometers read out with NTDs.
Cherenkov detectors~\cite{teo2_cher, teo2_cher2}, instead, uses the Cherenkov effect to discriminate between particles that are over the Cherenkov effect threshold (beta) and below the threshold (alpha and other nuclei).
Cherenkov detectors produce signals of much lower amplitude (typically a few tens of $eV/MeV$) and hence require more sensitive and complicated readout, using TESs~\cite{tes} or KIDs~\cite{kid}, or other signal amplification techniques like Neganov-Luke effect~\cite{luke,luke2}.
In addition, not all the \ndbd candidates are suitable for one or another technique.
Tellurium, for example, does not scintillate and PID can only be performed with Cherenkov light detection.

At the moment there is not a clear preferred solution, and several R\&D programs are trying to investigate which combination of \ndbd candidate and PID technique offers the best performance.

CUPID-0 (previously known as LUCIFER)~\cite{lucifer,cupid0} is the first large scale demonstrator that implements particle identification in a bolometric detector.
It uses scintillating $Zn^{82}Se$ crystals for the heat channel and germanium slabs for the detection of scintillating light, both read out with standard NTD thermistors.

The choice of $^{82}Se$ as \ndbd candidate also allow to further suppress the natural background since its Q-value of $2997\ keV$ is among the highest of the other \ndbd candidates.

The main problem of $^{82}Se$ comes from the rather low natural isotopic abundance which require an enrichment process that is particularly critical for costs and for radiopurity.

\begin{figure}
\begin{minipage}[b][][t]{.45\linewidth}
	\centering
	\includegraphics[width=\linewidth]{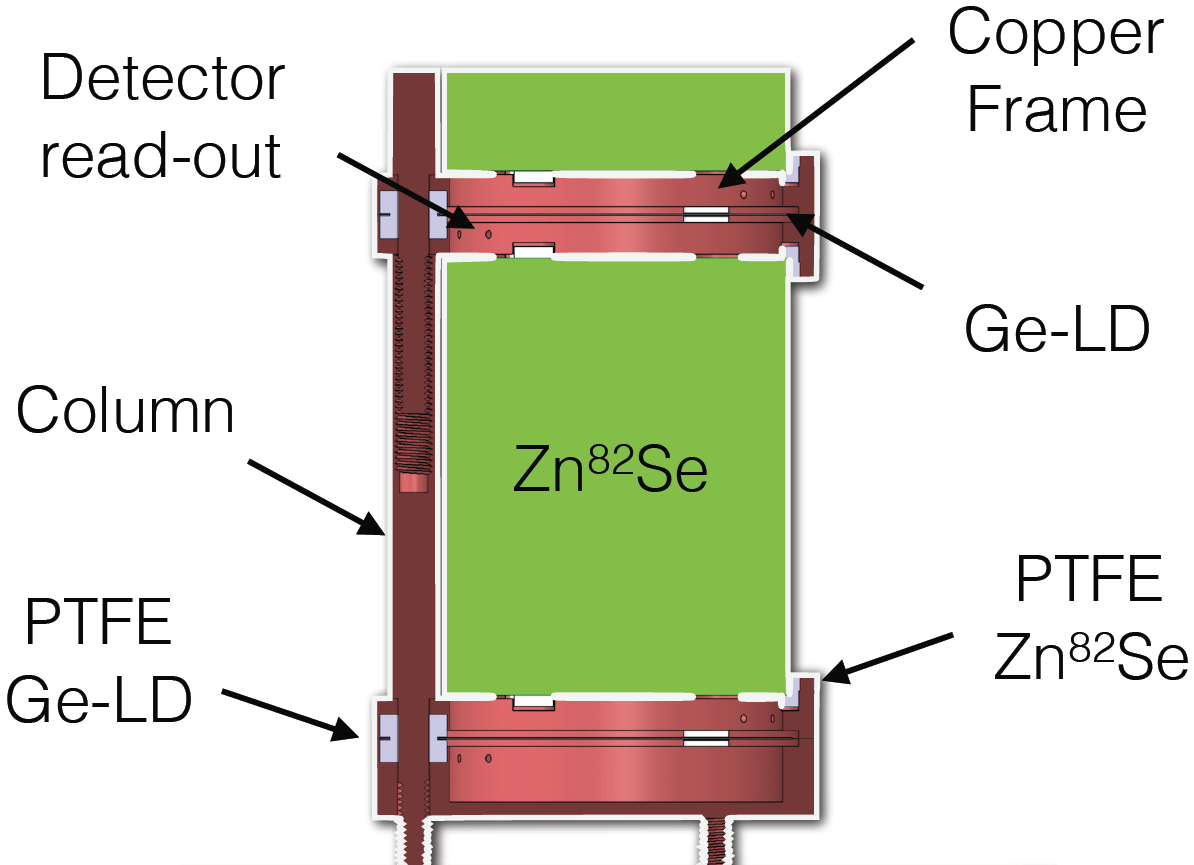}
\end{minipage}
\ \hspace{.1mm} \
\begin{minipage}[b][][t]{.23\linewidth}
	\centering
	\includegraphics[height=6 cm]{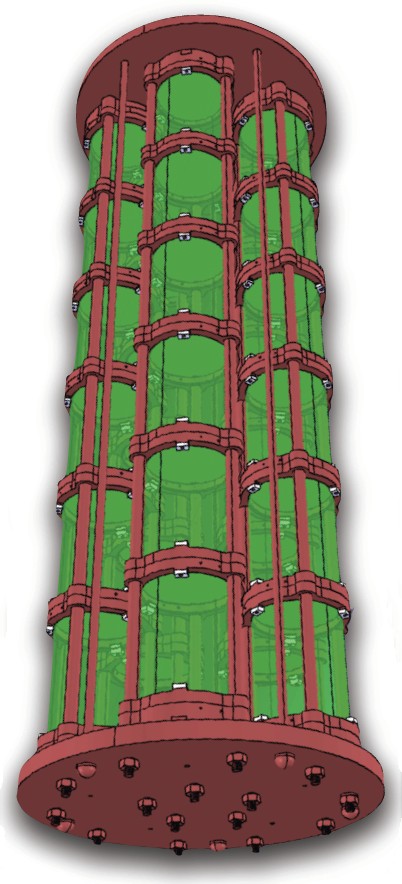} 
\end{minipage}
\ \hspace{.3mm} \
\begin{minipage}[b][][t]{.2\linewidth}
	\centering
	\includegraphics[height=6 cm]{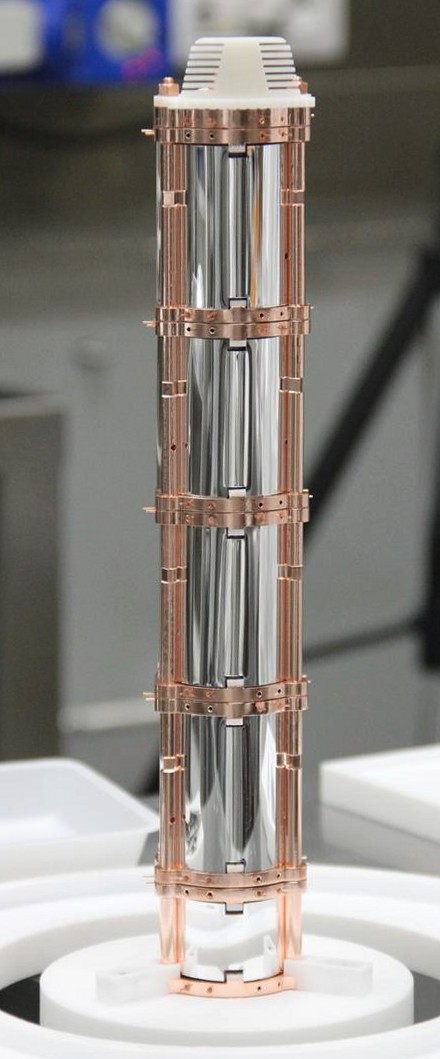} 
\end{minipage}
\caption{Schematic pictures and a photograph of the CUPID-0 detector. The photo on the right depicts a two-thirds complete tower, with the light-reflecting foil surrounding each crystal.}
\label{fig:cupid_disegni}
\end{figure}

Nevertheless, all these issues were overcome and the CUPID-0 is currently operational in the cryostat that was used by CUORE-0, next to the CUORE experiment.
CUPID-0 is made by 26 cylindric crystals (24 with 95.4\% enriched $^{82}Se$ and 2 with natural $Se$), arranged in 5 column of 6 crystal each.
The light detector is a thin germanium disc placed on the top and on the bottom of each crystal.
Figure~\ref{fig:cupid_disegni} shows two sketches and a photograph of the CUPID detector.
 
Particle identification is performed with pulse shape discrimination, since it was found to be more efficient than pure light yield~\cite{cupid0}.
The separation between the $\alpha$ and $\beta$ distributions at the $^{82}Se$ Q-value is more than $10\: \sigma$.

\section{The electronic readout system}

In this section I will provide an overview of the electronic readout system for the CUORE and CUPID-0 experiments.
Then, in the following sections I will describe the work performed in my Ph.D. for the qualification and optimization of the electronics sytems.
In the last section I will finally introduce the latest upgrades for some electronic components for the CUPID-0 experiment.

The electronics of the two experiments is exactly the same, with the only difference being the higher cut-off frequency of the antialiasing filters for the light channels of CUPID-0.

\subsection{Front-end boards}

The signal bandwidth in bolometric experiments is very small, a few tens of Hz.
As a consequence, the front-end (FE) electronics can be placed far from the detector at room temperature.
This configuration greatly simplifies the design of the readout electronics and it becomes mandatory for large array of macrobolometers, where the number of channels is very high.
Having a cold stage electronics, in fact, would imply a much more complex design of the cryogenic system due to the additional space needed by the cold stages and also the issues related to their thermalization and heat injection in the cryogenic system itself.
The cold readout approach has been used with success in the past, on smaller scale experiments~\cite{mare}.

For CUORE and CUPID-0 electronics, a fully differential DC-coupled voltage preamplifier was chosen.
The fully differential scheme allow to suppress common mode sources of noise like electromagnetic disturbances induced in the long links.
DC coupling is particularly convenient for the simpler bias circuitry and for DC baseline studies.

The amplifying system must have adjustable gain on a channel-by-channel basis, in order to match the large spread in the detector energy conversion coefficient an also exploit the full dynamic range of the data acquisition system.
The biasing of each bolometer has to be independently adjustable for each detector to optimize the working point and thus maximize signal amplitude.
Each bolometer, in fact, can show very different optimal working point and having the flexibility to adjust it channel by channel will greatly improve the readout performance. 

Finally, the noise of the electronic system must be adequately low and, given the rareness of the process under study and the typical long acquisition times, the electronic system has to exhibit a very high stability in both gain and offset with respect to external factors like temperature or power supplies.
The high stability is also very important since the monitoring of the bolometer baseline can be used to compensate changes in the energy conversion gain, combined with a pulser stabilization system.

\begin{figure}
	\centering
	\includegraphics[width=1\textwidth]{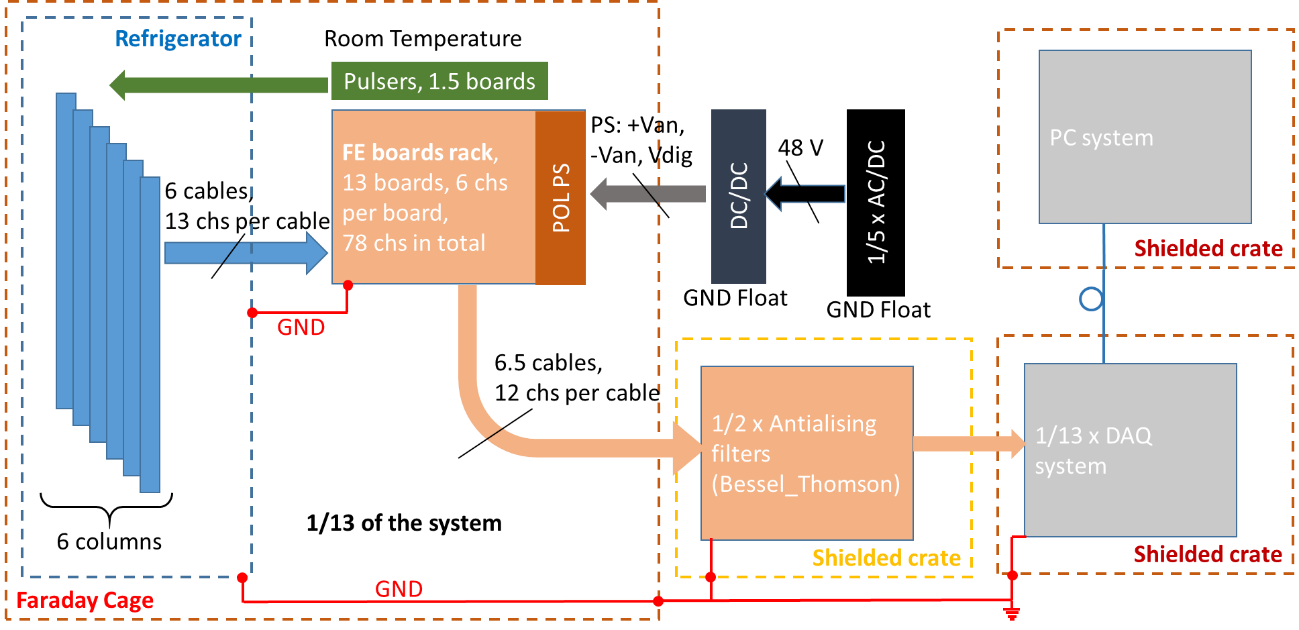}
	\caption{Simplified scheme of one module out of 13. Only the signal paths are shown. Every part of the system is in communication with the control room by means of optical fibers.}
	\label{fig:CUOREelectronicsSegmentation}
\end{figure}

In order to maximize the occupied space in the cryostat, the detectors are arranged in 19 towers, 4 columns each and 13 crystals per column.
This unusual arrangement forces the electronics to be segmented accordingly.
Figure~\ref{fig:CUOREelectronicsSegmentation} shows the block schematic of $1/13^{th}$ of the electronic system.
The signals are routed up to the mixing chamber on PEN ribbon cables.
From there a set of kapton boards route the signals on twisted cables that arrive at the top of the cryostat in a dedicated box.
From this box a 13-signal (26-wire) shielded cable connects the signals to the front-end electronics.
This cables were specifically selected and individually tested in order to have very high parasitic impedance between the pairs of much more than $200\ G\Omega$ (the limit of the electrometer used for the measurement).
The cables are also soft so that they can suppress the transmission of vibrational noise from the support structure and the cryostat.
Finally, cables are surrounded by an additional shielding mesh and by an insulating foam which again helps in minimizing unwanted vibrations which could induce microphonic noise on the links.

The front-end boards are arranged in 13 card cages for the readout of the detectors plus 1 card cages for other monitoring thermometers.
These 14 card cages are arranged into 7 racks, placed on top of the cryostat, on a Y-beam structure anchored to the main support plate (MSP).
The card cages are a standard 19'' 6 units height enclosure that houses 13 FE boards, or mainboards, each one hosting 6 differential voltage preamplifiers, and one power supply unit for the generation of the voltage references and analog supply voltages (point-of-load power supply, POL PS, in the picture).

On the top of these crates there is another rack, 19'' 3 units height, that houses the pulsers, a set of boards able to inject very precise and stable pulses to each detector.

The refrigerator~\cite{CUOREcryostat}, the 14 FE boards racks and the pulsers are placed inside a Faraday cage~\cite{CUOREfaraday}.
The power supply units are supplied by a two-stage AC/DC floating system~\cite{CUOREpowerSupply}.
Outside of the Faraday cage, five other shielded crates are placed: two host the anti-aliasing filters, two host the commercial data acquisition (DAQ) boards and the last one hosts the DAQ PC system.

The DAQ boards communicate with the DAQ PC system via optical fibers to suppress EMI and ground loop disturbances~\cite{CUOREdaq}.
Nevertheless the DAQ boards cannot be isolated from the main ground, and therefore we chose it as the main ground reference connection of our star-like layout, as shown in the red electrical link in Figure~\ref{fig:CUOREelectronicsSegmentation}.

\begin{figure}
	\centering
	\includegraphics[width=1\textwidth]{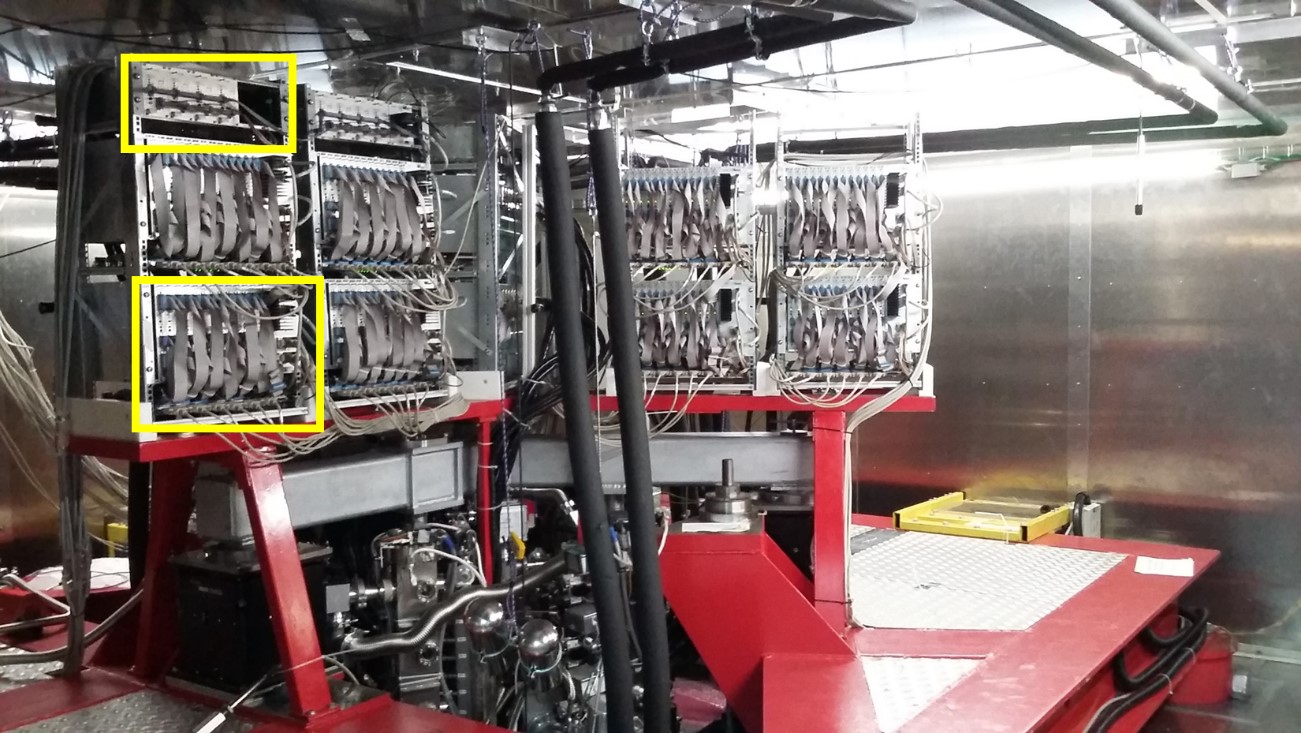}
	\caption{View of the electronic system on top of the CUORE cryostat. The yellow rectangles highlight a single front-end card cage (bottom) and pulser card cage (top).}
	\label{fig:CUOREracks}
\end{figure}

The pulser boards are used to send short, periodic, $\delta$-like (on the detector timing scale) voltage pulses to the heater resistors~\cite{CUOREstabilization}, glued on the crystals, with the aim to stabilize the detector response. These pulses emulate the energy release by particles in the crystals and have been designed to be very stable and precise~\cite{CUOREpulsers1,CUOREpulsers2}.

Figure~\ref{fig:CUOREracks} shows the photograph of the 14 card cages housing the FE mainboard and 6 card cages for the pulsers (the yellow rectangles highlight one of each type) in their final layout on the top of the refrigerator of CUORE.
The racks are organized in stacks of 2 FE mainboard card cages and one pulser card cage.

\begin{figure}
	\centering
	\includegraphics[width=.65\textwidth]{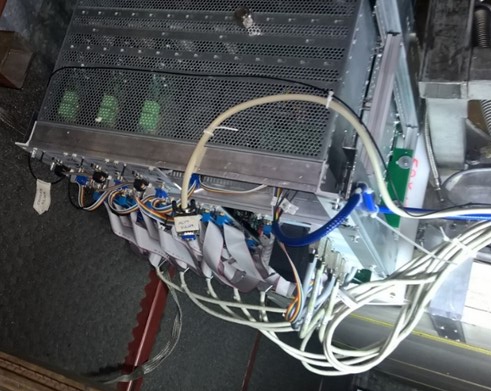}
	\caption{Picture of the front-end electronics of CUPID 0, consisting of one rack.}
	\label{fig:CUPIDracks}
\end{figure}

Figure~\ref{fig:CUPIDracks} shows the front-end electronics of CUPID-0 placed on top of its refrigerator.
All the channels of CUPID-0 are managed by a single rack of FE boards, while the remaining part of the set up is similar, except the channels number, to what already described for CUORE.

\begin{figure}
	\centering
	\includegraphics[width=.7\textwidth]{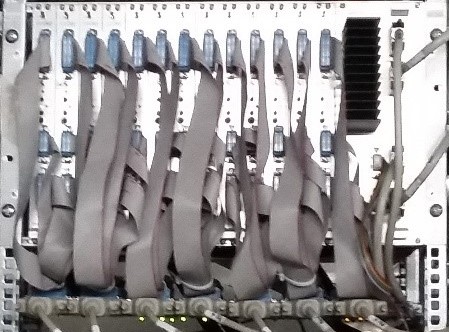}
	\caption{Photograph of a FE card cage. The 13 boards are on the left, while the POL PS (which can be recognized by the black heat sink) is on the right. Under the rack there is an underplane board that re-routes the signals exiting from the boards.}
	\label{fig:CUOREcardCage}
\end{figure}

Each FE card cage (Figure~\ref{fig:CUOREcardCage}) hosts 13 boards, $233\times280\ mm^2$, 8 layers.
On one side of the rack there is the Point-Of-Load Power Supply, POL PS, composed by 2 independent linear power supplies showing very low 1/f noise and high stability~\cite{CUORElps}.
The POL PSs also provide the reference voltages that are used by the boards for the generation of bias voltage and offset correction.

Each board hosts 6 channels and 6 complete detector columns can be managed with one card cage. The signal cables are connected to the back board of the rack, which distributes them to the FE mainboards.
The first 12 channels of each cable are connected to 2 adjacent boards, while the thirteenth channel is routed to the thirteenth board of the rack.
This layout tries to re-order the awkward arrangement of the towers, while maintaining a regularity in the connections.
On the back board only the signals from the detectors and their biases are present: no voltage supplies or digital signals are routed there to avoid disturbances.

\begin{figure}
	\centering
	\includegraphics[width=\textwidth]{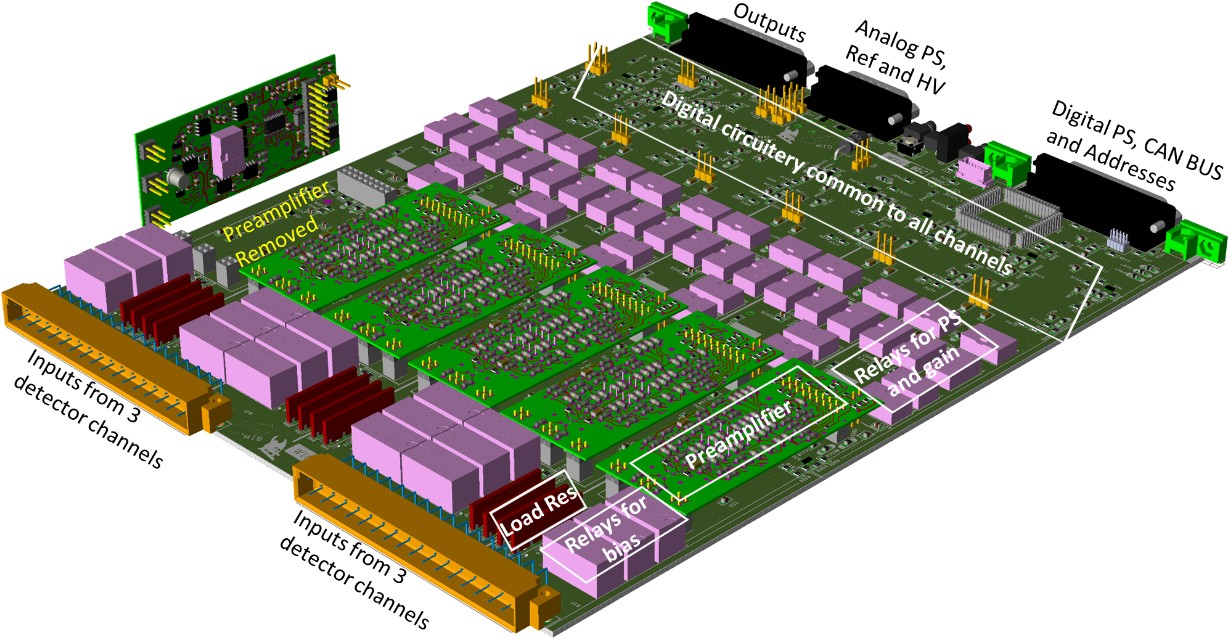}
	\caption{3D view of the top of the mainboard. The preamplifier on the left was removed and rotated.}
	\label{fig:CUOREmainboard3D}
\end{figure}

Figure~\ref{fig:CUOREmainboard3D} shows the 3D view of the mainboard.
On the bottom left there are the 2 input connectors.
Since the impedance of the thermistors can be very high, in the hundreds of $M\Omega$ range and up to $G\Omega$ in some cases, the parasitic impedance towards ground and towards adjacent detectors must be kept as large as possible.
The electrical links from the thermistors to the input connectors were extensively studied~\cite{CUORElink1,CUORElink2}.
They consist of three parts, two of which are inside the cryostat (the first from the detectors to the mixing chamber, the second from the mixing chamber to the top of the refrigerator)~\cite{CUORElink3,CUORElink4}, while the last is at room temperature and connects the top of the cryostat to the FE inputs.
The total length of the link is $5\ m$ on average.
The readout of each thermistor is differential and the parasitic capacitance in parallel with the pairs is about $500\ pF$, while the parasitic resistance is in the hundreds of $G\Omega$ range.

The load resistors are placed near the input connectors, together with the circuitry that generates the detector bias. In the same region, slightly towards the center of the boards, there are the preamplifiers, which will be described with more detail in section \ref{sec:CUOREpre}.

Going towards the output and power supply connectors of the board, we find the second stage of amplification, Programmable Gain Amplifier (PGA), located on the bottom layer, below the relays that enable and disable the power supply for each channel.

Closer to the output connectors, after the amplification has already taken place, there is a region where the digital section of the board is distributed.
This circuitry is common to all the channels.

The output and power supply connectors are 3.
The 6 differential outputs from the PGAs are routed to the upper left 25-pin female D-Sub connector.
The analog power supplies and the high voltage supplies for detector biasing are located on the 15-pin male D-Sub central connector: $V_{CC}$ and $V_{EE}$ ($\pm5\ V$) are the analog power supplies, $V_{REF+}$ and $V_{REF-}$ ($\pm5\ V$) are the reference stable and low noise power supplies, $V_{HV+}$ and $V_{HV-}$ are the power supplies used for the detector biasing, whose range is from $\pm5\ V$ to $\pm25\ V$. 3 pins are left for the analog ground. The digital power supply, CAN-bus signals for the communication with the control room, and the board address are routed on the 25-pin male D-Sub connector, located on the right.

Cryogenic detectors typically show a large spread in their characteristics: their energy conversion gain is strongly dependent on their biasing and it is not precisely known in design phase.
For this reason each detector channel has to be calibrated independently, starting with load curves for the determination of the dynamic impedance with respect to the input bias, and followed by a noise measurement and the pulse response.
With this process, which is totally automatized and managed by the DAQ, it is possible to determine the optimal working point with respect to the signal to noise ratio, SNR, as a function of the DC bias applied.

\begin{figure}
	\centering
	\includegraphics[width=\textwidth]{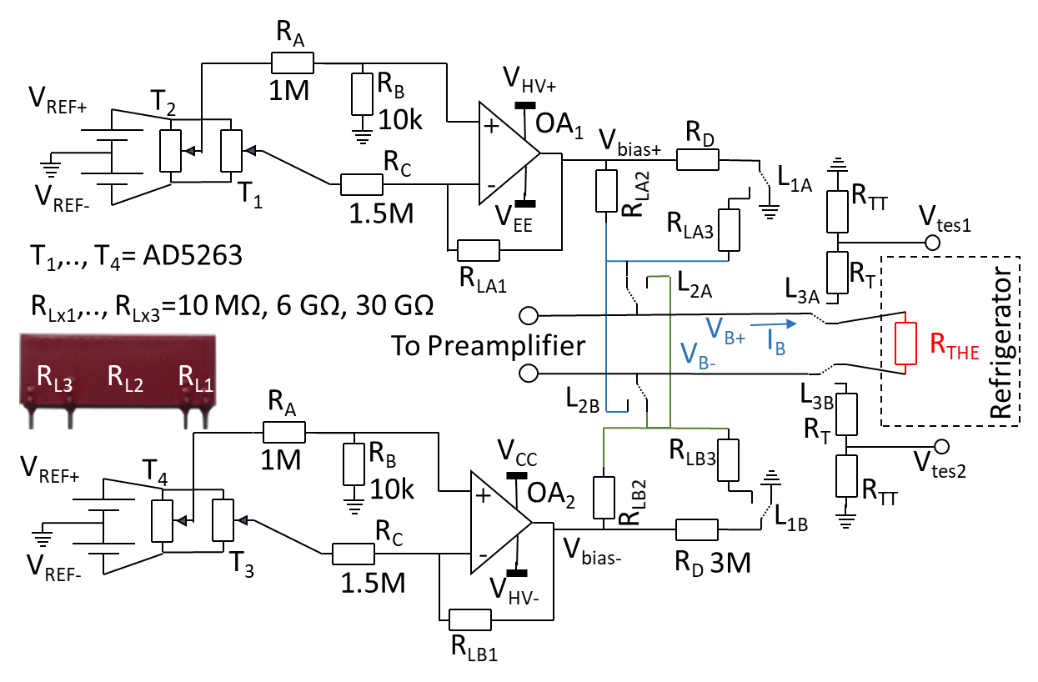}
	\caption{Scheme of the biasing circuit for the detector thermistor $R_{THE}$.}
	\label{fig:CUOREbiasCircuit}
\end{figure}

Figure~\ref{fig:CUOREbiasCircuit} shows the circuitry for detector biasing.
The thermistor is biased differentially.
Two symmetric voltages, $V_{bias+}$ and $V_{bias-}$, are generated with amplifiers $OA_1$ and $OA_2$.
The values of the 2 voltages is set by the 2 pairs of programmable trimmers, $T_1$-$T_2$ and $T_3$-$T_4$, all part of the same device, an Analog Devices 8-bit digital trimmer AD5263.
$T_1$ ($T_3$) sets the coarse value, while $T_2$ ($T_4$) tunes the fine value since they are attenuated by $100\ V/V$ with $R_A$ and $R_B$.
The range of output voltages is between $-5\ V$ and $25 V$ for $V_{bias+}$ and between $-25\ V$ and $5 V$ for $V_{bias-}$.
An asymmetric power supply is applied to $OA_1$ and $OA_2$ (Texas Instrument OP140) so that their absolute maximum range of $36\ V$ is not exceeded: $OA_1$ is powered between $V_{HV+}$ and $V_{EE}$, $25\ V$ and $-5\ V$, while $OA_2$ is powered between $V_{CC}$ and $V_{HV-}$, $+5\ V$ and $-25\ V$.

The actual bias voltages $V_{B+}$ and $V_{B-}$ applied to the detector are a fraction of $V_{bias+}$ and $V_{bias-}$ due to the partition coming from the load resistor system, $R_{LA2}$ and $R_{LB2}$.
In equation \ref{eq:LoadResNoise} we have shown that the noise due to the electronics is dominated by the noise of the load resistors.
Ideally this could be minimized by selecting very large value resistors.
However, for practical reasons coming from the maximum voltage bias available and limitations on parasitic impedance, we chose $R_L$ of the value of 60 $G\Omega$.
This characteristic could have been better if the load resistors were installed at a cold stage, within the cryostat, thus greatly reducing its Johnson noise.
This solution is known \cite{CUOREcold1, CUOREcold2} and has been pursued as a preliminary study for CUORE~\cite{CUOREpreliminarycoldstudy}, however it was discarded in favor of a room temperature operated front-end due to the easier implementation and for the fact that the electronic noise is not the dominant parameter that limits the resolution as mechanical friction has the larger contribution~\cite{CUOREvibration1, CUOREvibration2, CUOREvibration3}.

Since $R_L$ is much larger than $R_{THE}$, the biasing system behaves like a current generator:
\begin{equation}
\label{eq:biasi}
I_B = \frac{V_{REF+}-V_{REF-}}{R_C} \frac{R_{LA1}}{R_L}  \left[ \left(\alpha_3-\alpha_1\right) + \left(\alpha_2-\alpha_4\right) \left(\frac{R_B}{R_A+R_B}\right) \left(\frac{R_C}{R_{LA1}}+1\right)\right]
\end{equation}
where $\alpha_i$ is the $T_i$ trimmer setting fraction ranging from 0 to 1 (half-scale or $\alpha_i = 0.5$ corresponds to $I_B = 0\ A$). $R_{LB1}$ and $R_{LB2}$ have been considered equal to $R_{LA1}$ and $R_{LA2}$.

Voltages $V_{REF\pm}$ and trimmers $T_i$ are all stable at the order of a few $ppm/^\circ C$, while resistors $R_A$, $R_B$ and $R_C$ are metal film (mini-melf), with thermal drift of less than $10\ ppm/^\circ C$.

If relays $L_{1A}$ and $L_{1B}$ are switched (and assuming $R_{LA3}=R_{LB3}$), the expression still holds but load resistance is reduced, since $R_{LA3}$ is added in parallel to $R_{LA2}$.
$R_{LA3}$ and $R_{LB3}$ are both $6\ G\Omega$ resistors so that, when the switches are shorted, the total load resistance can be reduced from $60\ G\Omega$ to $10\ G\Omega$.

Resistors $R_{L1}$, $R_{L2}$ and $R_{L3}$ have very large values, thus they cannot be made with standard metal film resistors.
Nevertheless, they enter in equation \ref{eq:biasi} with their ratio, so that the important feature is their relative tracking.
We designed a custom array for the resistors $R_{Li}$ in order to optimize their matching.
The arrays are produced by two companies, Ohmcraft and SRT Resistor Technologies, using high voltage radial leaded resistors.
A photograph of one sample is included in Figure~\ref{fig:CUOREbiasCircuit}.
The tracking ratio between the resistors was measured as better than $10\ ppm/^\circ C$ for both the companies.
Thermal drift is not the only concern, but also the low frequency or 1/f noise is important when large bias voltages are applied.
When the resistor is long, as is the case for the large value resistor $R_{L2}$, which about $20\ mm$, the electric field per unit of length is minimized, helping in limiting the low frequency noise~\cite{CUOREloadresnoise}.

The settings of the relays shown in Figure~\ref{fig:CUOREbiasCircuit} are those intended for the normal running condition.
A number of other combinations are available and exploited during the characterization of the thermistors.
Each relay in the board is made by a pair of double-pole-single-throw (DPST) switches and is bi-stable.
The relay model chosen for $L_1$, $L_2$ and $L_3$ (DS2E-SL2-DC5V) has a distance of $5.08\ mm$ ($200\ mils$) between the terminals in order to maximize the value of the parasitic resistances.
In addition, several small cuts in the PCB have been made around all the terminals that need large parasitic impedances, for breaking possible conductive paths due to solder flux residues and dust.
 
$L_{2A}$ and $L_{2B}$ are used to reverse the polarity of the bias applied to the thermistor.
This is very useful during the characterization phase, since the difference between the measurements taken with opposite bias polarity is independent from any offset.

When $L_{3A}$ and $L_{3B}$ are switched in their alternative position, the pair of test resistors $R_T$ and $R_{TT}$ ($2\ M\Omega$ and $100\ k\Omega$ respectively) are connected in place of the detector.
With this feature it is possible to apply the bias to these known resistors, allowing the calibration of the bias resistors and of the entire biasing system.
Across resistors $R_{TT}$ it is possible to apply an external test signal (either DC or dynamic) through the backplane, again for calibration purposes.

\begin{figure}
	\centering
	\includegraphics[width=.8\textwidth]{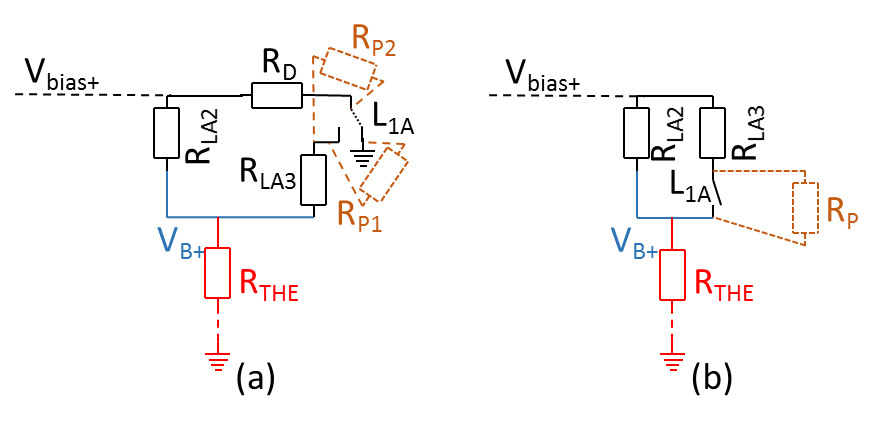}
	\caption{Detailed view of the biasing circuit. Configuration (a) is the solution adopted, while (b) shows a different and more canonical configuration.}
	\label{fig:CUORErelayconnection}
\end{figure}

It is worth to better explain the reason of the uncommon configuration used for $L_{A1}$ and $L_{B1}$, and highlighted in Figure~\ref{fig:CUORErelayconnection}.
Our first choice was the classical configuration (b), where $R_{LA3}$ has the switch $L_{1A}$ in series.
However, we noticed a large thermal drift when $L_{1A}$ was left open (a few hundreds of $ppm/^\circ C$), which is related to the presence of the parasitic resistance $R_P$ in parallel to the switch due to the insulating materials that compose the relay.

In configuration (a), adopted in our solution, the two parasitic resistances $R_{P1}$ and $R_{P2}$ are still present in parallel to the switch in open position, but this time $R_{LA3} + R_{P1} \parallel R_{P2}$ is connected between the detector and ground, and the voltage applied to it, $V_{B+}$, is much smaller than $V_{bias+}$ and, therefore, the effect of the thermal drift is greatly attenuated and made negligible.

The microcontroller on the mainboard manages the whole biasing system: voltages $V_{bias\pm}$ are read with a 24 bit $\Delta\Sigma$ ADC (Analog Devices AD7732) and the microcontroller is able to adjust the bias voltage to the desired value, compensating for the spread of the load resistor values, if necessary.

\subsection{Preamplifier}
\label{sec:CUOREpre}

The first stage of the amplification chain, usually referred to as preamplification, is fully differential.
This characteristic allow to suppress any common mode disturbance, either crosstalk between adjacent channels and common mode noise sources like microphonism or electromagnetic pickup on the long link from the cryostat.
Each detector is contacted by two wires that are twisted for the largest part of their path, so that any external signal injected on the wires, induces a similar effect on both.
This common mode signal is then canceled at the preamplifier inputs due to the differential configuration.

The preamplifier circuit configuration is similar to an instrumentation amplifier with only one pair of JFET transistors at its input, to minimize noise.
The JFETs are custom pairs selected from the NJ132 process from InterFET.
The gate to source pinch off voltage has been selected of small value in order to minimize the dropout voltage, hence the gate current~\cite{Pre1}, which was measured at about $30\ fA$ at room temperature, and around $100\ fA$ at $50\: ^\circ C$, close to the working condition inside the CUORE Faraday cage.
The parallel noise developed by the two input JFETs has a remarkable figure of less than $0.15\ fA/\sqrt{Hz}$, which is negligible with respect to the $0.5\ fA/\sqrt{Hz}$ noise coming from the load resistors.
The series white noise is about $3\ nV/\sqrt{Hz}$ and about $7\ nV/\sqrt{Hz}$ at 1 Hz, matching again the detector impedance.
Other fundamental characteristics featured by the preamplifier are a small thermal drift and high gain stability. Although the two JFETs in the pair are not from the same die and are only roughly matched, the preamplifier have some dedicated circuitry which can be used to compensated its voltage offset thermal drift down to less than $0.2\ \mu V/^\circ C$, as will be shown later with more detail.

\begin{figure}
	\centering
	\includegraphics[width=.8\textwidth]{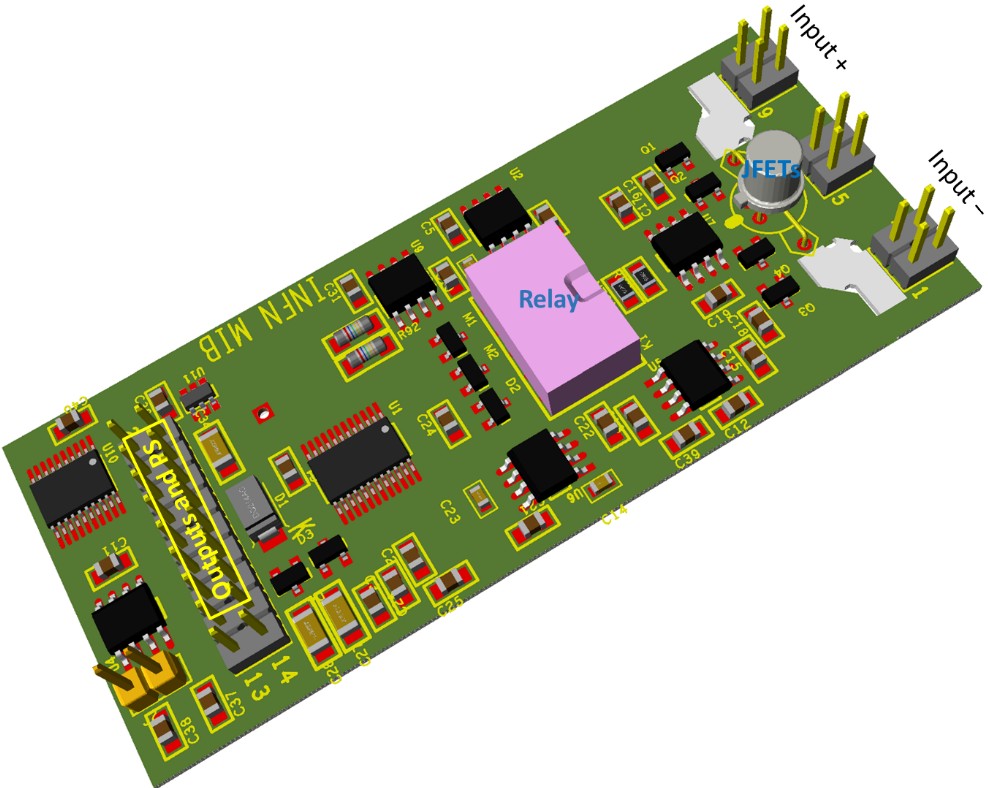}
	\caption{3D view of the top layer of the preamplifier, the side that faces the mainboard in Figure~\ref{fig:CUOREmainboard3D}.}
	\label{fig:CUOREpre3D}
\end{figure}

A 3D view of the preamplifier is shown in Figure~\ref{fig:CUOREpre3D}.
On the right side of the picture there are the two input connectors, close to the corners of the board.
The JFETs are enclosed in the same package, placed near the two connectors.
The leads of the gate connections are bent and routed far from the sources and drains.
The PCB area hosting the gate connections is well isolated from the rest of the circuit also by using cuts in the boards, in order to minimize any parasitic impedance.
The output connector is placed on the left side of the board.
On this connector, in addition to the amplified output signals, there is place also for the power supplies, the reference voltages and the digital communication signals.

\begin{figure}
	\centering
	\includegraphics[width=.9\textwidth]{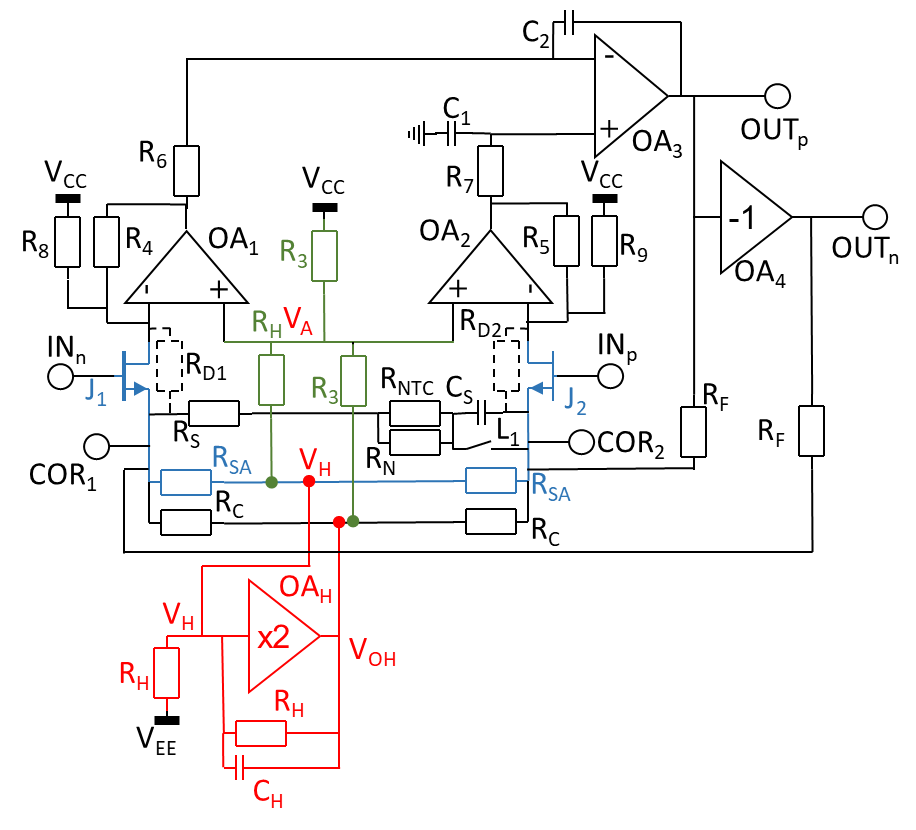}
	\caption{Schematic diagram of the preamplifier. Nodes $COR_1$ and $COR_2$ are where currents are injected for offset, thermal drift and CMRR compensations.}
	\label{fig:CUOREpreSchematic}
\end{figure}

The schematic circuit of this preamplifier is the latest evolution of previous designs~\cite{Pre2, Pre3, Pre4}.
Figure~\ref{fig:CUOREpreSchematic} shows the schematic of the preamplifier with the exception of the compensation circuitry, which will be described later.

The aim of the design is to operate the input JFETs, $J_1$ and $J_2$, at constant channel current and channel dropout voltage.
This is obtained by keeping perfect symmetry in the circuit schematic and using a slightly modified version of the Howland current generator, which was found adequate in our case where the frequency range of the signals is limited.
The Howland current generator is the network highlighted in red in Figure~\ref{fig:CUOREpreSchematic}, which includes the amplifier $OA_H$.
In a standard configuration, its output would be node $V_H$, but in our design we used also the output of $OA_H$.
Let's suppose an equilibrium condition for which $IN_n$ and $IN_p$ are equal.
The relay switch $L_1$ is closed in normal working condition.
Writing the balance of the currents at node $V_H$, we obtain ($C_H$ is a small compensating capacitance):
\begin{equation}
\label{eq:howland1}
\frac{V_A-V_H}{R_H} + 2\frac{V_{COR1}-V_H}{R_{SA}}+\frac{V_H}{R_H} = \frac{V_H-V_{EE}}{R_H} 
\end{equation}
with $V_{COR1}=V_{COR2}$.

The original idea of the Howland configuration is the presence of the term $\frac{V_H}{R_H}$ that is in the left side of the above equation.
This term, which comes from resistor $R_H$ connected between nodes $V_H$ and $V_{OH}$ (considering $V_{OH}=2 V_H$), simplifies with the corresponding term in the right side of the equation, forcing the sum of the 2 terms on the left to be constant.
Now, we write the equation at node $V_A$ (again using $V_{OH}=2 V_{H}$):
\begin{equation}
\frac{V_{CC}-V_A}{R_3} + \frac{2V_H-V_A}{R_3} = \frac{V_A-V_H}{R_H}
\end{equation}
and thus
\begin{equation}
\label{eq:howland2}
V_A-V_H = R_H I_{HA}
\end{equation}
where
\begin{equation}
I_{HA} = \frac{V_{CC}}{2R_H+R_3}
\end{equation}

The important result from equation \ref{eq:howland2} is that $V_A-V_H$ is constant.

Inserting this in equation \ref{eq:howland1} we can write
\begin{equation}
\frac{V_{COR1}-V_H}{R_{SA}} = \frac{I_H-I_{HA}}{2}
\end{equation}
where $I_H = -\frac{V_{EE}}{R_H}$. In this way the current through $R_{SA}$ is also constant.

The same is true also for the channel dropout voltage $V_A-V_{COR1\left(2\right)}$, once we consider that both the drains of $J_1$ and $J_2$ are at the same potential $V_A$, thanks to the local feedback of $OA_1$ and $OA_2$:
\begin{equation}
V_A-V_{COR1} = \frac{2R_H+R_{SA}}{2} I_{HA} - \frac{R_{SA}}{2} I_H \ .
\end{equation}

The inverting and non inverting terminals of amplifier $OA_3$ are kept at the same voltage due to its large gain and the consequence is that currents through $R_8$ and $R_9$ and through $R_4$ and $R_5$ are equal, and $OA_1$ and $OA_2$ behave as cascode current mirrors: current $I_{D1}$ equals current $I_{D2}$.
Capacitors $C_1$ and $C_2$ are $4.7\ nF$ compensating capacitors that limit the bandwidth to about $5\ kHz$.

The channel currents of both JFETs are forced equal by $OA_1$ and $OA_2$. Furthermore, thanks to the presence of the two resistors $R_C$, and under the condition $R_C = R_F$ while also assuming large impedances connected to nodes $COR_1$ and $COR_2$, they can be written as:
\begin{equation}
I_{D1(2)}=\frac{I_H-I_{HA}}{2} \left(1+\frac{2 R_{SA}}{R_F}\right) \ .
\end{equation}
In addition, nodes $OUT_p$ and $OUT_n$ are at zero potential.

To summarize, we have shown that, when the inputs are at the same voltage, the current through the input JFET channels and the voltage across them are constant, and that the outputs are at zero potential.

A differential input signal would determine a change of the output voltage given by (assuming $R_{NTC}\gg R_N$):
\begin{equation}
OUT_p - OUT_n = 2 \frac{R_{SE}+R_F}{R_{SE}} \left(V_{COR2}-V_{COR1}\right)
\end{equation}
with $R_{SE} = \left(R_S+R_N\right) \parallel 2 R_{SA}$.

In this case the channel dropout voltages of $J_1$ and $J_2$ become different by ($V_{COR1}-V_{COR2}$), while their currents remain equal.

The current through the drain resistors $R_{D1}$ and $R_{D2}$ undergoes a change when the drain to source voltage changes.
Since the current through the JFET is constant, it results that:
\begin{equation}
V_{COR1(2)}=\frac{g_{m1(2)}R_{D1(2)}}{1+g_{m1(2)}R_{D1(2)}} IN_{p(n)}
\end{equation}
where $g_{m1(2)}$ is the transconductance of the JFETs. Therefore we have:
\begin{equation}
OUT_p - OUT_n = 2 \: \frac{R_{SE}+R_F}{R_{SE}} \: \frac{g_{m1(2)} R_{D1(2)}}{1+g_{m1(2)} R_{D1(2)}} \left(IN_p-IN_n\right) \ .
\end{equation}
The term that depends on the drain resistors is close to unity since $g_m$ is $5\ mA/V$ and $R_{DS}$ is $12\ k\Omega$.

The thermal drift of the gain is minimized by adding the negative thermal coefficient (NTC) resistor $R_{NTC}$ in parallel to a small $5\ Omega$ metal film resistor.
The differential gain is about $210\ V/V$ with $R_F$ equal to $20\ k\Omega$ and $R_{SE}$ $190\ Omega$.
The channel current of each JFET is set at $0.5\ mA$, while $V_{DS}$ is $0.5\ V$.

When switch $L_1$ is set in the open position the DC gain is lowered to $20\ V/V$.
This working condition is adopted during the detector load curve characterization, to extend the preamplifier dynamic range at large thermistor biases.
Since at lower gain the stability becomes worse, we added capacitor $C_S$ ($4.7\ nF$), which ensures that the high frequency gain remains unaffected.

\begin{figure}
	\centering
	\includegraphics[width=\textwidth]{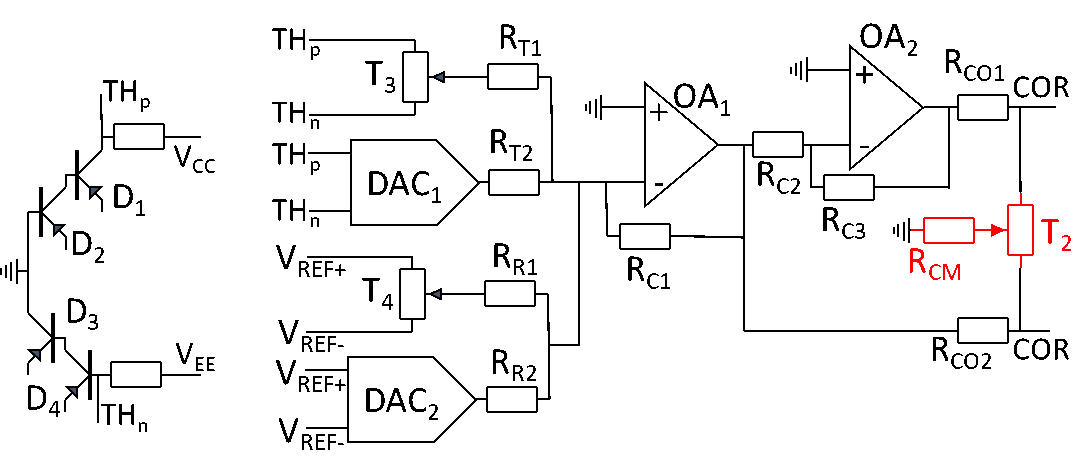}
	\caption{Offset, thermal drift and common mode correction circuitry for the preamplifier of Figure~\ref{fig:CUOREpreSchematic}.}
	\label{fig:CUOREpreCorrection}
\end{figure}

In the preamplifier description above, perfect matching between components was assumed.
This, however, is far from reality in particular for the input JFET pair.
In the real world, an input offset and a thermal drift are present but the preamplifier is able to compensate for both with the additional circuit shown in Figure~\ref{fig:CUOREpreCorrection}.

Offset voltage is adjusted with the 8-bit digital trimmer $T_4$ (from the 4 channel Analog Devices AD5263) and the 12-bit $DAC_2$ (from the two channel Analog Devices AD5415).
The voltage at the wiper terminal of $T_4$ is multiplied by about $-1$ by $R_{R1}$ and $R_{C1}$ in feedback to $OA_1$.
The output of $OA_1$ is made differential and applied to the 2 resistors $R_{CO1(2)}$, which are connected to the sources of the two input JFETs, at the nodes $COR_1$ and $COR_2$.
The output of $DAC_2$ follows the same path, except that resistor $R_{R2}$ is about 15 times larger than resistor $R_{R1}$ and therefore the weight of $DAC_2$ is 15 times smaller than that of $T_4$.
The reference voltage of $T_4$ and $DAC_2$ is $V_{REF\pm}$ ($\pm5\ V$), from the stable and low noise voltage generator described briefly above.
The maximum input compensation is about $\pm60\ mV$, more than enough to compensate for the maximum offset coming from the JFETs, which are matched at $\pm20\ mV$, and  the maximum bias foreseen across the detectors.
It is worth to note that, in contrast to previous assumptions, the resistors connected to the sources of the JFETs are not just $R_F$, and thus the value of resistors $R_C$ of Figure~\ref{fig:CUOREpreSchematic} is set to $R_F \parallel R_{CO1(2)}$.

The network for the offset thermal drift compensation uses the same path as the offset correction, except that it is based on trimmer $T_3$ and $DAC_1$ which have, as input references, the series of two pairs of forward biased diodes, $D_1$-$D_2$, and $D_3$-$D_4$, whose voltage has a dependence with temperature of $\pm4\ mV/^\circ C$.
Such range allows to compensate up to about $\pm40\ \mu V/^\circ C$ at the preamplifier input, with an accuracy of about $100\ nV/^\circ C$.
Drift compensation requires a calibration, which will be described with more detail in the section \ref{sec:thermal_cali}.

The common-mode rejection ratio, CMRR, is optimized with trimmer $T_2$ and resistor $R_{CM}$, by sinking asymmetric currents from nodes $COR_1$ and $COR_2$.
A good rejection is important not only for the suppression of common mode disturbances, but also to suppress parallel noise from the load resistors.
If we call $A_d$ the differential gain and $A_c$ the common mode gain of the amplifier of Figure~\ref{fig:CUOREpreModelNoise}, then, at the amplifier output, the noise contribution from the load resistors only is:

\begin{figure}
	\centering
	\includegraphics[width=.35\textwidth]{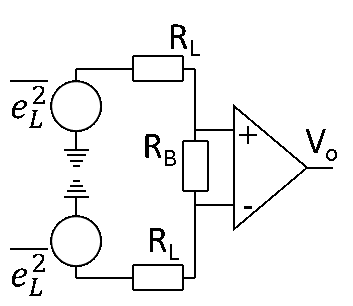}
	\caption{Model for the parallel noise at the preamplifier inputs.}
	\label{fig:CUOREpreModelNoise}
\end{figure}

\begin{equation}
\overline{V_o^2} = A_d^2 \left[ \frac{\overline{e_L^2}}{2 R_L^2} R_B^2 + \left(\frac{A_c}{A_d}\right)^2 \frac{\overline{e_L^2}}{2} \right] \ .
\end{equation}
If we require to the second term, coming from the common mode amplification, to be lower than $1\ nV/\sqrt{Hz}$, the ratio $\frac{A_c}{A_d}$ should be higher than $84\ dB$, for our $30\ G\Omega$ load resistors.

For the future CUPID experiment, we have developed a slightly upgraded version of the preamplifier described here, in order to reach a lower input series noise.
This new version will be presented in section~\ref{sec:CUPIDpre}.

\subsection{Second stage programmable gain amplifier}

The second stage amplifier is a programmable gain amplifier, PGA, which can be configured from $1\ V/V$ to $50\ V/V$.
This wide range is able to compensate the spread of the energy gain of the detectors.

\begin{figure}
	\centering
	\includegraphics[width=.65\textwidth]{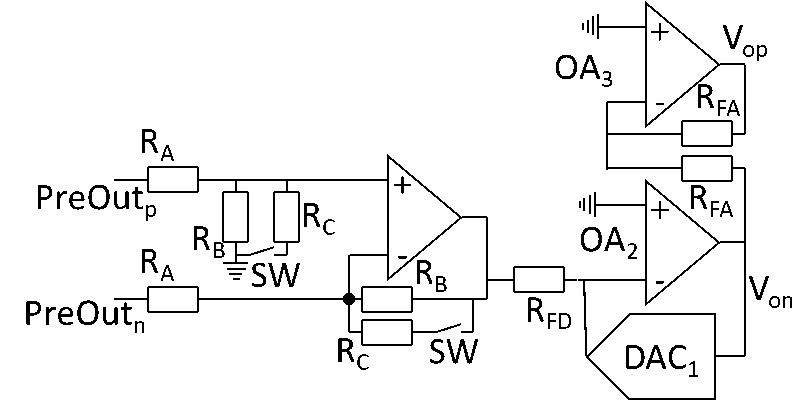}
	\caption{Second stage PGA.}
	\label{fig:CUOREpga}
\end{figure}

Its schematic circuit is shown in Figure~\ref{fig:CUOREpga}.
The PGA is made by two stages, with a differential to single ended configuration.
The first stage has a coarse gain that can be set to $10\ V/V$, with switches SW open, or $1\ V/V$, with switches SW closed.

The fine gain stage adds a gain with a range from $1\ V/V$ to $10\ V/V$, by means of the DAC in the feedback path of $OA_2$ (the same Analog Devices AD5415 used also in the offset compensation circuit).
The output is then converted back to differential by $OA_3$ in unitary inverting gain.
Resistors $R_{FA}$ and $R_{FD}$ are all internal to the DAC, with a relative stability of a few $ppm/^\circ C$, while resistors $R_A$, $R_B$ and $R_C$ are metal film.
The gain stability of this stage is better than $5\ ppm/^\circ C$.

\subsection{Antialiasing filter}

Before the digitization, the output signals are filtered by an active antialiasing filter implemented with a 6-pole Bessel-Thomson filter~\cite{CUOREbessel}.
This board features a roll off of $120\ dB/dec$ with 4 selectable bandwidth settings of $15\ Hz$, $35\ Hz$, $100\ Hz$ and $120\ Hz$.
The gain below the cut-off frequency is unitary.
The thermal noise of the resistors that are needed for the bandwidth selection, limits the input series noise, which is about $60\ nV/\sqrt{Hz}$ for the lower frequency bandwidth, where the larger input resistance is used.
The input RMS thermal drift is about $1.5\ \mu V/^\circ C$ as measured in the $40\ ^\circ C$ to $60\ ^\circ C$ temperature range.
Each antialiasing board has 12 channels and is equipped with a microcontroller that manages their operation following remote instructions from the DAQ.

An upgraded version with expanded functionalities for the CUPID experiment has been already designed and will be described in section~\ref{sec:CUPIDfilter}.

\subsection{Data acquisition}

The data acquisition system (DAQ) is built with an 18-bit commercial solution by National Instruments.
Data acquisition digitizes the signals without applying any on-line filtering.
The acquisition is continuous since the sampling rate is quite low ($1\ ksps$ for heat channels and $2\ ksps$ for the light channels of CUPID-0).
Triggering and off-line filtering for optimizing S/N ratio are performed in software with a dedicated software stack that implements a Wiener filter.

\subsection{Power supply}

The power supplies have been developed following the point-of-load (POL), approach.
With this approach the final stage of the power supply is distributed and located as close as possible to the load in order to guarantee better regulation and stability.

\begin{figure}
	\centering
	\includegraphics[width=.9\textwidth]{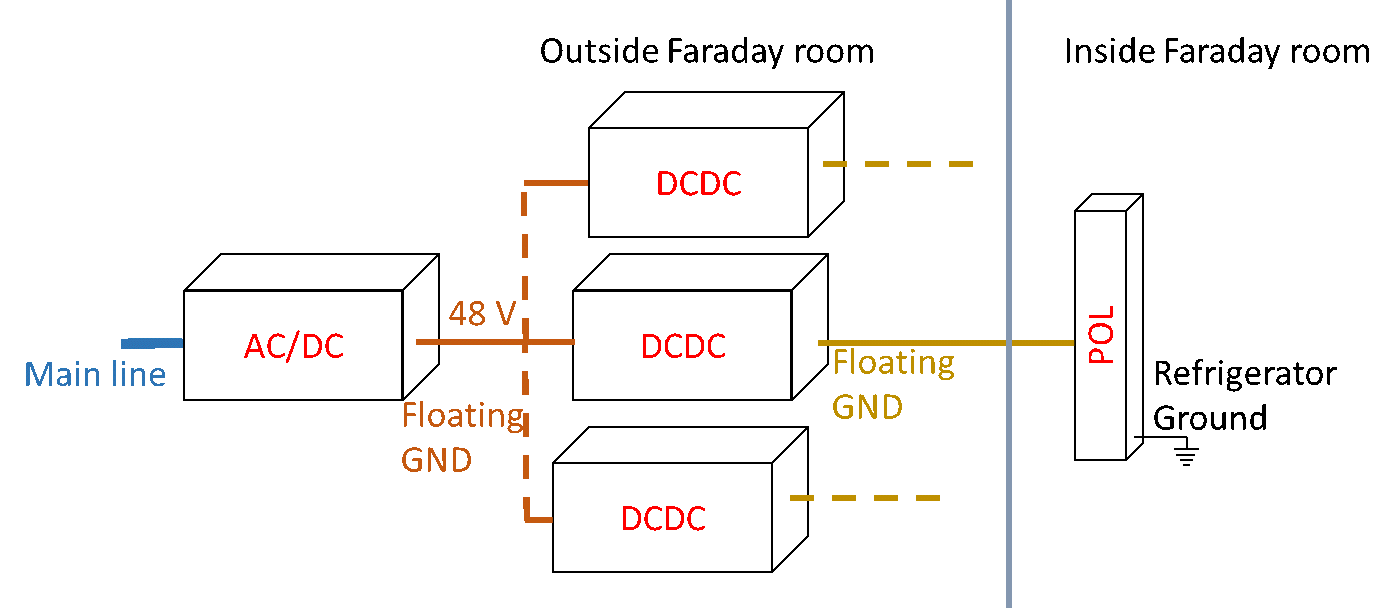}
	\caption{Diagram of the power supply system.}
	\label{fig:CUOREpowersupply}
\end{figure}

The scheme we used is shown in Figure~\ref{fig:CUOREpowersupply}.
A commercial AC/DC converter provides a $48\ V$ supply voltage for several DC/DC units~\cite{CUOREpowerSupply} having two analog outputs, adjustable between $\pm5\ V$ and $\pm15\ V$, and one digital output, adjustable from $2.5\ V$ to $8\ V$.
Each DC/DC unit powers the final stage which is a linear power supply generating $\pm5\ V$, with a very good regulation, low noise of less than $50\ nV/\sqrt{Hz}$ down to $1\ Hz$, and high stability at the level of $1\ ppm/^\circ C$~\cite{CUORElps}.
We required these characteristics as the final regulated voltage is used also as a reference for offset adjustment and for detector biasing.

Floating grounds on both the AC/DC and DC/DC regulator outputs allows to accomodate a single grounding node on the cryostat, suppressing ground loops.
Common grounding of the whole detector is at the DAQ input stage, and cannot be left floating for safety reasons.
The power dissipation of the whole front-end electronics inside the Faraday room is about $700\ W$.

\subsection{Communication system}

The communication link between the DAQ and the electronics system is very critical with respect to the possible injection of disturbances and creation of ground loops, as all the boards have to be connected together and to the DAQ through it.

\begin{figure}
	\centering
	\includegraphics[width=\textwidth]{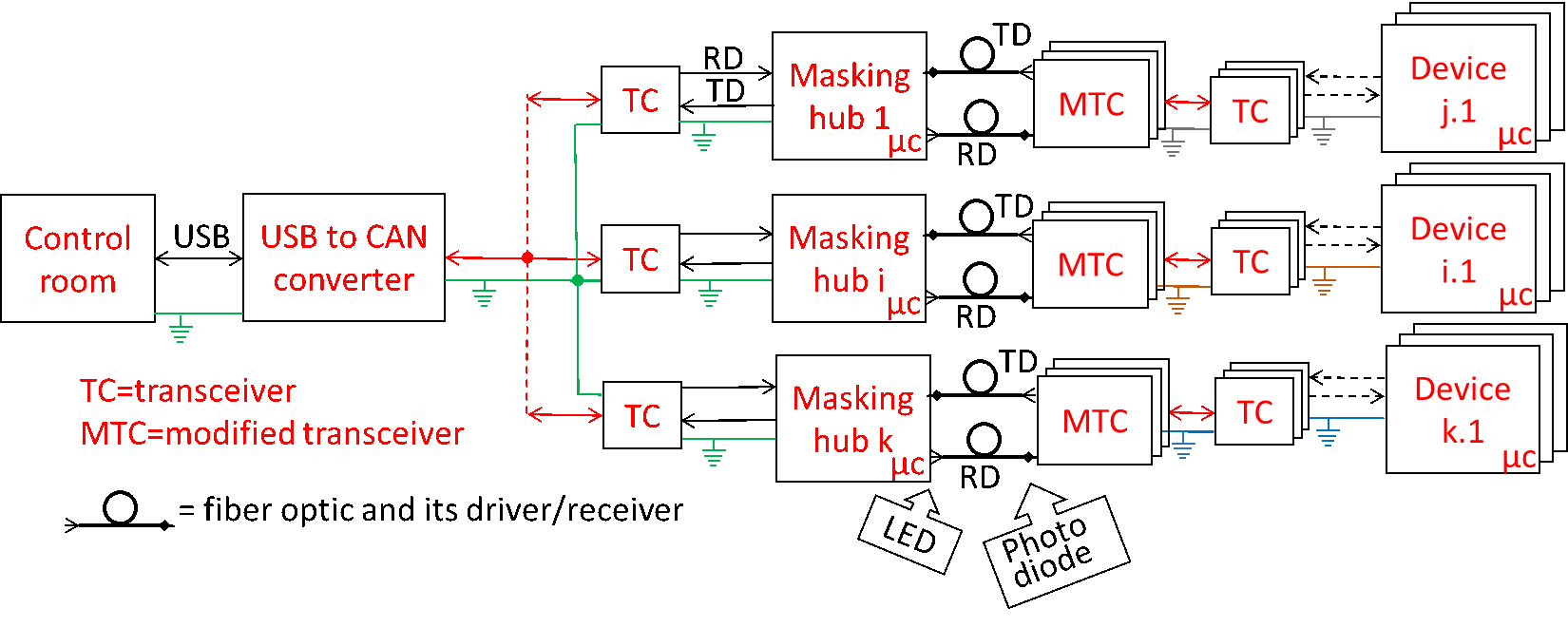}
	\caption{CAN bus connecting system based on fiber optic transmission.}
	\label{fig:CUOREcomm2}
\end{figure}

To avoid these effects we have modified the standard connection of the CAN-bus serial link by introducing fiber optics on the main paths, as shown in Figure~\ref{fig:CUOREcomm2}.
In addition to that, we reorganized the global system network into several virtual sub-networks in a similar fashion to what is done with VLANs in Ethernet switches.

A signal from/to the USB-to-CAN converter arrives to a series of trans-ceivers (TC), that translate it to the transmission (TD) and reception (RD) lines to the masking hubs.
A CAN packet is routed by the masking hub to the sub-network only if its address field includes the proper sub-address set into the masking hub itself.
In this way only a group of the full system receives the message, and hence is waken up, minimizing unnecessary digital activity (and related noise) in the front-end.

The masking hubs manage two separate CAN-bus domains.
The input CAN-bus is from the control room and the slow control system, while the output CAN-bus is converted to optical signals and distributed to the boards with optical fibers.
This allows to separate the ground connection between the different sub-groups.
The optical signals are converted back to electrical differential links by the MTC, a modified passive transceiver with a LED and a photodiode that convert signals back to the electrical domain needed by the boards.
The conversion from the electrical to the optical domain is easier for the TD and RD lines rather than with the standard CAN-bus differential link, since the two lines are mono-directional.

The transceivers that interface with the slow control are housed outside of the Faraday room, while the rest is located inside.
This solution helps further in minimizing the interferences close to the detector area.

\section{Measurements and compensations}

In this section we will describe the work that has been performed in the qualification phase of the CUORE and CUPID-0 electronics, followed by the performance results obtained.

All the boards produced were assembled in the final card cages configuration, as described in the previous sections, and we tested each channel for eventual production defects and component failures. For each channel we also performed the required calibrations in order to minimize the thermal drift and maximize common mode rejection. Finally, each channel was completely characterized in order to verify the compliance of its performance specifications.

\subsection{Thermal drift optimization}
\label{sec:thermal_cali}

The circuit for the offset voltage thermal drift compensation has been already described in section \ref{sec:CUOREpre}.
The aim of the optimization process is to find the required parameters that have to be set in the trimmer and DAC, in order to minimize the offset thermal drift.
This procedure has to be done on each channel independently, since each channel exhibit a different matching.

The optimization comprises two aspects: the characterization of the channel parameters, and the development of the firmware for controlling the setting of those parameters.

The first phase of the work was conducted in parallel.
A first characterization of the thermal drift served to understand the strategies to be adopted in the firmware algorithm.
In this phase we found that the coarse correction coefficient can be set independently of the input offset or preamplifier temperature, allowing to reach a residual drift just below the $\mu V/^\circ C$, as it will be shown later.
A further correction dependent on both input offset and preamplifier temperature, will refine this correction even further, taking into account second order effects like the quadratic drift.

The firmware was developed accordingly, before starting the final measurements for all the channels.
A lookup table, stored on an EEPROM memory, hosts all the required coefficients used by the algorithm.
In this way, at the end of the optimization process, it was possible to perform a final test using the definitive algorithm and, if this test is passed, the coefficients can be written into the preamplifier memory.
The thermal compensation is executed at user request, at the end of the offset correction command.

Due to the high number of channels involved, the measurement setup is completely automatized and remotely controllable.
The card cage under optimization is inserted in a V\"{o}tsch VC4018 climatic chamber and the output voltages are read out using three Keithley 2700 series 6.5 digit multimeters, for a total number of 60 channels.
Since the number of channel in one card cage is 78, the procedure had to be split in two halves for each card cage.
All the instrumentation was managed using a MATLAB graphical user interface (GUI).
The same GUI also managed the communication with the front-end electronics using a command library developed on purpose, and also controlled the entire process of the measurement and optimization phase, checking data quality and performing the required calculations.

The process needs several steps, consisting in a temperature sweep each.
In the first sweep spanning from $20\:\degC$ to $50\:\degC$ in $10\:\degC$ steps, the coarse corrector is varied and optimized in order to minimize the residual drift.
Then it is written into the EEPROM memory installed in each preamplifier.
A second temperature sweep, with finer $5\:\degC$ steps, allow to optimize the fine corrector.
In this phase the sweep is divided in 3 temperature intervals (centered at $25\:\degC$, $35\:\degC$ and $45\:\degC$) and for each interval a specific value for the fine corrector is found.
This allow to minimize the residual quadratic drift.
At the end, a third sweep is performed in order to verify the effectiveness of the process and the measurement of the final residual drift is stored in a database.
The thermal correction also depends on the input offset voltage and detector bias, so an external signal is applied during the sweeps and the algorithm is able to compensate also for this dependence. 

\begin{figure}
\centering
\includegraphics[width=.8\linewidth]{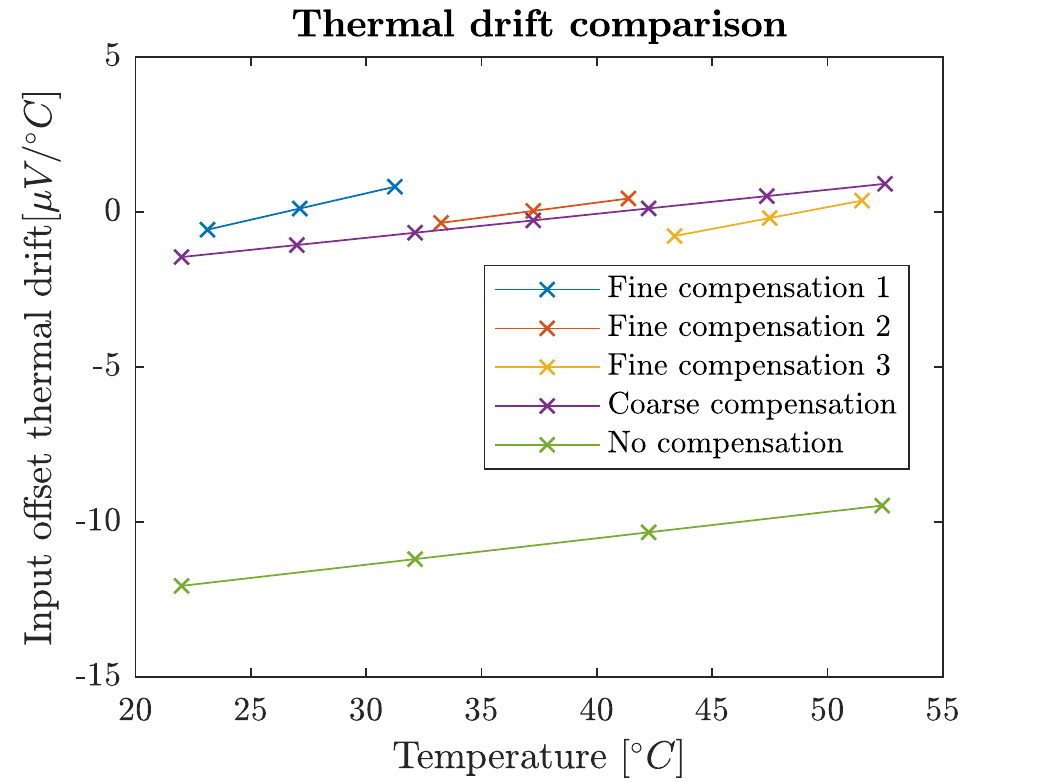}
\caption{Input offset thermal drift as a function of temperature for a typical channel at each different optimization step. The curves are not flat since there is a residual quadratic component. The closest to zero, the lowest the residual drift.}
\label{fig:temp_interval_comparison_curves}
\end{figure}

Figure~\ref{fig:temp_interval_comparison_curves} shows an example of the thermal drift in each phase for a typical channel.
The uncompensated drift amounts to about $-10\ \mu V/\degC$, while the coarse compensation is able to reduce it to about $\pm1 \ \mu V/\degC$.
The three sets of the finer compensation further reduce the residual drift and also better compensate the quadratic drift at the border of the interval.

\begin{figure}
\centering
\includegraphics[width=.9\linewidth]{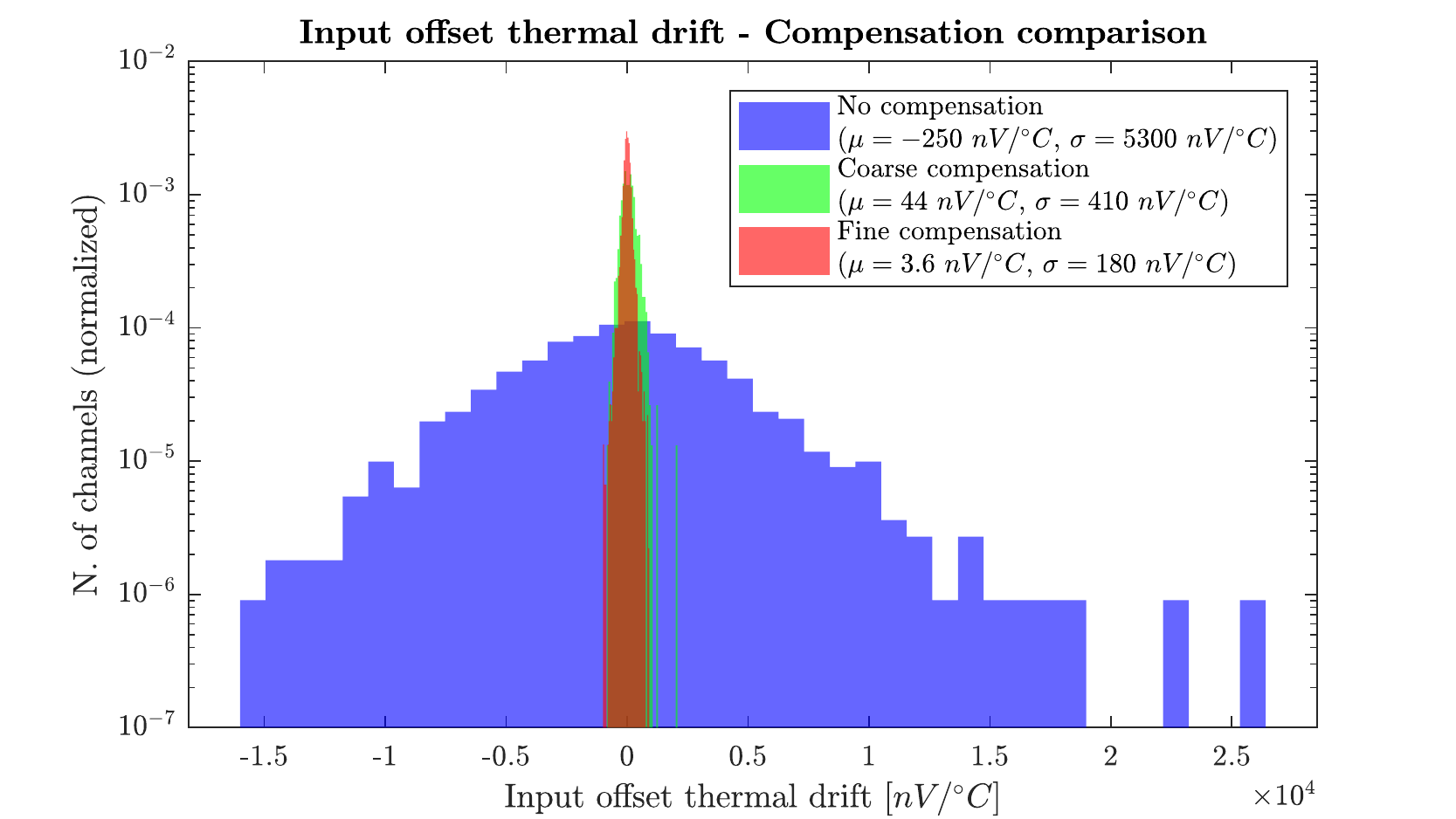}
\caption{Distribution of the residual thermal drift for all the channels at each step. The blue distribution represent the original uncompensated drift of the preamplifiers, while green and red represent the drifts after the coarse and fine compensation steps, respectively.}
\label{fig:temp_comparison}
\end{figure}

Figure~\ref{fig:temp_comparison} shows the distribution of the residual drifts at the end of the three phases.
The RMS value for the uncompensated drift is $5\ \mu V/\degC$.
It is reduced to $0.4\ \mu V/\degC$ with the coarse correction and, finally, $0.18\ \mu V/\degC$ with the finer compensation.

\begin{figure}
\centering
\includegraphics[width=.8\linewidth]{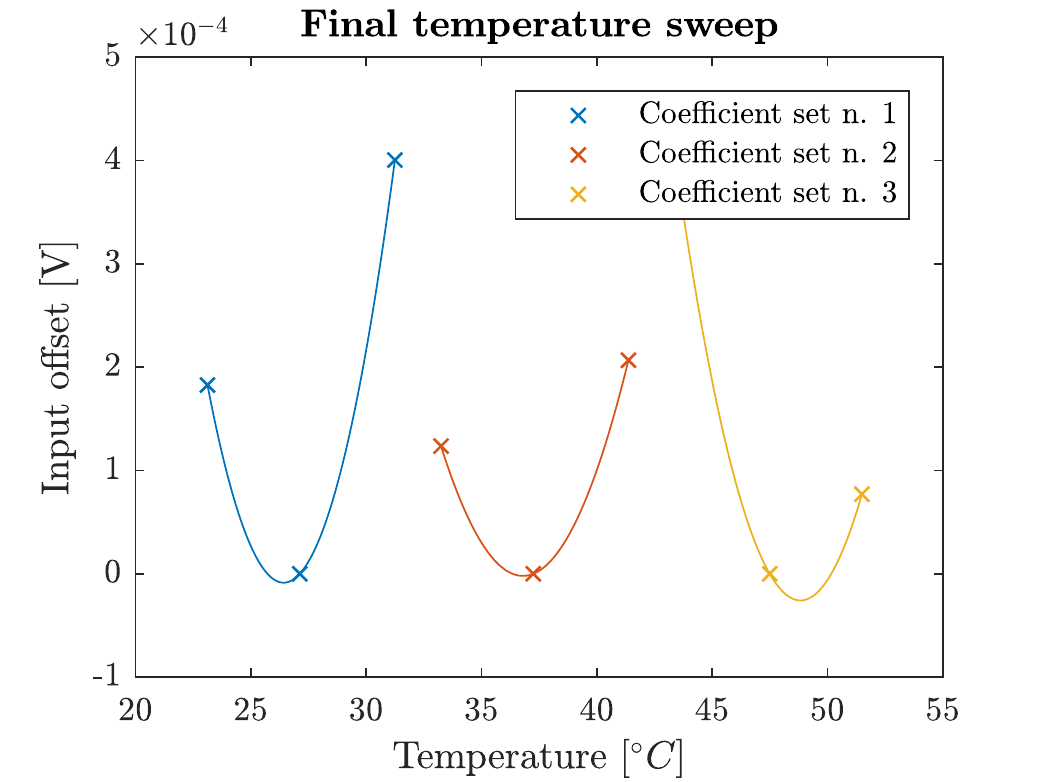}
\caption{This plot shows the offset voltage as a function of temperature during the final temperature sweep. It is possible to appreciate the residual quadratic drift and how beneficial it is to have different coefficient sets.}
\label{fig:temp_interval_comparison}
\end{figure}

In Figure~\ref{fig:temp_interval_comparison} it is possible to appreciate a detail of the final temperature sweep, where it is clear the benefit of having three separate set of corrections in order to minimize the quadratic thermal drift.

After the optimization phase was completed, we developed a further firm-ware upgrade which is able to improve even further the quadratic drift in the region between two intervals.
This is done by averaging the coefficients with a weight that depends from the distance of each interval center.
The effectiveness of this algorithm was verified using temperature variations between night and day, since the system was already mounted at the LNGS.

\begin{figure}
\centering
\definecolor{mycolor1}{rgb}{0.00000,0.44700,0.74100}%
\begin{tikzpicture}

\begin{axis}[%
width=3.536in,
height=1.916in,
at={(0.593in,0.415in)},
scale only axis,
xmin=-8,
xmax=8,
xlabel style={font=\color{white!15!black}},
xlabel={Offset voltage drift [$\mu V/^{\circ}C$]},
ymin=0,
ymax=200,
ylabel style={font=\color{white!15!black}},
ylabel={N. of channels},
axis background/.style={fill=white},
title style={font=\bfseries},
title={\textbf{Bessel filter offset voltage drift $\mathbf{(40\ ^\circ C - 60 \ ^\circ C )}$}},
legend style={legend cell align=left, align=left, draw=white!15!black}
]
\addplot[ybar interval, fill=mycolor1, fill opacity=0.6, draw=black, area legend] table[row sep=crcr] {%
x	y\\
-6.5	1\\
-6	0\\
-5.5	0\\
-5	4\\
-4.5	13\\
-4	43\\
-3.5	108\\
-3	173\\
-2.5	195\\
-2	137\\
-1.5	111\\
-1	88\\
-0.5	109\\
0	94\\
0.5	49\\
1	21\\
1.5	4\\
2	0\\
2.5	0\\
3	0\\
3.5	0\\
4	1\\
4.5	1\\
5	0\\
5.5	0\\
6	0\\
6.5	0\\
7	0\\
7.5	1\\
8	1\\
};
\addlegendentry{Normal distribution\\$\mu = -1.6 \ \mu V/^\circ C$\\$\sigma = 1.4 \ \mu V/^\circ C$}

\end{axis}
\end{tikzpicture}%
\caption{Offset voltage thermal drift of the antialiasing filter. In order to compare it to the drift of the front-end, it must be divided by the gain, which typically ranges from $1000\ V/V$ to $5000\ V/V$. So this contribution is completely negligible.}
\label{fig:Drift_Bessel_40_60_exported}
\end{figure}
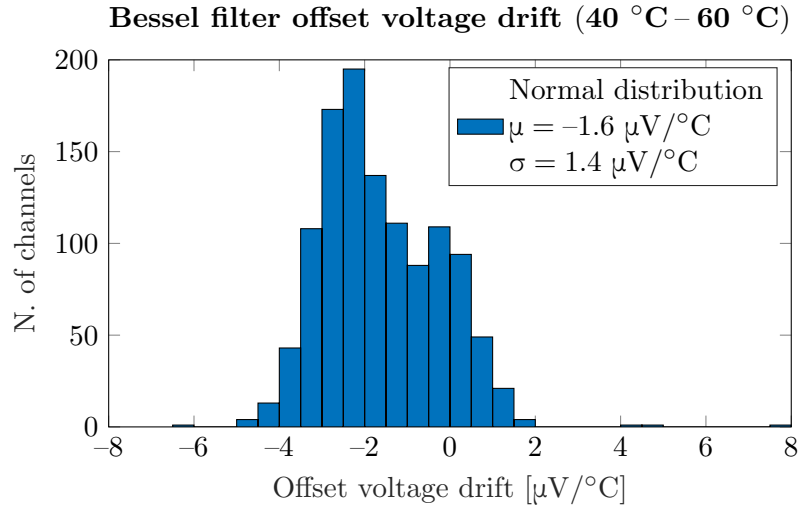

The offset thermal drift of the anti-aliasing boards was also tested.
The results are plotted in Figure~\ref{fig:Drift_Bessel_40_60_exported} and amounts to $-1.6\pm1.4\ \mu V/\degC$.
Since the gain of the amplification stage is typically \mytilde1000~V/V, this drift is negligible with respect to the contribution of the front-end.

\subsection{Gain measurements}

The card cages were then arranged on the racks and placed in a second test setup for further measurements. This setup adopts also the final power supply scheme described before.
This setup allow to perform gain, noise, and load resistors measurements in order to verify the compliance of each channel's characteristics with the specifications.
The test system is installed inside a Faraday cage and the readout is done with an 80 channel National Instruments PXI system.
Also in this system, a MATLAB GUI controls all the instrumentation, the data taking and the data analysis. 

\begin{figure}
\centering
\definecolor{mycolor1}{rgb}{0.00000,0.44700,0.74100}%
\begin{tikzpicture}

\begin{axis}[%
width=3.633in,
height=2.471in,
at={(0.609in,0.419in)},
scale only axis,
xmin=205.074,
xmax=206.526,
xlabel style={font=\color{white!15!black}},
xlabel={Preamplifier gain [V/V]},
ymin=0,
ymax=157.5,
ylabel style={font=\color{white!15!black}},
ylabel={N. of channels},
axis background/.style={fill=white},
title style={font=\bfseries},
title={\textbf{Preamplifier high gain distribution}},
legend style={legend cell align=left, align=left, draw=white!15!black}
]
\addplot[ybar interval, fill=mycolor1, fill opacity=0.6, draw=black, area legend] table[row sep=crcr] {%
x	y\\
205.14	1\\
205.206	1\\
205.272	8\\
205.338	22\\
205.404	33\\
205.47	61\\
205.536	91\\
205.602	120\\
205.668	150\\
205.734	136\\
205.8	134\\
205.866	113\\
205.932	83\\
205.998	45\\
206.064	25\\
206.13	13\\
206.196	3\\
206.262	3\\
206.328	1\\
206.394	1\\
206.46	1\\
};
\addlegendentry{Avg. gain\\$(205.8 \pm 0.2) \ V/V$}

\end{axis}
\end{tikzpicture}%
\caption{Gain distribution in preamplifier high gain mode.}
\label{fig:gain_distr_high}
\end{figure}
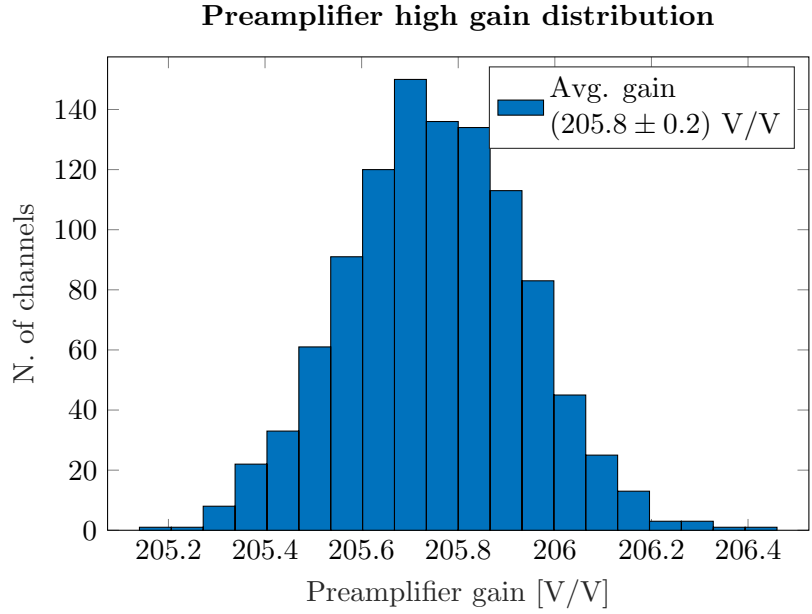

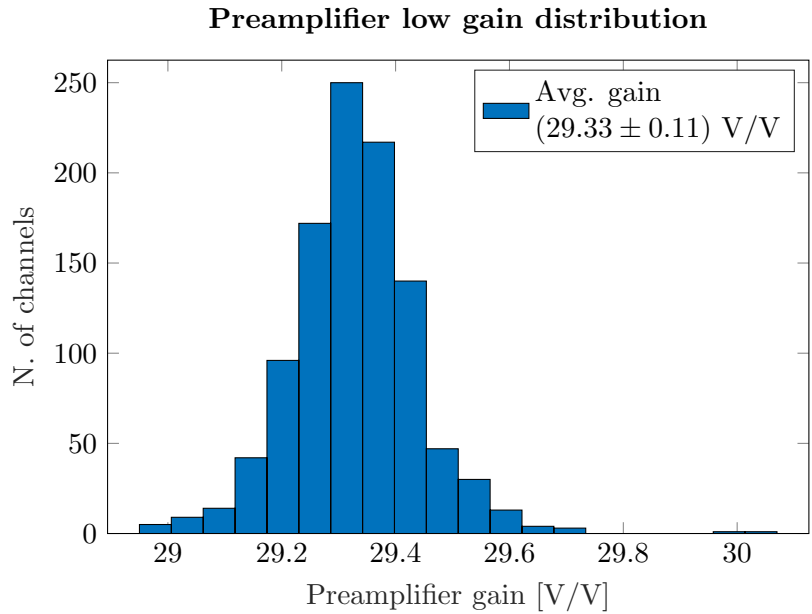
\begin{figure}
\centering
\definecolor{mycolor1}{rgb}{0.00000,0.44700,0.74100}%
\begin{tikzpicture}

\begin{axis}[%
width=3.633in,
height=2.471in,
at={(0.609in,0.419in)},
scale only axis,
xmin=28.894,
xmax=30.126,
xlabel style={font=\color{white!15!black}},
xlabel={Preamplifier gain [V/V]},
ymin=0,
ymax=262.5,
ylabel style={font=\color{white!15!black}},
ylabel={N. of channels},
axis background/.style={fill=white},
title style={font=\bfseries},
title={\textbf{Preamplifier low gain distribution}},
legend style={legend cell align=left, align=left, draw=white!15!black}
]
\addplot[ybar interval, fill=mycolor1, fill opacity=0.6, draw=black, area legend] table[row sep=crcr] {%
x	y\\
28.95	5\\
29.006	9\\
29.062	14\\
29.118	42\\
29.174	96\\
29.23	172\\
29.286	250\\
29.342	217\\
29.398	140\\
29.454	47\\
29.51	30\\
29.566	13\\
29.622	4\\
29.678	3\\
29.734	0\\
29.79	0\\
29.846	0\\
29.902	0\\
29.958	1\\
30.014	1\\
30.07	1\\
};
\addlegendentry{Avg. gain\\$(29.33 \pm 0.11) \ V/V$}

\end{axis}
\end{tikzpicture}%
\caption{Gain distribution in preamplifier low gain mode.}
\label{fig:gain_distr_low}
\end{figure}

Gain measurements verified the correct functionality of each element of the amplification chain.
Several input signals and PGA gain settings were used.
The distributions of the preamplifier gain are plotted in Figure~\ref{fig:gain_distr_high} and Figure~\ref{fig:gain_distr_low}, for the two gain settings.

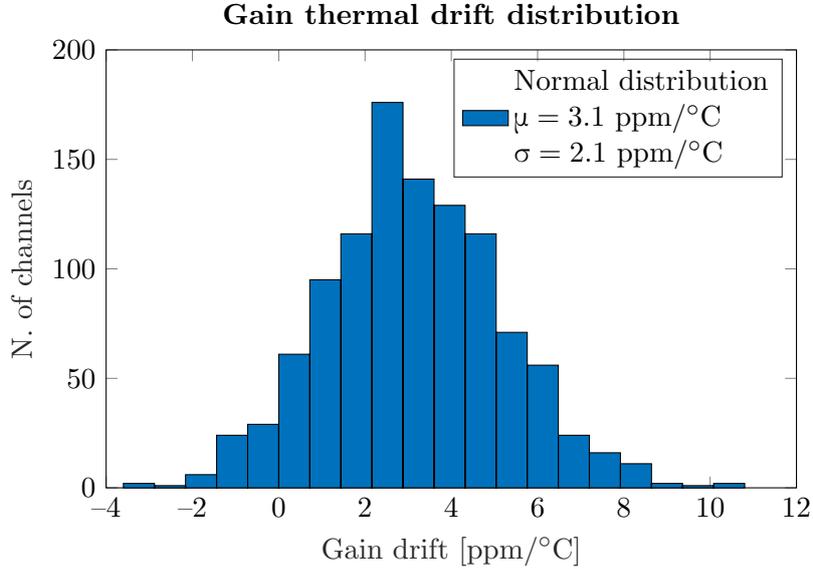
\begin{figure}
\centering
\definecolor{mycolor1}{rgb}{0.00000,0.44700,0.74100}%
\begin{tikzpicture}

\begin{axis}[%
width=3.576in,
height=2.286in,
at={(0.6in,0.419in)},
scale only axis,
xmin=-4,
xmax=12,
xtick={-4, -2,  0,  2,  4,  6,  8, 10, 12},
xminorticks=true,
xlabel style={font=\color{white!15!black}},
xlabel={Gain drift [$ppm/^\circ C$]},
ymin=0,
ymax=200,
ylabel style={font=\color{white!15!black}},
ylabel={N. of channels},
axis background/.style={fill=white},
title style={font=\bfseries},
title={\textbf{Gain thermal drift distribution}},
legend style={legend cell align=left, align=left, draw=white!15!black, fill=none}
]
\addplot[ybar interval, fill=mycolor1, fill opacity=0.6, draw=black, area legend] table[row sep=crcr] {%
x	y\\
-3.6	2\\
-2.88	1\\
-2.16	6\\
-1.44	24\\
-0.720000000000001	29\\
-8.88178419700125e-16	61\\
0.72	95\\
1.44	116\\
2.16	176\\
2.88	141\\
3.6	129\\
4.32	116\\
5.04	71\\
5.76	56\\
6.48	24\\
7.2	16\\
7.92	11\\
8.64	2\\
9.36	1\\
10.08	2\\
10.8	2\\
};
\addlegendentry{Normal distribution\\$\mu = 3.1 \ \mathrm{ppm/^\circ C}$\\$\sigma = 2.1 \ \mathrm{ppm/^\circ C}$}

\end{axis}
\end{tikzpicture}%
\caption{Distribution of the thermal drift of the preamplifier gain.}
\label{fig:Gain_drift_Paolo_exported}
\end{figure}

The thermal drift of the preamplifier gain has been already characterized in the past and are shown in Figure~\ref{fig:Gain_drift_Paolo_exported} for completeness.
These results are obtained from a temperature drift from $25\:\degC$ to $35\:\degC$.
An average gain drift of $3.1\ ppm/\degC$ with a sigma of $2.1\ ppm/\degC$ was found.

\subsection{Noise measurements}

The National Instruments PXI system allowed to measure the noise of each channel with a bandwidth of $31.25\ kHz$.
This is enough for our requirements since the bandwidth of the preamplifiers is limited to $5\ kHz$.
With this measurement we were able to detect channels with unexpectedly higher noise and substitute them.
We also corrected a few channels that had missing components.
Some channels showed an increase in the white noise injected through the offset correction circuit, so we added a $200\ nF$ capacitor in parallel to resistance $R_{C1}$ in Figure~\ref{fig:CUOREpreCorrection}, limiting its bandwidth to $8\ Hz$.
Lower bandwidth would have slowed the offset correction procedure too much.
Voltage noise, however, is not expected to be the limiting source of noise in the detector readout, as discussed before.

\begin{figure}
\centering
\includegraphics[width=.8\linewidth]{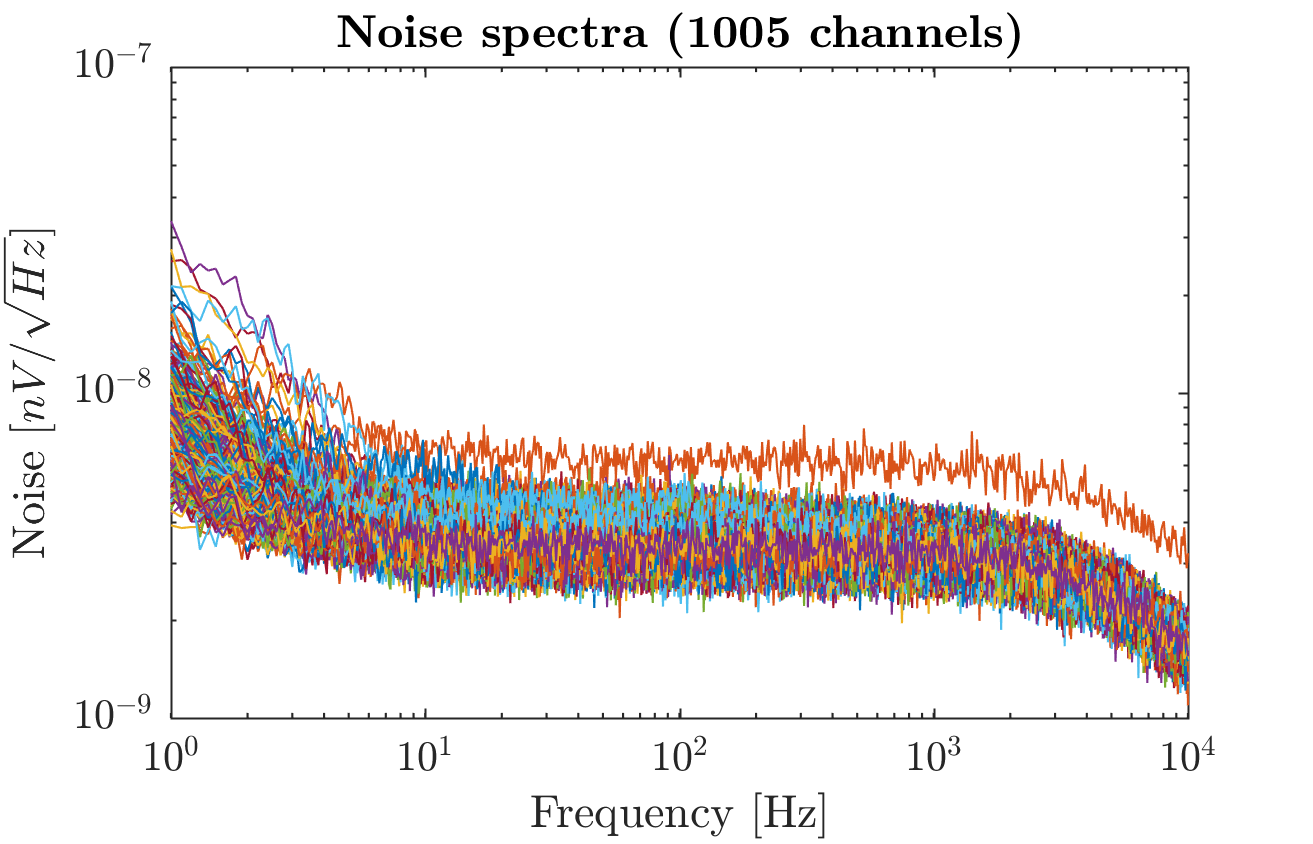}
\caption{Superposition of the noise spectra of all the measured channels.}
\label{fig:noise_spectra}
\end{figure}

\begin{figure}
\centering
\definecolor{mycolor1}{rgb}{0.00000,0.44700,0.74100}%
\begin{tikzpicture}

\begin{axis}[%
width=2.532in,
height=1.736in,
at={(0.485in,0.436in)},
scale only axis,
xmin=3,
xmax=5,
xlabel style={font=\color{white!15!black}},
xlabel={Noise [$nV/\sqrt{Hz}$]},
ymin=0,
ymax=300,
ylabel style={font=\color{white!15!black}},
ylabel={N. of channels},
axis background/.style={fill=white},
title style={font=\bfseries},
title={\textbf{White noise distribution}},
legend style={legend cell align=left, align=left, draw=white!15!black}
]
\addplot[ybar interval, fill=mycolor1, fill opacity=0.6, draw=black, area legend] table[row sep=crcr] {%
x	y\\
3.2	27\\
3.283	224\\
3.366	287\\
3.449	138\\
3.532	73\\
3.615	50\\
3.698	33\\
3.781	59\\
3.864	32\\
3.947	26\\
4.03	13\\
4.113	8\\
4.196	8\\
4.279	7\\
4.362	5\\
4.445	6\\
4.528	3\\
4.611	3\\
4.694	0\\
4.777	2\\
4.86	2\\
};
\addlegendentry{Average noise:\\3.53 $nV/\sqrt{Hz}$}

\end{axis}
\end{tikzpicture}%
\caption{White noise distribution.}
\label{fig:noise_White_Noise}
\end{figure}
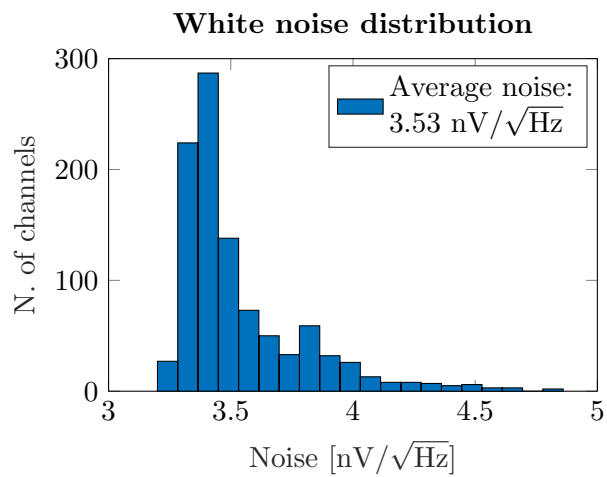

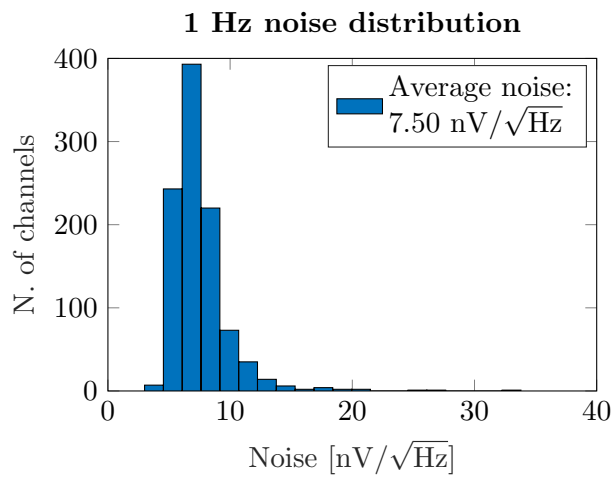
\begin{figure}
\centering
\definecolor{mycolor1}{rgb}{0.00000,0.44700,0.74100}%
\begin{tikzpicture}

\begin{axis}[%
width=2.532in,
height=1.736in,
at={(0.485in,0.436in)},
scale only axis,
xmin=0,
xmax=40,
xlabel style={font=\color{white!15!black}},
xlabel={Noise [$nV/\sqrt{Hz}$]},
ymin=0,
ymax=400,
ylabel style={font=\color{white!15!black}},
ylabel={N. of channels},
axis background/.style={fill=white},
title style={font=\bfseries},
title={\textbf{1 Hz noise distribution}},
legend style={legend cell align=left, align=left, draw=white!15!black}
]
\addplot[ybar interval, fill=mycolor1, fill opacity=0.6, draw=black, area legend] table[row sep=crcr] {%
x	y\\
3	7\\
4.54	243\\
6.08	393\\
7.62	220\\
9.16	73\\
10.7	35\\
12.24	14\\
13.78	6\\
15.32	2\\
16.86	4\\
18.4	2\\
19.94	2\\
21.48	0\\
23.02	0\\
24.56	1\\
26.1	1\\
27.64	0\\
29.18	0\\
30.72	0\\
32.26	1\\
33.8	1\\
};
\addlegendentry{Average noise:\\7.50 $nV/\sqrt{Hz}$}

\end{axis}
\end{tikzpicture}%
\caption{Low frequency ($1\ Hz$) noise distribution.}
\label{fig:noise_Hz1_Noise}
\end{figure}

\begin{figure}
\centering
\definecolor{mycolor1}{rgb}{0.00000,0.44700,0.74100}%
\begin{tikzpicture}

\begin{axis}[%
width=2.532in,
height=1.749in,
at={(0.485in,0.423in)},
scale only axis,
xmin=30,
xmax=55,
xlabel style={font=\color{white!15!black}},
xlabel={Noise [nV]},
ymin=0,
ymax=250,
ylabel style={font=\color{white!15!black}},
ylabel={N. of channels},
axis background/.style={fill=white},
title style={font=\bfseries},
title={\textbf{RMS noise distribution $\mathbf{(0.5\ Hz < f < 120\ Hz)}$}},
legend style={legend cell align=left, align=left, draw=white!15!black}
]
\addplot[ybar interval, fill=mycolor1, fill opacity=0.6, draw=black, area legend] table[row sep=crcr] {%
x	y\\
34.2	5\\
35.16	45\\
36.12	195\\
37.08	243\\
38.04	162\\
39	92\\
39.96	56\\
40.92	47\\
41.88	55\\
42.84	27\\
43.8	22\\
44.76	13\\
45.72	11\\
46.68	12\\
47.64	6\\
48.6	5\\
49.56	5\\
50.52	1\\
51.48	1\\
52.44	1\\
53.4	1\\
};
\addlegendentry{Average noise:\\39.0 nV}

\end{axis}
\end{tikzpicture}%
\caption{RMS noise distribution in the frequency range from $0.5\ Hz$ to $120\ Hz$ (the highest cut-off frequency of the antialiasing filter).}
\label{fig:noise_RMS_Noise_120Hz}
\end{figure}
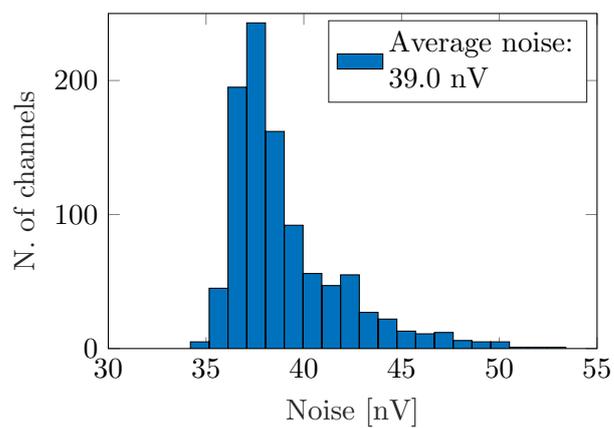

Figure~\ref{fig:noise_spectra} shows a superposition of the spectra of all the channels.
Input white noise is $3.5\ nV/\sqrt{Hz}$ on average and its distribution is shown in Figure~\ref{fig:noise_White_Noise}.
Average noise at $1\ Hz$ is $7.5\ nV/\sqrt{Hz}$, as in Figure~\ref{fig:noise_Hz1_Noise}.
Finally, the RMS noise evaluated in the bandwidth from $1\ Hz$ to $120\ Hz$ (the highest cut-off frequency of the anti-aliasing filter) was measured at $39\ nV$ RMS, as in Figure~\ref{fig:noise_RMS_Noise_120Hz}.

The input current noise of the preamplifier was not measured in this setup, since it has been previously characterized on the entire preamplifier pre-production and none of the devices showed a noise higher than $0.08 fA/\sqrt{Hz}$ at room temperature.

\subsection{CMRR optimization}

In an ideal differential amplifier, the output voltage does not depend on the common mode of its inputs and the output voltage can be written as:
\begin{equation}
\label{eq:ideal_ampli}
V_o = A_d\left(V_+ - V_-\right) \ ,
\end{equation}
where $A_d$ is the differential gain.
Real amplifiers, however, can have mismatches in the input differential pair or in the feedback network, that would cause also the input common mode voltage $\frac{V_+ + V_-}{2}$ to be amplified with a gain $A_{cm}$. 
Then, equation \ref{eq:ideal_ampli} becomes:
\begin{equation}
\label{eq:ampli_cm}
V_o = A_d\left(V_+-V_-\right) + \frac{A_{cm}}{2} \left(V_+ + V_-\right) \ .
\end{equation}
The common mode rejection ratio (CMRR) quantifies the common mode gain with respect to the differential gain and it can be written as:
\begin{equation}
\label{eq:cmrr}
CMRR = 20\log\left(\frac{A_d}{A_{cm}}\right) \ .
\end{equation}

CMRR depends on the frequency of the input signal, however, since the frequencies of interest in this application are very low, dynamic and static CMRR are considered equal.

The measurement of this parameter consists in applying a purely common mode signal ($V_+ = V_- = V_{cm}$, so that $V_+-V_- = 0$) and measuring the output voltage $V_o$ which is given only by the common mode gain $A_{cm}$.
With our very flexible front-end system we can exploit the bias circuitry in order to generate an asymmetric voltage.
This measurement was performed on site at LNGS, when the detectors were still hot and therefore having a very low impedance, completely negligible with respect to the load resistance of $60\ G\Omega$.
Asymmetric bias ($1.5\ V$) was applied in both polarities, and the average of the two output voltages was considered for the calculation of the CMRR, in order to cancel any contribution due to offset.

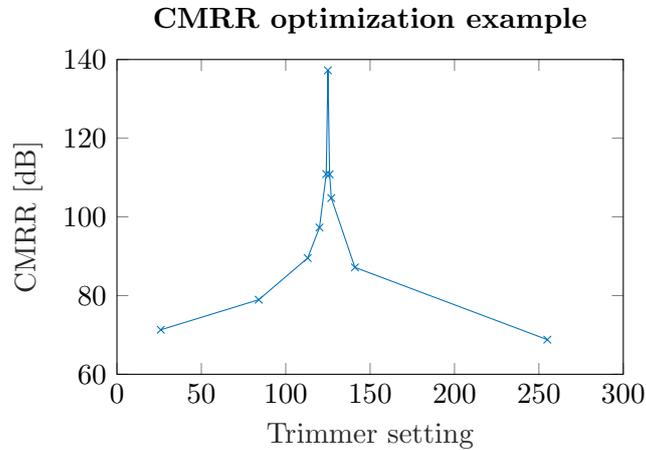
\begin{figure}
\centering
\definecolor{mycolor1}{rgb}{0.00000,0.44700,0.74100}%
\begin{tikzpicture}

\begin{axis}[%
width=2.623in,
height=1.643in,
at={(0.488in,0.422in)},
scale only axis,
xmin=0,
xmax=300,
xlabel style={font=\color{white!15!black}},
xlabel={Trimmer setting},
ymin=60,
ymax=140,
ylabel style={font=\color{white!15!black}},
ylabel={CMRR [dB]},
axis background/.style={fill=white},
title style={font=\bfseries},
title={\textbf{CMRR optimization example}}
]
\addplot [color=mycolor1, mark=x, mark options={solid, mycolor1}, forget plot]
  table[row sep=crcr]{%
26	71.3078086370103\\
84	78.9469453195093\\
113	89.5531414887137\\
120	97.3082269768499\\
124	110.893935008891\\
125	137.251852456642\\
126	110.823870513276\\
127	104.79101353141\\
141	87.1687915909949\\
255	68.7963735246286\\
};
\end{axis}
\end{tikzpicture}%
\caption{Example of the CMRR optimization process for a typical channel. The maximum is found with a bisection algorithm, starting from the lowest and highest trimmer setting that are do not saturate the outputs.}
\label{fig:cmrr_opt_example}
\end{figure}

As already described in section \ref{sec:CUOREpre}, the CMRR of the preamplifier can be compensated by acting on a specific trimmer (T2 in Figure~\ref{fig:CUOREpreCorrection}).
We developed a completely automatized algorithm that was able to optimize the CMRR with an iterative process, using the internal ADC for the read out of the output voltages.
An example of the procedure is shown in Figure~\ref{fig:cmrr_opt_example}, where the CMRR as a function of the trimmer setting is plotted.
The algorithm starts with the minimum and maximum trimmer voltages that does not cause saturation (32 and 255 in this specific example) and then uses an adaptive process based on bisection, in order to minimize the number of measurements and thus the total time, which would have been unacceptably long for 1000 channels, even if they are measured in parallel, whenever it is possible.
The trimmer setting that maximizes CMRR is saved on the EEPROM of each preamplifier and can be recalled with a specific command.

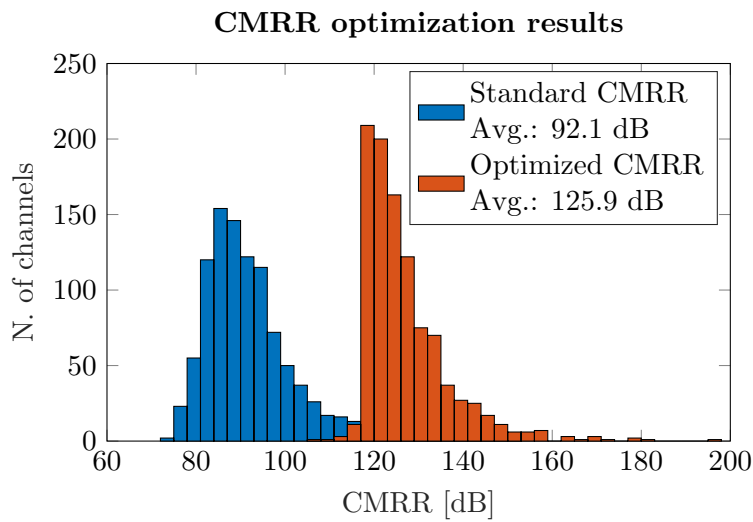
\begin{figure}
\centering
\definecolor{mycolor1}{rgb}{0.00000,0.44700,0.74100}%
\definecolor{mycolor2}{rgb}{0.85000,0.32500,0.09800}%
\begin{tikzpicture}

\begin{axis}[%
width=3.229in,
height=1.971in,
at={(0.542in,0.414in)},
scale only axis,
xmin=60,
xmax=200,
xlabel style={font=\color{white!15!black}},
xlabel={CMRR [dB]},
ymin=0,
ymax=250,
ylabel style={font=\color{white!15!black}},
ylabel={N. of channels},
axis background/.style={fill=white},
title style={font=\bfseries},
title={\textbf{CMRR optimization results}},
legend style={legend cell align=left, align=left, draw=white!15!black}
]
\addplot[ybar interval, fill=mycolor1, fill opacity=0.6, draw=black, area legend] table[row sep=crcr] {%
x	y\\
72	2\\
75	23\\
78	55\\
81	120\\
84	154\\
87	146\\
90	122\\
93	115\\
96	72\\
99	50\\
102	37\\
105	26\\
108	17\\
111	16\\
114	13\\
117	13\\
120	6\\
123	2\\
126	4\\
129	3\\
132	4\\
135	2\\
138	0\\
141	0\\
144	0\\
147	0\\
150	1\\
153	1\\
};
\addlegendentry{Standard CMRR\\Avg.: 92.1 dB}

\addplot[ybar interval, fill=mycolor2, fill opacity=0.6, draw=black, area legend] table[row sep=crcr] {%
x	y\\
105	1\\
108	1\\
111	3\\
114	11\\
117	209\\
120	200\\
123	163\\
126	122\\
129	75\\
132	70\\
135	37\\
138	27\\
141	25\\
144	17\\
147	11\\
150	6\\
153	6\\
156	7\\
159	0\\
162	3\\
165	1\\
168	3\\
171	1\\
174	0\\
177	2\\
180	1\\
183	0\\
186	0\\
189	0\\
192	0\\
195	1\\
198	1\\
};
\addlegendentry{Optimized CMRR\\Avg.: 125.9 dB}

\end{axis}
\end{tikzpicture}%
\caption{Distribution of the CMRR before (blue) and after (orange) the optimization process.}
\label{fig:cmrr_opt_histo}
\end{figure}

\begin{figure}
\centering
\definecolor{mycolor1}{rgb}{0.00000,0.44700,0.74100}%
\begin{tikzpicture}

\begin{axis}[%
width=3.229in,
height=1.971in,
at={(0.542in,0.414in)},
scale only axis,
xmin=-20,
xmax=120,
xlabel style={font=\color{white!15!black}},
xlabel={CMRR improvement [dB]},
ymin=0,
ymax=200,
ylabel style={font=\color{white!15!black}},
ylabel={N. of channels},
axis background/.style={fill=white},
title style={font=\bfseries},
title={\textbf{CMRR improvement after optimization}},
legend style={legend cell align=left, align=left, draw=white!15!black}
]
\addplot[ybar interval, fill=mycolor1, fill opacity=0.6, draw=black, area legend] table[row sep=crcr] {%
x	y\\
-5	1\\
0	38\\
5	14\\
10	19\\
15	42\\
20	91\\
25	111\\
30	173\\
35	191\\
40	142\\
45	73\\
50	48\\
55	24\\
60	16\\
65	9\\
70	2\\
75	3\\
80	3\\
85	1\\
90	1\\
95	0\\
100	0\\
105	0\\
110	1\\
115	1\\
};
\addlegendentry{Avg. improvement:\\34.8 dB ($=55.2 \: \times$)}

\end{axis}
\end{tikzpicture}%
\caption{Distribution of the CMRR improvement for each channel.}
\label{fig:cmrr_opt_histo_impr}
\end{figure}
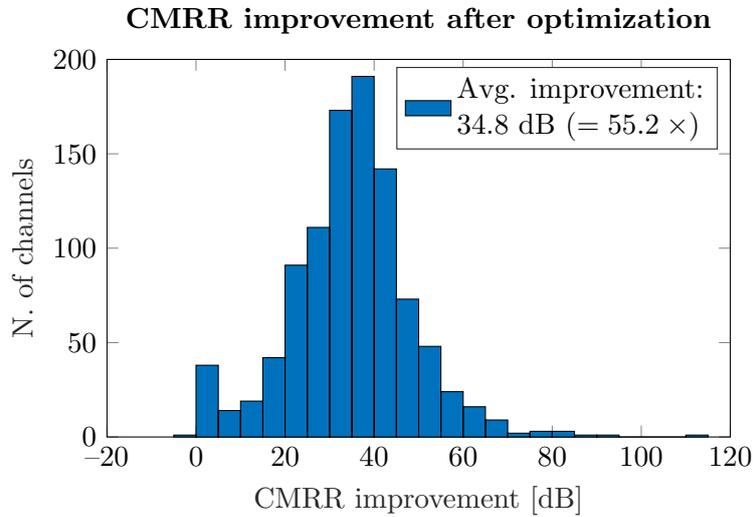

In Figure~\ref{fig:cmrr_opt_histo} we show the distribution of the CMRR values before the optimization (with trimmer setting at mid-scale) and after the optimization process.
The unoptimized CMRR is $92.1\ dB$ on average, while the optimized CMRR is $125.9\ dB$.
Figure~\ref{fig:cmrr_opt_histo_impr} shows the distribution of the CMRR improvement (difference between uncompensated and compensated CMRR for each channel), which gives an average improvement of $34.8\ dB$, or 55.2 times in linear scale.
A few channels that failed the optimization, but already had CMRR higher than 110 dB were left with the trimmer setting at half-scale and comprise the small peak at $0\ dB$.

\begin{figure}
\centering
\includegraphics[width=.75\linewidth]{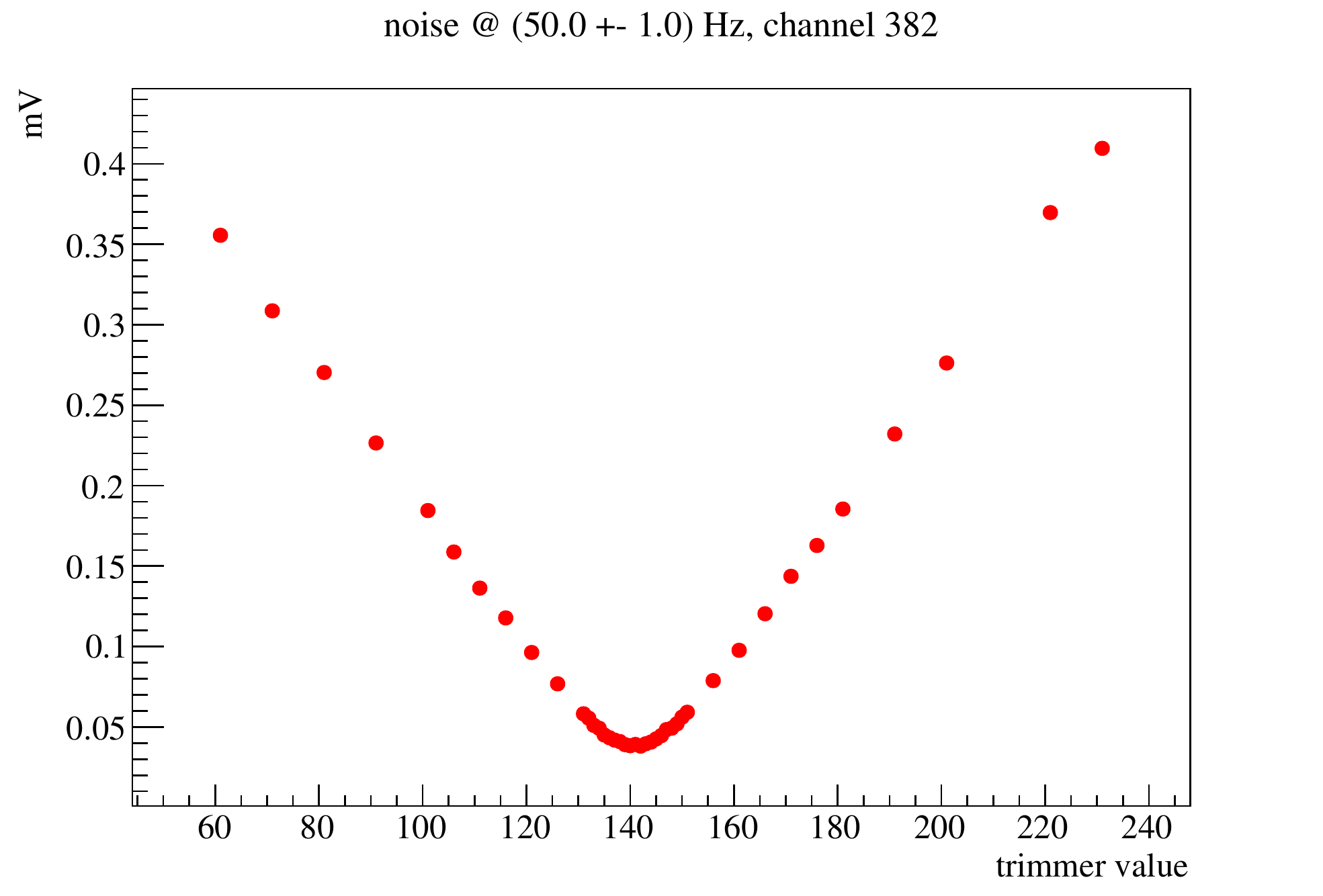}
\caption{Amplitude of the $50\ Hz$ peak from the detector spectrum, as a function of the CMRR trimmer value. This plot was acquired during several noise runs, while scanning the trimmer settings around the optimum value.}
\label{fig:cmrr_ver_50hz}
\end{figure}

\begin{figure}
\centering
\includegraphics[width=.75\linewidth]{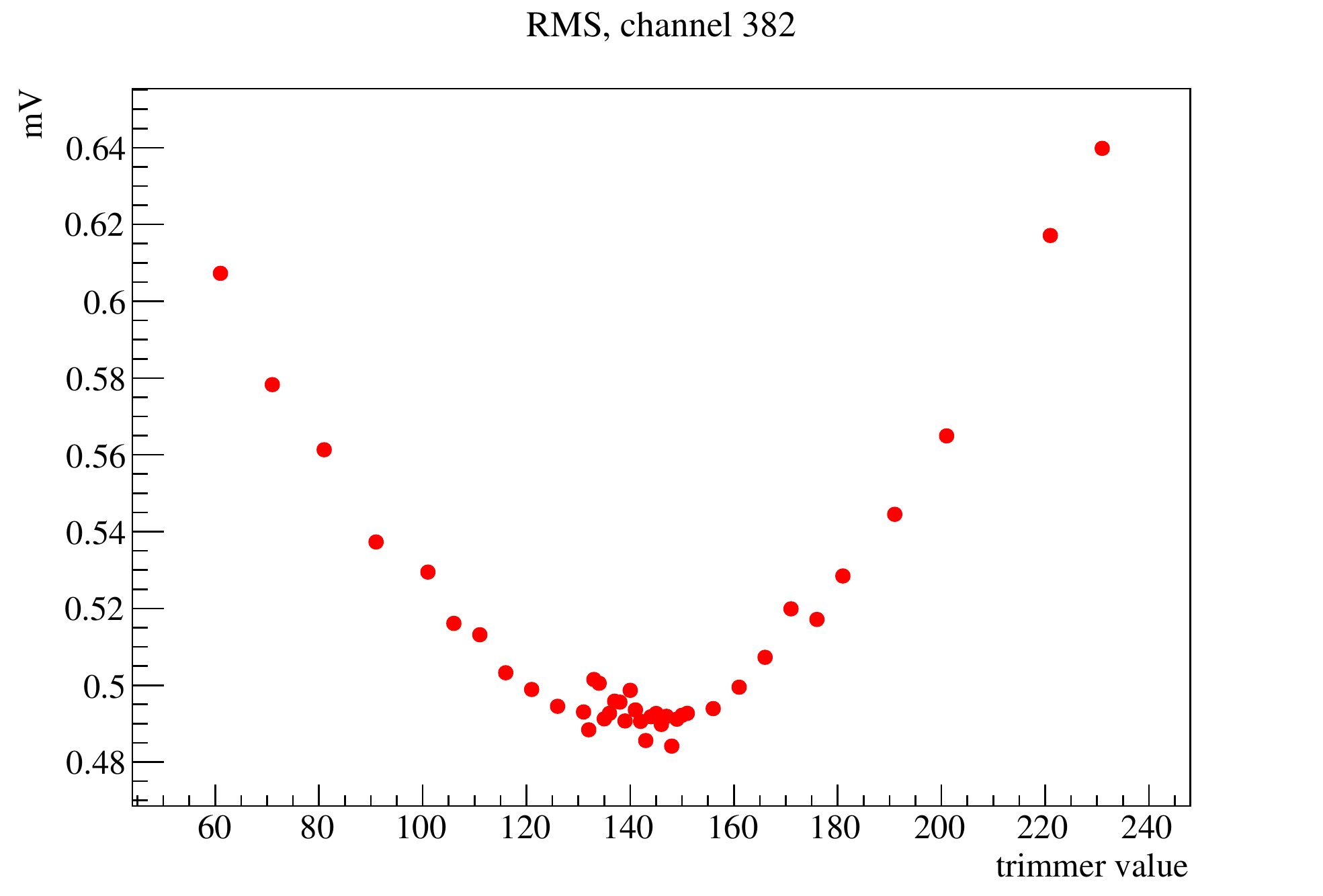}
\caption{Same as the previous plot, but showing the output voltage RMS noise.}
\label{fig:cmrr_ver_rms}
\end{figure}

As a further confirmation of the optimization procedure, when the detector reached the base temperature, several noise runs were acquired, while scanning the different trimmer settings around the value that was calculated with the process described previously.
Figure~\ref{fig:cmrr_ver_50hz} and Figure~\ref{fig:cmrr_ver_rms} show, respectively, the output RMS noise around the $50\ Hz$ peak, and the total output RMS noise on the entire bandwidth, for a typical channel.
As it can be seen, when approaching the optimized value, the contribution coming from the common mode becomes negligible and the curve reaches a minimum in correspondence of value calculated during optimization.
This measurement confirmed the effectiveness of our procedure.

\section{Installation}

We dedicated a great effort also in the installation of the entire electronic system for both the CUORE and CUPID-0 experiments.

In CUORE, the front-end electronics is installed on top of the cryostat, standing on the main support plate (MSP) and not on the cryostat, in order to minimize the induced vibrations.
This layout also minimizes the length of the link from the top of the refrigerator, which is beneficial for both the stray capacitance and noise.
The cables are wrapped in soft insulation foam to reduce vibrations and fixed to the MSP structure.

\begin{figure}
\centering
\includegraphics[width=.8\linewidth]{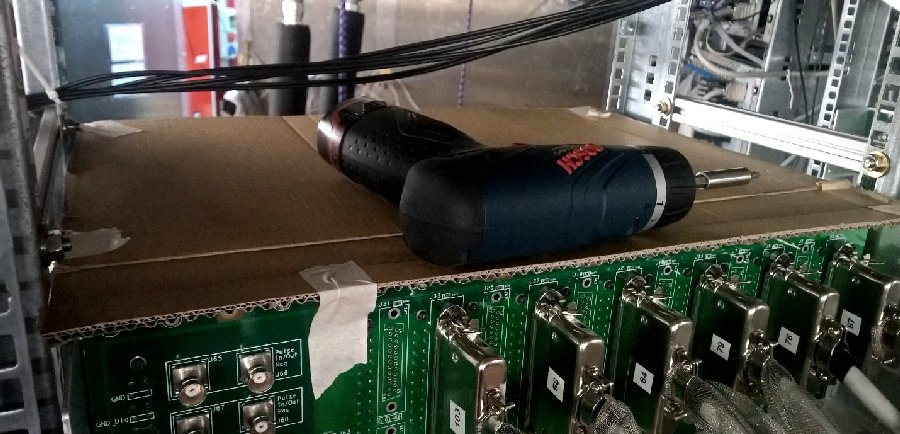}
\caption{Photograph of the provisional cardboard cover that avoids the airflow in the front-end card cages. Later on it was substituted by a metal cover, designed on purpose.}
\label{fig:cover}
\end{figure}

After the installation at LNGS, we found that the very low frequency noise (below $1\ Hz$) was more than one order of magnitude worse than what was measured in our test system.
The issue was caused by the air flow in the area of the JFETs, which induced thermocouple effects at the inputs.
The phenomenon was more evident on channel 6 of each mainboard, which is the channel closer to the bottom of the board, where the air flow is more turbulent.
In our test system this issue was not visible since the Faraday cage was much smaller and thus the air flow was suppressed.
We were able to solve the problem by covering the top and bottom openings of the card cages.
This solution completely stopped the heat convection and thus the air flow.
In Figure~\ref{fig:cover} it is possible to see this provisional cover made with cardboard during one of the first trials.
For the final configuration we designed metal panels that did not cover completely the entire opening but only the most sensitive area where the preamplifiers are placed, reaching the best compromise between noise minimization and temperature of the mainboard.

The electronic system is powered up and it has been running stably in both CUORE and CUPID-0 experiments for several months, without any issue.


\section{Electronics upgrades}

For the final CUPID experiment we studied a further improvement for two elements of the read out chain: the preamplifier and the antialiasing filter.
For both these two boards, the circuital design is completed and the performance evaluation on test bench has started.
In the following sections I will describe with more details these two upgraded boards.

\subsection{CUPID preamplifier} \label{sec:CUPIDpre}

During a preliminary run of the CUPID-0, the cryostat was unable to reach the base temperature and the detectors were operating at $20\ mK$.
In this working condition the NTD resistances were in the order of $10\ M\Omega$ or less and their intrinsic noise decreased down to about $3\ nV/\sqrt{Hz}$, comparable to the input series noise of the preamplifier.

In order to improve the preamplifier performance even in such working conditions, we decided to select a new JFET that could be used to upgrade the present preamplifier for the CUPID experiment.
The new FET must have lower series noise, both white and flicker, hence its area must be larger than the current one since 1/f noise is inversely proportional to device area.
Lower series white noise is achieved by increasing the bias current, increasing the transconductance of the transistor.

As a first test of the new preamplifier, an actual CUORE preamplifier was equipped with the new input pair, based on two InterFET IF9030 JFET transistors, and the values of the bias components were changed in order to set a similar transistor working point, but with bias current increased by a factor 5, from $0.5\ mA$ to $2.5\ mA$.
The input capacitance for one JFET increased by a factor 2, up to about $80\ pF$, which is still reasonably small with respect to the typical parasitic capacitance of the link (about $200\ pF$ in the case of CUPID-0, but probably more in the final CUPID detector with CUORE-like dimensions).
Anyway, at these smaller detector impedances the effect of the capacitance has a negligible effect on the signal bandwidth.
For instance $100\ pF$ limits the bandwidth of a $10\ M\Omega$ detector to $150\ Hz$, outside the signal bandwidth.

\begin{figure}
\centering
\includegraphics[width=.75\linewidth]{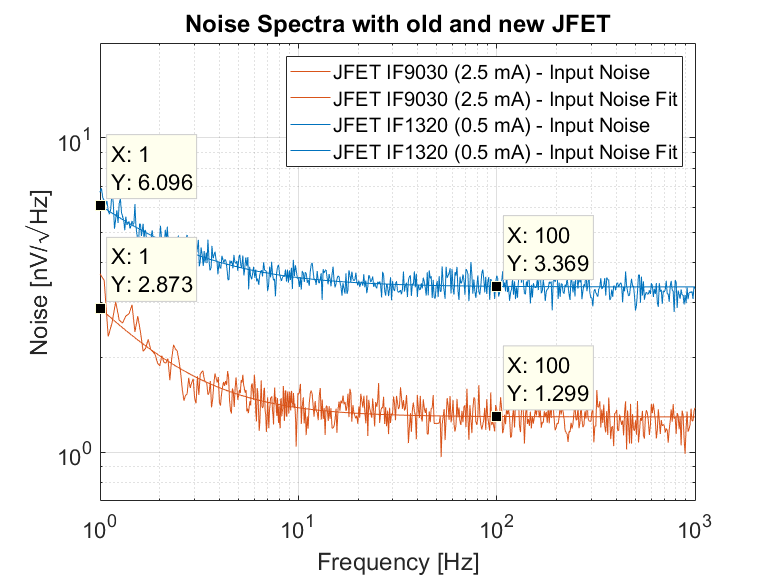}
\caption{Comparison of the input noise of the CUORE preamplifier (blue) and of the modified preamplifier with the larger JFETs (orange).}
\label{fig:pre_cupid_noise}
\end{figure}

Figure~\ref{fig:pre_cupid_noise} shows the plots of the noise spectra for the CUORE preamplifier equipped with the old IF1320 and for the modified preamplifier with the new JFETs.
Clearly, both the 1/f and white noise improved.
Low frequency noise at $1\ Hz$ was reduced by a factor 4.5 (quadratic) from $6.1\ nV/\sqrt{Hz}$ to $2.9\ nV/\sqrt{Hz}$, while white noise was reduced by a factor 6.7 (quadratic) from $3.7\ nV/\sqrt{Hz}$ to $1.3\ nV/\sqrt{Hz}$.
As it can be seen, white noise is reduced by more than the ratio of the bias currents since the new JFET also has higher transconductance at equal bias current with respect to the old one.

The new preamplifier design implements also other minor modifications.
The main one consists in the removal of the DAC for the fine correction of the voltage offset and its thermal drift.
In this new version the corrections are performed using only the 4-channel trimmer: two channels are dedicated to the coarse and fine offset corrections, one channel is dedicated to the CMRR compensation and one channel is dedicated to the thermal drift correction.
The new preamplifier will be able to reach a similar thermal drift specification thanks to an optimized range of the corresponding corrector.
Removing the DAC will have the advantage to lower the quadratic drift, which was contributed also by the DAC.

The new preamplifier has been already produced and it is currently being assembled.

\subsection{CUPID antialiasing filter and DAQ} \label{sec:CUPIDfilter}

As already said, bolometers exhibit a large spread of their characteristic from channel to channel.
This is particularly evident on the energy conversion factor, which lead us to design the mainboard with programmable gain and independent bias generation.
For dual readout experiments, like CUPID, the thermal dynamics of heat and light detectors can be even more spread due to the completely different characteristics of the two detectors.
Light detectors like the Ge discs used so far in CUPID-0, in fact, are much lighter than the bolometric crystals and their coupling to the heat sink can also differ.
For these reasons, the bandwidth of the light channels is usually one order of magnitude higher than heat channels.

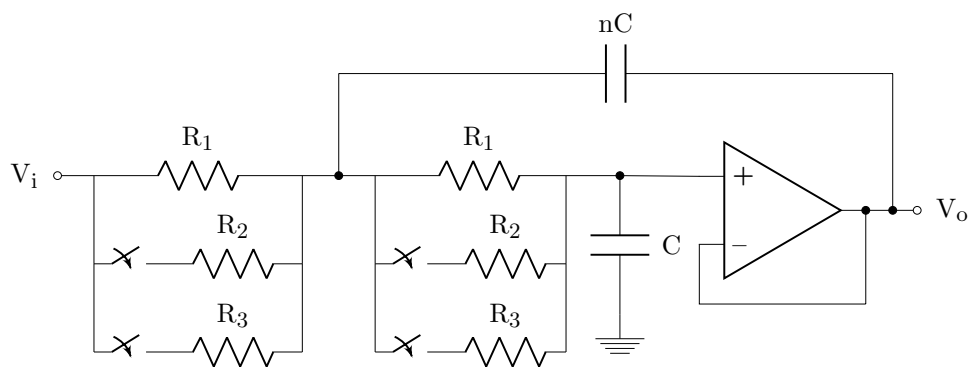
\begin{figure}
\centering
\resizebox{\textwidth}{!}{%
\begin{tikzpicture}
	\node 			(vi)		[label=left:$V_\mathrm{i}$] {};
	\node			(a)			[right=3.75 of vi] {};
	\node			(b)			[right=3.75 of a] {};
	\node			(r1a)		[right=0.25 of vi] {};
	\node			(r1b)		[left=0.25 of a] {};
	\node			(r1aa)		[below=1 of r1a] {};
	\node			(r1bb)		[below=1 of r1b] {};
	\node			(r1aaa)		[below=1 of r1aa] {};
	\node			(r1bbb)		[below=1 of r1bb] {};
	\node			(r2a)		[right=0.25 of a] {};
	\node			(r2b)		[left=0.5 of b] {};
	\node			(r2aa)		[below=1 of r2a] {};
	\node			(r2bb)		[below=1 of r2b] {};
	\node			(r2aaa)		[below=1 of r2aa] {};
	\node			(r2bbb)		[below=1 of r2bb] {};
	\node[ground]	(gnd)		at ($(b)+(0,-2)$) {}; 
	\node[op amp]	(oa)        [right=1 of b, yshift=-5mm, yscale=-1] {};
	\node			(vo)        [right=0.6 of oa.out] [label=right:$V_\mathrm{o}$] {};
	\node			(ff)        [right=0.25 of oa.out] {};
	\node			(a')        [above=1.2 of a] {};
	
	\draw			(vi)		to[R=$R_{1}$, o-]		(a)
								to[R=$R_{1}$]			(b)
								to[short]			(oa.+);
	\draw			(a)			to[short, *-]		(a')
								to[C=$nC$]			(a' -| ff)
								to[short, -*]		(ff);
	\draw			(b)			to[C, l^=$C$, *-]	(gnd);
	\draw			(oa.out)	to[short, -o]		(vo);
	\draw			(oa.out)	to[short, *-]		+(0, -1.35)
								to[short]			+(0,0) -| (oa.-);
	\draw			(r1a)		to[short]			(r1aa);
	\draw			(r1aa)		to[short]			(r1aaa);
	\draw			(r1b)		to[short]			(r1bb);
	\draw			(r1bb)		to[short]			(r1bbb);
	\draw			(r1aa)		to[spst]			+(1,0)
								to[R=$R_{2}$]		(r1bb);
	\draw			(r1aaa)		to[spst]			+(1,0)
								to[R=$R_{3}$]		(r1bbb);
	\draw			(r2a)		to[short]			(r2aa);
	\draw			(r2aa)		to[short]			(r2aaa);
	\draw			(r2b)		to[short]			(r2bb);
	\draw			(r2bb)		to[short]			(r2bbb);
	\draw			(r2aa)		to[spst]			+(1,0)
								to[R=$R_{2}$]		(r2bb);
	\draw			(r2aaa)		to[spst]			+(1,0)
								to[R=$R_{3}$]		(r2bbb);
\end{tikzpicture}
}%
\caption{Circuit schematic of a single Sallen-Key cell of the existing CUORE and \mbox{CUPID-0} antialiasing filter. The switches allow to connect resistances in parallel and increase the cut-off frequency.}
\label{fig:bessel_attuale}
\end{figure}

In the current CUORE and CUPID-0, the bandwidth of the antialiasing Bessel filter is selectable among 4 values only, and the cut-off frequency spans less than one order of magnitude.
Frequencies are set by acting on a digital switch that connects different resistances within the Sallen-Key cell, as shown in Figure~\ref{fig:bessel_attuale}.

In CUORE and on the heat channel of CUPID-0 the cut-off frequencies are $15\ Hz$, $35\ Hz$, $100\ Hz$ and $120\ Hz$, while for the light channels of CUPID-0 we modified the resistors in order to have slightly higher cut-off frequencies of $15\ Hz$, $100\ Hz$, $140\ Hz$ and $200\ Hz$.

To overcome this limitation, it could be possible to substitute the switches and resistances with a digital trimmer. However, this solution has several challenges:
\begin{itemize}
\item \emph{High voltage.} In Sallen-Key cells, signals must be converted from differential to single ended. In our system, operating with $\pm5\ V$ differential signals, this means that the power supplies of the components of the cell have to be in excess of $\pm 10\ V$. A high voltage trimmer is thus required and unfortunately they are not common since most of the digital trimmers are intended for low voltage or single supply applications. 
\item \emph{Resistance precision.} High resistance precision or at least a good matching between the two resistances of the Sallen-Key cell, are required for adequate frequency response.
\item \emph{Low parasitics.} Parasitic capacitances must be negligible with respect to the capacitances of the cell, otherwise the filter response could be worsened by the large spreads of such parasitics or due to their thermal drift.
\item \emph{High resolution.} Being able to change the cut-off frequency over two order of magnitudes with adequate granularity requires at least 8-bit or 10-bit components.
\item \emph{High value resistances.} The lowest cut-off frequency must be about $20\ Hz$ and the capacitors cannot be chosen too large due to the reduced space for each channel. This means that the resistances of the filter must be in the $100\ k\Omega$ range.
\end{itemize}

Indeed the design of the upgraded antialiasing filter was made possible by the commercialization of a digital trimmer that suits perfectly the requirements of this project.
The Analog Devices AD529x is a 8- or 10-bit digital trimmer that can operate up to $\pm 16.5\ V$ and features a proprietary solution in order to reach 3\% resistance precision over almost 95\% of the range.
It is available in $100\ k\Omega$, $50\ k\Omega$ and $20\ k\Omega$, but the higher resistance one has the best precision specification.
Parasitics capacitances on the terminals are not so large for such a device, $85\ pF$ on the opposite terminals and $65\ pF$ on the wiper terminal.

\begin{figure}
\centering
\resizebox{\textwidth}{!}{%
\begin{tikzpicture}
	\node 			(vi)		[label=left:$V_\mathrm{i}$] {};
	\node			(a)			[right=3.75 of vi] {};
	\node			(b)			[right=3.75 of a] {};
	\node			(r1a)		[right=0.25 of vi] {};
	\node[ground, color=black!20!white]	(gnd1a)		at ($(r1a)+(0,-1.5)$) {};
	\node			(r1b)		[left=0.25 of a] {};
	\node[ground, color=black!20!white]	(gnd1b)		at ($(r1b)+(0,-1.5)$) {};
	\node			(r2a)		[right=0.25 of a] {};
	\node[ground, color=black!20!white]	(gnd2a)		at ($(r2a)+(0,-1.5)$) {};
	\node			(r2b)		[left=0.5 of b] {};
	\node[ground, color=black!20!white]	(gnd2b)		at ($(r2b)+(0,-1.5)$) {};
	\node[ground]	(gnd)		at ($(b)+(0,-2)$) {}; 
	\node[op amp]	(oa)        [right=1 of b, yshift=-5mm, yscale=-1] {};
	\node			(vo)        [right=0.6 of oa.out] [label=right:$V_\mathrm{o}$] {};
	\node			(ff)        [right=0.25 of oa.out] {};
	\node			(a')        [above=1.2 of a] {};
	
	\draw			(vi)		to[vR=$R$, o-]	(a)
								to[vR=$R$]		(b)
								to[short]			(oa.+);
	\draw			(a)			to[short, *-]		(a')
								to[C=$nC$]			(a' -| ff)
								to[short, -*]		(ff);
	\draw			(b)			to[C, l^=$C$, *-]	(gnd);
	\draw			(oa.out)	to[short, -o]		(vo);
	\draw			(oa.out)	to[short, *-]		+(0, -1.35)
								to[short]			+(0,0) -| (oa.-);
	\draw[gray]		(r1a)		to[/tikz/circuitikz/bipoles/length=1cm,C, l^=$C_B$, color=black!30!white]	(gnd1a);
	\draw[gray]		(r1b)		to[/tikz/circuitikz/bipoles/length=1cm,C, l_=$C_A$, color=black!20!white]	(gnd1b);
	\draw[gray]		(r2a)		to[/tikz/circuitikz/bipoles/length=1cm,C, l^=$C_B$, color=black!20!white]	(gnd2a);
	\draw[gray]		(r2b)		to[/tikz/circuitikz/bipoles/length=1cm,C, l_=$C_A$, color=black!20!white]	(gnd2b);
\end{tikzpicture}
}%
\caption{Schematic of a single Sallen-Key cell of the upgraded antialiasing filters. The resistors are implemented with a 10-bit $100\ k\Omega$ high voltage digital trimmer. The parasitic capacitances are shown in gray. $C_B$ is typically $85\ pF$, while $C_A$ is $150\ pF$.}
\label{fig:bessel_nuovo}
\end{figure}
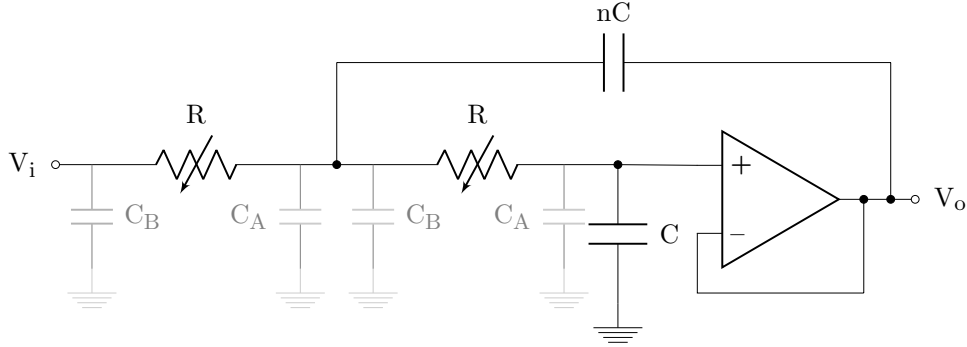

Figure~\ref{fig:bessel_nuovo} shows the modified Sallen-Key cell, adopting the digital trimmer in place of the input resistors.

These digital trimmers are only available in single channel configuration in a TSSOP package, so they occupy a large area on the channel.
The choice of the filter capacitors is thus limited by this space constraint, which requires the adoption of SMD components only.
Furthermore, since stability over temperature is an important parameter, only high quality C0G dielectric capacitors can be used, which are only available up to $100\ nF$.

The Sallen-Key transfer function for our case where resistances $R$ are equal and capacitors have a ratio of $n$, is given by
\begin{equation}
\label{eq:sk_ideal}
H(s)=\frac{1}{1+2sRC+s^2nR^2C^2} \ .
\end{equation}
The cut-off frequency is therefore
\begin{equation}
f_{c} = \frac{1}{2\pi RC\sqrt{n}}
\end{equation}
and the Q-factor
\begin{equation}
Q=\frac{\sqrt{n}}{2} \ .
\end{equation}
Table~\ref{tab:bessel_spec} lists the characteristics and component values of the 3 Sallen-Key cells that make the 6-pole Bessel-Thomson transfer function.
As it can be seen in the table, cut-off frequency spans from about $24\ Hz$ up to $2.4\ kHz$, values that are perfectly suitable for both slower heat channels and faster light channels.

\begin{table}
\centering
{\tabulinesep=1.2mm
\begin{tabu}{c c c c c P{15mm} P{12.5mm} P{11mm}} 
				& $\bm{Q_{bessel}}$	& $\bm{C}$		& $\bm{nC}$ 		& $\bm{Q}$	& $\bm{f_{cut}}$ $\bm{(100\ k\Omega)}$	& $\bm{f_{c}}$ $\bm{(10\ k\Omega)}$	& $\bm{f_{c}}$ $\bm{(1\ k\Omega)}$ 	\\ 
				&  				& [nF]		& [nF]		& 		& [Hz]	& [Hz]	& [Hz]	\\ 
\hline
\textbf{Cell 1} & 1.023 		& $16.8$	& $69.8$	& 1.019	& 24.4	& 243.7	& 2437	\\
\textbf{Cell 2}	& 0.611 		& $32$ 		& $47$		& 0.606	& 24.3	& 242.7	& 2427	\\
\textbf{Cell 3}	& 0.510 		& $39.8$ 	& $41.2$	& 0.509	& 24.5	& 244.7	& 2447	\\
\end{tabu}}
\caption{Summary of the specs for each of the three cell of the new antialiasing board.}
\label{tab:bessel_spec}
\end{table}

For the component choice adopted in this configuration, the parasitic capacitances are negligible and they account for less than the capacitor tolerance which is usually 5\% for C0G capacitors.

Things change in case the filter is configured to operate at higher frequencies.
For example, if cut-off frequency range is increased by a factor 10 ($240\ Hz$-$24\ kHz$), the value of the capacitors must be reduced by a factor 10 too, and parasitic capacitances becomes about 10\% of the value of the discrete components.
In this case, the solution shown in equation \ref{eq:sk_ideal} is not valid anymore and the exact solution must be calculated.
From the scheme in Figure~\ref{fig:bessel_nuovo} we can calculate the new transfer function, also considering $C_B \approx C_A/2 \approx 78\ pF$:
\begin{equation}
\label{eq:sk_parasitics}
H(s)=\frac{1}{1+2sRC\left(1+\frac{7C_B}{2C}\right)+s^2nR^2C^2\left[1+\frac{C_B}{nC}\left(2n+3\right)+\frac{6C_B^2}{nC^2}\right]} \ .
\end{equation}
The term $\frac{6C_B^2}{nC^2}$ can be neglected, since it accounts for less than 1\%. 

From the previous equation we obtain
\begin{equation}
f_{c} = \frac{1}{2\pi RC\sqrt{n}\sqrt{1+\frac{C_B}{nC}\left(2n+3\right)}}
\end{equation}
and
\begin{equation}
Q = \frac{\sqrt{n}\sqrt{1+\frac{C_B}{nC}\left(2n+3\right)}}{2\left(1+\frac{7C_B}{2C}\right)} \ .
\end{equation}

The components for the higher bandwidth version are listed in Table~\ref{tab:bessel_spec_hf}.

\begin{table}
\centering
{\tabulinesep=1.2mm
\begin{tabu}{c c c c c P{15mm} P{12.5mm} P{11mm}} 
				& $\bm{Q_{bessel}}$	& $\bm{C}$		& $\bm{nC}$ 		& $\bm{Q}$	& $\bm{f_{cut}}$ $\bm{(100\ k\Omega)}$	& $\bm{f_{c}}$ $\bm{(10\ k\Omega)}$	& $\bm{f_{c}}$ $\bm{(1\ k\Omega)}$ 	\\ 
				&  				& [nF]		& [nF]		& 		& [kHz]	& [kHz]	& [kHz]	\\ 
\hline
\textbf{Cell 1} & 1.023 		& $1.62$	& $5.93$	& 1.019	& 0.25	& 2.53	& 25.27	\\
\textbf{Cell 2}	& 0.611 		& $3.03$ 	& $4.12$	& 0.614	& 0.25	& 2.53	& 25.30	\\
\textbf{Cell 3}	& 0.510 		& $3.57$ 	& $3.9$		& 0.510	& 0.25	& 2.53	& 25.28	\\
\end{tabu}}
\caption{Summary of the specs for each of the three cell of the new antialiasing board, in the high frequency option, also taking into account trimmer parasitic capacitances.}
\label{tab:bessel_spec_hf}
\end{table}

For the values obtained here, an eventual change of the parasitic capacitance $C_B$ would not be too critical for both $f_c$ and $Q$, since we can evaluate
\begin{equation}
\frac{1}{f_c}\frac{df_c}{dC_B} = -0.077\: \%/pF
\end{equation}
and
\begin{equation}
\frac{1}{Q}\frac{dQ}{dC_B} = -0.108\: \%/pF \ ,
\end{equation}
hence, for a parasitic capacitance change of 25\%, $f_c$ and $Q$ change only by 1.5\% and 2.2\%, respectively.

\begin{figure}
\centering
\includegraphics[width=\linewidth]{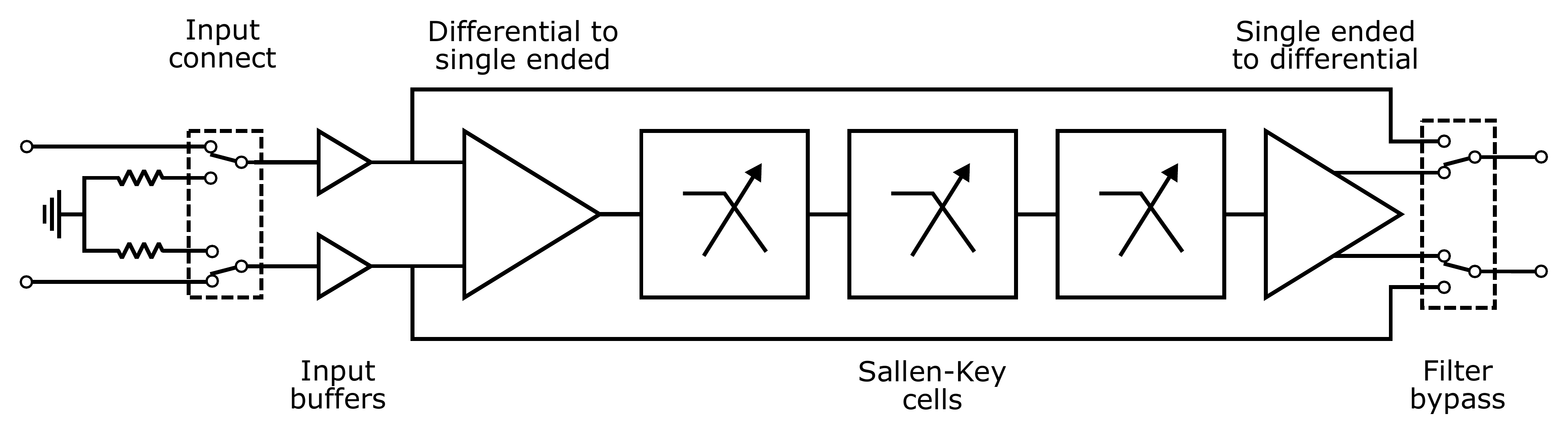}
\caption{Scheme of one analog channel of the antialiasing filter.}
\label{fig:bessel_nuovo_schema_analog}
\end{figure}

Proceeding in the bottom-top analysis of the upgraded board, Figure~\ref{fig:bessel_nuovo_schema_analog} shows the complete scheme of one analog channel.
One board hosts 12 analog channels, like the actual version.

The inputs signals can be disconnected from the filter by means of an analog switch, which also connects the inputs of the analog block to ground using two resistors.
This configuration is useful for debugging purposes, when calibrating the offset of the channel or during other measurements (noise, PSRR, etc.).
When the inputs are disconnected, the values of the trimmer resistances can be individually measured by applying a reference voltage to specific nodes of each cell, by means of a multiplexer.

The inputs are then fed to a unitary gain buffer with high input impedance.
These buffers are implemented using a Texas Instruments OPA2188, an auto-zero, zero-drift operational amplifier by Texas Instruments, in non-inverting buffer configuration.
The bufferized differential signals are then converted to a single ended signal with a unitary gain summing stage, and then filtered by 3 Sallen-Key cells described above.
All these 4 stages adopts Analog Devices AD8622 opamps. 
The filtered output signal is converted back to differential.
A relay allow to completely bypass the filter, while still having the inputs buffered.
The gain of the whole chain is unitary.

The noise at lowest cut-off frequency is dominated by the Johnson noise of the $100\ k\Omega$ series resistors, that contribute with $100\ nV/\sqrt{Hz}$ for all the three cells.
Since high source impedances are present, the amplifier input current noise can also become quite high.
The AD8676 used in the actual version, in fact, has a current noise of $0.3\ pA/\sqrt{Hz}$ and would contribute with an additional $100\ nV/\sqrt{Hz}$.
The AD8622, on the contrary, features a lower current noise of $0.15\ pA/\sqrt{Hz}$, or $50\ nV/\sqrt{Hz}$.
The AD8622 also has other advantages: much lower supply current which would allow to consume 1/5 of the power for each channel, better compatibility of the inputs to the rails which allow to save an additional 15\% of power since the boards can be supplied at $\pm12\ V$ instead of $\pm 13\ V$, and lower input offset current, which is also important for precision measurements with high source impedance.
The only drawback of this operational amplifier is the higher white noise of $11\ nV/\sqrt{Hz}$ with respect to $2.8\ nV\sqrt{Hz}$ of the AD8676, which increases the overall noise at higher cut-off frequencies, where the trimmer resistance is lower and its contribution vanishes.

\begin{figure}
\centering
\includegraphics[width=\linewidth]{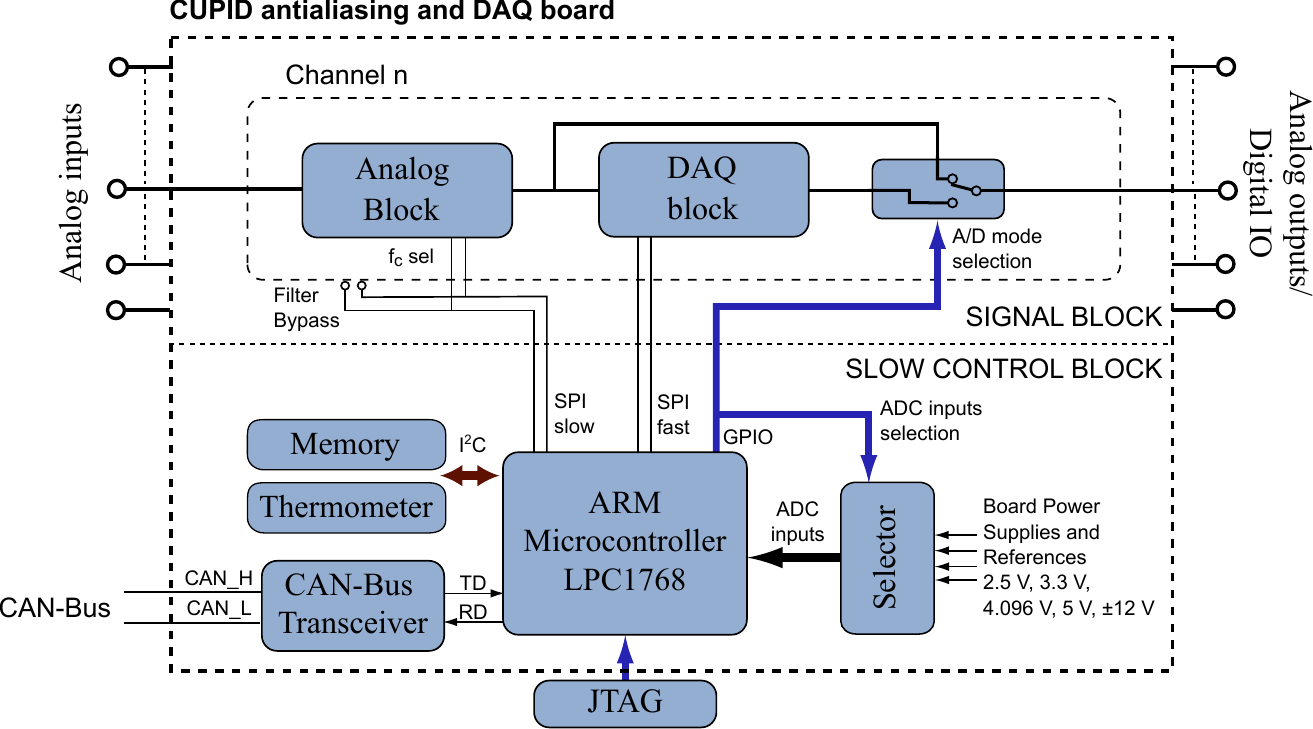}
\caption{Block schematic of the whole antialiasing and DAQ board.}
\label{fig:bessel_nuovo_schema}
\end{figure}

Finally, in Figure~\ref{fig:bessel_nuovo_schema} the overall block schematic of the board is presented. 

As it can be seen, the board is made by two main blocks, the signal block and the slow control block.
The latter is equipped with a ARM Cortex M3 microntroller LPC1768, which takes care of the analog block settings (cut-off frequency, filter bypass, and input disconnect), as well as many other functions like thermometer readout, EEPROM memory interface and power supply voltages readings.
The board receives commands through CAN-bus, like the present version.

One of the other main characteristics of the board is the presence of the DAQ block.
The board, in fact, can work also as a complete DAQ system, thanks to a set of ADCs that read out each filtered output.
The ADCs can be accessed either by the microcontroller or externally, from the output connector.
The board is able to work either in DAQ mode or in analog mode, since both the signals (filtered analog signals or digital signals required to interface the ADCs) can be routed to the output connector by the A/D mode selection block.
This makes the board fully compatible with the current analog-only scheme, using external commercial DAQ, but will also enable the use of the board as of a fully integrated DAQ system.
The mode selection block is made by a set of high voltage SPDT switches (Analog Devices ADG1434).

The DAQ block is composed by a fully-differential input buffer, Analog Devices AD8475, that drives a 2-channel 24-bit $\Delta\Sigma$ ADC, Analog Devices AD7175-2.
The input buffer is required since it attenuates and offsets the signals to the voltage specifications required by the ADC ($\pm4\ V$ differential signals with common mode of $2.5\ V$).
We chose the adoption of an all-in-one solution like the AD8475 due to space constraints, even if its noise specifications, especially at low frequency, are not the best ($2.5\ \mu V$ peak-to-peak from $0.1\ Hz$ to $10\ Hz$) with respect to the performance of the analog block.
Such noise, however, corresponds to a peak-to-peak resolution of 21.7 bit, which is still better than what is achievable by the AD7175-2 at most data rates.

\begin{figure}
\centering
\includegraphics[width=.65\linewidth]{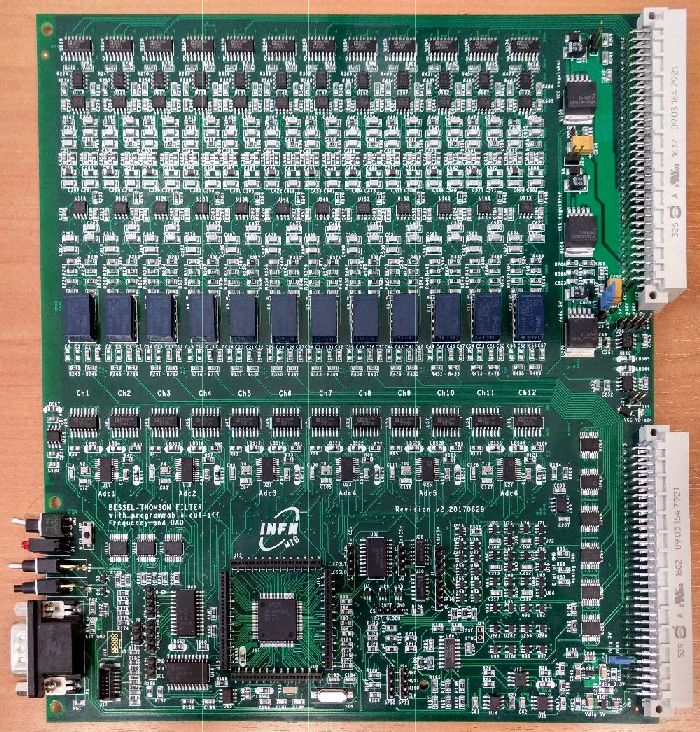}
\caption{Photograph of the upgraded antialiasing filter with integrated DAQ.}
\label{fig:bessel_new}
\end{figure}

The AD7175-2 has two channels, so one device is shared between neighboring channels.
The maximum data rate when only one channel is active is $250\ kHz$, and it reduces to $25\ kHz$ in dual channel mode since each input signal fully settles in $20\ \mu s$ ($50\ kHz$).
As a comparison, the data rates used in CUPID-0 are $1\ kHz$ for heat channels and $2\ kHz$ for light channels.
The high acquisition rate provided by this board would allow its use also in other kind of experiments which adopt faster detectors.
The AD7175-2 is a $\Delta\Sigma$ ADC~\cite{candy1962oversampling}, hence its resolution improves at lower data rates due to the oversampling technique adopted by this class of devices.
At $250\ kHz$ (single channel) and $25\ kHz$ (dual channel), the effective RMS resolution is 19.8~bits, or $8.7\ \mu V$ RMS at the ADC input.
At $2.5\ kHz$ (dual channel), resolution improves to 21.4~bits or $3.1\ \mu V$ RMS, also taking into account the contribution of the buffer stage. The noise of the analog part, in comparison, amounts to less than $0.8\ \mu V$ at the input of the ADC, due to the attenuation performed by the AD8475.

The ADC can be configured by the microcontroller using a fast $20\ MHz$ SPI.
During data acquisition, however, the microcontroller and its CAN-bus interface to the control system are not fast enough to manage the high data rate achievable by the system, so the SPI signals can also be routed externally, by means of the switches described above, so that they can be managed by a faster device, like, for example an FPGA.
We chose not to install an FPGA onboard, because it would be more convenient to share it between multiple boards and thus to install it on the backplane.

Figure~\ref{fig:bessel_new} shows a photograph of the upgraded antialiasing and DAQ board.
The form-factor is the same of the old version, a 6U $220\ mm$ and the pinout is also completely retrocompatible, so that the new version can be installed without changing anything of the present infrastructure.

\begin{figure}
\centering
\includegraphics[width=.7\linewidth]{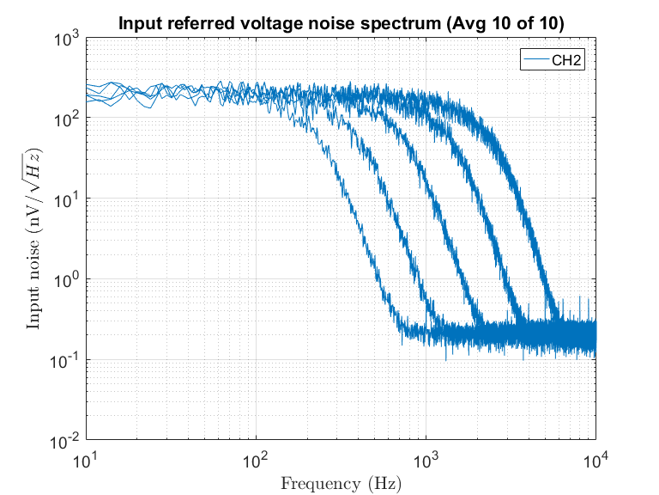}
\caption{Test of different cut-off frequency settings.}
\label{fig:bessel_new_cutoff}
\end{figure}

The board is fully operational: preliminary tests on the first sample boards verified the complete operation of all the functionalities.
The analog mode was tested by connecting the board in our lab to a CUORE front-end card cage with detector inputs grounded.
Several spectra were acquired at different cut-off frequencies, as depicted in the plot of Figure~\ref{fig:bessel_new_cutoff}.
In this measurement the input referred noise is dominated by the noise of the $2\ M\Omega$ resistors that connect the inputs of the front-end board to ground.

\begin{figure}
\begin{minipage}[b][][t]{.47\linewidth}
\centering
\includegraphics[width=\linewidth]{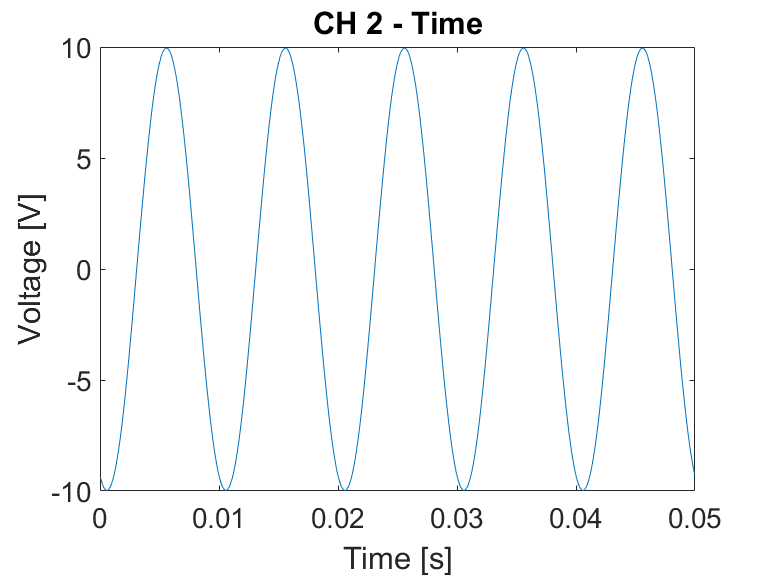}
\caption{Test of the data acquisition with a $\pm10\ V$ peak-to-peak signal at $100\ Hz$.}
\label{fig:bessel_new_segnale}
\end{minipage}
\ \hspace{.3mm} \
\begin{minipage}[b][][t]{.47\linewidth}
\centering
\includegraphics[width=\linewidth]{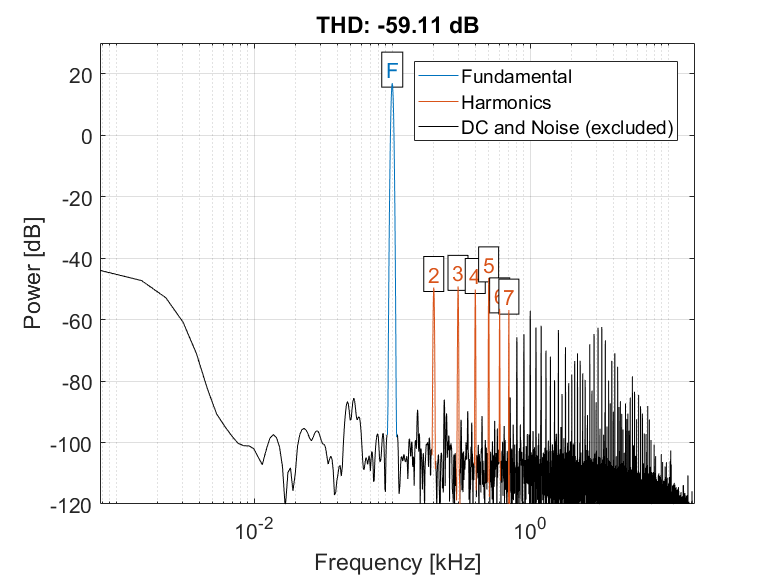}
\caption{Total harmonic distortion measurement of the signal acquired with the integrated DAQ system.}
\label{fig:bessel_new_thd}
\end{minipage}
\end{figure}

The board was also tested in DAQ mode.
A $100\ Hz$, $10\ V$ peak-to-peak signal was injected at the input of one channel with cut-off frequency set at $2.4\ kHz$ and the corresponding output was read out using an external FPGA.
The acquired data were stored in a buffer within the RAM of the FPGA and downloaded to the PC using JTAG direct access to the RAM.
The acquired signal is shown in Figure~\ref{fig:bessel_new_segnale}.
Even if the signal amplitude is at the maximum specification, the total harmonic distortion (THD) is still dominated by the distortion of the wave generator, at about $-60\ dB$, as depicted in Figure~\ref{fig:bessel_new_thd}.

The final back-end infrastructure of the DAQ system will be based on a UDP protocol over 10 Gbps Ethernet interface.
The expected maximum data rate for an entire card cage of 16 boards (96 channel in fast $250\ kHz$ mode) is 1.5 Gbps for 64-bit packets (24 bits of ADC data, 8 bits of channel ID and control data and 32 bit of timestamp), which is a manageable data rate for an hardware-implemented UDP protocol.


\chapter*{Conclusions}
\addcontentsline{toc}{chapter}{Conclusions}
\label{Chapter02}
\thispagestyle{empty}

The electronic instrumentations described in this dissertation have reached a very high level of maturity and they are currently already operating or about to be installed in their respective experiments.

The optoelectronic readout chain for the upgraded RICH detector at LHCb passed all the Production Readiness Reviews and the final production, quality assurance and assembly are ongoing.
The complete system is due to be installed starting from Q2 2019, when the LHC machine will be shut down for the LS2 stop period.
Testbeams proved the effectiveness of the system in reading out successfully Cherenkov photons produced in a solid radiator with excellent performance from the photodetector sensors and front-end electronics.
The ALDO project will be a fundamental backup solution in case the tests with the final PDMDB digital board at full $40\ MHz$ will show unexpected behavior.
The first ALDO prototype described in this work posed the base for further developments for the High Luminosity era at LHC (HL-LHC), where noise, stability and radiation hardness requirements will be pushed even beyond the present level.
In particular, the CMS barrel timing layer group showed a great interest for the adoption of an ALDOv2, in order to improve the power supply stability and filtering for their front-end ASIC, TOFHiR, used for precise timing measurements with SiPMs.
Perspectives for this project are thus promising.

The electronic instrumentations developed for the CUORE and CUPID-0 experiments are currently up and running since late Q2 2017.
Both the experiments were able to present background data and first physics results in summer 2017 with a total exposure of about $85\ kg\: y$ for the CUORE experiment and $1\ kg\: y$ for CUPID-0.
CUORE was able to set a new limit on the neutrinoless double beta decay in $^{130}Te$ of $1.5\cdot 10^{25}\ y$, which restricts the effective Majorana neutrino mass to $m_{\beta\beta}<140\ meV-400\ meV$.
In this dissertation I presented a detailed overview of the complete electronic system.
My contribution was predominant in the calibration, qualification and optimization of the entire front-end systems for both the experiments and in the development of the upgraded instrumentations for next-generation experiments.
Next-generation experiments such as CUPID will greatly increase the sensitivity to this ultra-rare decay thanks to the active rejection of background sources.
For this experiment the upgraded preamplifier and antialiasing filter will further expand the performance and flexibility of the instrumentation.
The DAQ capability of the new antialiasing filter is particularly promising as it would represent a great cost saving compared to typical commercial solutions, allowing to double or even quadruple the number of channels at the same cost of the present solution, with a fraction of the actual power consumption.
An R\&D project dedicated to direct search of dark matter, named COSINUS, expressed its interest in the adoption of such system in the faster configuration, which was also described in this work.

Hopefully, the next-generation of experiments, either at high energy accelerators or in the neutrino field, will mark the observation of New Physics beyond the Standard Model, a moment that is long-awaited by the scientific community.
As electronic designers we want to be sure that our instruments are among the firsts to witness such an incredible event.

\vfill

\noindent A warm thank you to everybody who contributed, directly or indirectly, to the realization of this dissertation.

%




\bibliography{BibMain}
\bibliographystyle{paolo}

\cleardoublepage

\end{document}